\documentclass[twocolumn]{aastex631}

\begin{document}

\title{Characterizing the Extended Molecular Hydrogen Winds in Protoplanetary Disks from the JWST Disk Infrared Spectroscopic Chemistry Survey}

\correspondingauthor{Mayank Narang}
\email{mayank.narang@jpl.nasa.gov}
\author[0000-0002-0554-1151]{Mayank Narang}
\affiliation{Jet Propulsion Laboratory, California Institute of Technology, 4800 Oak Grove Drive, Pasadena, CA 91109, USA}

\author[0000-0001-7552-1562]{Klaus M. Pontoppidan}
\affiliation{Jet Propulsion Laboratory, California Institute of Technology, 4800 Oak Grove Drive, Pasadena, CA 91109, USA}

\author[0000-0003-3682-6632]{Colette Salyk}
\affil{Department of Physics and Astronomy, Vassar College, 124 Raymond Avenue, Poughkeepsie, NY 12604, USA}

\author[0000-0003-2631-5265]{Nicole Arulanantham}
\affiliation{Astrophysics \& Space Institute, G, New York, NY 10011, USA}

\author[0000-0003-0787-1610]{Geoffrey A. Blake}
\affiliation{Division of Geological and Planetary Sciences, California Institute of Technology, MC 150-21, Pasadena, CA 91125, USA}

\author[0000-0003-4335-0900]{Andrea Banzatti}
\affiliation{Department of Physics, Texas State University, 749 N Comanche Street, San Marcos, TX 78666, USA}

\author[0000-0002-5758-150X]{Joan Najita}
\affiliation{NSF’s NOIRLab, 950 N. Cherry Avenue, Tucson, AZ 85719, USA}

\author[0000-0001-7962-1683]{Ilaria Pascucci}
\affil{Department of Planetary Sciences, University of Arizona, 1629 East University Boulevard, Tucson, AZ 85721, USA}

\author[0000-0001-6947-6072]{Jane Huang}
\affiliation{Department of Astronomy, Columbia University, 538 W. 120th Street, Pupin Hall, New York, NY 10027, USA}

\author[0000-0002-3291-6887]{Sebastiaan Krijt}
\affil{Department of Physics and Astronomy, University of Exeter, Exeter, EX4 4QL, UK}

\author[0000-0001-8798-1347]{Karin \"{O}berg}
\affil{Center for Astrophysics | Harvard \& Smithsonian, 60 Garden St., Cambridge, MA 02138, USA}

\author[0000-0003-4853-5736]{Giovanni Rosotti}
\affiliation{Dipartimento di Fisica, Universit`a degli Studi di Milano, Via Celoria 16, 20133 Milano, Italy}

\author[0000-0001-8240-978X]{Till Kaeufer}
\affil{Department of Physics and Astronomy, University of Exeter, Exeter, EX4 4QL, UK}

\author[0000-0003-2985-1514]{Emma Dahl}
\affiliation{Division of Geological and Planetary Sciences, California Institute of Technology, MC 150-21, Pasadena, CA 91125, USA}

\author[0000-0003-2076-8001]{L. Ilsedore Cleeves}
\affil{Astronomy Department, University of Virginia, Charlottesville, VA 22904, USA}

\author[0000-0002-0661-7517]{Ke Zhang}
\affil{Department of Astronomy, University of Wisconsin-Madison, Madison, WI 53706, USA}

\author[0000-0003-1665-5709]{Joel Green}
\affiliation{Space Telescope Science Institute, 3700 San Martin Drive, Baltimore, MD 21218, USA}

\author{The JDISCS Collaboration}
\affiliation{}

\begin{abstract}

We present a comprehensive analysis of extended H$_2$ emission from 34 protoplanetary disks observed with the JWST Disk Infrared Spectroscopic Chemistry Survey (JDISCS), supplemented by archival data. We investigated the morphology, kinematics, excitation conditions, and mass dynamics of H$_2$. Extended emission {from  pure rotational H$_2$ lines} is found to be common, with 16 sources exhibiting clear signatures of disk winds. These include monopolar and bipolar structures in inclined disks and ring-like or bubble-like morphologies in face-on systems features indicative of wide-angle disk winds. Our analysis shows that the H$_2$ is consistent with slow {(4.2$^{+6.7}_{-3.0}$ km s$^{-1}$)} MHD driven winds. {For ten disks, we model the wind morphology and find a median half-opening angle of $45\arcdeg^{+5}_{-4}$ and a characteristic power-law index of $  \alpha \sim$ 1.6.} Excitation analysis yields a median  gas temperature of  { 624 $\pm$ 130 K and a column density of $ \log(N_{\mathrm{\rm tot}} [\mathrm{cm}^{-2}]) = 18.6 \pm 0.6$.} {The median wind mass-loss rate, { ${\rm log_{10}}(\dot{\rm M}_{\rm wind}^{\rm tot}) = -9_{-0.4}^{+0.8}\,{\rm M_\odot\,yr^{-1}}$}, implies that, if molecular winds are the dominant mechanism responsible for disk dispersal, a typical disk with a mass {of $2-3\,M_{\rm Jup}$ would dissipate on a $\sim$2-3 Myr} timescale, consistent with observed disk lifetimes.} The $\dot{\rm M}_{\mathrm{\rm wind}}^{\rm tot}$ span a relatively narrow range ($\sim$2 dex) and do not correlate strongly with accretion rates onto the star, suggesting that the mass loss rate and the accretion rates are probing different timescales. Our findings demonstrate that spatially extended warm H$_2$ emission is a widespread and reliable tracer of molecular disk winds in protoplanetary systems.

\end{abstract}


\section{Introduction} \label{sec:intro}

The dissipation of protoplanetary disks dictates the timescale available for the formation of planetesimals,  planetary migration, and the accretion of giant planet atmospheres, thereby playing a crucial role in shaping planetary system architectures \citep[e.g.,][]{Haisch01, Alexander14, 2017ApJ...839...16C,2017RSOS....470114E, 2020A&A...633A...4K, 2021ApJ...909...75T,2022EPJP..137.1357E, 2022Natur.604..643L}. Understanding the physical mechanisms that drive disk dispersal is thus fundamental to advancing our knowledge of planet formation. Disk evolution is governed by both internal and external processes that act to remove gas and dust \citep[e.g.,][]{2000prpl.conf..401H, Alexander06, 2016SSRv..205..125G, 2017RSOS....470114E,2023ASPC..534..567P}, ultimately determining how much matter remains available for planet building at any given time.

Once the protostellar envelope is depleted, there are four major sinks for the gas available for further planet formation: (i) accretion onto the star from the innermost disk \citep[e.g.,][]{1996ApJ...457..355G,2013ApJ...767..112I,2016ARA&A..54..135H}, (ii) launched as a high-velocity jet formed as a consequence of the interaction of the rotating stellar magnetic field and the stellar accretion flow \citep[e.g.,][]{2002ApJ...576..222B,2007ApJ...663..350C,2018A&A...609A..87N}, (iii) accretion onto giant planets  \citep[e.g.,][]{2009MNRAS.393...49A}, and (iv) or launched as low-velocity wind   (either photoevaporative or MHD in origin) from the disk surface \citep[e.g,][]{1982MNRAS.199..883B,2000prpl.conf..759K,2009ApJ...702..724P,2013ApJ...772...96B,2013ApJ...772...60R, 2016ApJ...831..169S, 2019ApJ...870...76B}.

Among these mechanisms, low-velocity disk winds have been particularly challenging to characterize observationally \citep{2001MNRAS.326..524D,2011A&A...527A..13P}. They are typically inferred from the presence of a low-velocity component (LVC) in forbidden atomic and ionic emission lines \citep[e.g.,][]{1983PASP...95..883J, 1984A&A...141..108A, 2013ApJ...772...60R, 2016ApJ...831..169S, 2019A&A...631A..44G, 2019ApJ...870...76B}. Only  the blue-shifted emission is detected due to the obscuration of the red-shifted emission by the disk \citep[][]{1983PASP...95..883J,1984A&A...141..108A,1987ApJ...321..473E}. There are also hints of the influence of disk winds on molecular emission lines from CO and, sometimes, H$_2$O \citep{2011ApJ...733...84P,2011A&A...527A.119B, 2013ApJ...770...94B, 2022AJ....163..174B}. 

Theoretical models propose two dominant mechanisms for launching winds from protoplanetary disks: photoevaporation (PE) \citep[e.g.,][]{1994ApJ...428..654H, 2001MNRAS.328..485C, 2009ApJ...699.1639E,2009ApJ...690.1539G,2025arXiv251100515N}, in which high-energy radiation (from the central star or an external source) ionizes the disk surface and drives thermal outflows through electron heating, and magnetohydrodynamical (MHD) winds \citep{1993ApJ...410..218W, 1997A&A...319..340F, 2014prpl.conf..411T, 2015ApJ...801...84G, 2021A&A...650A..35L}, in which rotating magnetic fields extract angular momentum and accelerate gas outward. By removing angular momentum, the latter also drives accretion onto the star, while the former does not. In younger, embedded sources, X-winds can also be launched \citep[e.g.,][]{2000prpl.conf..789S,2008ApJ...672..489C,2020ApJ...905..116S}, which also removes angular momentum from the disk. In the X-wind model, material is launched from a narrow inner disk region, close to the co-rotation radius, with the help of the open stellar magnetic field lines that thread the inner disk.

However, significant tension remains between these models, as they predict different mass-loss rates, wind kinematics, and dependence on disk properties. Disentangling the relative contributions of PE and MHD processes is crucial for refining our understanding of disk dispersal and its implications for planet formation.
 
Recent JWST observations have uncovered large, extended ($\gtrsim$100\,au) molecular disk winds in isolated protoplanetary disks.  \citep{2025NatAs...9...81P,2025ApJ...980..148S,2024ApJ...965L..13A,2026arXiv260412242C}. Most of these detections are of edge-on disks, where the central system’s light is suppressed, allowing the extended wind emission to be more readily detected. For three edge-on disks that have dispersed their natal envelope, \cite{2025NatAs...9...81P} find that the launch radius (determined via the H$_2$ S(9) emission) is smaller that what is predicted by PE winds (which can only launch winds from outside the gravitational radius), leaving MHD winds as the likely launching mechanism.  {Further,  \citet{2025arXiv251100515N} have recently shown that photoevaporative-wind models can reproduce several key features of observed H$_2$ winds, including their nested morphology and overall fluxes, although their models tend to underpredict the higher-J emission from more extended H$_2$ features. These models also predict relatively high molecular fractions in disk winds. }

A key remaining question is whether typical protoplanetary disks harbor observable molecular disk winds, and how the wind properties are linked to observables more commonly detected in moderate-inclination disks, such as mass accretion onto the central star. 

We present an analysis of data from the JWST Disk Infrared Spectroscopic Chemistry Survey \citep[JDISCS; ][]{Pontoppidan24,2025arXiv250507562A} augmented by archival data from the Mid INfrared Disk Survey \citep[MINDS;][]{2024PASP..136e4302H}, focusing on extended emission structures traced by pure rotational lines of H$_2$. These structures are observed using the Medium Resolution Spectrometer \citep[MRS; ][]{Wells15} on the Mid-infrared Instrument \citep[MIRI;][]{Rieke15,Wright23} aboard the James Webb Space Telescope \citep[JWST;][]{Gardner23}.

Our observations reveal that extended emission from rotational H$_2$ lines is widespread and not isolated to a few extreme cases. We carry out a uniform analysis of the excitation conditions, the kinematics and the morphology of these wide angled H$_2$ flows. In section \ref{sec:obs}, we describe the sample selection and methods for extracting line maps from the three-dimensional cubes. Section \ref{sec:analysis} discusses the morphology and physical parameters of the molecular winds, while section \ref{sec:discussion}  discusses the implications for our results. Finally, we summarize our findings in Section 5.

\begin{deluxetable*}{lccccccccccc}
\tablecaption{\label{tab: sample} Properties of the disk sample.}
\tablehead{
\colhead{Target} & \colhead{Program ID} & \colhead{Distance} & \colhead{SpT} & \colhead{$L_\star$ } & \colhead{$M_\star$ } & \colhead{$\log_{10}\dot{M}_{acc}$ } & \colhead{$r_{\rm dust}$ } & \colhead{$i_{\rm disk}$} & \colhead{PA} & \colhead{Ref} \\
\colhead{} & \colhead{} & \colhead{(pc)} & \colhead{} & \colhead{ ($L_\odot$)} & \colhead{($M_\odot$)} & \colhead{($M_\odot$/yr)} & \colhead{(au)} & \colhead{(\arcdeg)} & \colhead{(\arcdeg)} & \colhead{}
}
\startdata
AS 205 N & 1584 & 132 & K5 & 1.3 & 0.87 & -7.4 & 60 & 20 & 114 & 1 \\
AS 209 & 2025 & 121 & K5 & 1.4 & 0.83 & -7.3 & 139 & 35 & 86 & 1 \\
CI Tau & 1640 & 160 & K7 & 0.8 & 0.65 & -7.5 & 174 & 50 & 11 & 1 \\
DoAr 25 & 1584 & 138 & K5 & 0.9 & 0.62 & -8.9 & 165 & 67 & 111 & 1 \\
DoAr 33 & 1584 & 142 & K4 & 1.5 & 0.69 & -9.6 & 27 & 67 & 111 & 1 \\
DR Tau & 1282 & 193 & K6 & 0.63 & 0.93 & -6.7 & 54 & 5 & 3 & 2 \\
Elias 20 & 1584 & 138 & M0 & 2.6 & 0.88 & -6.7 & 64 & 49 & 153 & 1 \\
Elias 24 & 1584 & 139 & K5 & 6.8 & 1.1 & -6.3 & 136 & 29 & 46 & 1 \\
Elias 27 & 1584 & 110 & M0 & 1.5 & 0.63 & -7.1 & 254 & 56 & 119 & 1 \\
FZ Tau & 1549 & 129 & M0 & 1.0 & 0.51 & -6.5 & 15 & 26 & 30 & 1 \\
GK Tau & 1640 & 129 & K7 & 0.9 & 0.58 & -8.3 & 13 & 40 & 120 & 1 \\
GM Aur & 2025 & 158 & K7 & 1.0 & 0.69 & -8.0 & 220 & 53 & 57 & 1 \\
GO Tau & 1640 & 142 & K5 & 0.2 & 0.36 & -9.5 & 144 & 54 & 21 & 1 \\
GQ Lup & 1640 & 154 & K7 & 1.4 & 0.61 & -7.4 & 22 & 61 & 346 & 1 \\
GW Lup & 1282 & 155 & M1.5 & 0.33 & 0.46 & -9.0 & 105 & 39 & 38 & 3, 4 \\
HD 142666 & 1584 & 146 & A8 & 9.1 & 1.23 & -8.4 & 59 & 62 & 162 & 1 \\
HD 143006 & 1584 & 167 & G7 & 3.9 & 1.48 & -7.7 & 82 & 19 & 169 & 1 \\
HD 163296 & 2025 & 101 & A1 & 17.0 & 2.04 & -7.4 & 169 & 47 & 133 & 1 \\
HP Tau & 1640 & 171 & K0 & 1.1 & 0.84 & -10.3 & 21 & 18 & 57 & 1 \\
HT Lup & 1584 & 153 & K2 & 5.1 & 1.32 & -8.1 & 33 & 48 & 166 & 1 \\
IQ Tau & 1640 & 132 & M0.5 & 1.0 & 0.42 & -7.9 & 96 & 62 & 42 & 1 \\
IRAS 04385+2550 & 1640 & 160 & M0.5 & 0.5 & 0.56 & -8.1 & 32 & 60 & 162 & 1 \\
MWC 480 & 1584 & 156 & A2 & 22.0 & 3.58 & -7.0 & 105 & 36 & 148 & 1 \\
MY Lup & 2025 & 157 & K0 & 0.9 & 1.2 & -8.0 & 87 & 73 & 59 & 1 \\
RU Lup & 1584 & 158 & K7 & 1.5 & 0.55 & -7.0 & 63 & 19 & 121 & 1 \\
RY Lup & 1640 & 153 & K2 & 1.9 & 1.27 & -8.1 & 80 & 67 & 109 & 1 \\
SR 4 & 1584& 135 & K7 & 1.2 & 0.61 & -6.9 & 31 & 22 & 18 & 1 \\
SY Cha &1282 & 181 & K5 & 0.6 & 0.7 & -9.2 & $\leq 180$ & 51 & 345 & 5, 6 \\
Sz 114 & 1584 & 157 & M5 & 0.2 & 0.16 & -9.1 & 58 & 21 & 165 & 1 \\
Sz 129 & 1584 & 160 & K7 & 0.4 & 0.73 & -8.3 & 76 & 34 & 151 & 1 \\
TW Cha & 1549 & 183 & K7 & 0.4 & 0.7 & -8.6 & 97 & 27 & 121 & 1 \\
TW Hya & 1282 & 60 & K6 & 0.3 & 0.9 & -8.6 &  96 & 5 & 152 & 7, 8, 9 \\
VZ Cha & 1549 & 191 & K7 & 0.5 & 0.5 & -7.1 & 47 & 19 & 30 & 1 \\
WSB 52 & 1584 & 135 & M1 & 1.7 & 0.55 & -7.9 & 32 & 54 & 138 & 1 \\ \enddata
\tablecomments{(1) Stellar properties and accretion rates from the compiled table in \cite{2023ASPC..534..539M}.  {The distances are based on Gaia DR3.} The disk radii based on flux at 1.3 mm from \citealt{2018ApJ...869L..42H}, \cite{2018ApJ...869...17L}, and \citealt{2019ApJ...882...49L} as formatted and presented in  \citealt{2025arXiv250507562A}, (2) \citealt{2019ApJ...882...49L}, (3) \citealt{2018ApJ...869L..42H}, (4) \citealt{2023ApJ...947L...6G}, (5) \citealt{2024ApJ...962....8S}, (6) \citealt{2023PASJ...75..424O}, (7) \citealt{2018ApJ...852..122H}, (8) \citealt{2023ApJ...956..102H}, (9) \citealt{d24}. \label{tab:Table1} }
\end{deluxetable*} 

\section{Observations} \label{sec:obs}
\subsection{Sample}
We use observations of 30 protoplanetary disks from the JDISCS Cycle 1 sample \citep{2025arXiv250507562A}. The data were obtained from four observation programs, namely: PID 1549 (PI Pontoppidan), PID 1584 (PI Salyk), PID 1640 (PI Banzatti), and PID 2025 (PI \"{O}berg). To expand our study, we incorporate previously published archival data of four additional protoplanetary disks from the MINDS program (PID 1282, PI Henning, \citealt{2024PASP..136e4302H,2024ApJ...962....8S,2023ApJ...947L...6G,2024A&A...686A.117T}). By incorporating this archival data set, we aim to determine whether the observed prevalence of extended winds is a characteristic feature specific to the JDISCS sample or if it represents a more widespread phenomenon among protoplanetary disks, allowing for a more comprehensive assessment of disk wind properties across different disk populations. In total, we analyzed 34 protoplanetary disks. The details of the full sample are provided in Table \ref{tab: sample}.

\subsection{Observations and data reduction}\label{sec:reduction}

For all disks analyzed in this work, observations were obtained with the four MIRI MRS channels, each comprising short, medium, and long sub-bands, covering a wavelength range of $4.90$--$27.9\,\mu\mathrm{m}$. 
The spectral resolving power varies across the MRS bands, from 
$R = \lambda / \Delta \lambda \sim 3{,}500$ at $5\,\mu\mathrm{m}$ 
to $\sim 1{,}500$ at $27.9\,\mu\mathrm{m}$ \citep{Pontoppidan24}\footnote{\url{https://jwst-docs.stsci.edu/jwst-mid-infrared-instrument/miri-observing-modes/miri-medium-resolution-spectroscopy}}. 
The spatial resolution of MIRI MRS also depends on wavelength, ranging from $0\farcs67$ for the H$_2$~S(1) line at $17.035\,\mu\mathrm{m}$ 
to $0\farcs33$ for the H$_2$~S(5) line at $6.910\,\mu\mathrm{m}$ \citep{2023AJ....166...45L}.

The JDISCS observations uses the point-source dither pattern, providing access to the largest possible field of view with a single MRS pointing. Relatively deep exposures of 500\,s to several 1000\,s on-source time per sub-band provide access to faint, extended line emission. The details of the observations with date, program id and exposure time for the JDISCS sample is presented in the overview paper by \cite{2025arXiv250507562A}.  For the MINDS sample the details of the observations are given in the individual papers \citealt{2024PASP..136e4302H,2024ApJ...962....8S,2023ApJ...947L...6G,2024A&A...686A.117T}. 
 All targets in the JDISCS sample are dominated by a bright point(-like) source, necessitating careful subtraction of the unresolved component to accurately isolate any extended emission (which we describe in \S 2.3). 

{The data were processed using the JWST Calibration Pipeline \citep[version 1.20.2, CRDS context 1464; ][]{bushouse_2025_17515973} only to produce rectified and calibrated data cubes. We then used the JDISCS methodology (specifically version 9.1)} to extract calibrated 1-dimensional spectra of the central point sources and make line images \citep{Pontoppidan24,2025arXiv250507562A}.

\begin{deluxetable}{cccccc}
    \tablewidth{0pt}
    \tablecaption{Observed rotational H$_2$ lines. The wavelength, Einstein A coefficient $A_{ul}$, upper state energy and g$_{up}$ are from \cite{2022JQSRT.27707949G}. }
    \label{TableH2}
    \tablehead{
    \colhead{Wavelength} &  \colhead{Name} &   \colhead{$A_{ul}$} & \colhead{$E_{up}$} & \colhead{g$_{up}$} \\
    \colhead{($\micron$)} & \colhead{} &  \colhead{(s$^{-1}$)}   &  \colhead{(K)} & \colhead{} \\
    }
    \startdata
17.035	 & 	H$_2$ 0-0 S(1)	 		 & 	4.8 $\times 10^{-10}$ 	 & 		1015	&21	\\
12.279	 & 	H$_2$ 0-0 S(2)	 		 & 	  2.8 $\times  10^{-9}$	 & 		1682	& 9	 \\
9.665	 & 	H$_2$ 0-0 S(3)	   		 & 	 9.8 $\times 10^{-9}$ 	 & 		2504	& 33	 \\
8.025	 & 	H$_2$ 0-0 S(4)           & 	 2.6 $\times 10^{-8}$ 	 & 		3475	& 13	 \\ 
6.910	 & 	H$_2$ 0-0 S(5)	 	  		 & 	 5.9 $\times 10^{-8}$ 	 & 		4586	& 45	 \\
\enddata
\end{deluxetable}

\subsection{PSF correction and line maps}

Beginning with the level 3 data cubes, we generated emission line maps for each disk by extracting both the integrated line flux and radial velocity per spaxel. For each spaxel we integrate the observed  emission line to derive the line fluxes. We also fit a Gaussian profile to the emission line using the {\tt astropy.modeling} package, allowing us to measure both the line flux and the velocity shift. The fits were used to measure the flow velocity. 

To isolate the extended emission from the bright unresolved central source, we subtracted a model of the point spread function (PSF), constructed from wavelength channels immediately adjacent to the line. Specifically, we averaged 10 spectral planes (five on each side of the line) to create a noise-suppressed model of the PSF. This model was then scaled to match the peak point-source line emission at the line center before being subtracted from the line image. {For each spaxel, we estimate the error on the line flux as the RMS of the PSF planes: $\rm RMS/\sqrt{n_{\rm line\, planes}}$, where $n_{\rm line\, planes}$ is the number of wavelength planes over which the line flux is summed.} 

Many of the disks in our sample exhibit strong continuum emission as well as molecular line emission from species such as H$_2$O, which can blend with the rotational H$_2$ lines. These features contribute to wavelength-dependent scattered light from the central source, complicating the detection of faint extended emission. The scaled PSF subtraction mitigates these effects by removing spatially unresolved line emission originating from the disk itself. Crucially, this correction enhances the imaging contrast and allows for the detection of fainter, spatially extended features that would otherwise be masked by the central emission.

\section{Analysis}\label{sec:analysis}

\subsection{H$_2$ emission from typical protoplanetary disks}

\begin{figure*}
\centering
\includegraphics[width=18cm]{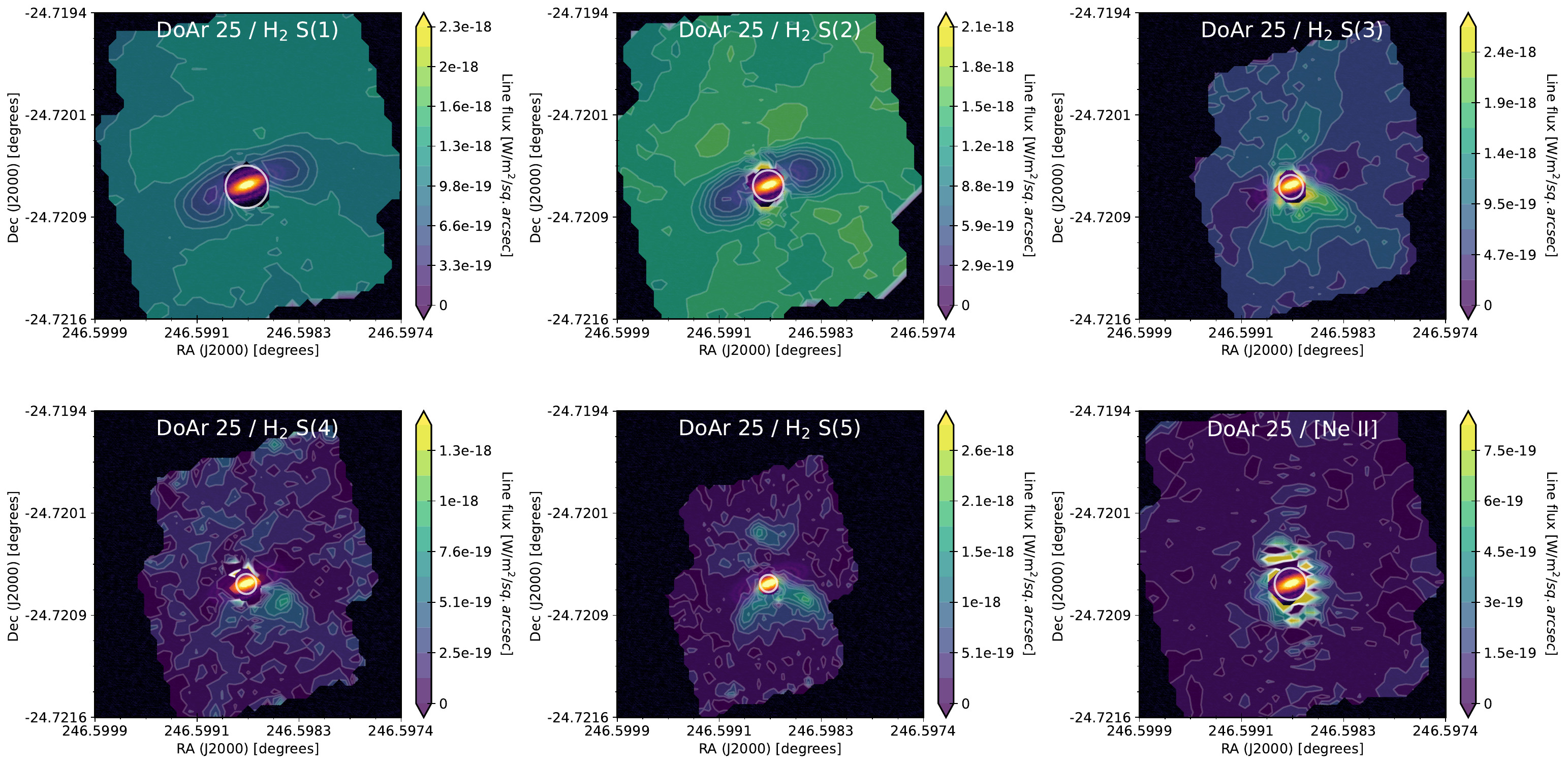}
\caption{The  H$_2$ and [Ne II] emission detected towards DoAr 25 on top of the ALMA 240 GHz continuum emission (shown in the background as a reddish-yellow image). The white central circle shows the inner working angle from JWST. In this case, the disk itself is seen in absorption against line emission from the Ophiuchus PDR for H$_2$ S(1)-S(3) (17.03-9.66\,$\mu$m). \label{fig:doar25} }
\end{figure*}

We analyzed the PSF-subtracted images of the H$_2$~0--0 S(1) to S(5) rotational transitions (see Table~\ref{TableH2}) and the [Ne~II] line at 12.81~$\mu$m; higher-lying H$_2$ transitions are generally not detected in an extended component. In Figure \ref{fig:doar25} we show the detected H$_2$ emission from DoAr 25 as an example. The complete galleries of the detected H$_2$ emission overlaid on ALMA continuum images are shown in the Appendix \ref{Notes} (Figures~\ref{fig:as205}–\ref{fig:wsb52}),{ with detailed notes on individual sources}.

In many disks, the morphology of the S(3) and S(5) lines clearly reveals signatures of wide-angle flows and extended emission. We classify all extended H$_2$ emission as evidence of disk winds (see Section \ref{sec:prevalence}). Several sources exhibit wide-angle conical winds, observed in both monopolar (e.g., GM Aur; Figure~\ref{fig:gmaur}, SY Cha; Figure~\ref{fig:sycha}) and bipolar (e.g., DoAr 25; Figure~\ref{fig:doar25}, IRAS 04385+2550; Figure~\ref{fig:iras04385}, IQ Tau; Figure~\ref{fig:iqtau}, MY Lup Figure~\ref{fig:mylup}) configurations. These morphologies provide strong evidence for the presence of disk winds when the system is viewed at high inclinations, but we find more symmetric emission from less inclined systems, consistent with wide-angle structures. The morphology generally matches the known inclinations of the disks. A detailed analysis of these conical wind structures is presented in \S~\ref{sec:OA}.

Most face-on disks (inclination $\lesssim$ 30\arcdeg), show bubble-like (VZ Cha: Figure~\ref{fig:vzcha}) or ring-like (e.g., Elias 24: Figure~\ref{fig:elias24}, FZ Tau: Figure~\ref{fig:fztau}, TW Cha: Figure~\ref{fig:twcha}, TW Hya: Figure~\ref{fig:twhya}) morphologies. For face-on systems, such circular emission patterns are consistent with predictions from disk wind models \citep[e.g.,][]{2022A&A...668A..78D}. The observed ring-like structures may trace temporal or spatial variations in the wind mass-loss rate. In the face-on or close to face-on disks we only detected one side of the emission since the emission from the other side is obscured by the disk \citep[e.g.,][]{{1983PASP...95..883J,1984A&A...141..108A,1987ApJ...321..473E}}.

In many cases (e.g.,  DoAr 25, Figure~\ref{fig:doar25}; Elias 27, Figure \ref{fig:elias27}; SR 4 Figure \ref{fig:sr4}), the lower-excitation H$_2$ transitions, particularly S(1) and S(2), show that the MIRI MRS fields of view are often filled by emission likely originating from foreground or background photodissociation regions (PDRs) or the exposed surfaces of nearby molecular clouds. In some systems, we also detect disk silhouettes of absorption against  extended backgrounds H$_2$ emission (DoAr 25, Figure~\ref{fig:doar25}; Elias 27, Figure \ref{fig:elias27}). We also detect extended [Ne II] emission in a subset of targets (see Section \ref{sec:Ne}).

\begin{figure*}
\centering
\includegraphics[width=16cm]{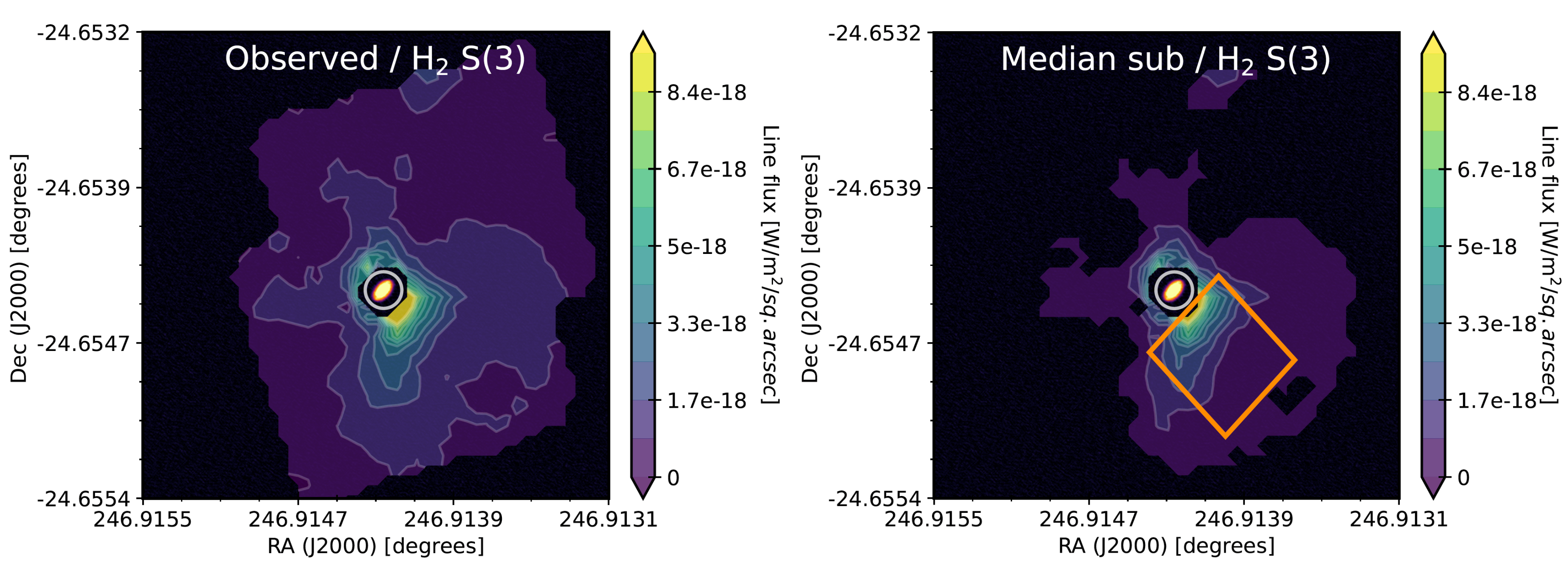}
\caption{The observed and median subtracted H$_2$ S(3) lime maps towards {WSB 52} on top of the ALMA 240 GHz continuum emission (shown in the background as a reddish-yellow image). The white central circle delineates the inner working angle of the observation. The median subtraction removes the ambient fore/background H$_2$ emission. {The orange square represents the $250\,\mathrm{au} \times 250\,\mathrm{au}$ aperture from which the line intensities are extracted.} \\
\label{fig:median_sub} }
\end{figure*}

\subsection{Excitation analysis of H$_2$ lines in extended winds} \label{sec:rot}

{The winds show varied geometries; e.g., the ring-like geometry of FZ Tau (Figure \ref{fig:fztau}) or the tadpole-like geometry in HP Tau (Figure \ref{fig:hptau}). We experimented extensively with different methods for determining the region of interest, and found that anything more complex than a standard box size tended to add error (both statistical and systematic). The simplest, and most stable, solution was a single box size for every source and line.} Therefore we measured the surface line intensities using a square aperture with 250 au sides, centered on the bright part of the wind with high S/N. {To limit the influence of PSF residuals, we mask a region within $1.5 \times 1.22 \lambda/D$ of the central source. We further mask pixels with S/N$<3$ of the integrated line flux.} For  bipolar sources the apertures were placed along the outflow such that they encompass the majority of the significant H$_2$ emission from one lobe. This fixed-size of the physical aperture size allows for a uniform comparison of the excitation conditions and kinematics of the extended H$_2$ emission across different sources. We list the coordinates for the apertures in Table \ref{tab:H2}. In most cases, the aperture has an position angle of 0\arcdeg; the exceptions to this are marked in Table \ref{tab:H2}, where we used a position angle of 45\arcdeg.

In several disks, the H$_2$ emission associated with the system is superimposed on background or foreground contributions. To account for this, we subtracted the median H$_2$ line flux, computed from all spaxels in the image (after performing PSF correction and masking the region inside the inner working angle). {The median background line intensities are provided in Table \ref{tab:H2}.} Figure \ref{fig:median_sub} presents an illustrative example of this procedure. {The uncertainty in the aperture-integrated line flux is estimated as the quadratic sum over the rms image within the same aperture. Finally the aperture integrated line fluxes are divided by the aperture solid angle to yield average surface line intensities.} Table \ref{tab:H2} lists the average surface line intensities. {To ensure conservative estimates, line intensities within the aperture yielding S/N $< 3\sigma$ are designated as upper limits.}  

\begin{figure*}
\centering
\includegraphics[width=7cm]{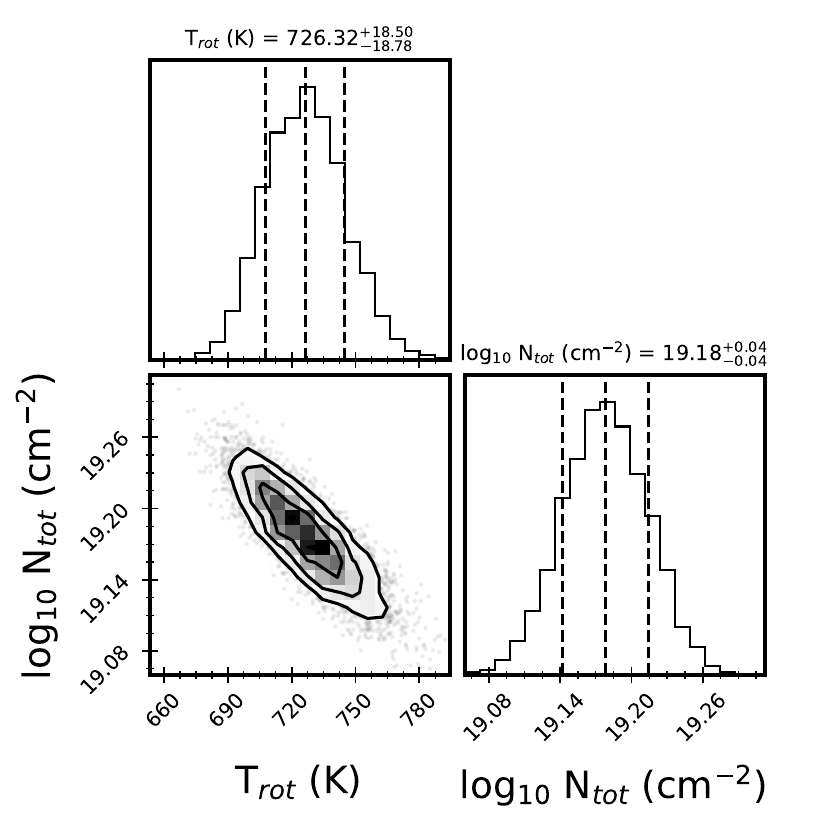}\includegraphics[width=7cm]{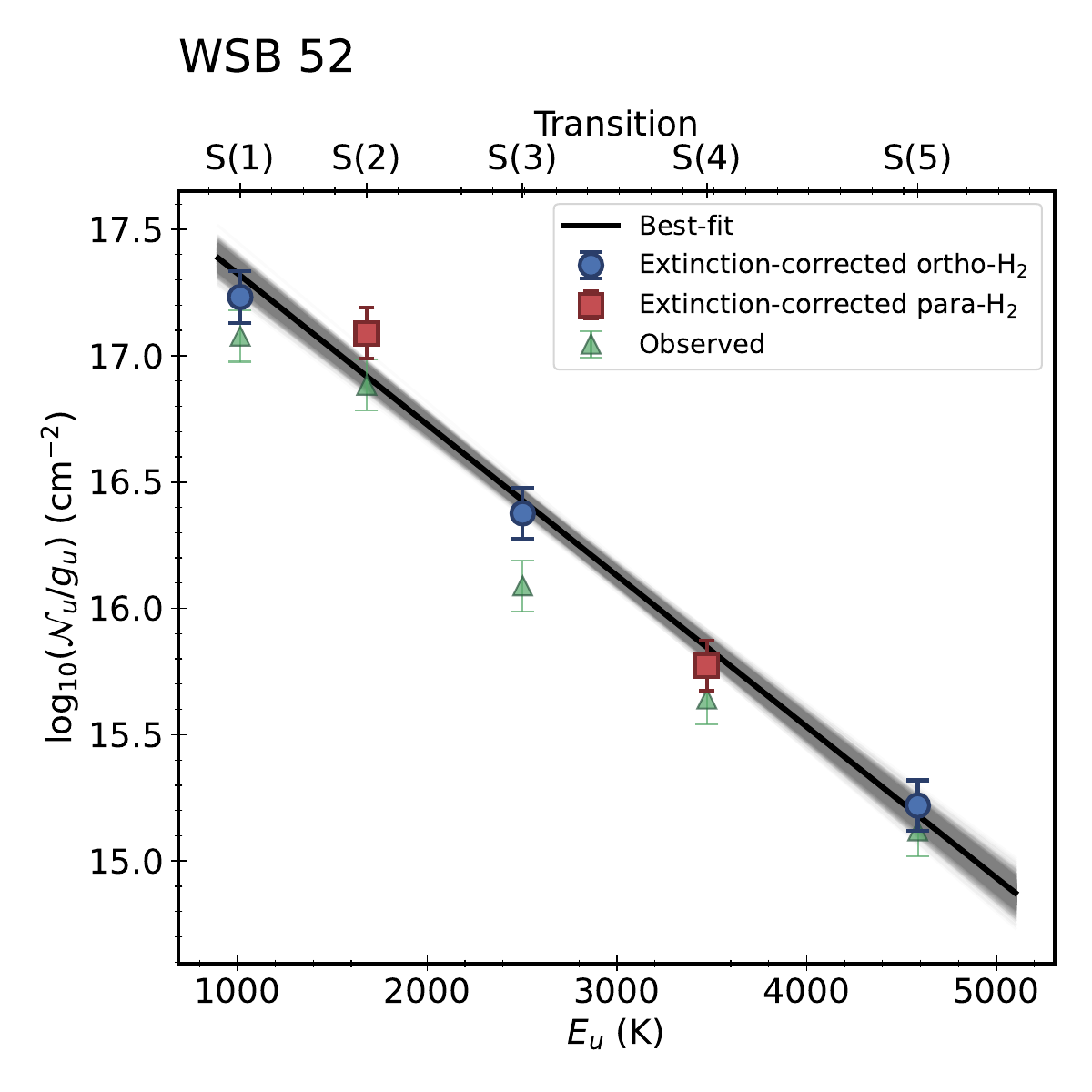}
\caption{(Left) Corner plot showing the posterior distributions and best-fit parameters from the MCMC fit to the rotation diagram of {WSB 52}, presented as a representative example. (Right) Rotation diagram with observed data points in green and extinction-corrected points in blue for ortho-H$_2$ and red for para-H$_2$. The black line denotes the best-fit model, and the shaded grey region represents the uncertainty range derived from the MCMC fitting.}
\label{fig:rot}
\end{figure*}

We used rotational analysis to estimate the column density $N({\rm H}_2)$ and excitation temperature $T(\rm{H}_2)$ of the emitting gas (Figure \ref{fig:rot}). We construct a rotation diagram by plotting $\ln (\mathcal{N}_u/g_u) $, the natural logarithm of the column density of H$_2$ molecules in the upper rotational state $ u $ divided by its degeneracy $ g_u $, against the upper state energy $ E_u$ \citep[e.g.,][]{2006ApJ...649..816N,2010ApJ...724...69N}. The degeneracy is given by $ g_u = g_{J_u} \times g_I $, where $ g_{J_u} = 2J_u + 1 $ accounts for rotational degeneracy, and $ g_I = 2I + 1 $ is the spin degeneracy, with $ I = 1 $ for ortho-H$_2$ (odd J) and $ I = 0 $ for para-H$_2$ (even J). 

For optically thin H$_2$ emission, the column density of H$_2$ in the (upper) $ u^{\rm th} $ rotational state is given by:
\begin{equation}
    \mathcal{N}_u = \frac{4~\pi~I_{u,l}~\lambda_{u,l}}{h~c~A_{u,l}},
\end{equation} 
where $h$, $c$, $I_{u,l}$, $\lambda_{u,l}$ and $A_{u,l}$ are the Planck's constant, speed of light in vacuum, line intensity, wavelength, and Einstein A-coefficient, in that order, corresponding to the transition from an upper state $u$ to a lower state $l$. 

\begin{deluxetable*}{lccccccccccccccc}
\rotate
\tabletypesize{\scriptsize}
\tablewidth{0pt}
\tablecaption{\label{tab:H2}
{Background-subtracted line intensities for the H$_2$ S(1)–S(5) transitions are presented, together with their 1$\sigma$ uncertainties and the median background line intensities used in the subtraction.  Also listed are the observed velocity of the S(1) line in the telescope frame ($\rm V_{H_2}$), the systemic velocity of the source ($\rm V_{\rm sys}$), and the inclination-corrected wind velocity ($\rm V_{\rm wind}$). Upper limits are 3$\sigma$. Sources denoted by $^*$ correspond to apertures with non-zero position angles (see below)}}
\tablehead{
\colhead{Target} & 
\multicolumn{2}{c}{Aperture Coordinates} & 
\multicolumn{5}{c}{Background-subtracted line intensity} & 
\multicolumn{5}{c}{Background line intensity} & 
\multicolumn{3}{c}{Velocities} \\ 
& \colhead{RA} & \colhead{Dec} 
& \colhead{S(1)} & \colhead{S(2)} & \colhead{S(3)} & \colhead{S(4)} & \colhead{S(5)} 
& \colhead{S(1)} & \colhead{S(2)} & \colhead{S(3)} & \colhead{S(4)} & \colhead{S(5)} 
& \colhead{$\rm V_{H_2}$} & \colhead{$\rm V_{ sys}$} & \colhead{$ \rm V_{ wind}$} \\ 
\colhead{} & 
\multicolumn{2}{c}{\centering [deg]} & 
\multicolumn{5}{c}{\centering [$10^{-9}$ W\,m$^{-2}$\,sr$^{-1}$]} & 
\multicolumn{5}{c}{\centering [$10^{-9}$ W\,m$^{-2}$\,sr$^{-1}$]} & 
\multicolumn{3}{c}{\centering [ km s$^{-1}$]} 
}
\startdata
AS 205 & 242.8801 & -18.6408 & 7.28 $\pm$ 0.90 & 35.88 $\pm$ 1.75 & 114.77 $\pm$ 7.03 & 39.32 $\pm$ 4.21 & 82.48 $\pm$ 1.30 & 8.92 & 15.29 & 56.77 & 30.63 & 60.36 & -9.91 $\pm$ 7.16 & -6.03 $\pm$ 3.38 & -4.13 $\pm$ 7.92 \\
AS 209 & 252.3141 & -14.3693 & 3.13 $\pm$ 0.44 & 52.11 $\pm$ 0.95 & 25.48 $\pm$ 1.34 & $<$3.30 & 11.24 $\pm$ 0.47 & 7.85 & 12.85 & 41.14 & 24.93 & 41.59 & -3.32 $\pm$ 5.63 & -23.24 $\pm$ 22.85 & 24.32 $\pm$ 23.53 \\
CI Tau & 68.4664 & 22.8418 & 5.82 $\pm$ 0.06 & 7.36 $\pm$ 0.20 & 14.21 $\pm$ 0.29 & 7.95 $\pm$ 0.38 & 11.87 $\pm$ 0.29 & 3.17 & 2.30 & 8.04 & 3.61 & 5.94 & 13.32 $\pm$ 3.09 & 11.48 $\pm$ 1.93 & 2.86 $\pm$ 3.64 \\
DoAr 25 & 246.5984 & -24.721 & 1.37 $\pm$ 0.05 & 1.74 $\pm$ 0.07 & 17.13 $\pm$ 0.04 & 5.40 $\pm$ 0.06 & 12.04 $\pm$ 0.05 & 49.42 & 56.28 & 26.78 & 6.65 & 8.14 & -8.64 $\pm$ 1.78 & -7.20 $\pm$ 5.37 & -3.69 $\pm$ 5.66 \\
DoAr 33 & 246.9128 & -23.9718 & 3.18 $\pm$ 0.04 & 4.96 $\pm$ 0.17 & 5.14 $\pm$ 0.08 & 0.33 $\pm$ 0.08 & 1.06 $\pm$ 0.06 & 13.67 & 6.44 & 5.49 & 1.81 & 3.04 & -9.22 $\pm$ 1.95 & -19.31 $\pm$ 7.26 & 25.83 $\pm$ 7.52 \\
DR Tau & 71.7759 & 16.9782 & 5.19 $\pm$ 0.44 & 65.70 $\pm$ 0.80 & 58.49 $\pm$ 2.48 & 35.67 $\pm$ 1.57 & 6.20 $\pm$ 1.07 & 6.43 & 8.76 & 35.60 & 18.86 & 37.52 & --- & 22.14 $\pm$ 2.41 & --- \\
Elias 20 & 246.5782 & -24.4725 & 0.98 $\pm$ 0.07 & 3.46 $\pm$ 0.39 & 20.42 $\pm$ 0.13 & 13.15 $\pm$ 0.32 & 41.68 $\pm$ 0.17 & 115.11 & 90.25 & 88.76 & 21.61 & 32.79 & -5.68 $\pm$ 1.80 & -13.29 $\pm$ 5.92 & 11.61 $\pm$ 6.19 \\
Elias 24 & 246.6007 & -24.2702 & 14.88 $\pm$ 0.45 & 28.01 $\pm$ 1.11 & 51.41 $\pm$ 0.58 & 20.76 $\pm$ 0.78 & 70.13 $\pm$ 0.57 & 36.97 & 33.79 & 52.01 & 21.76 & 63.29 & -4.83 $\pm$ 3.50 & -7.82 $\pm$ 3.39 & 3.42 $\pm$ 4.87 \\
Elias 27 & 246.6876 & -24.3852 & 0.81 $\pm$ 0.06 & 4.95 $\pm$ 0.14 & 5.38 $\pm$ 0.11 & 1.83 $\pm$ 0.12 & 3.17 $\pm$ 0.10 & 37.48 & 32.64 & 20.34 & 7.15 & 10.02 & -5.76 $\pm$ 4.28 & -11.88 $\pm$ 4.01 & 10.95 $\pm$ 5.86 \\
FZ Tau & 68.1318 & 24.3334 & 6.28 $\pm$ 0.02 & 10.94 $\pm$ 0.03 & 39.36 $\pm$ 0.18 & 24.40 $\pm$ 0.16 & 7.13 $\pm$ 0.07 & 7.80 & 10.79 & 32.27 & 10.73 & 19.62 & 11.92 $\pm$ 3.88 & 14.63 $\pm$ 3.35 & -3.02 $\pm$ 5.12 \\
GK Tau & 68.3945 & 24.3515 & 1.14 $\pm$ 0.08 & 7.91 $\pm$ 0.28 & 5.11 $\pm$ 0.35 & 2.25 $\pm$ 0.33 & 5.86 $\pm$ 0.10 & 7.39 & 5.46 & 15.17 & 5.22 & 8.24 & 16.17 $\pm$ 5.19 & 5.52 $\pm$ 5.87 & 13.90 $\pm$ 7.83 \\
GM Aur & 73.7961 & 30.3662 & 1.02 $\pm$ 0.04 & 1.77 $\pm$ 0.04 & 10.34 $\pm$ 0.10 & 1.89 $\pm$ 0.07 & 3.38 $\pm$ 0.03 & 1.21 & 0.63 & 1.68 & 1.28 & 1.71 & 8.79 $\pm$ 5.01 & 29.42 $\pm$ 37.02 & -34.27 $\pm$ 37.36 \\
GO Tau & 70.7628 & 25.3388 & 0.12 $\pm$ 0.02 & 0.89 $\pm$ 0.03 & 0.37 $\pm$ 0.05 & $<$0.22 & 0.37 $\pm$ 0.05 & 1.00 & 0.52 & 2.08 & 2.35 & 2.47 & 8.14 $\pm$ 4.60 & 17.40 $\pm$ 16.08 & -15.75 $\pm$ 16.73 \\
GQ Lup & 237.2998 & -35.6515 & 1.64 $\pm$ 0.05 & 1.48 $\pm$ 0.10 & 6.92 $\pm$ 0.09 & $<$0.26 & 4.16 $\pm$ 0.10 & 6.90 & 2.93 & 8.18 & 2.82 & 5.00 & -3.84 $\pm$ 3.62 & -3.86 $\pm$ 4.07 & 0.03 $\pm$ 5.45 \\
GW Lup & 236.6867 & -34.5099 & 0.63 $\pm$ 0.03 & 1.52 $\pm$ 0.08 & 2.13 $\pm$ 0.06 & 0.41 $\pm$ 0.07 & 1.15 $\pm$ 0.04 & 2.62 & 1.64 & 3.79 & 2.16 & 2.80 & -2.46 $\pm$ 3.00 & -3.36 $\pm$ 5.20 & 1.15 $\pm$ 6.01 \\
HD 142666 & 239.1667 & -22.0275 & $<3$ & 11.00 $\pm$  $<3.2$ &  $<3.5$ &  $<3$ &  $<1.5$ & 4.35 & 9.39 & 35.51 & 22.51 & 32.27 & --- & --- & --- \\
HD 143006 & 239.6538 & -22.9547 & $<1.7$ &  $<1.9$ &  $<1$ &  $<1$ &  $<0.7$ & 1.60 & 2.26 & 8.85 & 6.63 & 9.93 & --- & --- & --- \\
HD 163296 & 269.0886 & -21.9563 &  $<9$ &  $<19$ &  $<50$ &  $<27$ &  $<15.2$ & 11.23 & 18.56 & 99.61 & 40.25 & 58.77 &--- & --- & --- \\
HP Tau & 68.9702 & 22.9061 & 6.44 $\pm$ 0.40 & 39.36 $\pm$ 0.49 & 34.64 $\pm$ 0.24 & 42.16 $\pm$ 0.73 & 72.37 $\pm$ 0.22 & 6.12 & 5.60 & 19.03 & 9.13 & 17.23 & 13.31 $\pm$ 8.60 & 7.85 $\pm$ 10.85 & 5.74 $\pm$ 13.85 \\
HT Lup & 236.3031 & -34.2919 & 1.72 $\pm$ 0.25 & 8.73 $\pm$ 0.85 & 33.59 $\pm$ 0.44 & 14.02 $\pm$ 0.53 & 44.31 $\pm$ 0.28 & 6.81 & 8.08 & 27.44 & 14.37 & 35.60 & -2.31 $\pm$ 5.53 & -0.56 $\pm$ 13.22 & -2.61 $\pm$ 14.33 \\
IQ Tau & 67.4651 & 26.1122 & 2.47 $\pm$ 0.06 & 7.18 $\pm$ 0.14 & 7.75 $\pm$ 0.14 & 10.92 $\pm$ 0.95 & 21.87 $\pm$ 0.21 & 1.50 & 1.38 & 6.33 & 3.05 & 5.54 & 8.11 $\pm$ 6.60 & 13.69 $\pm$ 1.05 & -11.89 $\pm$ 6.68 \\
IRAS 04385 & 70.4121 & 25.941 & 2.96 $\pm$ 0.15 & 6.37 $\pm$ 0.18 & 13.06 $\pm$ 0.10 & 5.73 $\pm$ 0.13 & 23.06 $\pm$ 0.09 & 2.88 & 2.68 & 9.83 & 6.39 & 12.13 & 12.76 $\pm$ 3.41 & 8.00 $\pm$ 7.72 & 9.52 $\pm$ 8.44 \\
MWC 480 & 74.6928 & 29.8431 & $<3.2$ &  $<9.1$ &  $<12$ &  $<4$ &  $<3$ & 2.70 & 5.23 & 27.87 & 13.13 & 17.16 & --- & --- & --- \\
MY Lup & 240.1855 & -41.925 & 4.30 $\pm$ 0.08 & 4.53 $\pm$ 0.22 & 4.80 $\pm$ 0.09 & 2.20 $\pm$ 0.18 & 3.21 $\pm$ 0.19 & 3.37 & 1.14 & 3.22 & 2.13 & 2.69 & -5.51 $\pm$ 4.06 & -5.34 $\pm$ 15.95 & -0.59 $\pm$ 16.46 \\
RU Lup & 239.1762 & -37.8214 & 4.67 $\pm$ 0.90 & 43.57 $\pm$ 1.40 & 21.25 $\pm$ 2.03 & 4.91 $\pm$ 1.21 & 6.92 $\pm$ 0.72 & 9.83 & 10.62 & 36.81 & 20.47 & 33.92 & -0.12 $\pm$ 2.54 & -5.38 $\pm$ 4.85 & 5.56 $\pm$ 5.47 \\
RY Lup & 239.8688 & -40.3644 & 3.07 $\pm$ 0.07 & 6.35 $\pm$ 0.09 & 10.48 $\pm$ 0.15 & 1.77 $\pm$ 0.11 & 2.09 $\pm$ 0.11 & 4.35 & 2.52 & 15.38 & 5.09 & 8.61 & -0.76 $\pm$ 3.56 & -16.90 $\pm$ 7.49 & 41.30 $\pm$ 8.29 \\
SR 4 & 246.4844 & -24.3464 & 24.36 $\pm$ 0.20 & 24.63 $\pm$ 0.15 & 22.34 $\pm$ 0.14 & 5.38 $\pm$ 0.10 & 5.02 $\pm$ 0.11 & 288.79 & 189.09 & 219.33 & 33.77 & 37.97 & -4.88 $\pm$ 3.46 & -13.06 $\pm$ 5.83 & 8.82 $\pm$ 6.78 \\
SY Cha & 164.1262 & -77.194 & 2.71 $\pm$ 0.05 & 3.37 $\pm$ 0.19 & 15.40 $\pm$ 0.08 & 8.05 $\pm$ 0.11 & 27.57 $\pm$ 0.05 & 1.87 & 1.23 & 5.82 & 2.78 & 6.63 & 10.20 $\pm$ 4.19 & 12.86 $\pm$ 5.91 & -4.23 $\pm$ 7.25 \\
Sz 114 & 242.2573 & -39.0868 & 0.73 $\pm$ 0.02 & 2.00 $\pm$ 0.09 & 2.36 $\pm$ 0.09 & 1.56 $\pm$ 0.16 & 2.39 $\pm$ 0.07 & 5.32 & 1.89 & 3.19 & 2.09 & 3.60 & 0.47 $\pm$ 1.44 & -3.28 $\pm$ 3.07 & 4.02 $\pm$ 3.40 \\
Sz 129 & 239.8182 & -41.9528 & 2.07 $\pm$ 0.01 & 1.84 $\pm$ 0.10 & 3.65 $\pm$ 0.06 & 0.48 $\pm$ 0.07 & 1.23 $\pm$ 0.04 & 2.98 & 0.93 & 2.41 & 1.44 & 2.88 & 1.05 $\pm$ 1.93 & 1.86 $\pm$ 3.52 & -0.98 $\pm$ 4.02 \\
TW Cha & 164.754 & -77.3783 & 1.27 $\pm$ 0.05 & 1.43 $\pm$ 0.19 & 3.79 $\pm$ 0.36 & 1.25 $\pm$ 0.16 & 2.67 $\pm$ 0.10 & 1.02 & 0.70 & 2.52 & 1.89 & 3.45 & 10.87 $\pm$ 8.81 & 11.43 $\pm$ 1.72 & -0.63 $\pm$ 8.98 \\
TW Hya & 165.4659 & -34.7048 & 8.16 $\pm$ 0.08 & 2.27 $\pm$ 0.14 & 6.49 $\pm$ 0.07 & 2.18 $\pm$ 0.06 & 4.18 $\pm$ 0.06 & 4.47 & 2.51 & 6.86 & 5.84 & 7.98 & 8.57 $\pm$ 4.72 & 14.88 $\pm$ 18.25 & -6.34 $\pm$ 18.85 \\
VZ Cha & 167.3473 & -76.3892 & 9.50 $\pm$ 0.10 & 17.32 $\pm$ 0.25 & 56.12 $\pm$ 0.29 & 13.64 $\pm$ 0.25 & 36.56 $\pm$ 0.18 & 2.95 & 2.44 & 8.06 & 4.05 & 7.63 & 23.81 $\pm$ 2.64 & 14.60 $\pm$ 2.56 & 9.74 $\pm$ 3.68 \\
WSB 52 & 246.914 & -24.6548 & 11.09 $\pm$ 0.20 & 24.38 $\pm$ 0.35 & 65.16 $\pm$ 0.09 & 29.64 $\pm$ 0.13 & 79.72 $\pm$ 0.11 & 14.43 & 15.12 & 29.92 & 11.46 & 26.35 & -2.45 $\pm$ 3.22 & -8.28 $\pm$ 1.17 & 9.92 $\pm$ 3.43 \\
\enddata

\tablecomments{ DoAr 25 45\arcdeg{}; DoAr 33 45\arcdeg{}; GM Aur 45\arcdeg{}; HP Tau 45\arcdeg{}; IQ Tau 45\arcdeg{}; MY Lup 45\arcdeg{};  Sz 114 45\arcdeg{};  Sz 129 45\arcdeg{}; WSB 52 45\arcdeg }

\end{deluxetable*}

\begin{deluxetable*}{cccccccc}
\tabletypesize{\footnotesize}
\tablecaption{\label{tab:wind} The derived $A_V$, column density ${\rm (N(H_2))}$ and rotational temperature {$T_{rot}({\rm H}_2)$} for the gas  within the aperture as well as the correction factor $C_{\rm apert}$, and the total mass of gas ${M}_{ \rm {wind}}^{\rm tot}$ and the total mass loss rate   $\dot{M}_{ \rm {wind}}^{\rm tot}$. $^\dagger$ Bipolar sources. }
\tablehead{\colhead{Disk} & \colhead{$A_V$} & \colhead{${\rm log_{10}(N(H_2))}$} & \colhead{$T_{\rm rot}({\rm H}_2)$} & $C_{\rm apert}$ &  \colhead{${\rm log_{10}}({M}_{ \rm {wind}}^{\rm tot}$)} & \colhead{${\rm log_{10}}(\dot{M}_{\mathrm{wind}}^{\rm tot}$)} \\
& & $\rm cm^{-2}$  & K & & $\rm M_{\odot}$ & $\rm M_{\odot}\,\rm yr^{-1}$ }
\startdata
AS 205 & 1.7 & 19.17 $\pm$ 0.05 & 734 $\pm$ 22 & 2.60 $\pm$ 0.16 & -5.68 $\pm$ 0.05 & -8.34$^{+0.48}_{-0.60}$ \\
AS 209 & 1.6 & 19.33 $\pm$ 0.05 & 507 $\pm$ 11 & 3.02 $\pm$ 0.06 & -5.46 $\pm$ 0.05 & -8.15$^{+0.47}_{-0.59}$ \\
CI Tau & 1.2 & 18.69 $\pm$ 0.04 & 615 $\pm$ 14 & 3.35 $\pm$ 0.07 & -6.05 $\pm$ 0.04 & -8.77$^{+0.46}_{-0.58}$ \\
DoAr 25 & 3.2 & 18.18 $\pm$ 0.04 & 815 $\pm$ 24 & 3.42 $\pm$ 0.01 & -6.86 $\pm$ 0.04 & -9.58$^{+0.45}_{-0.57}$ \\
DoAr 33 & 3.7 & 18.81 $\pm$ 0.04 & 442 $\pm$ 8 & 3.51 $\pm$ 0.05 & -5.91 $\pm$ 0.04 & -8.64$^{+0.46}_{-0.58}$ \\
DR Tau & 0.3 & 19.28 $\pm$ 0.04 & 574 $\pm$ 16 & 3.97 $\pm$ 0.18 & -5.39 $\pm$ 0.05 & -8.14$^{+0.47}_{-0.59}$ \\
Elias 20 & 6.8 & 18.31 $\pm$ 0.04 & 1109 $\pm$ 47 & 3.24 $\pm$ 0.04 & -6.45 $\pm$ 0.04 & -9.16$^{+0.45}_{-0.57}$ \\
Elias 24 & 7.5 & 19.43 $\pm$ 0.04 & 646 $\pm$ 15 & 3.01 $\pm$ 0.04 & -5.36 $\pm$ 0.04 & -8.05$^{+0.45}_{-0.58}$ \\
Elias 27 & 2.7 & 18.31 $\pm$ 0.04 & 590 $\pm$ 14 & 2.68 $\pm$ 0.06 & -6.53 $\pm$ 0.04 & -9.20$^{+0.46}_{-0.58}$ \\
FZ Tau & 2.8 & 19.04 $\pm$ 0.04 & 577 $\pm$ 11 & 3.57 $\pm$ 0.02 & -5.68 $\pm$ 0.04 & -8.41$^{+0.45}_{-0.57}$ \\
GK Tau & 1.2 & 18.32 $\pm$ 0.04 & 629 $\pm$ 15 & 6.46 $\pm$ 0.45 & -6.14 $\pm$ 0.05 & -9.00$^{+0.48}_{-0.60}$ \\
GM Aur & 0.0 & 18.03 $\pm$ 0.04 & 653 $\pm$ 16 & 2.76 $\pm$ 0.03 & -6.80 $\pm$ 0.04 & -9.47$^{+0.46}_{-0.58}$ \\
GO Tau & 3.4 & 17.73 $\pm$ 0.06 & 529 $\pm$ 16 & 3.90 $\pm$ 0.13 & -6.95 $\pm$ 0.06 & -9.70$^{+0.48}_{-0.60}$ \\
GQ Lup & 0.1 & 18.06 $\pm$ 0.04 & 655 $\pm$ 17 & 4.79 $\pm$ 0.07 & -6.53 $\pm$ 0.04 & -9.32$^{+0.46}_{-0.58}$ \\
GW Lup & 0.7 & 17.87 $\pm$ 0.04 & 570 $\pm$ 13 & 6.18 $\pm$ 0.19 & -6.61 $\pm$ 0.04 & -9.46$^{+0.46}_{-0.59}$ \\
HP Tau & 4.4 & 19.12 $\pm$ 0.04 & 749 $\pm$ 21 & 4.58 $\pm$ 0.03 & -5.48 $\pm$ 0.04 & -8.27$^{+0.45}_{-0.57}$ \\
HT Lup & 1.5 & 18.43 $\pm$ 0.05 & 914 $\pm$ 36 & 3.85 $\pm$ 0.06 & -6.25 $\pm$ 0.05 & -9.00$^{+0.46}_{-0.58}$ \\
IQ Tau & 1.4 & 18.34 $\pm$ 0.03 & 787 $\pm$ 22 & 3.04 $\pm$ 0.06 & -6.74 $\pm$ 0.04 & -9.44$^{+0.45}_{-0.58}$ \\
IRAS 04385+2550 & 3.4 & 18.51 $\pm$ 0.04 & 732 $\pm$ 19 & 4.57 $\pm$ 0.05 & -6.40 $\pm$ 0.04 & -9.18$^{+0.45}_{-0.58}$ \\
MY Lup & 1.3 & 18.59 $\pm$ 0.04 & 515 $\pm$ 10 & 3.92 $\pm$ 0.08 & -6.39 $\pm$ 0.04 & -9.14$^{+0.46}_{-0.58}$ \\
RU Lup & 0.0 & 19.42 $\pm$ 0.06 & 454 $\pm$ 11 & 3.02 $\pm$ 0.12 & -5.36 $\pm$ 0.06 & -8.06$^{+0.48}_{-0.61}$ \\
RY Lup & 0.8 & 18.68 $\pm$ 0.04 & 491 $\pm$ 9 & 6.33 $\pm$ 0.10 & -5.78 $\pm$ 0.04 & -8.64$^{+0.46}_{-0.58}$ \\
SR 4 & 1.1 & 19.49 $\pm$ 0.04 & 432 $\pm$ 7 & 4.18 $\pm$ 0.04 & -5.15 $\pm$ 0.04 & -7.92$^{+0.46}_{-0.58}$ \\
SY Cha & 1.5 & 18.27 $\pm$ 0.03 & 851 $\pm$ 25 & 3.46 $\pm$ 0.03 & -6.45 $\pm$ 0.03 & -9.18$^{+0.45}_{-0.57}$ \\
Sz 114 & 0.0 & 17.83 $\pm$ 0.04 & 647 $\pm$ 16 & 5.80 $\pm$ 0.24 & -6.67 $\pm$ 0.04 & -9.50$^{+0.47}_{-0.59}$ \\
Sz 129 & 0.5 & 18.27 $\pm$ 0.04 & 500 $\pm$ 10 & 3.37 $\pm$ 0.07 & -6.47 $\pm$ 0.04 & -9.18$^{+0.46}_{-0.58}$ \\
TW Cha & 0.5 & 18.00 $\pm$ 0.05 & 619 $\pm$ 15 & 5.80 $\pm$ 0.56 & -6.51 $\pm$ 0.06 & -9.34$^{+0.50}_{-0.62}$ \\
TW Hya & 0.3 & 18.60 $\pm$ 0.04 & 519 $\pm$ 10 & 1.09 $\pm$ 0.01 & -6.63 $\pm$ 0.04 & -9.10$^{+0.46}_{-0.58}$ \\
VZ Cha & 0.2 & 18.93 $\pm$ 0.04 & 661 $\pm$ 15 & 3.66 $\pm$ 0.03 & -5.77 $\pm$ 0.04 & -8.50$^{+0.45}_{-0.57}$ \\
WSB 52 & 4.3 & 19.18 $\pm$ 0.04 & 726 $\pm$ 19 & 2.67 $\pm$ 0.01 & -5.96 $\pm$ 0.04 & -8.63$^{+0.45}_{-0.57}$ \\
\enddata
\tablecomments{ DoAr 25,  IRAS 04385+2550,  IQ Tau, MY Lup, and WSB 52 are bipolar sources with $C_{\rm obsc}=1$}.
{For AS 209, GO Tau and RU Lup we used the S(2) line to compute $C_{\rm apert}$; for DoAr 33 and MY Lup we used the S(1) line.}
\end{deluxetable*}

\vskip -0.685in

Our analysis focuses on weaker, more diffuse emission, where many sources exhibit significantly fainter and more spatially patchy structures, particularly after median-flux subtraction. In our sample, extended H$_2$ emission is typically detected up to the S(5) line; therefore, for consistency, we limit our analysis to transitions up to H$_2$ S(5). {Because only 3–5 reliable lines are available to constrain the rotational temperature, the fit has limited degrees of freedom.}  We thus model the rotation diagram with a single temperature component, characterized by a column density $N$(H$_2$) and rotational temperature $T{_\mathrm{rot}}$(H$_2$).  {As a result, our analysis may not be sensitive to the hotter components {(1400 - 2000 K)} of the flow, such as those reported in previous studies \citep[e.g.,][]{2024ApJ...966L..22N, 2025A&A...694A.174F, 2025ApJ...980..148S,2025A&A...703A.139S,2025ApJ...983..110T,2025ApJ...995..199N}, which arise from higher-excitation lines.} The line properties are summarized in Table~\ref{TableH2}.

Our targets are optically visible and not deeply embedded. {Previous studies have reported low extinction values for these sources, typically $A_V \lesssim 4$ mag \citep[e.g.,][]{2011ApJS..195....3F,2010ApJS..188...75M,2011ApJS..193...11M,2014ApJ...786...97H,2024A&A...692A..69F}.} However, these values were derived using different methods and are not uniform across the literature. To obtain a consistent estimate, we used Gaia DR3 data \citep{2023A&A...674A...1G} to compute color excesses and derive extinction values uniformly. {We did not use the H$_2$ emission lines directly to measure extinction due to the limited number of detected transitions (degree's of freedom), which prevents a robust determination.} Instead, we used Gaia observed $G_{BP}-G_{RP}$, and the intrinsic colors were adopted from the online table based on \citet{2013ApJS..208....9P}\footnote{\url{https://www.pas.rochester.edu/~emamajek/EEM_dwarf_UBVIJHK_colors_Teff.txt}}
This table is regularly updated to reflect the latest calibrations of photometric filters, spectral types, and standards. The median differences in ${G_{BP}}$, and ${G_{RP}}$ between Gaia DR2 and DR3 are only 0.02, and 0.01 mag, respectively \citep{2023PhDT........17N,2024AJ....168....7B} . Therefore, the intrinsic colors from this table are suitable for use with \textit{Gaia} DR3 photometry. We list the derived $A_V$ values in Table \ref{tab:wind}. We then used the KP5 extinction law \citep{2024RNAAS...8...68P} to derive the extinction and de-redden the flux.  {After applying this extinction correction, we find that in most sources the H$_2$ S(3) line, which is most affected by extinction due to its location within the 10~$\mu$m silicate feature, now fits well within the rotation diagram, and any remaining deviation from the linear fit can be attributed to low S/N.
}

We also adopt a fixed ortho-to-para ratio (OPR) of 3, as expected under LTE conditions at the typical rotational temperatures of $\sim$650 K (see Table~\ref{tab:wind}) found for our disks \citep{1998ApJ...498..246T, 2000A&A...356.1010W,2006ApJ...649..816N}.

We used the MCMC sampler from the python package \textit{lmfit} \citep{2023zndo...8145703N} to fit the rotation diagram and compute the rotation temperature and the total column density along with the associated uncertainties. {We have also added a systematic error of 10\% from the relative spectro-photometric precision of MIRI MRS \citep{2023A&A...675A.111A} to the measured line fluxes. }  In Figure \ref{fig:rot} we show the result from the MCMC fit as well as the best fit rotation diagram for WSB 52 as a representative fit.  The rest of the rotation diagrams are shown in Figures \ref{rotfig1} to \ref{rotfig10}. We use the partition function from \cite{1996AJ....111.2403H} to compute the total column density of H$_2$, $N_{\rm tot}$:

\begin{equation}
    Z(T_{\rm rot}) = \frac{0.0247T_{\rm rot}}{1 - \exp\left(-\frac{6000}{T_{\rm rot}}\right)}
\end{equation}

Table \ref{tab:wind} lists the derived column density and rotational temperature for the sample. The median temperature of the emitting gas is $T_{\rm rot}\sim  624 \pm 130\,$ K,  and the median  column density $\log_{10}(N_{\rm tot}) \sim 18.6 \pm 0.6$ (cm$^{-2}$).  Figures \ref{fig:rot} (and  Figures \ref{rotfig1} -- \ref{rotfig10}) demonstrate that most disk winds are best fit with a single warm temperature component and specifically do not require a second hot component. The excitation conditions of H$_2$ for one disk SY Cha has previously been studied in literature \citep{2025ApJ...980..148S,2025ApJ...991..232S}. While compact H$_2$ emission from the central point source has is best fit by  two temperature components, this study also found that the extended component (radii greater than 0\farcs5) of  H$_2$ is best fit by a single component with $T_{\rm rot} = 1061\pm39$\,K and $N(\rm H_2) = 4.31\pm 0.72\times 10^{17} {\rm cm}^{-2}$. {These values differ from our measurements of $T_{\rm rot} = 851 \pm 25$ K and $N(\mathrm{H}_2) = (1.9 \pm 0.1)\times10^{18},\mathrm{cm}^{-2}$.} This discrepancy may arise because \cite{2025ApJ...980..148S,2025ApJ...991..232S} do not include emission from the warmer S(1) and S(2) transitions, which can bias the derived rotational temperature and column densities. Additionally, their analysis does not account for extinction correction, which can further impact the inferred column densities and temperatures. We repeated the analysis allowing the OPR to vary as a free parameter and find a median {OPR of 2.2$\pm 0.3$ (1$\sigma$)}.

\subsection{Wind velocities} 

The H$_2$ velocity is measured for each spaxel by fitting a Gaussian profile to the emission line and put on a precise reference frame by comparing to nearby water emission lines from the central disk, where available. Although the MIRI MRS instrument has a spectral resolving power $R \sim100\,{\rm km s}^{-1}$, the S/N of the observations allows for precise measurements of the central wavelengths, with uncertainties as low as a few km~s$^{-1}$ \citep{Pontoppidan24}. For aggregate velocities, we use the same 250 au square apertures previously adopted for line intensity measurements. We use the H$_2$ S(1) line, which typically has high S/N and is surrounded by numerous strong water lines. The nearby water lines makes it easy to compute the systemic velocity (see the following discussion).

\begin{figure}[h]
\centering
\includegraphics[width=8.5cm]{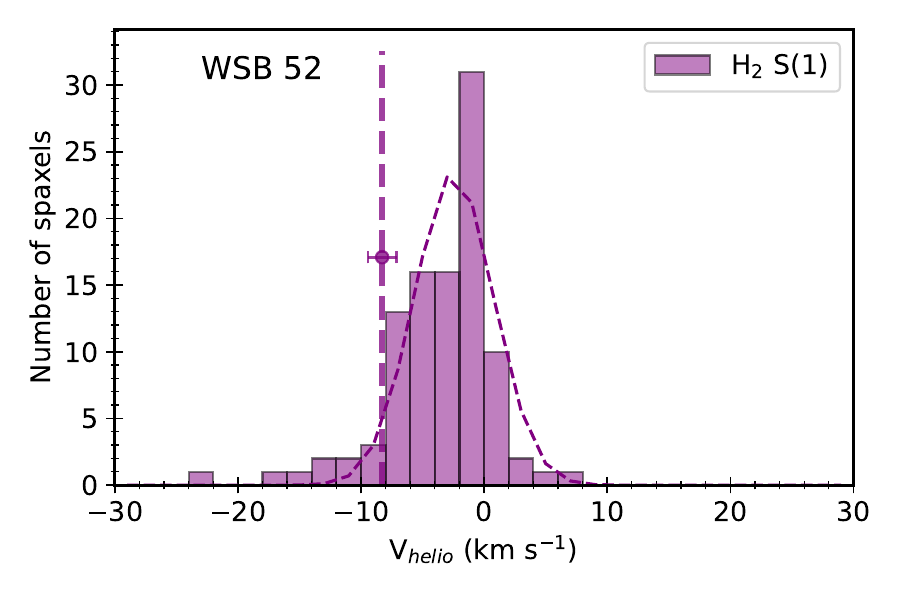}
\caption{ The velocity distribution for H$_2$ S(1) line in {WSB 52} from the spaxels within the 250 au square aperture used to measure the line fluxes in Section \ref{sec:rot}. Also shown is the best fit gaussian to the velocity distribution. The systemic velocity of the star as measured using H$_2$O lines is shown as the dashed vertical line with the error bar showing the uncertainty in determining the systemic velocity.  \label{fig:fztau_vel}}
\end{figure}

As an illustrative example, Figure~\ref{fig:fztau_vel} shows the velocity distribution of the H$_2$ S(1) emission for all spaxels within the 250 au aperture for {WSB 52}. We visualized the velocity distribution as a histogram with 30 bins spanning the velocity range of $\pm\,30 \,\rm km~s^{-1}$. We fit a Gaussian to the velocity distribution, where the mean $V_{\rm H_2}$, represents the velocity of the emitting gas in the barycenter frame (with the associated telescope wavelength calibration uncertainties). To determine the absolute (true) flow velocity, we correct for the systemic velocity of the star, $V_{\rm sys}$ in the barycentric frame (with the associated telescope wavelength calibration uncertainties). The systemic velocity is measured by measuring the central velocity of nearby H$_2$O lines, under the assumption that they originate from the disk. In Table \ref{tab:H2} we have listed the $V_{\rm H_2}$ and V$_{\rm sys}$ for all targets.

In Figure~\ref{fig:fztau_vel}, the dashed vertical line indicates this systemic velocity. Subtracting it from the mean velocity of the H$_2$ emission yields the intrinsic velocity of the molecular flow. We then correct for inclination by dividing by the cosine of the inclination angle to derive the true flow velocity $V_{\rm wind}$ (see Table \ref{tab:H2}). We also compute the error on the V$_{\rm wind}$ by adding the error in $V_{\rm H_2}$ and $V_{\rm sys}$ in quadrature. 

{Using this method, what we have measure is V$_{\rm wind}$ or the component of the flow velocity projected along the flow axis $z$, $V_{\rm z}$, under the assumption of axisymmetry \citep{2022A&A...668A..78D}. The true poloidal velocity of the flow is then obtained as }

\begin{equation}
    V_{\rm p} = \frac{V_{\rm z}}{\cos\theta},
\end{equation}

{where $\theta$ is the opening angle of the flow. For a typical opening angle of $\sim 45^\circ$ (see Table \ref{tab:OA}), this projection correction increases the inferred flow velocity by a factor of $\sim 1.4$.}

There is a spread in the measured velocities of sources due to large relative errors for some sources, as shown in Table \ref{tab:H2}. {Therefore, for sources with reliable measurements ($\sigma(V_{\rm wind}) \leq 10~{\rm km~s}^{-1}$ {and $|V_{\rm wind}|\leq25~{\rm km~s}^{-1}$)}, we compute the median velocity of the wind by taking the absolute value of the velocity. The median wind velocity is  4.2$^{+6.7}_{-3.0}$ km s$^{-1}$, which is comparable to the $5\,\mathrm{km\,s^{-1}}$ wind velocity adopted by \citep{2025ApJ...980..148S}. We use this median velocity (corrected for inclination) to compute the dynamical timescale (see Section 4.2 and Eq \ref{tdy}). }

\subsection{Opening angle and wind launch radius} \label{sec:OA}

\begin{figure*}
\centering
\includegraphics[width=16cm]{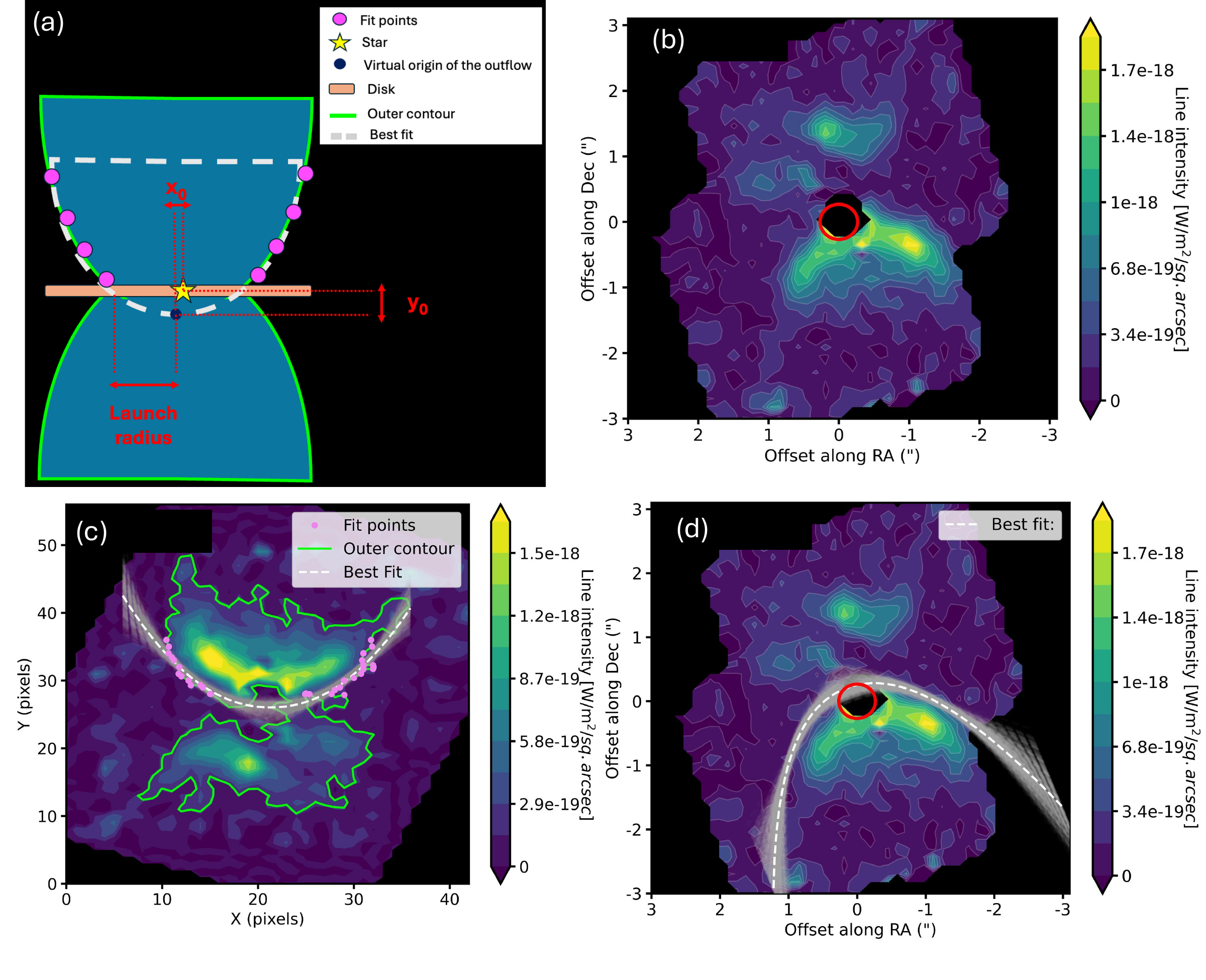}
\caption{(a) Schematic illustrating the fitting procedure used to derive the opening angle of the H$_2$ wind.   (b) The wide-angle H$_2$ wind emission as observed for DoAr 25. (c) The image rotated to align the wind axis vertically. The outer contour tracing the edge of the flow is shown in lime green, and the magenta points indicate the data selected for model fitting. (d) The resulting best-fit power-law model is shown in white and the shaded grey region is the uncertainty range derived from MCMC fitting. The red circle marks the inner working angle.   }
\label{Fig:OA}
\end{figure*}

The angular resolution of JWST/MIRI enables us to extract geometrical parameters such as the opening angles and launch radii of the winds. Since the wind morphology displays significant curvature, we fit disks with sufficiently high S/N with a polynomial of the form \citep[also see][]{2021ApJ...911..153H}:
\begin{equation}
     y(x) = y_0 + A |x - x_0|^{\alpha} 
\end{equation}
where $ y(x) $ denotes the height of the wind at a distance $ x $ from the central axis $ x_0 $. {Since the wind is launched from a finite radius, extrapolating the flow leads to a virtual origin located behind the disk, represented by the parameter $ y_0 $ {(see Figure \ref{Fig:OA}(a))}. }

The S(5) emission provides the best balance between S/N and angular resolution to determine the morphological properties of the wind, {except for CI Tau and Elias 27, where we use the S(3) line and IQ Tau where S(1) line was used due to insufficient S/N of the S(5) image}. If both lobes of the wind are visible, we fit them separately.

Figure \ref{Fig:OA}, shows an example of the fitting procedure using the SW component of the DoAr 25 SW flow (Figure \ref{Fig:OA}b). To fit the wind morphology, we first then rotate the H$_2$ image so that the flow axis is aligned along the $x$–$y$ plane, with the direction of wind along the $y$-axis (see Figure \ref{Fig:OA}c). A similar appoach is followed in \cite{2021ApJ...911..153H}. We then identify the edge of the wind by measuring the standard deviation, $\sigma$, of the background and applying a threshold for each flow to delineate the wide-angle wind structure (Figure \ref{Fig:OA}(c)).  We limit the curvature index $\alpha$ to 1--4, the horizontal offset $x_0$ to within $\pm1$ spaxel from the JWST/MIRI photo-center along the $x$-axis, and the vertical offset $y_0$ to between $-5$ and 0 pixels from the JWST/MIRI photo-center along the $y$-axis. 

In Figure \ref{Fig:OA}(d), we show the fit to the observed flow. The launch radius of the wind is assumed to be the half-width of the flow, $x_d$, measured at the JWST/MIRI photo-center along the $y$-axis (Figure \ref{Fig:OA}(a)). Since we do not fit a straight line to the wind edges, we cannot directly measure the half-opening angle, as it depends on the outer radius of the flow (see \cite{2021ApJ...911..153H}). Instead, we determine the slope of the fitted power-law between 2 pixels ($\sim$30 au) and 12 pixels ($\sim$180 au). The effective half-opening angle relative to the vertical is then calculated as the cotangent inverse of this slope.

We analyzed in detail the wind morphology of ten sources. The remaining targets were primarily close to face-on disks, which exhibited bubble-like or disrupted morphologies, or had insufficient S/N to reliably isolate the wind structure. From the {ten} analyzed sources, we identified clear bipolar winds in {five} cases. In the remaining {five}, we only detect the wind in the near side of the disk, which is likely due to extinction from the disk itself.  Figure~\ref{fig:wind_fit} displays the best-fit power-law models for each wind structure, with the corresponding fit parameters summarized in Table~\ref{tab:OA}. {The wind shapes are well characterized by a median power-law index of $\alpha$ = {1.6}$^{+0.5}_{-0.4}$. The median half-opening angle is $45\arcdeg^{+5}_{-4}$.}
Similar results have been derived for other protoplanetary disks as well \citep{2025NatAs...9...81P,2025ApJ...980..148S,2024ApJ...965L..13A}. These opening angles are significantly wider than the typical $29\arcdeg$ observed in less evolved Class 0 protostars \citep{2024MNRAS.533.3828D}, and are comparable to the $43\arcdeg$–$45\arcdeg$ half-opening angles characteristic of more evolved  Class I or flat-spectrum protostellar sources \citep{2024MNRAS.533.3828D}. Since protostellar outflows are entrained by envelope material, the unconstrained wind morphology observed from protoplanetary disks with little or no residual envelope may be a closer representation of the true wind opening angles.

The radii we derive for the wide-angle flows exceed some values reported for other protoplanetary disks observed with NIRSpec in the H$_2$ 0–0 S(9) line \citep{2025NatAs...9...81P}. Several factors may contribute to this discrepancy: (1) The higher spatial resolution of NIRSpec can trace emission features nearer the central source, which may influence the inferred flow geometry and give a tighter bound for the launch radii;
(2) The flows exhibit curvature rather than resembling idealized straight cones, and our use of a power-law fit tends to yield larger inferred radii compared to linear fits; (3) Our fitting routine assumes a finite flow radius at the location of the disk, resulting in the flow's geometric origin being placed behind the disk, which also broadens the inferred launch region;
(4) H$_2$ outflows from protostars frequently show a nested structure, with higher-excitation lines appearing more collimated (\citealt{2026arXiv260209837N}; H. Tyagi et al., in prep, V. Pathak in prep), and a similar stratification may be present in our observations. As expected for winds these launch radii are still significantly smaller than the gas disk radii derived from ALMA observations (see Table~\ref{tab:OA}).

\section{Discussion} \label{sec:discussion}

\subsection{Prevalence of disk winds in our sample}
\label{sec:prevalence}

We generally attribute observed extended H$_2$ emission centered on the source to disk winds. Several lines of evidence support this attribution. First, moderately inclined disks ($i\gtrsim 50$\,\arcdeg) exhibit wide-angle conical wind structures (Section~\ref{sec:OA}) predicted by disk wind models \citep[][]{2022A&A...668A..78D,2025NatAs...9...81P}. We identify ten such systems: CI Tau (Figure~\ref{fig:citau}), Elias 20 (Figure~\ref{fig:elias27}), Elias 27 (Figure~\ref{fig:elias27}), GM Aur (Figure~\ref{fig:gmaur}), SY Cha (Figure~\ref{fig:sycha}), WSB 52 (Figure \ref{fig:wsb52}), DoAr 25 (Figure~\ref{fig:doar25}), IRAS 04385+2550 (Figure~\ref{fig:iras04385}), IQ Tau (Figure \ref{fig:iqtau}), and MY Lup (Figure~\ref{fig:mylup}). The opening angles are consistent with those inferred for edge-on disks, where the detection of disk winds is more straightforward due to the favorable geometry \citep[e.g.,][]{2024ApJ...965L..13A,2025ApJ...980..148S,2025NatAs...9...81P}. The agreement across a range of inclinations supports the interpretation that the observed wide-angled emission is indeed driven by disk winds. A lot more sources e.g., GQ Lup and HT Lup also so tentative evidence for a wide-angle wind. 

The next most commonly observed morphology is associated with nearly face-on disks ( $i\lesssim$ 30\arcdeg), which often exhibit either bubble-like or ring-like extended emission structures. Ring-like morphologies are seen in Elias 24 (Figure~\ref{fig:elias24}), FZ Tau (Figure~\ref{fig:fztau}),  HP Tau (Figure~\ref{fig:hptau}), TW Cha (Figure~\ref{fig:twcha}), and TW Hya (Figure~\ref{fig:twhya}),  while bubble-like emission is seen in VZ Cha (Figure~\ref{fig:vzcha}). These radially symmetric emission patterns are naturally explained by wide-angle disk-driven winds viewed close to face-on \citep[e.g.,][]{2022A&A...668A..78D}. {While \citet{2022A&A...668A..78D} explicitly present results for a single inclination, their underlying formalism is general and can be modified for any arbitrary inclination angle\footnote{https://github.com/Alois-deValon/Axoproj/tree/main}. }

A few systems in our sample, especially AS 205, HT Lup, and GK Tau, are known components of wide stellar binaries (separations $\gtrsim$1$''$), while GQ Lup has a confirmed massive sub-stellar companion \citep{2024ApJ...966L..21C}. This may naturally influence the observed morphology of the extended emission. Indeed, the presence of a stellar or sub-stellar companion can significantly alter the dynamics and structure of the disk winds. In binaries, tidal interactions may truncate the disk or introduce asymmetries that distort the otherwise symmetric wind structure, leading to deviations from the expected morphologies in both face-on and inclined systems. 

\begin{figure*}
\centering
\includegraphics[width=6cm]{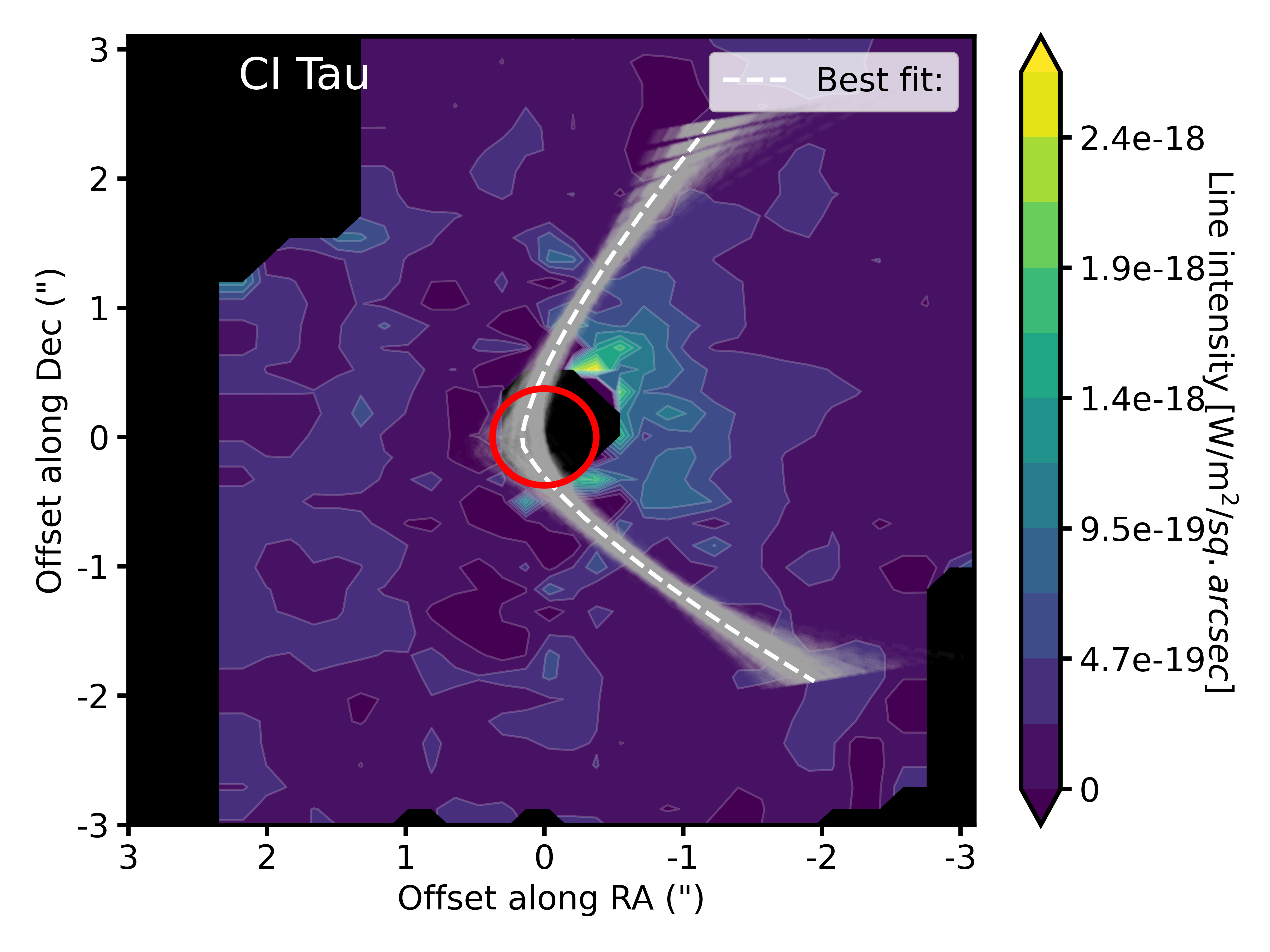}\includegraphics[width=6cm]{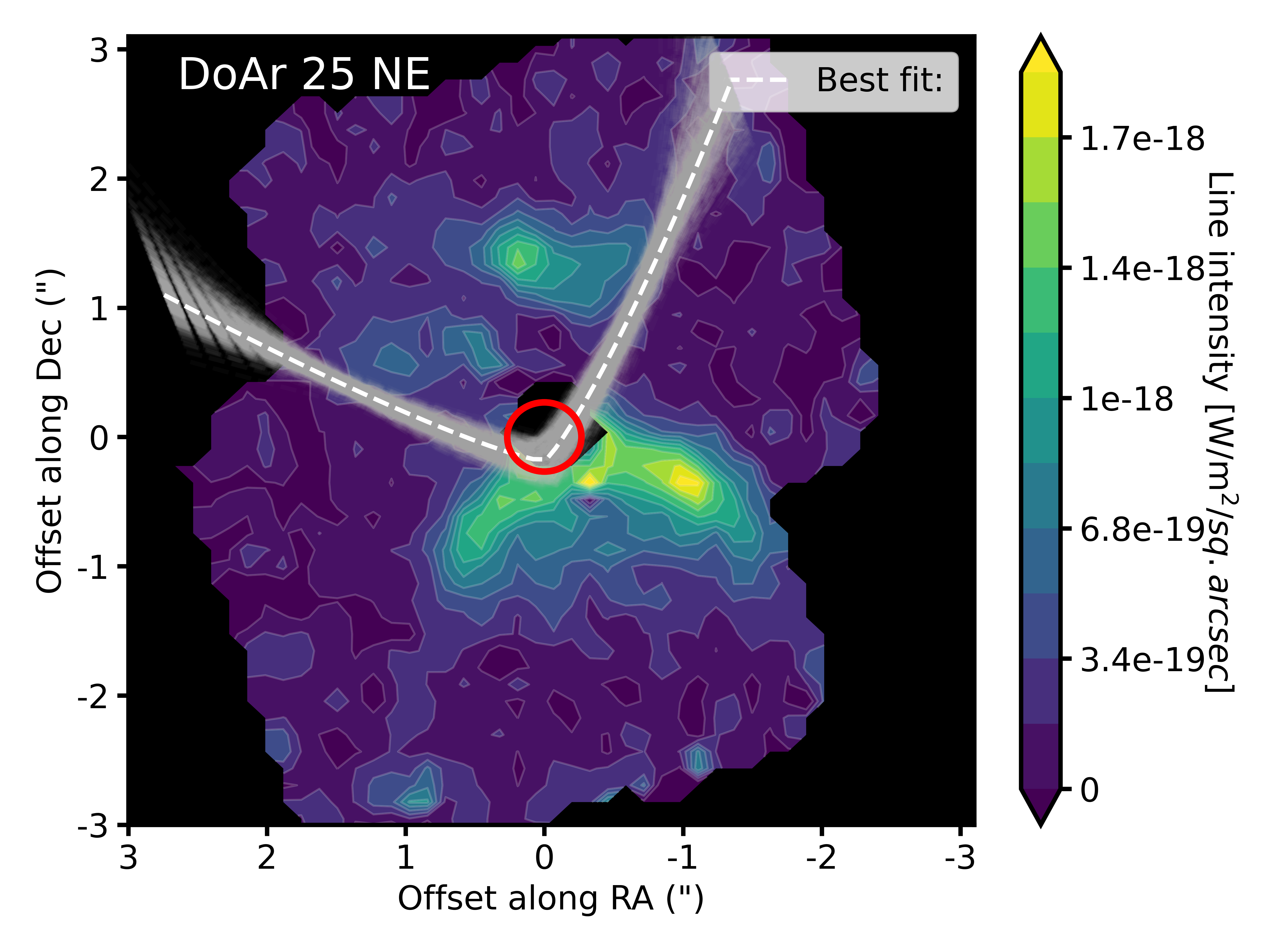}\includegraphics[width=6cm]{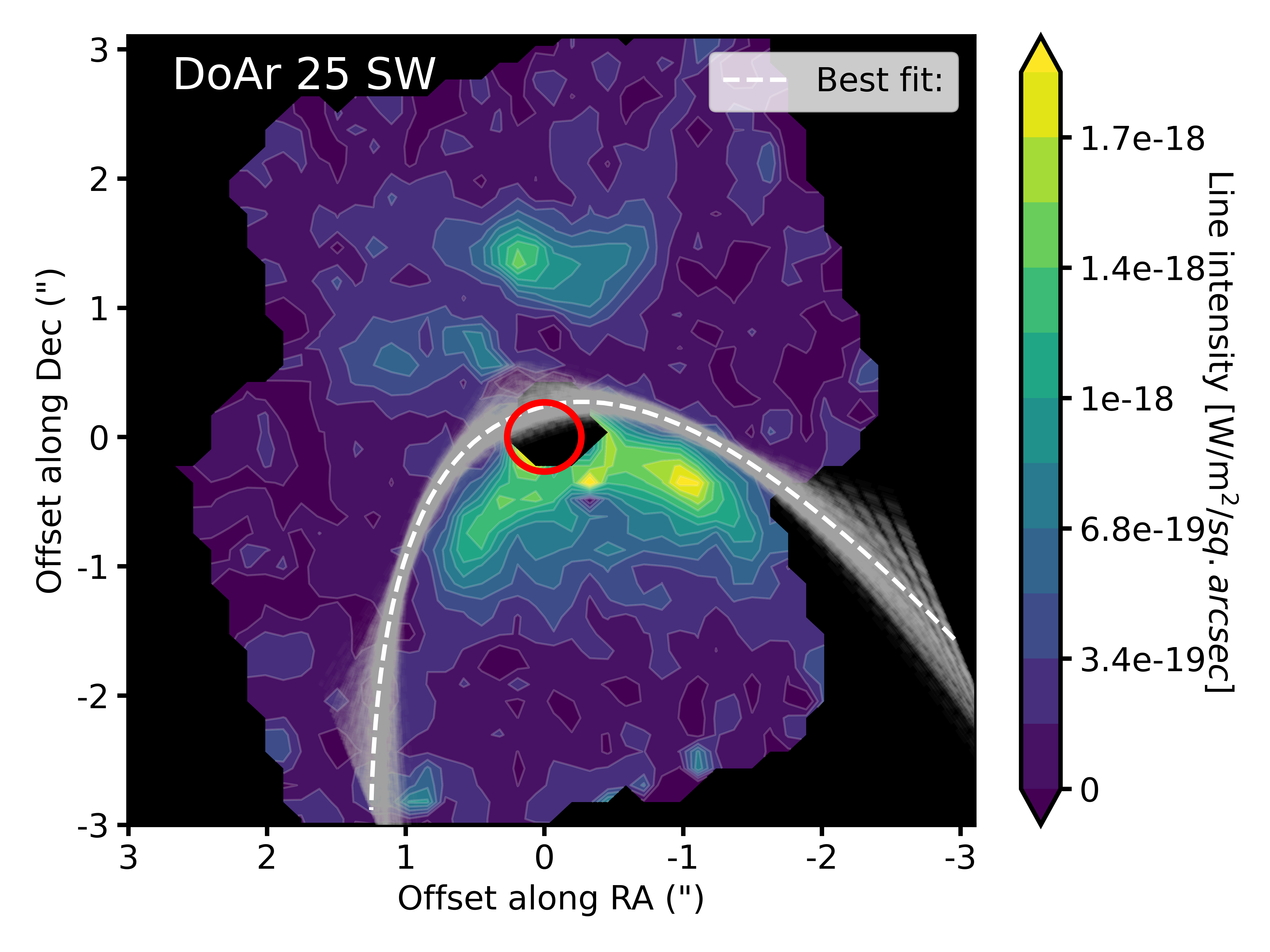}

\includegraphics[width=6cm]{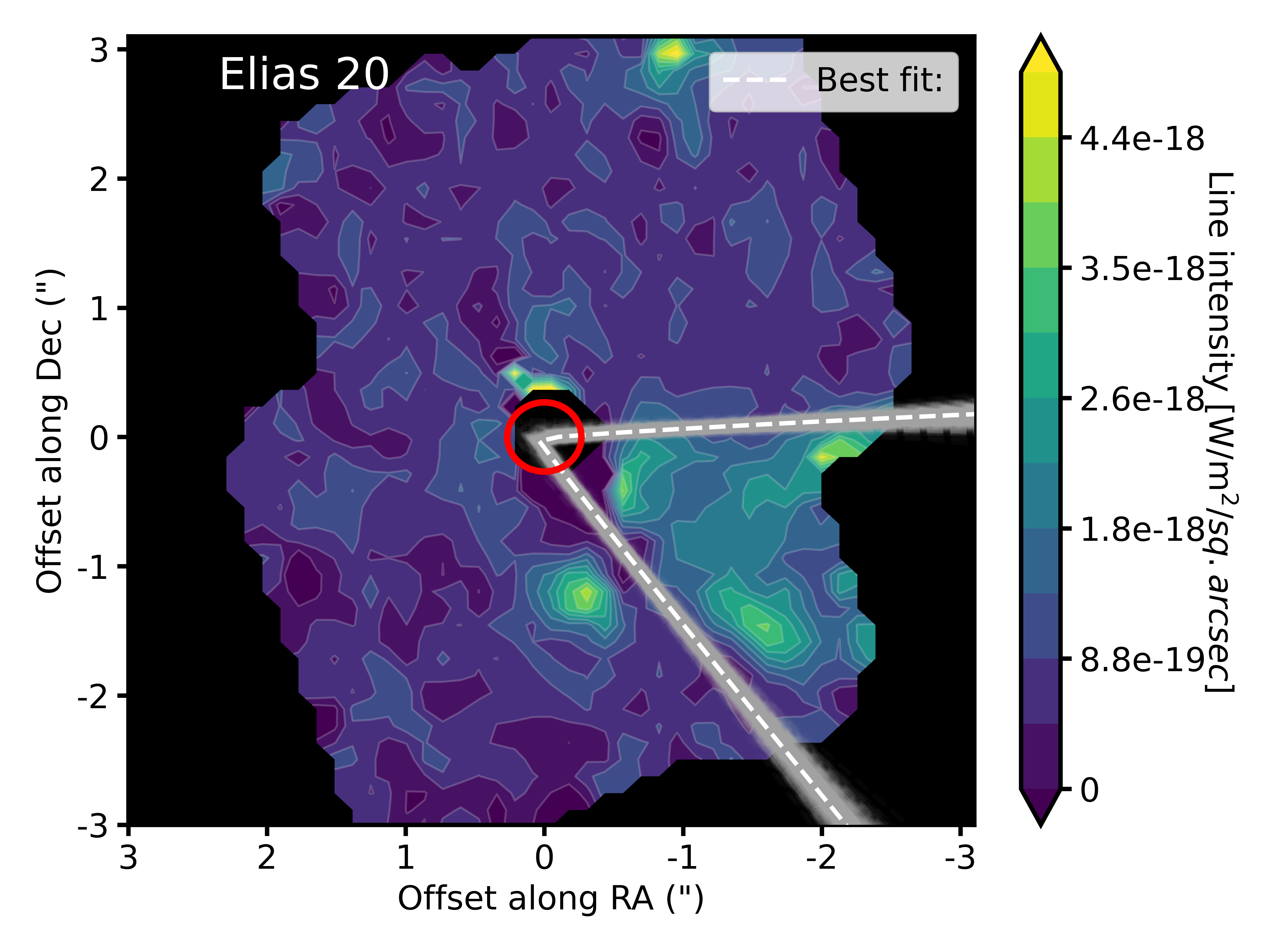}\includegraphics[width=6cm]{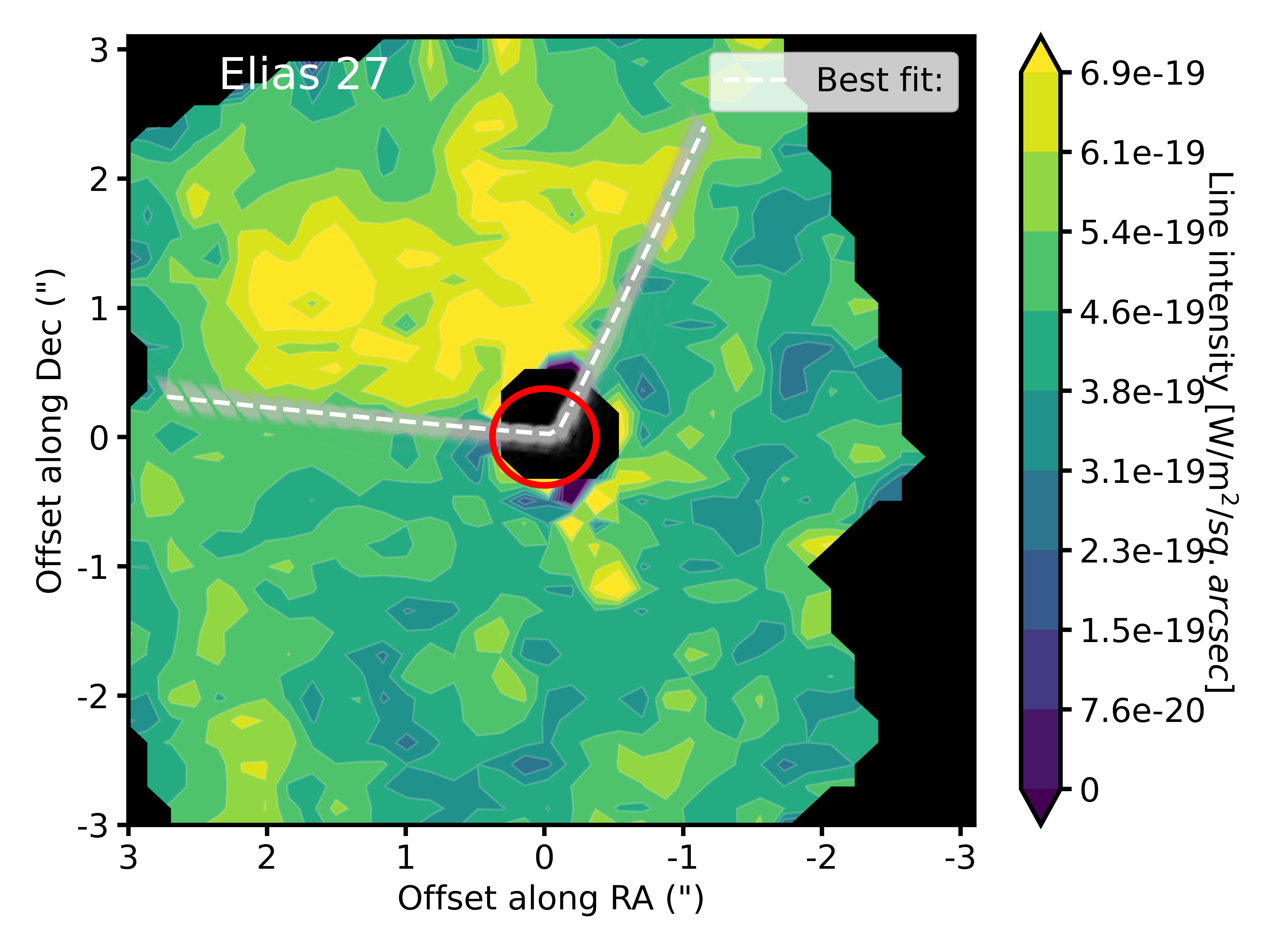}\includegraphics[width=6cm]{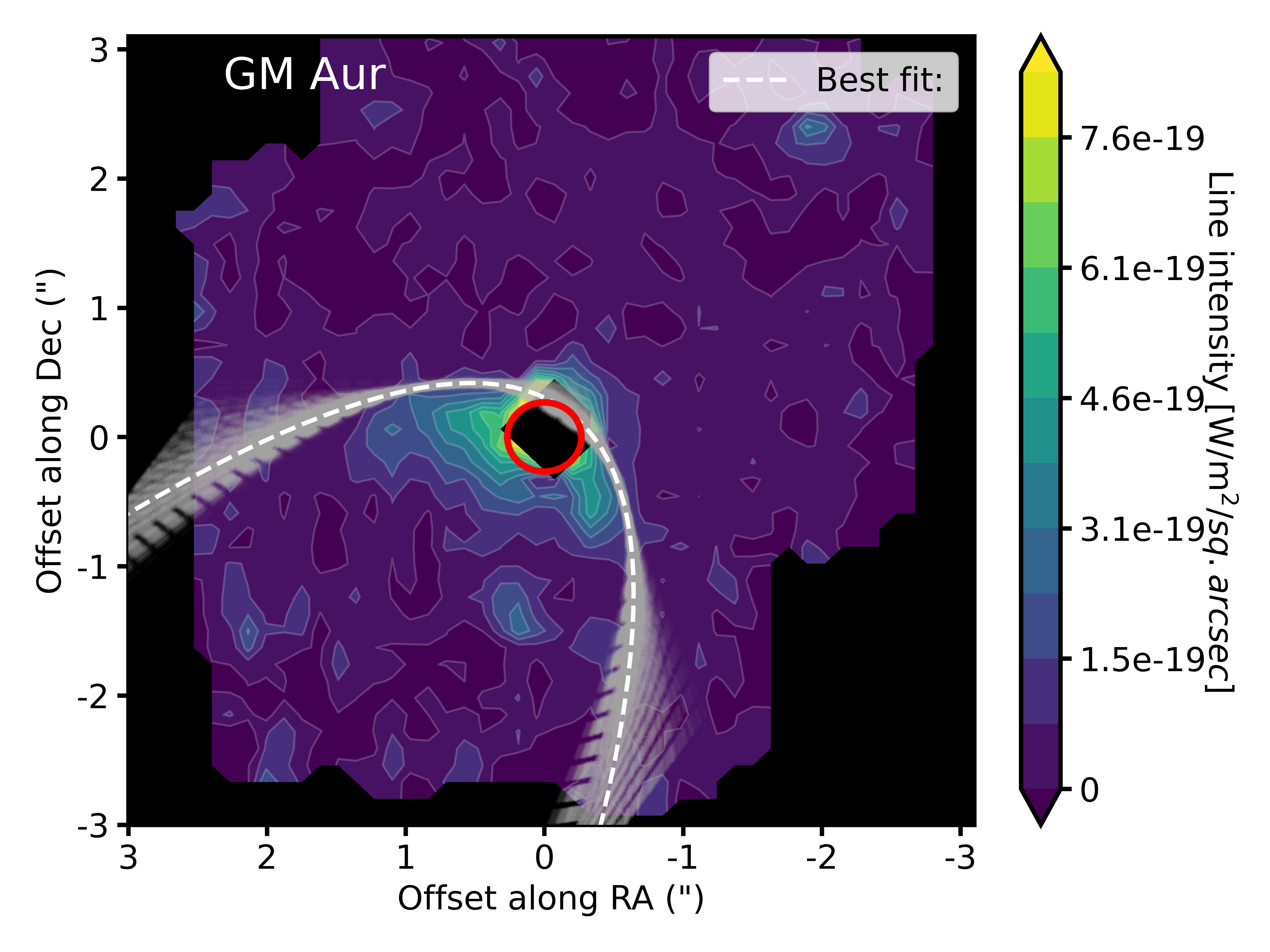}

\includegraphics[width=6cm]{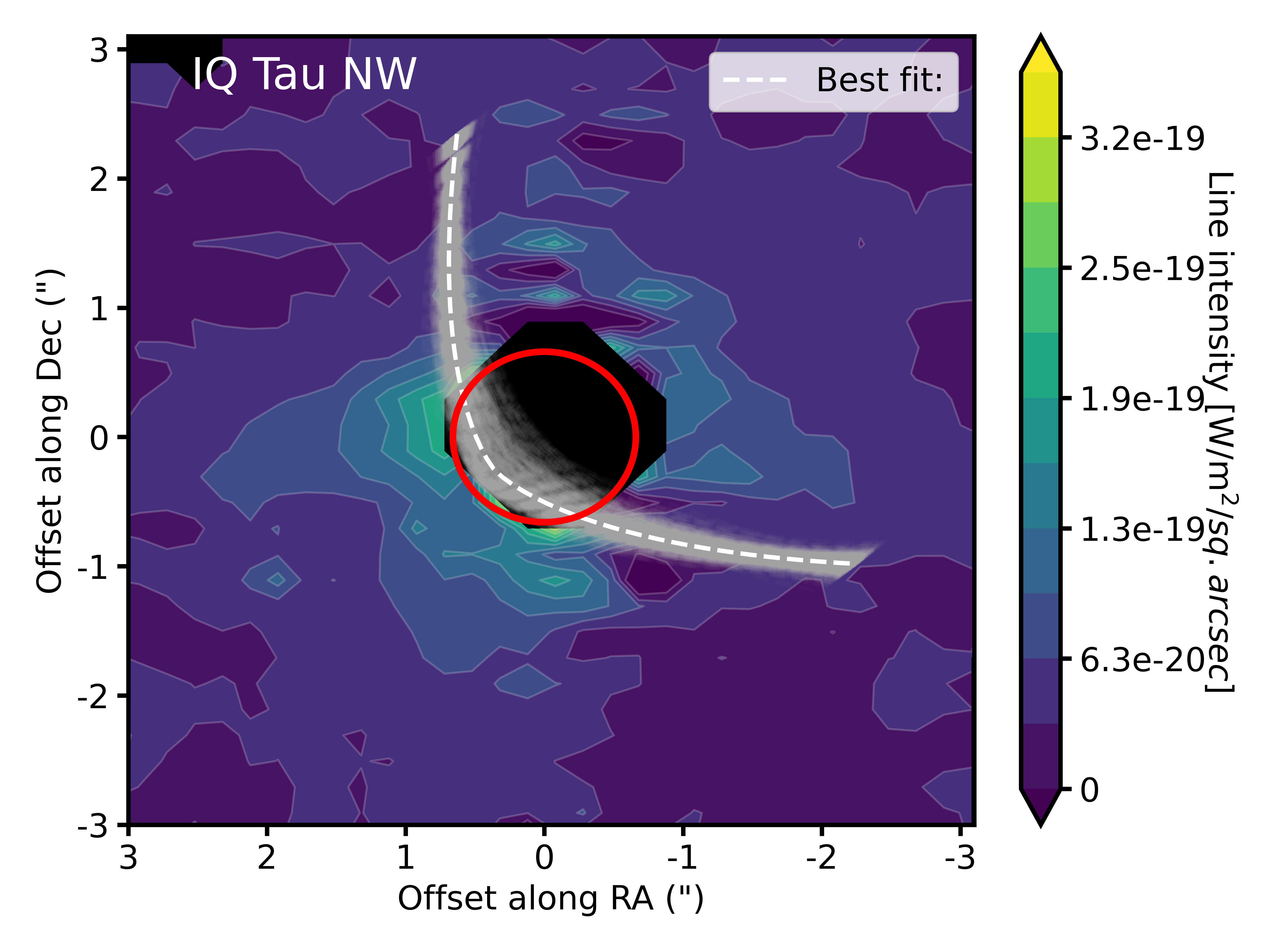}\includegraphics[width=6cm]{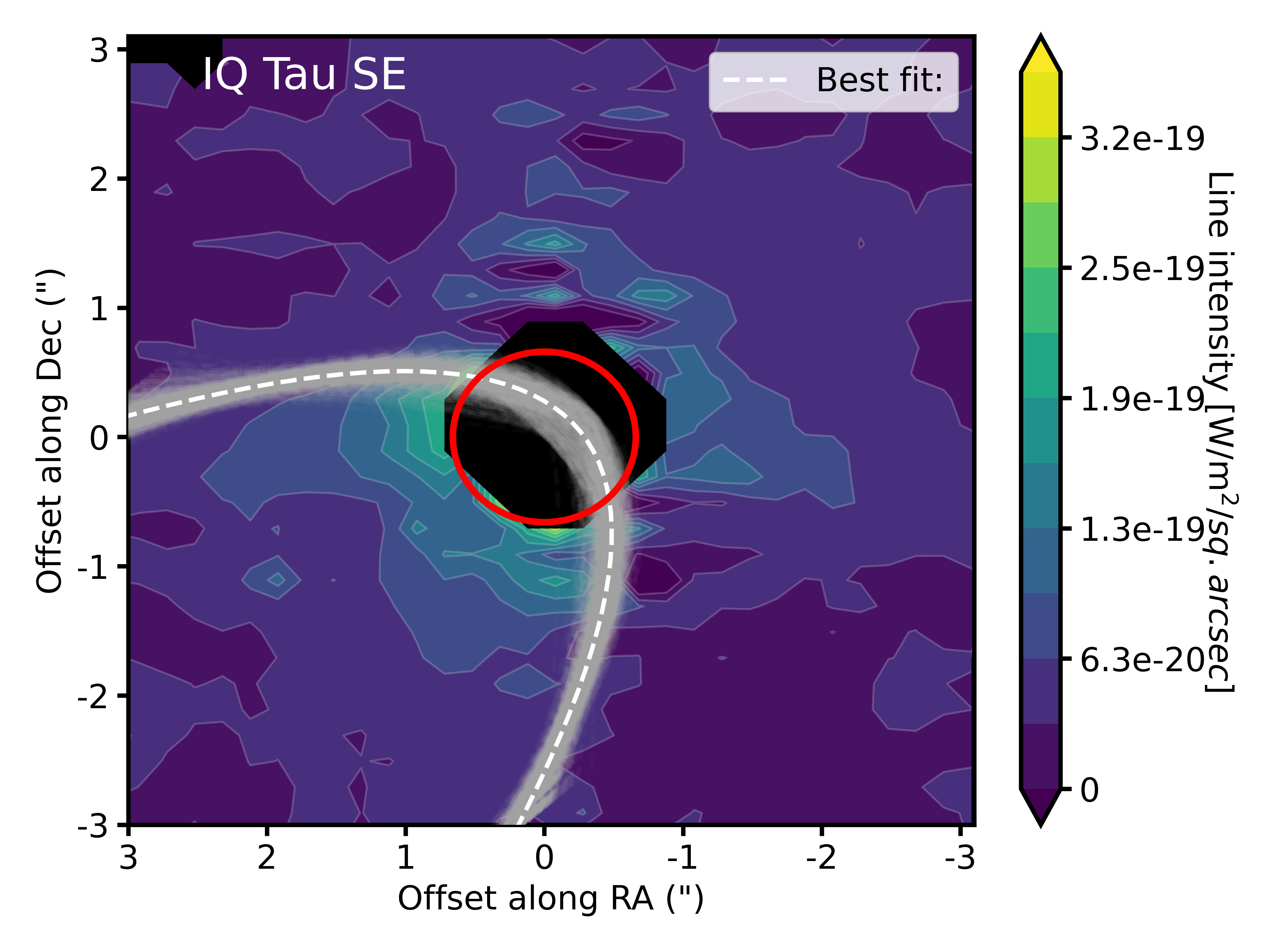}\includegraphics[width=6cm]{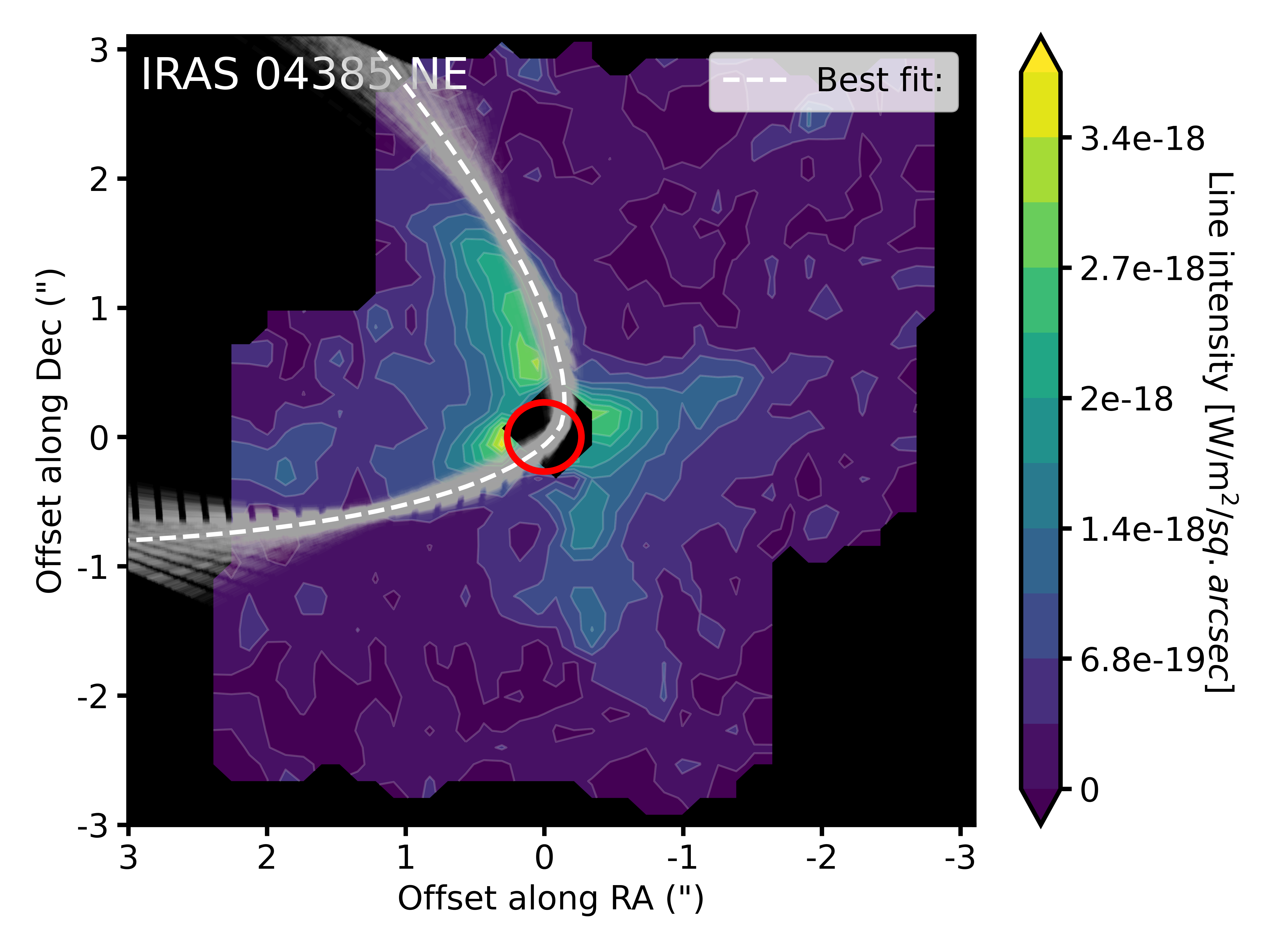}

\includegraphics[width=6cm]{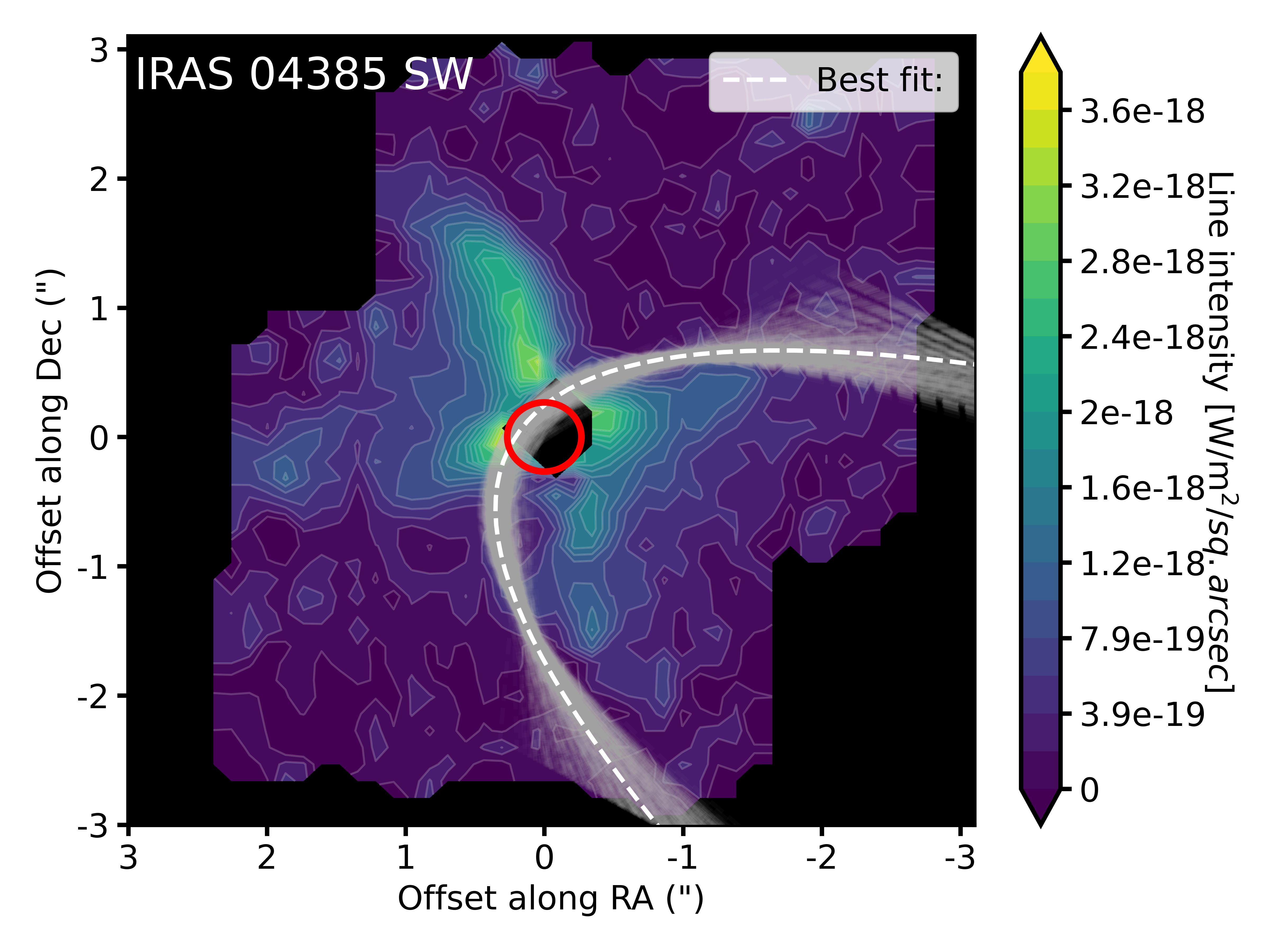}\includegraphics[width=6cm]{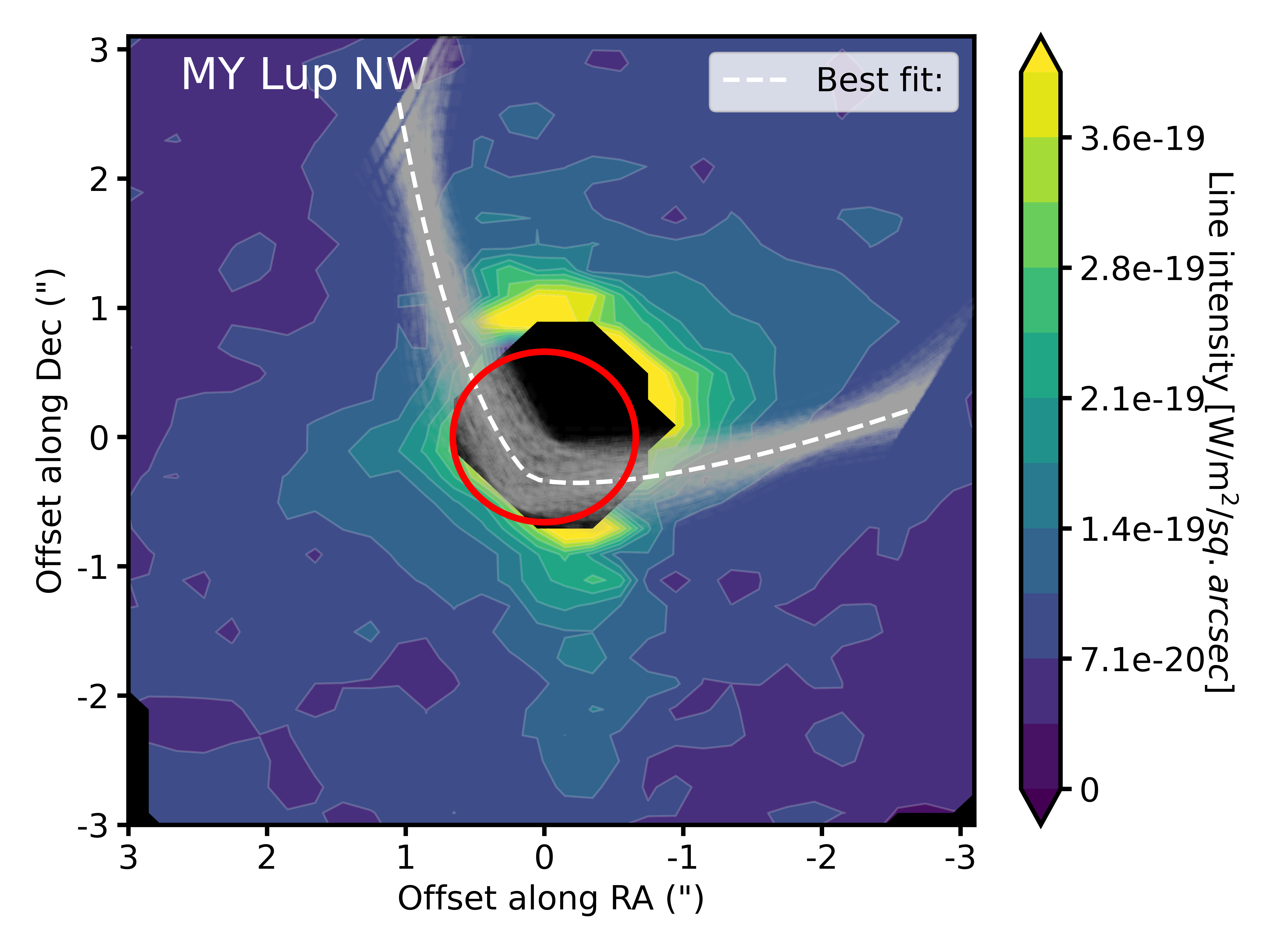}\includegraphics[width=6cm]{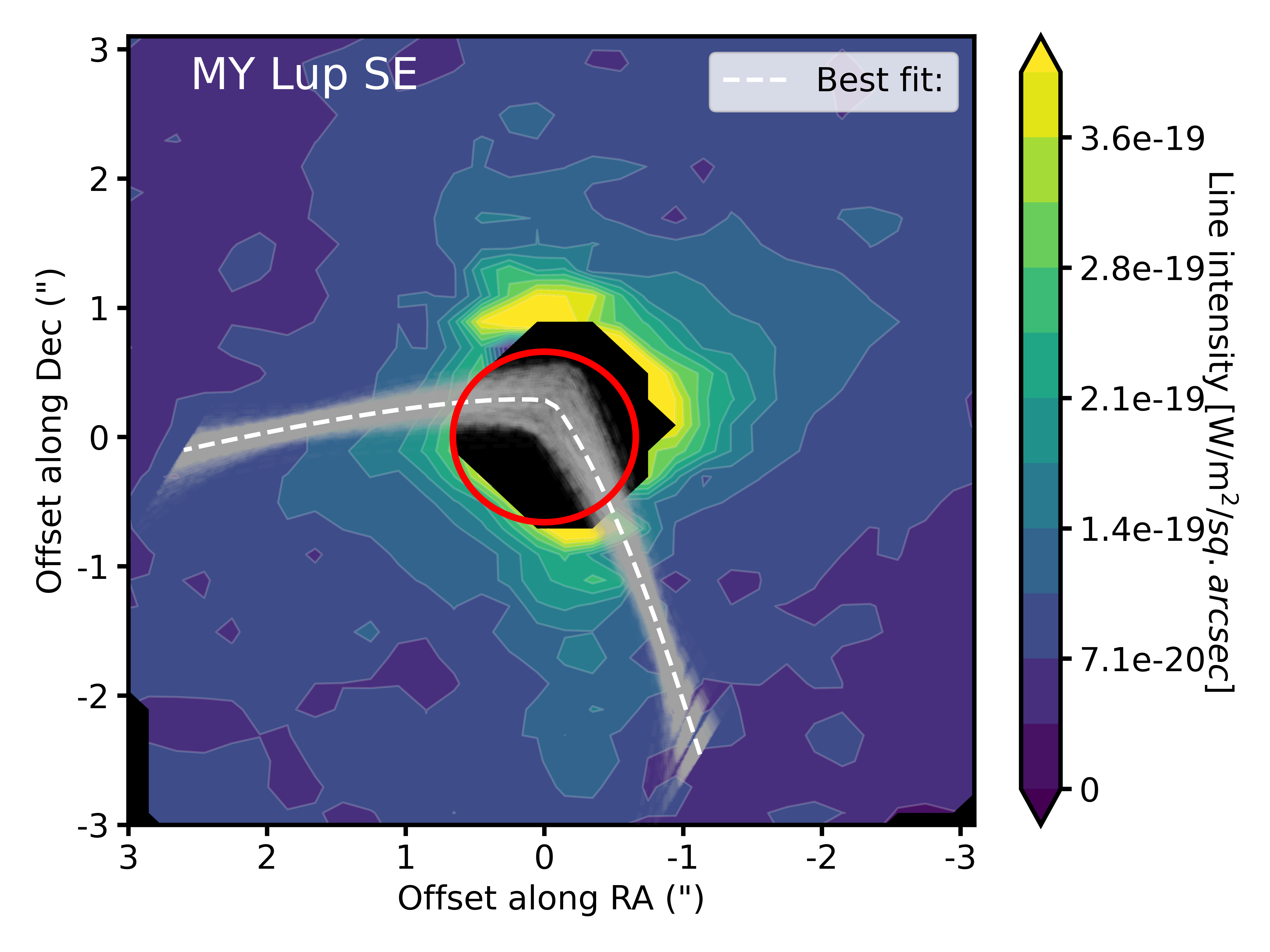}

\includegraphics[width=6cm]{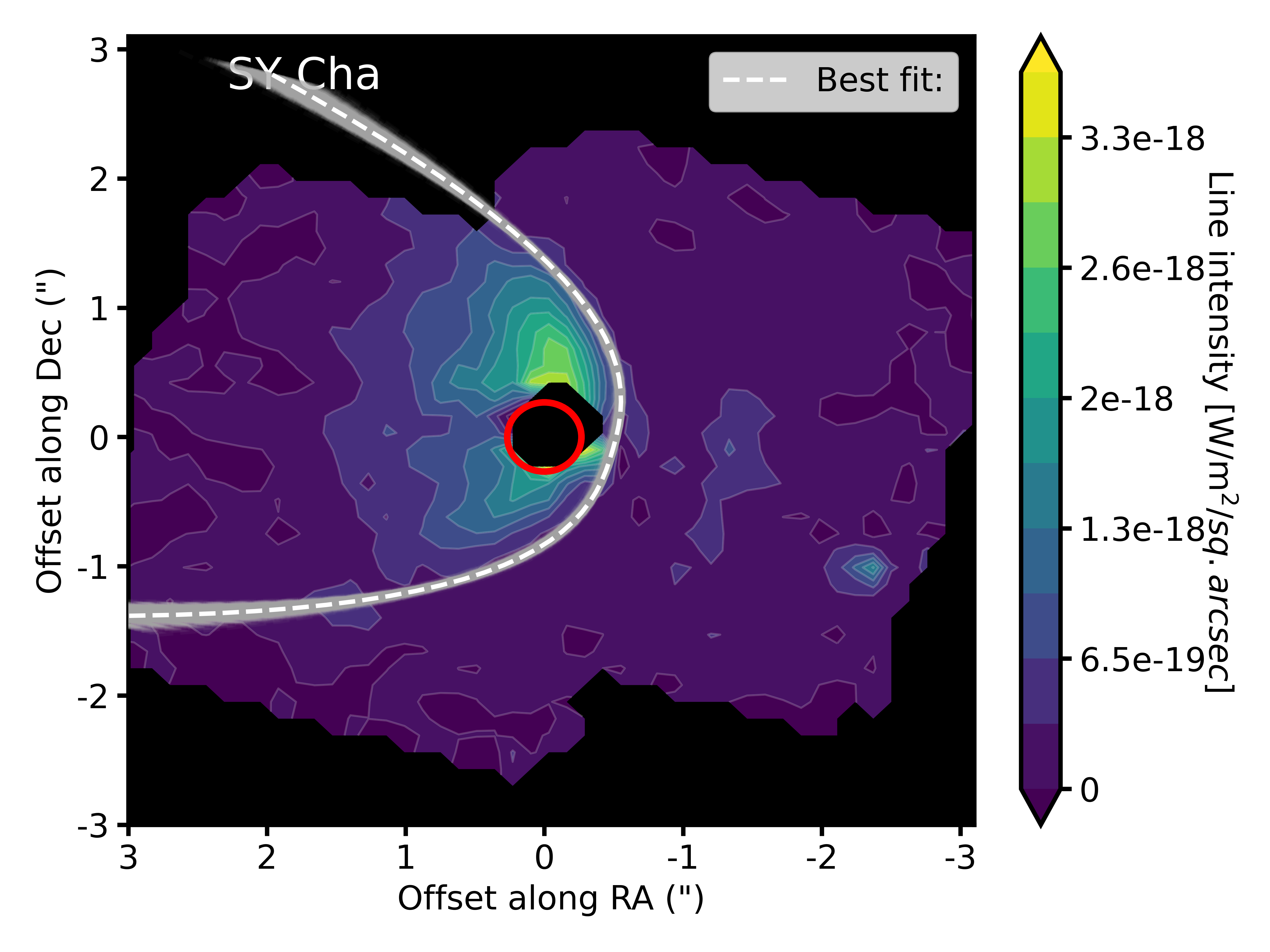}\includegraphics[width=6cm]{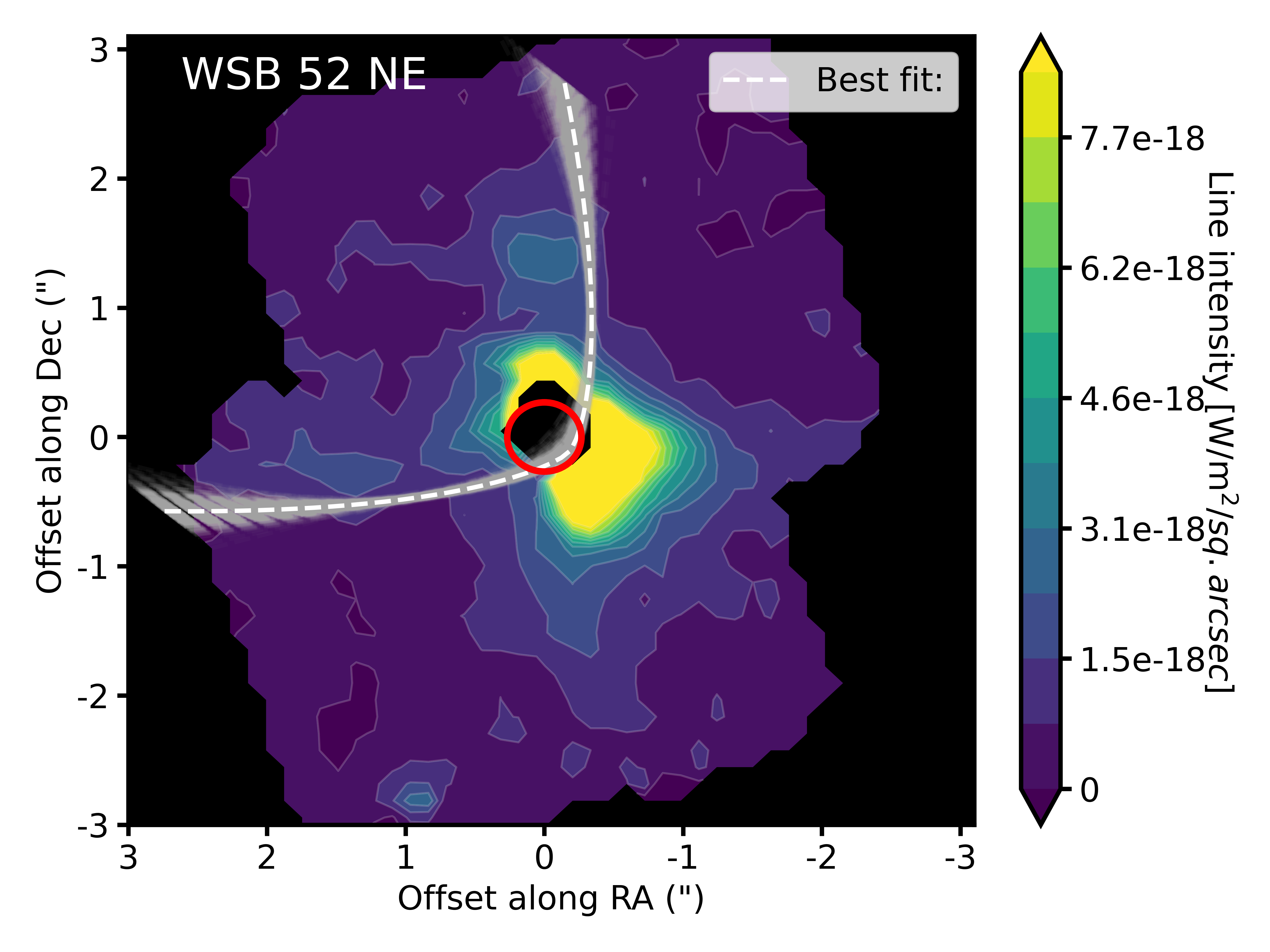}\includegraphics[width=6cm]{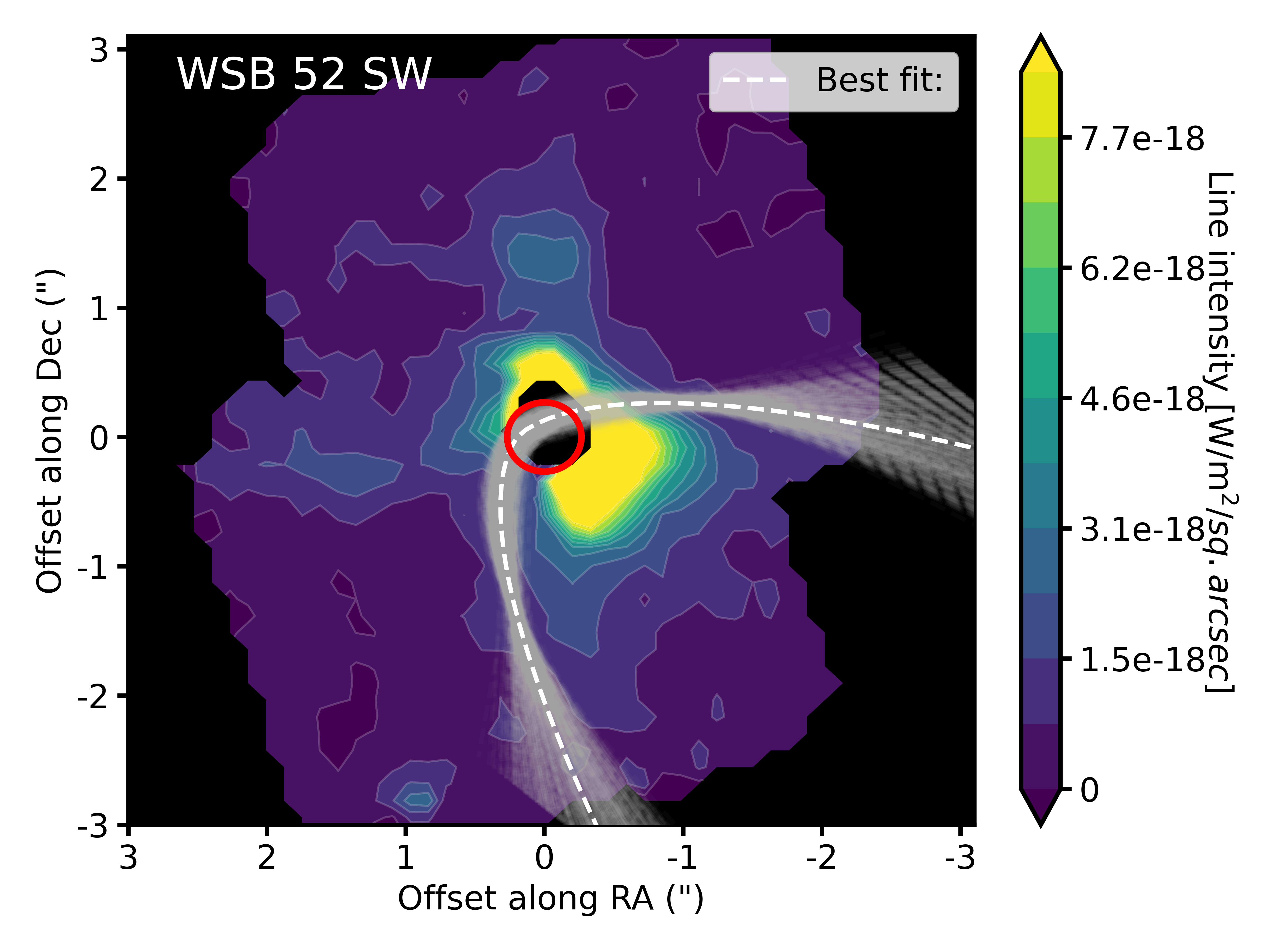}

\caption{The ten sources with detected wide angle flows. {The color scale is the H$_2$ line image.}  The dashed line is the best fit curve and the grey lines show the range of fits from the MCMC. The red circle delineates the inner working angle. }
\label{fig:wind_fit}
\end{figure*}

Of the 34 sources in our sample, 16 exhibit unambiguous morphological signatures of disk winds. These include ten systems with well-defined monopolar or bipolar outflows, typically associated with inclined disks, and six sources showing ring-like or bubble-like structures consistent with face-on geometries. Among the remaining 18 sources, four are known binaries, which may explain their dispersed or disrupted emission. Additionally, four targets (HD 142666, HD 143006, HD 163296, and MWC 480) are Herbig Ae/Be stars, which do not exhibit any H$_2$ emission, even in the central spectra. This is consistent with previous findings that Herbig Ae/Be stars generally lack detectable H$_2$ emission \citep[e.g.,][]{2009ApJ...695.1302M, 2010A&A...516A.110M, 2011A&A...533A..39C}, and these stars are sufficiently distinct from the rest of the sample to warrant exclusion from a homogeneous analysis.

\begin{deluxetable*}{cccccc}
\tablecaption{The best-fit values for the power-law exponent $\alpha$, half-opening angle, and launch radius.  $^\dagger$ Used H$_2$ S(3) transition for fitting. $^{\dagger\dagger}$ Used H$_2$ S(1) transition for fitting. {$^\star$ The derived radius corresponds to the outermost streamline of the wind, although the wind may also be launched from smaller radii. } (1) \cite{S24}; (2) \cite{2022ApJ...931....6L}; (3) \cite{2018MNRAS.477.1004H}; (4) \cite{2021ApJS..257...19H}; (5) updated from \cite{2009ApJ...701..698S}; (6) \cite{2023PASJ...75..424O} }
\label{tab:OA}
\tablehead{
    \colhead{Target} & \colhead{$\alpha$} & \colhead{Half-opening} & \colhead{Launch} & \colhead{ALMA $^{12}$CO gas} & \colhead{Ref} \\
    \colhead{} & \colhead{} & \colhead{Angle (\arcdeg)} & \colhead{radius (au)$^\star$} & \colhead{disk radius (au)} & \colhead{}\\
}
\decimals
\startdata
CI Tau$^\dagger$ & 1.3$^{+0.4}_{-0.3}$ & 48.3$^{+3.5}_{-3.9}$ & 61.2$^{+16.3}_{-24.9}$& 518 $\pm$ 1 & 1 \\
DoAr 25 NE &  1.2$^{+0.2}_{-0.1}$ & 44.4$^{+3.7}_{-3.1}$ & 35.2$^{+15.1}_{-20.2}$  & 233 $\pm$ 6 & 2 \\
DoAr 25 SW & 2.1$^{+0.4}_{-0.6}$ & 46.1$^{+2.7}_{-2.5}$ & 95.6$^{+8.6}_{-11.4}$ & 233 $\pm$ 6 & 2 \\
Elias 20 & 1.0$^{+0.0}_{-0.0}$ & 27.9$^{+1.1}_{-1.5}$ & 3.3$^{+4.2}_{-2.4}$ & --- & ---  \\
Elias 27$^\dagger$ & 1.0$^{+0.0}_{-0.0}$ & 53.6$^{+1.0}_{-1.3}$ & 4.6$^{+6.9}_{-3.4}$  & $>$300 &   3 \\
GM Aur & 2.0$^{+0.4}_{-0.4}$ & 44.6$^{+3.4}_{-3.6}$ & 109.5$^{+2.2}_{-2.5}$  & 550 & 4 \\
IQ Tau$^{\dagger\dagger}$ NW & 1.5$^{+0.3}_{-0.3}$ & 51.7$^{+3.5}_{-3.0}$ & 128.0$^{+15.3}_{-22.3}$ & 203 $\pm$ 4 & 1 \\
IQ Tau$^{\dagger\dagger}$ SE & 2.0$^{+0.4}_{-0.4}$ & 34.3$^{+4.2}_{-3.7}$ & 82.2$^{+18.8}_{-24.3}$ & 203 $\pm$ 4 & 1 \\
IRAS 04385+2550 NE & 1.6$^{+0.4}_{-0.3}$ & 42.2$^{+4.1}_{-3.8}$ & 34.7$^{+16.6}_{-17.3}$ & 343 $\pm$ 11 & 5 \\
IRAS 04385+2550 SW & 2.0$^{+0.5}_{-0.5}$ & 44.7$^{+5.5}_{-4.1}$ & 88.5$^{+17.6}_{-26.6}$ & 343 $\pm$ 11 & 5 \\
MY Lup NW & 1.4$^{+0.5}_{-0.3}$ & 43.9$^{+6.1}_{-7.4}$ & 90.5$^{+38.4}_{-47.2}$ & 192 $\pm$ 7 & 2 \\
MY Lup SE & 1.2$^{+0.2}_{-0.2}$ & 48.7$^{+4.6}_{-4.3}$ & 73.2$^{+25.6}_{-39.8}$ & 192 $\pm$ 7 & 2 \\
SY Cha & 2.5$^{+0.1}_{-0.1}$ & 43.6$^{+0.8}_{-0.7}$ & 187.9$^{+2.7}_{-3.0}$  & $<$ 650 & this work (also see 6) \\
WSB 52 NE &1.4$^{+0.2}_{-0.1}$ & 50.3$^{+1.6}_{-1.5}$ & 60.8$^{+3.7}_{-5.6}$ & $<200$  & this work \\
WSB 52 SW &1.6$^{+0.5}_{-0.4}$ & 40.3$^{+6.6}_{-6.4}$ & 46.2$^{+16.5}_{-17.7}$ &  $<200$  & this work  \\
\enddata
\end{deluxetable*} 
Excluding these four Herbig Ae/Be stars, over 50\% of our sample shows clear extended H$_2$ emission attributable to winds. If we further exclude the four known binaries, the fraction increases to 61\%. Thus, even in cases where well-defined morphological structures are absent, the widespread presence of extended emission supports our working assumption: that all extended H$_2$ emission in our sample originates from disk winds.

The ubiquitous nature of these wide-angle, slow-moving winds may help to explain a puzzling observation of cool molecular hydrogen toward older systems, including debris disks, observed in large apertures by the Infrared Space Observatory \citep{Thi2001}. Any momentum conserving component of these winds will be centered near the systemic velocity but present on large scales, and with even modest amounts of dust in the extended wind the reformation of molecular hydrogen should compete with loss via UV photodissociation.

\subsection{Wind mass loss rate vs mass accretion rate}

{
We use the total column density $N_{\rm tot}$ derived from the rotation diagram to compute the mass within the aperture ($M_{\rm wind}^{\rm apert}$) following:
\begin{equation}
    M_{\rm wind}^{\rm apert} = \mathrm{N}(H_2) \times \Psi \times \mu \times m_p,
\end{equation}
where $\Psi$ is physical area of the aperture (250\,au $\times$ 250\,au), $\mu=2.3$ is the mean molecular gas weight and $m_p$ is the proton mass. By integrating the line emission within a large 250 au square aperture (about $2\arcsec\times 2\arcsec$), we measure a solid-angle averaged column density. While this aperture typically accounts for a large fraction of winds, we account for line emission that falls outside the aperture using a correction factor, $C_{\rm apert}$, defined as the ratio of the total background-subtracted line flux measured over the full MRS field of view, $F_{\rm tot}$, to the line flux measured within the aperture, $F_{\rm apert}$. 

Due to obscuration from the disk, we often detect only one side of the wind {(blue-shifted side)}, except in a few cases where a bipolar structure is clearly seen (i.e., DoAr 25 in Figure~\ref{fig:doar25}; IQ Tau in Figure \ref{fig:iqtau}; IRAS 04385+2550 in Figure~\ref{fig:iras04385}; MY Lup in Figure~\ref{fig:mylup}, and WSB 52 in Figure~\ref{fig:wsb52}). To account for obscured emission from the other side of the disk, we introduce an additional correction factor, $C_{\rm obsc}$.

Thus, the total correction factor based on the H$_2$ S(3) line is ($C = C_{\rm apert} \times C_{\rm obsc}$). {We used the S(3) line as it has high S/N most often, however in a few sources the S(1) or S(2) lines are used when there is not a clear detection of the S(3) line. The median $C_{\rm apert}$ across the sample is {3.5}. The values of $C_{\rm apert}$ along with the errors are listed in Table~\ref{tab:wind}.}

We calculate the total molecular wind mass as $M_{\rm wind}^{\rm tot} = C_{\rm obsc} \times C_{\rm apert} \times M_{\rm wind}^{\rm apert}$ in Table \ref{tab:wind}, where $C_{\rm obsc}=1$ for the bipolar sources (i.e., DoAr 25,  IRAS 04385+2550,  IQ Tau, MY Lup, and WSB 52) and 2 for the rest of the sample.

The total wind mass-loss rate is:
\begin{equation}
    \dot{M}_{\rm wind}^{\rm tot} = \frac{M_{\rm wind}^{\rm tot}}{t_{ \rm dyn}},
\end{equation}
where $t_{\rm dyn}$ is the dynamical timescale of the flow. 

Approximating the flow as being approximately isotropic (spherically symmetric), the dynamical timescale is related to the characteristic length, $L$, of the emitting area of the aperture, $A$, as:
\begin{equation}
    t_{\rm dyn}^{\rm apert} = \frac{L_{\rm apert}}{v_{\rm wind}} \sim \frac{\sqrt{A}}{v_{\rm wind}},
    \label{tdy}
\end{equation}

Since $C_{\rm apert}$ traces the ratio of emitting areas between the total wind emission and that inside the aperture (assuming uniform emission), the dynamical timescale for the total wind scales as 

\begin{equation}
    t_{\rm dyn} = \frac{L_{\rm tot}}{V_{\rm wind}} \sim \frac{\sqrt{C_{\rm apert}\times A}}{V_{\rm wind}} = \sqrt{C_{\rm apert}} \,  t_{\rm dyn}^{\rm apert},
\end{equation}
where $V_{\rm wind}$ is the median wind {velocity  of  4.2$^{+6.7}_{-3.0}$~km~s$^{-1}$. Based on the median velocity, the $t_{\rm dyn}^{\rm apert}$ is 284$_{-175}^{+686}$. We propagate the error in $t_{\rm dyn}$ based on the errors in $t^{\rm apert}_{\rm dyn}$ and C$_{\rm apert}$}

{The median wind mass loss rate from the full sample is ${\rm log_{10}}(\dot{M}{\rm _{\rm wind}^{\rm tot}}) = -9_{-0.4}^{+0.8},{\rm M_\odot\,yr^{-1}}$}. Assuming that the observed molecular winds are the dominant mechanism responsible for disk dispersal, we can estimate a characteristic dissipation timescale, or disk lifetime. Assuming a constant wind mass loss rate, a disk with a typical mass {of $\sim 2-3\,M_{\rm Jup}$ \citep{2025ApJ...989....1Z} is depleted in $\sim 2-3$ Myr,} roughly consistent with current estimates for disk lifetimes. 

Figure~\ref{fig:massloss}(a) compares $\dot{M}{_{\rm wind}^{\rm tot}}$ with the mass accretion rate onto the central star ($\dot{M}{_\mathrm{acc}}$) from the literature \citep[see Table~\ref{tab:Table1} and][]{2023ASPC..534..539M}. We separate the sources based on their morphology: Ten sources with wide-angle wind morphology (see Section~4.1 and Figure~\ref{fig:wind_fit}) and six sources with ring- or bubble-like morphologies (see Section~4.1). The remaining 14 sources do not fit into either of these categories. {We find no strong correlation between $\dot{M}{_\mathrm{\rm wind}^{\rm tot}}$ and $\dot{M}{_\mathrm{acc}}$. The wind mass-loss rates remain within $\sim$2 dex, despite spanning $\sim${4 dex} in accretion rate, with the average ratio being consistent with $\dot{M}{_\mathrm{\rm wind}} \sim 0.1\, \times \dot{M}{_\mathrm{acc}}$}. Further, there is no apparent relation to wind morphology. 

The dominant source of uncertainty in $\dot{M}_{\rm wind}^{\rm tot}$ is the dynamical timescale. Our mass-loss estimates do not include contributions from any H$_2$ component too cold to be traced by MIRI  or a hot component potentially traced by higher S(J) transitions. A colder component would require observation of the H$_2$ S(0) line at 28.29~\micron{}, which is not currently possible. However, a significant cold component would imply implausible disk lifetimes shorter than 1 Myr. Based on non-detections in our sample, the contribution of a hotter component ($>$1000~K) must be at least one to two orders of magnitude lower than that of the observed warm component ($\sim$650~K), consistent with previous studies \citep[e.g.,][]{2024ApJ...966L..22N,2025A&A...694A.174F,2025ApJ...980..148S,2025A&A...703A.139S}. Strictly, the total molecular mass and the total molecular mass-loss rates derived from H$_2$ lines represent a lower limit of the total mass loss owing to the unknown molecular fraction in the wind. For photoevaporative winds, \cite{2025arXiv251100515N} have shown that the molecular fraction of the wind is highly dependent on from where the winds are launched. Nevertheless, the large wind mass and mass-loss rates suggest that the molecular fraction is likely near unity. 


{The wind mass-loss rate inferred from H$_2$ emission represents a time-averaged value over the dynamical timescale of the outflow (typically a few hundred years), whereas mass accretion rates are derived from instantaneous tracers such as UV excess or H-I line emission. Some of the dispersion in the observed accretion--mass-loss rate relation may therefore arise from the intrinsically stochastic and variable nature of mass accretion in YSOs, which can fluctuate on timescales ranging from days to years \citep{2014MNRAS.440.3444C,2022ApJ...924L..23Z,2023ApJ...956..102H,2025ApJ...985..224T}. In Figure \ref{fig:massloss}(b), we compare $\dot{M}{_{\rm wind}^{\rm tot}}$ with the mass accretion rate onto the central star ($\dot{M}{_\mathrm{acc}}$) derived from JWST MIRI observations of H-I lines. For most of our sources, we used the accretion rates determined by \cite{2026A&A...708A..22S}. However, for three sources (AS 209, Elias 24, and GW Lup) absent from the \cite{2026A&A...708A..22S} sample, we used the accretion luminosities from \cite{2025ApJ...985..224T} and computed the mass accretion rates following the methodology of \cite{2026A&A...708A..22S}. Similar to Figure \ref{fig:massloss}(a), we do not find a strong correlation between $\dot{M}{_\mathrm{\rm wind}^{\rm tot}}$ and $\dot{M}{_\mathrm{acc}}$, though the average ratio remains consistent with $\dot{M}{_\mathrm{\rm wind}} \sim 0.1\, \times \dot{M}{_\mathrm{acc}}$. Thus, part of the observed dispersion may reflect differences in the wind launching mechanisms and the radii from which these winds are launched (see Section~4.5 for details).}

The molecular wind mass-loss rates traced by H$_2$ emission has not been previously quantified in a systematic way. In contrast, ionic and atomic mass-loss rates from jets have been extensively derived using optical forbidden lines, in particular the high-velocity component (HVC) of the [O\,I]~6300\,\AA\ line \citep{1995ApJ...452..736H}. These studies report a strong correlation between $ \dot{M}_{\rm jet}$ and $\dot{M}_{\rm acc}$, consistent with the frequently quoted empirical scaling  $ \dot{M}_{\rm jet} \sim 0.1\,\dot{M}_{\rm acc}$ (\citealt{2018A&A...609A..87N}, also see \citealt{2026ApJ...999..264W}). However, this relation is established for a different physical process of highly collimated jets launched from within $\lesssim 1$~au of the central star, and not from an extended disk wind more likely to drive disk dispersal.

\begin{figure}[h]
\centering
\includegraphics[width=8.5cm]{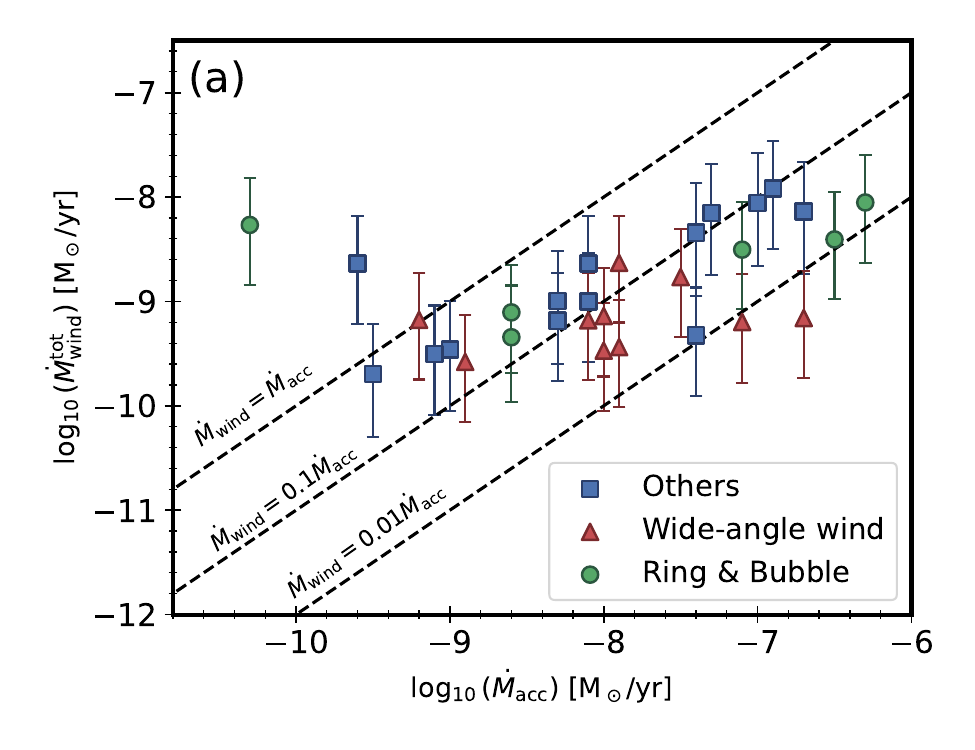}
\includegraphics[width=8.5cm]{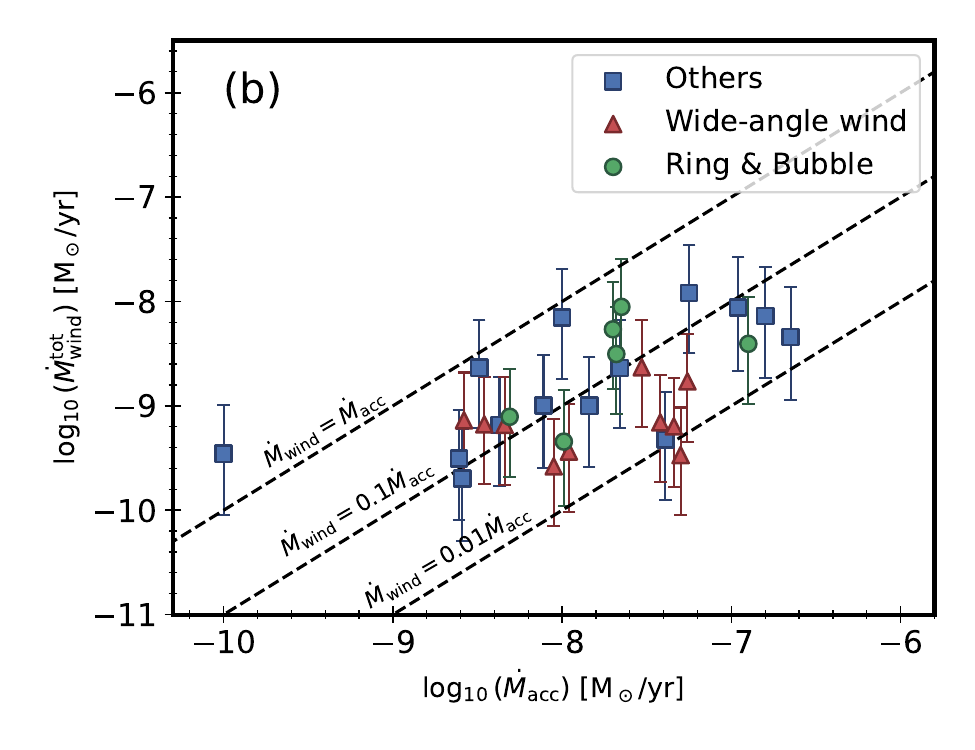}
\vskip -0.1in
\caption{The measured total wind mass-loss rates ($\dot{M}{_{\rm wind}^{\rm tot}}$) compared against stellar mass accretion rates obtained from {(a) previous literature and (b) HI lines simultaneously observed by MIRI MRS.} Sources are color-coded according to their morphology: the ten sources that show clear wide-angle wind structures are shown as red triangles, the six sources that exhibit ring- or bubble-like morphologies are shown as green circles, and the remaining sources are shown as blue squares. }
\label{fig:massloss}
\end{figure}

\subsection{Comparison between H$_2$ winds from protoplanetary disks and protostellar}

Extended H$_2$ emission has long been known to be a classic characteristic of protostellar outflows, primarily originating from outflows driven by accretion \citep{Pudritz86,Noriega-Crespo04,Neufeld09}. More recently, they have been studied in unprecedented detail using JWST \citep[e.g.,][]{2023arXiv231014061N, 2024A&A...687A..36T, 2024ApJ...966...41F,2024arXiv240319400D,2024arXiv240916061C,2025arXiv250200394N}. These outflows exhibit distinct morphological and kinematic properties, setting them apart from the H$_2$ emission observed in protoplanetary disks. In particular, in protostars, clear signatures of bipolar outflows with an hourglass-like morphology, often displaying a nested structure \citep[e.g.,][]{2024A&A...687A..36T,2024arXiv240916061C} are observed. Kinematically, the protostellar outflows show both blueshifted as well as redshifted emission with typical velocities of $\sim 10 - 30 $~km~s$^{-1}$  (\citealt{2024arXiv240916061C}; \citealt{2026arXiv260209837N}; H. Tyagi et al. in prep).  

In terms of excitation conditions, simple rotational analysis of the H$_2$ emission from protostars suggests that a two-component fit, comprising a warm and a hot component, best describes the rotational diagram \citep[e.g.,][]{2024ApJ...966L..22N,2025ApJ...995..199N}. The warm component typically has temperatures ranging from $\sim$400 to 1000 K, similar to the temperature derived for protoplanetary disks. While the warm H$_2$ emission from protostars shares a comparable temperature range with protoplanetary disks, the column densities in protostars is $\gtrsim$1-2 orders of magnitude higher than those derived for protoplanetary disks.        

In contrast, the hot component exhibits significantly higher temperatures, around $\sim$1000 to 3500 K \citep[e.g.,][H. Tyagi et al., in prep]{2024ApJ...966L..22N,2025ApJ...995..199N}{}. These temperature ranges appear to be consistent across the entire mass spectrum of protostars, from very-low-mass to high-mass objects \citep[e.g.,][]{2024arXiv240916061C,2023A&A...679A.108G}.

\subsection{Comparison between [Ne II] and H$_2$ emission} \label{sec:Ne}

{Among the 34 disks analyzed in this study, we detect clear extended jet-like [Ne II] emission in 8 sources: CI Tau, Elias 24, Elias 27,  HP Tau, SR 4, SY Cha,  VZ Cha, and WSB 52.} In most of these systems, the [Ne II] emission is spatially aligned with the wide-angle flow traced by H$_2$ emission, suggesting a common origin. It is important to note, however, that this list may not be exhaustive. Some sources in our sample could host weak or marginally extended [Ne II] emission {(see Appendix \ref{Notes})} that remains undetected due to limitations in sensitivity, spatial resolution, or image artifacts. In particular, the process of PSF subtraction can introduce significant residuals.  

The [Ne II] at 12.81 \micron{} emission traces ionized gas and given the collimated nature of the extended emission, the [Ne II] is most-likely associated with an atomic jets launched near the stellar surface \citep[e.g.,][]{2023arXiv231014061N,2024arXiv240916061C,2024A&A...687A..36T, 2026arXiv260109587F}. We do find not a strong correlation between the presence of extended [Ne II] emission and either mass accretion rates or wind mass-loss rates from H$_2$. This lack of correlation may arise because [Ne II] emission, H$_2$ emission, and accretion rates each trace different timescales: accretion rates reflect near-instantaneous conditions, H$_2$ flows represent dynamical timescales of several hundred years, and [Ne II] emission likely traces intermediate timescales of a few tens of years. 

\subsection{Wind launching mechanism}

{Our analysis reveals that, in many cases, the H$_2$ emission exhibits wide-angle structures. We clearly detect wind signatures in 16 sources, while several others show tentative evidence for similarly wide-angled winds. These winds are slow, with characteristic velocities of 4.2$^{+6.7}_{-3.0}$ km s$^{-1}$, and have large opening angles of $\sim 45\arcdeg$. They 
have a median excitation temperature of $624 \pm 130$ K and a median total column density of $\log_{10}(N_{{\rm tot}}~[\mathrm{cm}^{-2}]) \sim 18.6 \pm 0.6$. The inferred mass-loss rate is also significant, with ${\log_{10}}(\dot{M}_{\rm wind\,tot}) = -9_{-0.4}^{+0.8}\,\mathrm{M_\odot\,yr^{-1}}$. 

{In MHD disk wind theory, the ratio $\dot{M}_{\rm wind}/\dot{M}_{\rm acc}$ is not universal, but instead depends on the magnetic lever arm parameter $\lambda$ and on the radial extent of the launching region (the ratio of outer radius $r_{\rm e}$ to the inner radius $r_{\rm i}$; see Eq.~17 of \citealt{2006A&A...453..785F}). For jet-like solutions with $\lambda \sim 10$, characteristic of inner-disk streamlines, the expected ratio of $\dot{M}_{\rm wind}/\dot{M}_{\rm acc}$ is naturally of order 0.1, thereby reproducing the empirical jet scaling. In contrast, extended disk winds launched from larger disk radii can exhibit substantially lower lever arms ($\lambda \sim 2$--3; \citealt{2019ApJ...874...90W,2021A&A...650A..35L}). Such low-$\lambda$ solutions are more heavily mass-loaded and can yield wind mass-loss rates comparable to the stellar accretion rate \citep{2019ApJ...874...90W,2021A&A...650A..35L,2022A&A...668A..78D}. Thus,  part of the scatter observed in Fig. \ref{fig:massloss} could be indicative of molecular disk winds that originate at larger disk radii than the atomic jet component. This is consistent with the wide-angle morphology we observe for a large fraction of our sample. }
}

The relatively high accretion rates ($\geq 10^{-8}\,\mathrm{M_\odot\,yr^{-1}}$) measured for some of our targets may challenge a photoevaporative origin of the observed winds, since the strong UV radiation produced by accretion could efficiently dissociate H$_2$ \citep[e.g.,][]{1985ApJ...291..722T,2007ApJ...661..334N}. {However,} recent theoretical work has shown that warm H$_2$ can survive in photoevaporative winds \citep{2018ApJ...865...75N,2024A&A...690A.296S}. \citet{2025arXiv251100515N} show that photoevaporative wind models may reproduce several key observational characteristics of the H$_2$ flows, including the low-J line fluxes. However, these models tend to underpredict the high-$J$ H$_2$ emission associated with the more spatially extended H$_2$ structures. {Therefore, a photoevaporative disk wind interpretation remains plausible. However, robust discrimination requires quantitative comparisons with both photoevaporative and MHD wind models. Such comparisons are currently limited by the lack of detailed H$_2$ modeling in MHD disk winds.} 

\section{Summary} \label{sec:summary}

We present the results of our analysis of H$_2$ emission from 34 protoplanetary disks, utilizing data from the JDISCS program (PID: 1549, 1584, 1640) and the GTO program (PID: 1282). Our study examines the morphology, kinematics, excitation conditions, and mass dynamics of the H$_2$ emission. Our key findings are:

\begin{enumerate}

    \item Extended H$_2$ emission structures are widespread in protoplanetary disks, with morphologies consistent with slow { (4.2$^{+6.7}_{-3.0}$ km s$^{-1}$)} MHD driven winds. Of the 34 sources in our sample, 16 show clear morphological evidence for disk winds. This includes ten systems with monopolar or bipolar winds seen in inclined disks and six sources exhibiting ring-like or bubble-like morphologies typical of face-on geometries.  Accounting for binary stars and removing the Herbig Ae/Be stars from our sample we find that more than 60\% of the sources in our sample show H$_2$ morphology consistent with disk winds.  
    
    \item The low-J transitions [S(1)-S(2)] in some sources show emission consistent with background or foreground PDRs. In two sources (e.g., DoAr 25, Figure~\ref{fig:doar25}; Elias 27, Figure \ref{fig:elias27}) we observe a disk shadow in silhouette against this bright emission.   

    \item We model the spatial extent of the wind in ten disks using a power-law description of the wind edges. The median power-law index is{ $\alpha \approx$ {1.6}, with a median half-opening angle of $45\arcdeg^{+5}_{-4}$, }suggesting relatively wide-angle flows.

    \item  Rotational analysis yields a median excitation temperature of {$\sim$ 624 $\pm$ 130 K and a median total column density of  $\log_{10}(N_{\mathrm{\rm tot}}~[\mathrm{cm}^{-2}]) \sim 18.6 \pm 0.6$}, assuming local thermodynamic equilibrium and an ortho-to-para ratio of 3.
    
    \item  {The median mass-loss rate, {${\rm log_{10}}(\dot{M}_{\rm wind}^{\rm tot}) = -9_{-0.4}^{+0.8}\,{\rm M_\odot\,yr^{-1}}$,} implies that, if molecular winds are the dominant mechanism responsible for disk dispersal, a typical disk with a mass of {$2-3 \, M_{\rm Jup}$} would dissipate on a {$\sim$2-3 Myr} timescale, consistent with observed disk lifetimes.}
    
    \item We do not find a strong correlation between   $\dot{M}_{\mathrm{wind}}^{\rm tot}$ with the mass accretion rates  of $\dot{M}_{\mathrm{acc}}$,  instead finding a small ($\sim$2 dex) scatter in $\dot{M}_{\mathrm{wind}}^{\rm tot}$ for a large range {($\sim$4 dex)} of accretion rates.

    \item  {Compared to protostellar (Class 0/I) outflows, traced in both pure rotational and ro-vibrational H$_2$, more evolved, protoplanetary disk winds, primarily detected in pure rotational H$_2$, exhibit lower velocities (of order a few km s$^{-1}$), wider opening angles, and lower column densities. In contrast, protostellar outflows are characterized by highly collimated bipolar morphologies with velocities reaching up to $\sim$30 km s$^{-1}$, along with additional extended, hotter H$_2$ components that are not detected in most Class II disks. } 

    \item  {We interpret the derived wind properties as evidence that MHD winds are a primary driver of disk dispersal during the first few Myr of evolution. However, a photoevaporative origin cannot be ruled out. Definitive discrimination requires quantitative comparisons with both photoevaporative and MHD wind models, presently hindered by the absence of detailed H$_2$ emission modeling in MHD disk winds.}
    
    \item Our findings establish that spatially extended warm H$_2$ emission is a widespread and robust tracer of molecular disk winds in protoplanetary systems. This emission offers direct observational evidence of slow, wide-angle winds originating from the disk surface, which play a critical role in redistributing angular momentum and regulating mass loss from the disk.

\end{enumerate}

\facilities{JWST(MIRI), ALMA}

{Dataset:} All of the JWST data presented in this article were obtained from the Mikulski Archive for Space Telescopes (MAST) at the Space Telescope Science Institute. The specific observations analyzed can be accessed via \dataset[doi:10.17909/r2y3-az93]{https://doi.org/10.17909/r2y3-az93}.

\section{Acknowledgment}
This research was carried out at the Jet Propulsion
Laboratory, California Institute of Technology, under a contract with the National Aeronautics and Space Administration (80NM0018D0004). This work is based on observations made with the NASA/ESA/CSA James Webb Space Telescope. The data were obtained from the Mikulski Archive for Space Telescopes at the Space Telescope Science Institute, which is operated by the Association of Universities for Research in Astronomy, Inc., under NASA contract NAS 5-03127 for JWST. These observations are associated with programs 1282, 1549, 1584, 1640, 2025. This paper makes use of the following ALMA data. ALMA is a partnership of ESO (representing its member states), NSF (USA), and NINS (Japan), together with NRC (Canada), MOST and ASIAA (Taiwan), and KASI (Republic of Korea), in cooperation with the Republic of Chile. The Joint ALMA Observatory is operated by ESO, AUI/NRAO, and NAOJ. \\

© 2025. All rights reserved

\bibliographystyle{aasjournal}
\bibliography{JDISCS}{}
\appendix

\section{Individual sources \label{Notes}}

\subsection{Notes on individual sources}
This appendix provided details about the H$_2$ and [Ne II] morphology for individual sources.

\begin{enumerate}

\item {AS 205} shows prominent and spatially extended H$_2$ emission, particularly in the strong ortho H$_2$ lines, consistent with a wide-angle wind. The winds appears to originate primarily from the northern component, AS 205N. Given the close separation of the two components, the wind may also interact with the secondary star, potentially influencing the observed spatial structure of the emission. In the S(2) image, a potential artifact is present to the northeast of the primary source.

\item {AS 209} is a moderately inclined disk. The observed H$_2$ morphology is consistent with this viewing geometry, with emission symmetric around the central source. Residual features associated with PSF subtraction are present in the immediate vicinity of the central source, but the extended emission beyond this region remains clearly detectable mainly in S(1), S(2), and S(3) lines. 

\item {CI Tau} exhibits a clear wide-angle wind. Only the blue-shifted side of the wind is detected, likely because the red-shifted emission is obscured by the disk. The wind is detected in the S(1), S(2), S(3), and S(5) lines. The S(4) line is faint and is dominated by PSF artifacts close to the center. The monopolar wind shows has a large opening angle. Complementing the H$_2$ emission, the [Ne II] emission reveals a collimated, jet-like structure that is aligned with the wide-angle molecular wind, suggesting a layered outflow system with a fast, ionized core surrounded by a slower, molecular component.

\item {DoAr 25} stands out in the sample because its dust disk is observed in absorption against bright background H$_2$ emission, particularly in the S(1), S(2), and S(3) lines. The uniform background line emission originates from the photon-dominated region (PDR) near the Ophiuchus core. Superimposed on the background emission, the H$_2$ data show clear signatures of a bipolar wind in the S(3), S(4), and S(5) lines (tentatively in the S(1) and S(2) lines). The [Ne II] emission provides tentative evidence for a jet component, indicating the presence of collimated jet alongside the wide-angle wind. 

\item {DoAr 33} shows strong, uniform background H$_2$ emission in the S(1) and S(2) lines. We detect extended H$_2$ emission associated with the source in the S(1), S(2), and S(3) lines. The S(4) line shows little to no evidence of wind-like emission, while the S(5) line is affected by residal instrumental artifacts and PSF subtraction effects, making detailed interpretation of this transition less certain.

\item {DR Tau} is a nearly face-on disk, which exhibits faint extended H$_2$ emission, and a symmetric morphology around the central source. This is consistent with the face-on orientation of the system. However, the faint nature of the extended component and limited S/N prevents further characterization of the spatial structure of the emission.

\item {Elias 20} is superimposed on bright background H$_2$ line emission associated with the the Ophiuchus core PDR. The background H$_2$ emission exhibits a gradient that increases toward the northeast, reflecting the structured nature of the PDR environment. The source displays a monopolar outflow, particularly prominent in the higher-$J$ transitions (S(3)--S(5)), where the wind emission appears more distinct from the background PDR. In addition, the [Ne II] emission appears to be slightly extended beyond the disk. 

\item {Elias 24} shows a ring-like morphology with a bright knot to the northeast in all H$_2$ lines. The ring is asymmetric, extending more along the northeast–southwest direction and less along the east–west direction. The bright northeastern knot stands out as a localized enhancement in H$_2$ emission, suggesting a region of higher excitation, density, or possible interaction with a localized outflow or wind feature. The source displays a [Ne II] jet oriented perpendicular to the plane of the dust disk, indicating the presence of an ionized outflow component alongside the extended molecular structure. The ring morphology may indicate episodic outflow activity.

\item {Elias 27} is observed in silhouette against strong background H$_2$ line emission from the Ophiuchus PDR in the S(1) and S(2) lines due to the dust in the disk. In the S(3) and S(5) lines, a monopolar outflow extending from the disk along the northeast side is observed. Extended [Ne II] emission is observed, indicating the presence of a collimated, ionized jet in addition to a molecular wind. The [Ne II] emission exhibits a prominent knot along the southwest side, suggesting localized regions of enhanced excitation or density in the ionized jet.

\item {FZ Tau} is a moderately inclined system. It represents one of the clearest examples of ring-like H$_2$ emission in our sample, with an average radius of $\sim$2.4\arcsec{} (310 au) and a width of about 1.15\arcsec{} (148 au) \citep{Pontoppidan24}. A localized region of enhanced emission is visible toward the southwest portion of the ring, suggesting possible asymmetry in the gas distribution or localized excitation. MIRI's field of view in the S(4) and S(5) lines is smaller than the full extent of the ring, and a significant fraction of the emission falls outside the observed region in these lines. As a result, the morphology of the ring is best traced in the lower-energy transitions. The presence of a ring-like structure in the H$_2$ emission may be indicative of episodic outflow activity or past ejection events that have created a shell of shocked molecular gas expanding away from the central source. The [Ne II] appears to be slightly extended in the source. 

\item {GK Tau} shows extended H$_2$ emission in the S(1), S(2), S(3), and S(5) lines. The emission is diffuse, without a clear preferred orientation. There are no obvious morphological signatures indicative of a collimated wind. The lack of structure suggests that any molecular wind, if present, is intrinsically weak with a wide opening angle.

\item {GM Aur} exhibits a monopolar outflow oriented perpendicular to the dust disk plane detected in all observed H$_2$ lines. The emission traces a blue-shifted wide-angle structure extending away from the central source, consistent with a molecular wind emerging from the disk surface, while the red-shifted emission is obscured by the dust disk. GM Aur one of the clearest examples of a well-structured molecular outflow in this sample.

\item {GO Tau} shows weak extended emission in the S(3), S(4), and S(5) lines. In contrast, the S(1) line reveals more discernible structure, suggestive of a wide-angle outflow and a disk silhouette with major axis oriented north-east/south-west.

\item {GQ Lup} shows tentative extended emission with a wide-angle morphology in the S(1), S(2), and S(3) lines. Some PSF subtraction artifacts are present, particularly in the S(5) line, where a bright artifact is visible to the north of the source. 

\item {GW Lup} shows bright H$_2$ emission surrounding the central source in the S(1) and S(2) lines. The extended emission is concentrated around the star. Some of the bright features seen in the S(2) line near the central source may be caused by PSF subtraction artifacts.

\item {HD 142666, HD 143006, HD 163296, and MWC 480} are Herbig Ae stars and show no extended H$_2$ emission in any of the observed lines. The apparent bright emission surrounding the star is consistent with PSF subtraction artifacts.

\item {HP Tau} exhibits a striking ring-like H$_2$ morphology with a tail extending toward the west. The tail is prominent and coherent, suggesting material flowing or being entrained along this direction. In addition, the [Ne II] emission is also spatially extended along the western tail, indicating that both warm molecular and ionized gas trace the same outflow or wind structure. This morphology may reflect a combination of disk winds and jet activity shaping the surrounding molecular environment.

\item {HT Lup} is a triple system with bright continuum emission. The primary component is associated with extended H$_2$ emission, although not with an classical wind-like morphology. A tentative wide-angle wind component becomes more discernible in the S(3), S(4), and S(5) lines. This suggests that the primary hosts a molecular wind that is partially disrupted by PSF artifacts from the companion sources.

\item {IQ Tau} is an inclined source that exhibits a clear bipolar H$_2$ outflow oriented perpendicular to the dust disk plane. The northwestern lobe of the outflow has lower S/N in lines higher than S(1) and S(2). The [Ne II] emission appears slightly extended along the outflow direction, suggesting that both warm molecular and ionized gas trace the outflow structure. 

\item {IRAS 04385+2550} is moderately inclined ($i = 60\arcdeg$) and presents an unambiguous example of wide-angle bipolar H$_2$ emission, reminiscent of emission seen in protostars, though with a significantly wider opening angle. The bipolar structure is well-defined across the observed H$_2$ lines. This demonstrates that bipolar winds can be visible even in sources that are not close to edge-on. The [Ne II] emission appears slightly extended along the same outflow direction.

\item {MY Lup} exhibits a clear bipolar morphology in the H$_2$ S(1) line, with tentative evidence for the S(2) line. The higher-$J$ transitions have lower S/N. Weak bipolar emission is detected in the S(3) line, while the S(4) line shows little to no detectable emission. The [Ne II] emission is marginally extended.

\item {RU Lup} shows bright, but compact, emission close to the central source in the S(1), S(2), and S(3) transitions, suggesting that any extended molecular wind is confined to small spatial scales. The symmetric structure is consistent with the nearly face-on orientation of this system. 

\item {RY Lup} is unique within the sample, exhibiting H$_2$ emission both parallel and perpendicular to the disk plane. The perpendicular component likely traces a molecular outflow or wind emerging from the disk surface, whereas the emission aligned with the disk may originate from cooler, quiescent disk gas. This dual morphology provides a rare view of both the disk and outflow components in a single system, highlighting complex gas dynamics in the inner regions of the protoplanetary disk. The [Ne II] emission is also extended perpendicular to the disk, coincident with the perpendicular H$_2$ emission, suggesting that the ionized and molecular gas trace the same outflowing structure.

\item {SR 4} is dominated by bright background H$_2$ emission originating from a nearby PDR in the Ophiuchus core, showing a clear spatial gradient increasing from the southwestern to the northeastern side. No definitive signatures of outflows or winds are detected in the H$_2$ emission, except in the S(5) transition. Extended [Ne II] emission is centered on the source, suggesting a possible low-level outflow traced by the ionized gas from the face-on system.

\item {SY Cha} exhibits a conical monopolar H$_2$ outflow detected in all observed transitions, oriented toward the east \citep{2025ApJ...980..148S}. Emission from the western side is observed, albeit significantly weaker. Additionally, SY Cha shows an extended [Ne II] jet on both sides of the disk, aligned along the wide-angle H$_2$ wind, indicating the presence of ionized, collimated jet.

\item {Sz 114} is a nearly face-on source ($i = 20\arcdeg$). The extended emission is concentrated around the central source, consistent with a face-on disk wind, and is detected in all observed H$_2$ lines. 

\item {Sz 129} is a low-inclination source ($i = 34\arcdeg$). Extended emission from S(1), S(2), and S(3) H$_2$ lines is concentrated around the central source, consistent with a face-on disk wind. The [Ne II] emission is also marginally extended.

\item {TW Cha} displays a centrally concentrated emission from the S(1), S(2), S(3), and S(5) lines, distributed symmetrically around the central source, consistent with a face-on disk wind. The [Ne II] emission also appears extended, albeit more compact.

\item {TW Hya} shows a clear, symmetric, extended H$_2$ morphology in all observed lines, consistent with a nearly face-on disk wind. The emission appears symmetric around the central source and likely traces the wide-angle disk wind. The [Ne II] emission also appears extended, albeit more compact.

\item {VZ Cha} exhibits an unusual H$_2$ outflow morphology characterized by a bubble-like structure on the northwestern side of the source. The bubble has a brighter rim and in the inner region is less bright. The bubble component appears offset from the central source, possibly indicating a past episodic ejection event. The [Ne II] emission clearly reveals a jet along the same side of the outflow, suggesting that the ionized gas traces a more collimated component embedded within the broader molecular wind.

\item {WSB 52} exhibits a wide-angle H$_2$ outflow. The S(1) transition is dominated by bright, extended emission, while the higher-excitation H$_2$ lines reveal a weaker but discernible bipolar morphology. The southwestern side of the outflow appears brighter than the northeastern side, indicating asymmetry in the emission or excitation conditions. The source also shows extended, jet-like [Ne II] emission aligned with the direction of the outflow.

\end{enumerate}

\newpage

\subsection{Line images for individual sources}

This appendix shows the H$_2$ and [Ne II] line images for our sources.

\begin{figure*}[h]
\centering
\includegraphics[width=18cm]{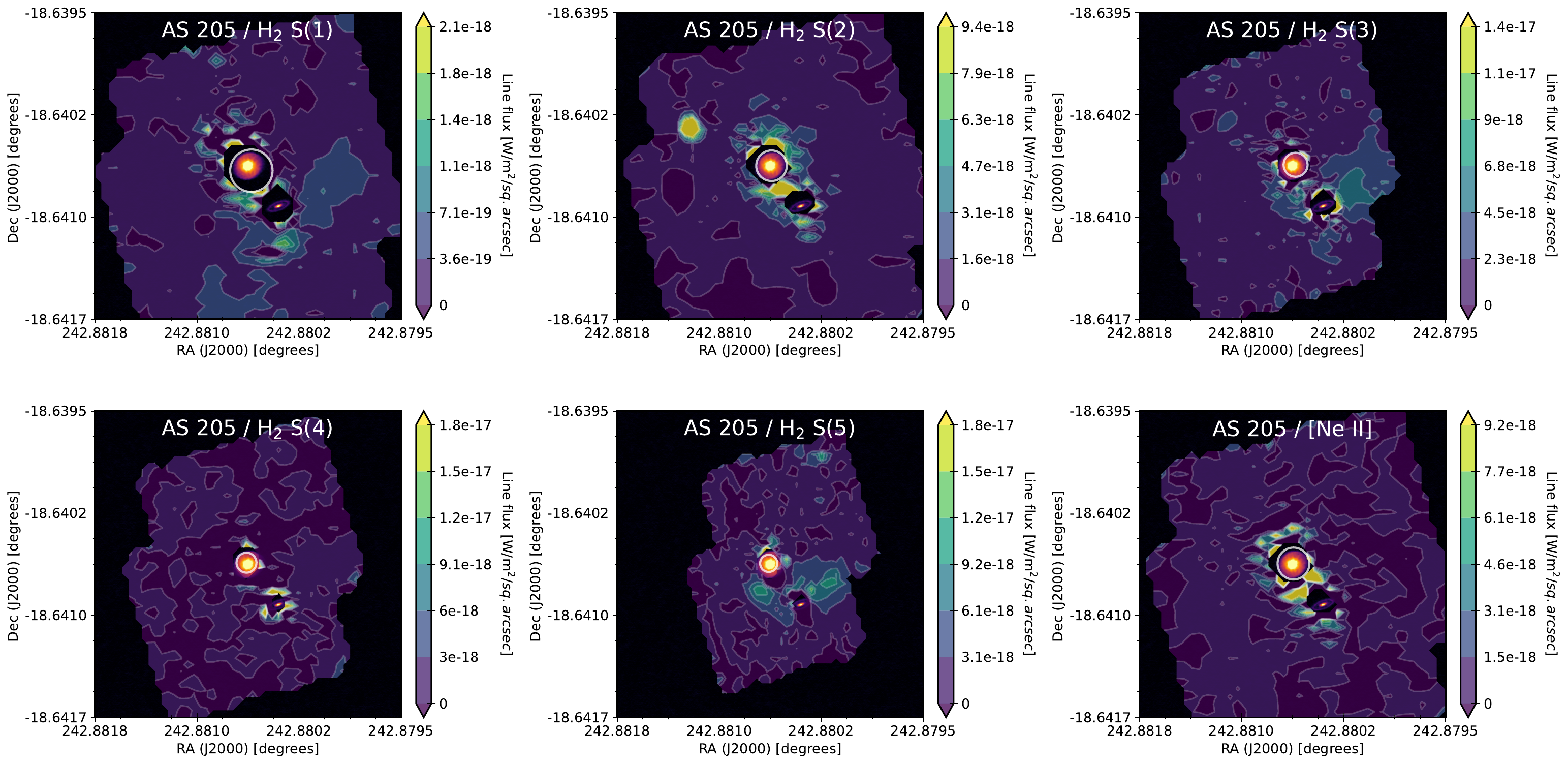}
\caption{\label{fig:as205}}
\end{figure*}

\begin{figure*}[h]
\centering
\includegraphics[width=18cm]{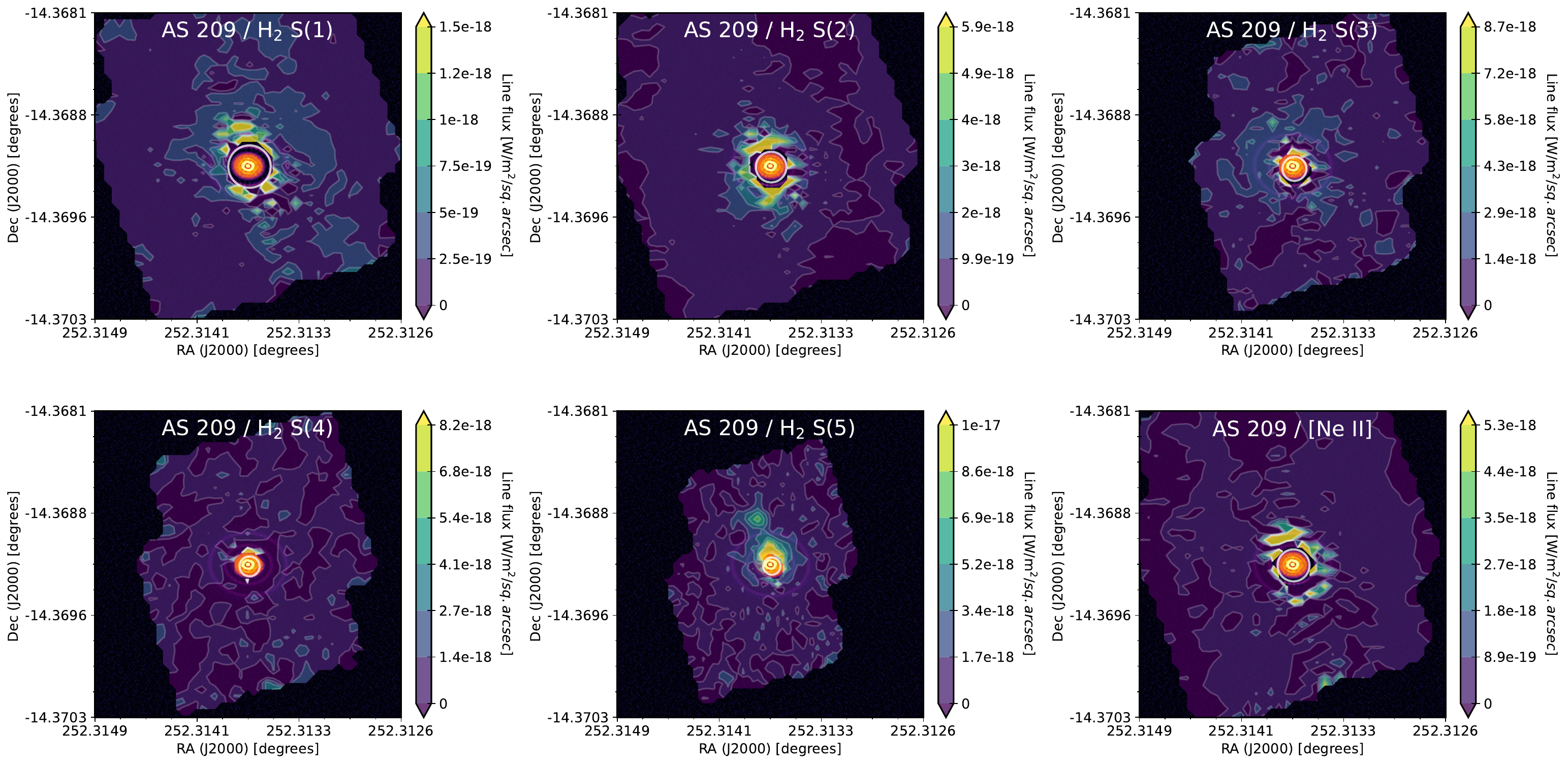}
\caption{The  H$_2$ and [Ne II] emission detected towards AS 209 on top of the ALMA 240 GHz continuum emission (shown in the background as a reddish-yellow image). The white central circle shows the inner working angle from JWST. The same plotting scheme is followed for all the following images in the gallery, and for AS 205 above.  \label{fig:as209}}
\end{figure*}

\newpage

\begin{figure*}[h]
\centering
\includegraphics[width=18cm]{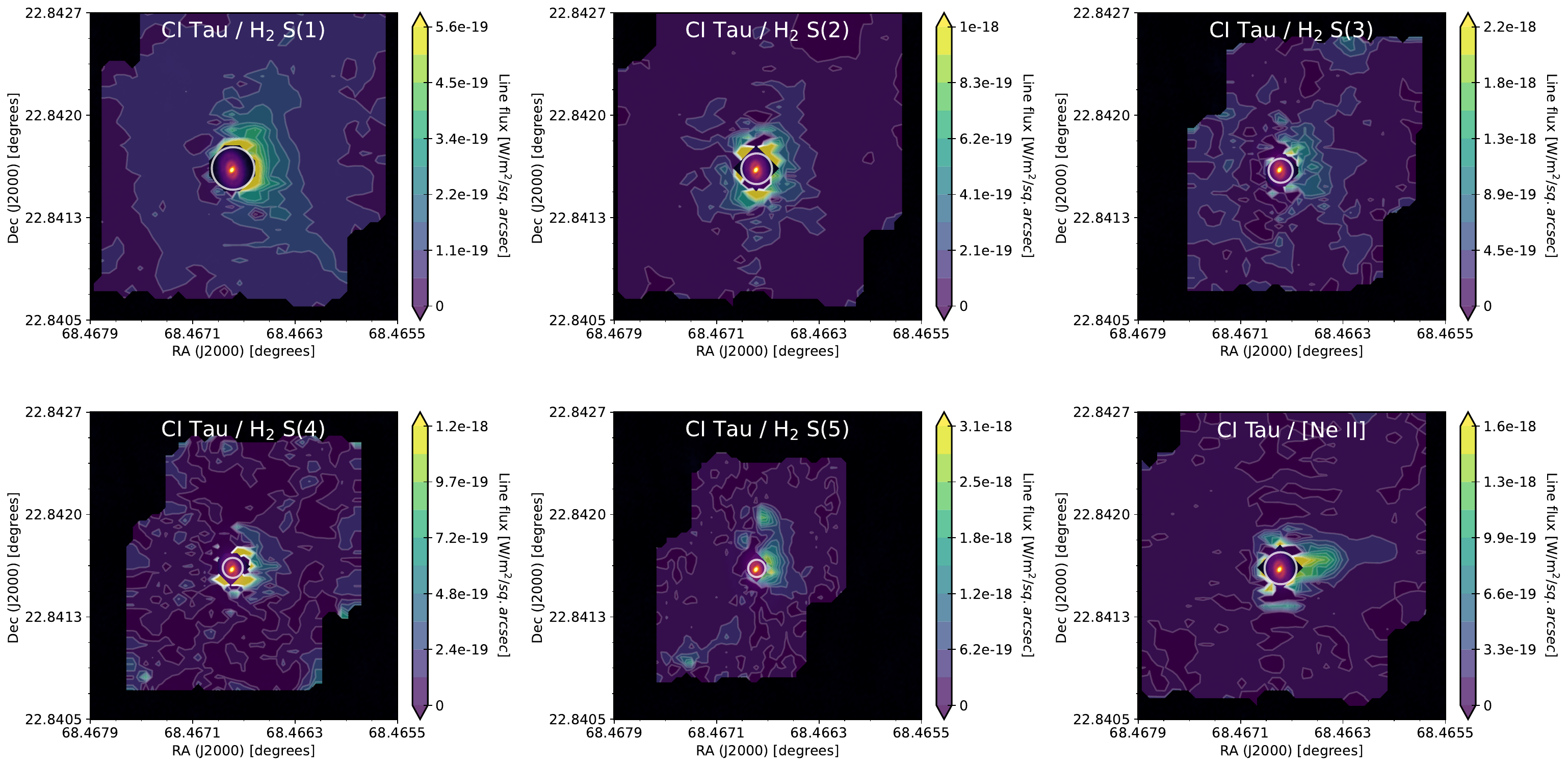}
\caption{\label{fig:citau}}
\end{figure*}

\begin{figure*}[h]
\centering
\includegraphics[width=18cm]{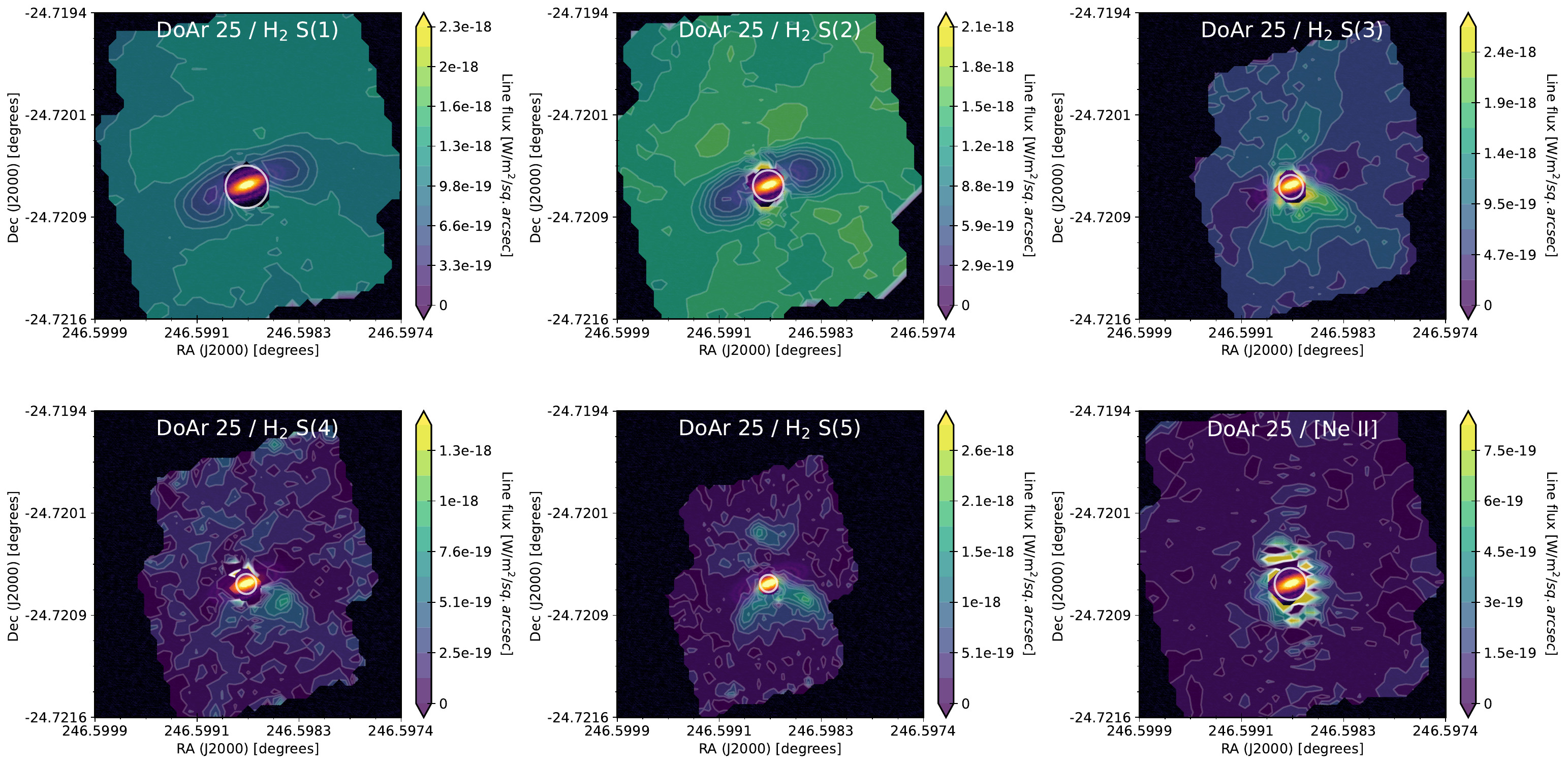}
\caption{\label{fig:doar25b}}
\end{figure*}

\newpage

\begin{figure*}[h]
\centering
\includegraphics[width=18cm]{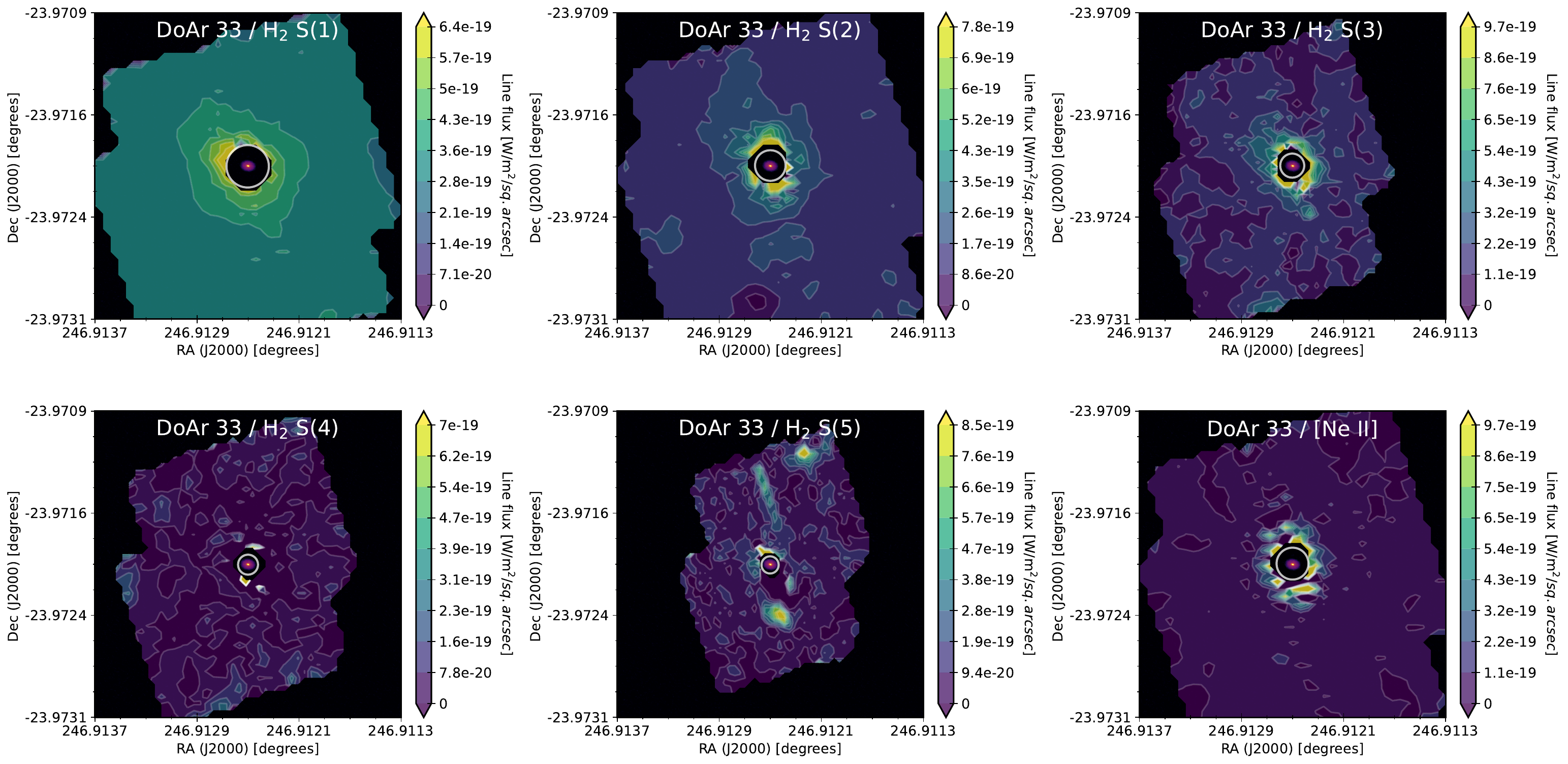}
\caption{\label{fig:doar33}}
\end{figure*}

\begin{figure*}[h]
\centering
\includegraphics[width=18cm]{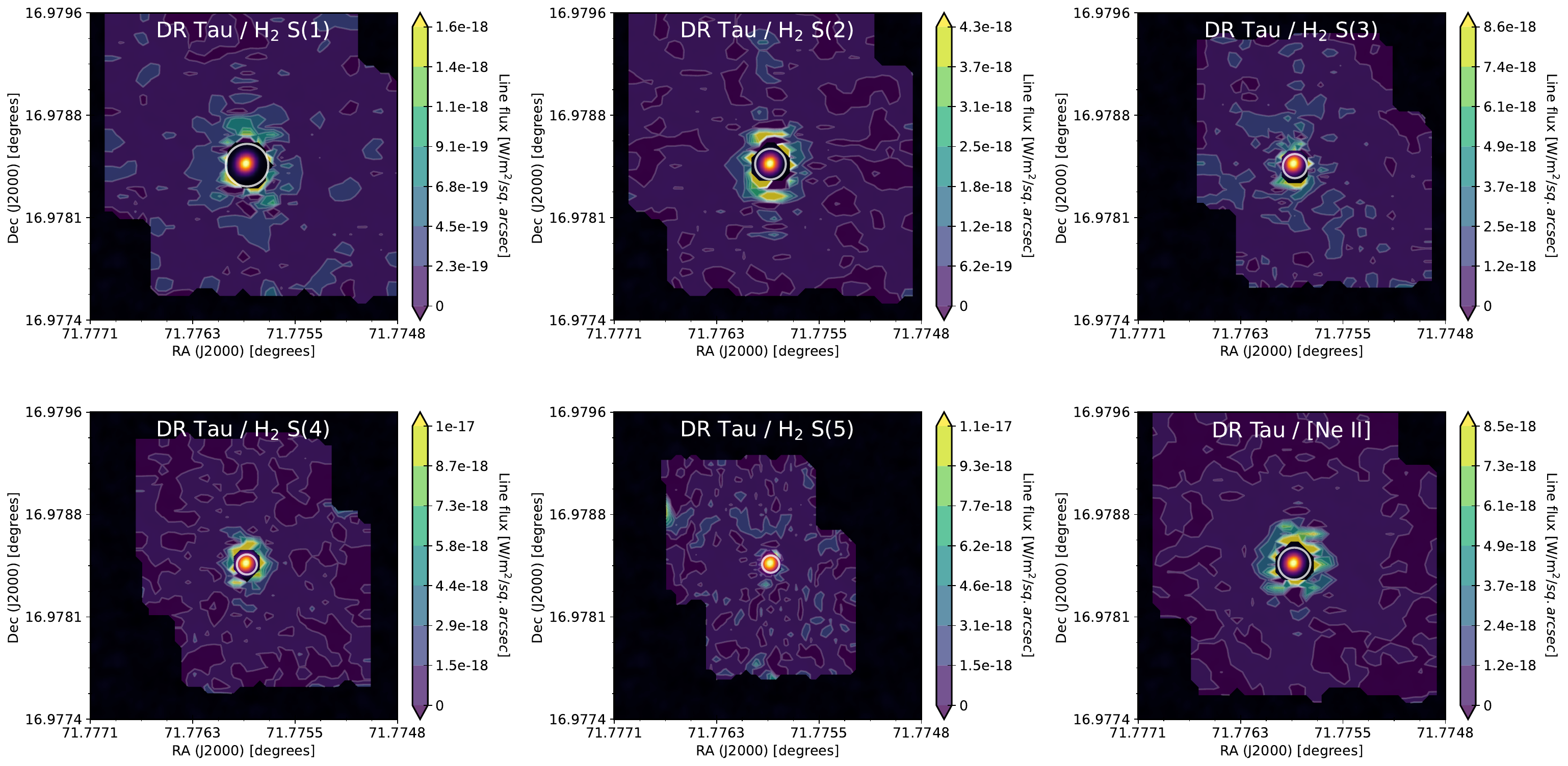}
\caption{\label{fig:drtau}}
\end{figure*}

\newpage

\begin{figure*}[h]
\centering
\includegraphics[width=18cm]{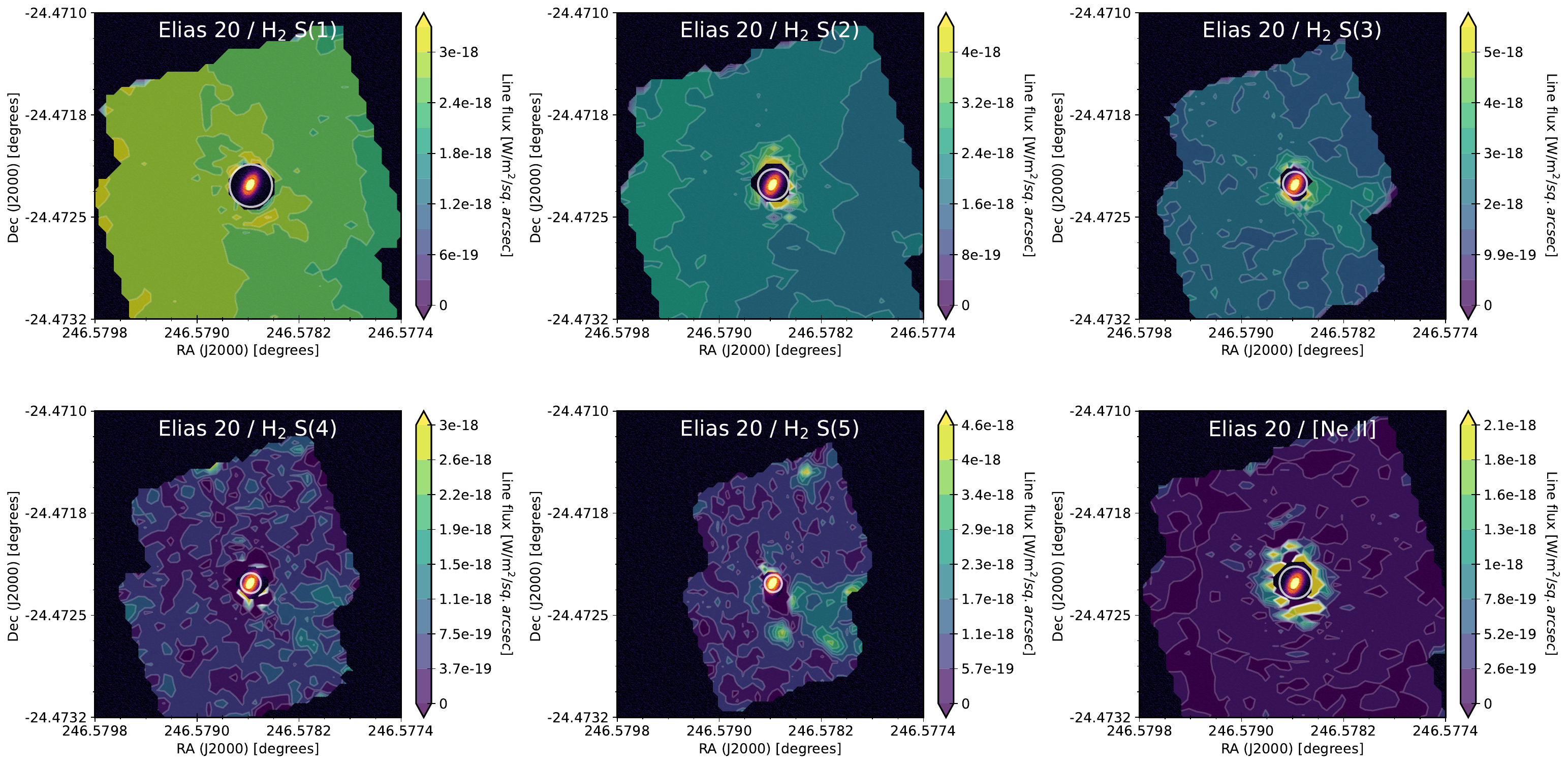}
\caption{\label{fig:elias20}}
\end{figure*}

\begin{figure*}[h]
\centering
\includegraphics[width=18cm]{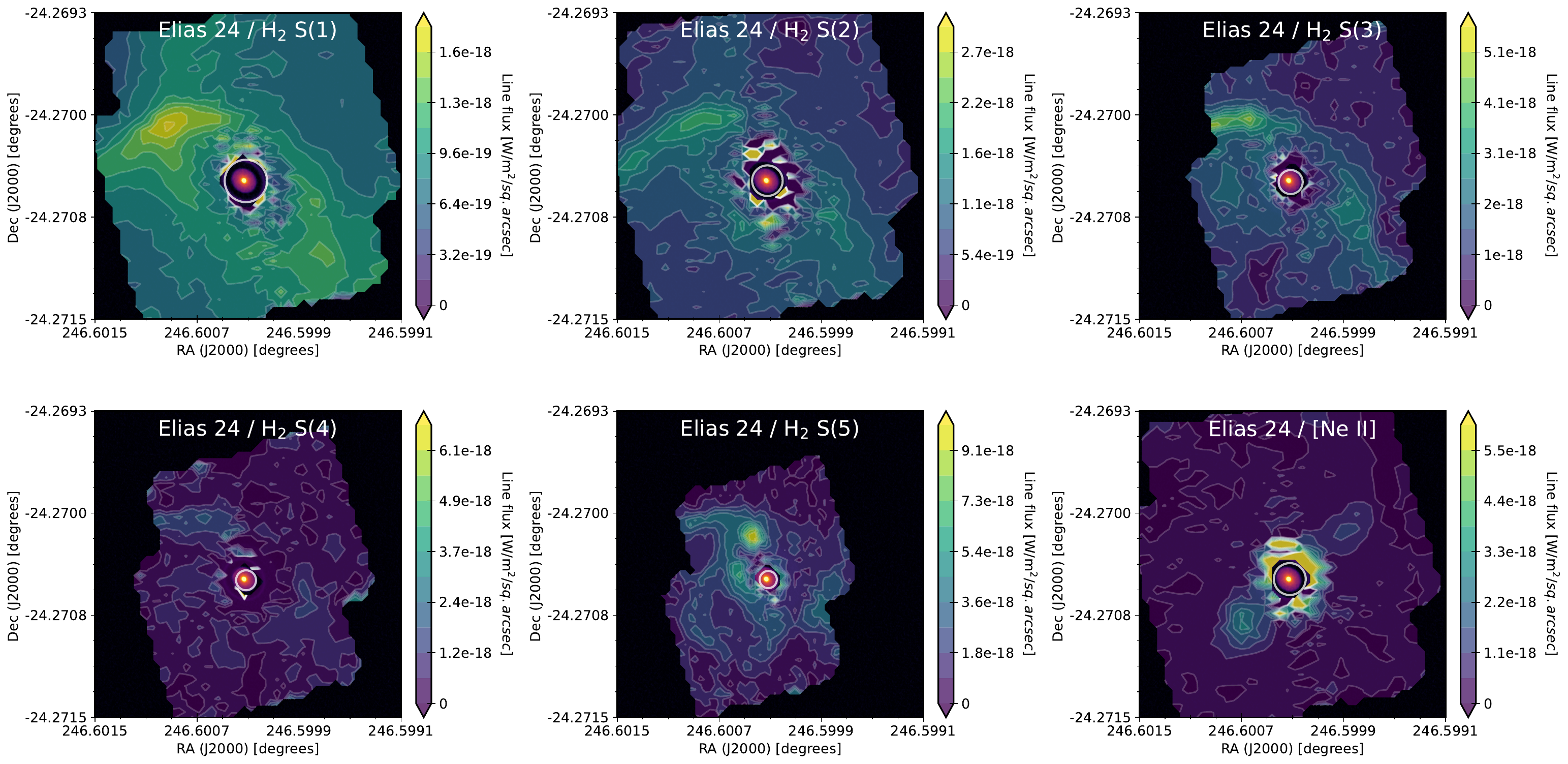}
\caption{\label{fig:elias24}}
\end{figure*}

\newpage

\begin{figure*}[h]
\centering
\includegraphics[width=18cm]{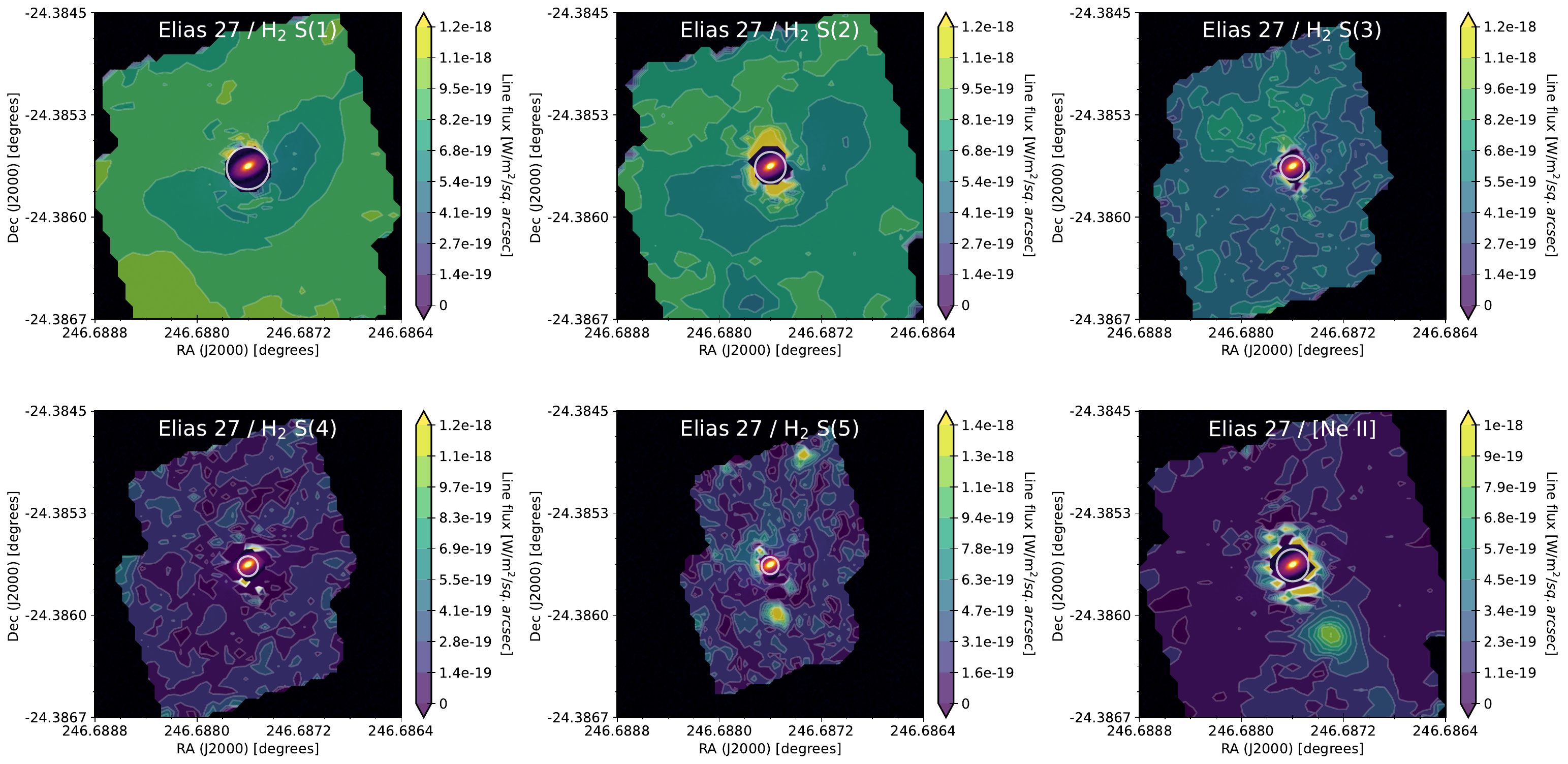}
\caption{\label{fig:elias27}}
\end{figure*}

\begin{figure*}[h]
\centering
\includegraphics[width=18cm]{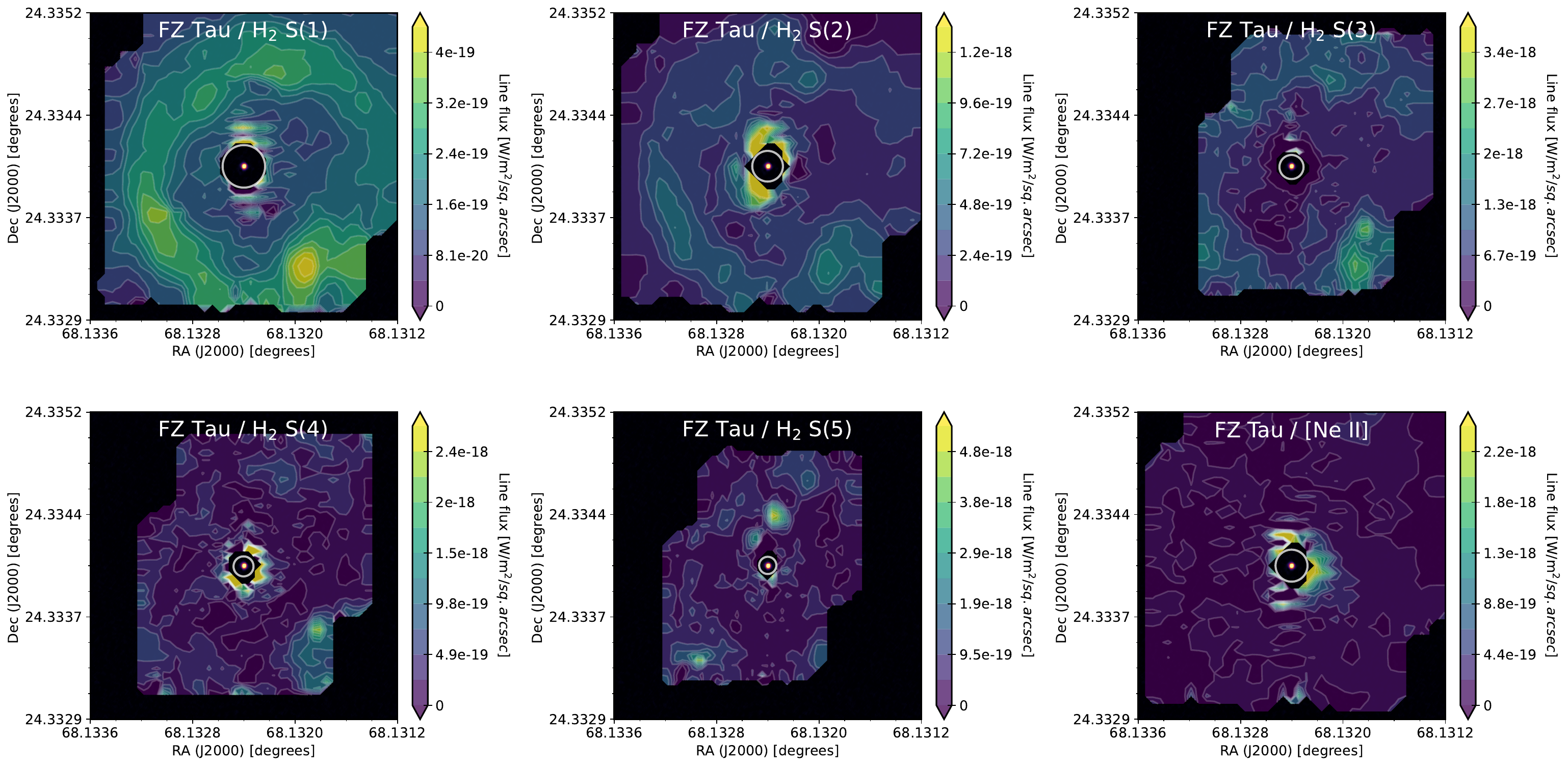}
\caption{\label{fig:fztau}}
\end{figure*}

\newpage

\begin{figure*}[h]
\centering
\includegraphics[width=18cm]{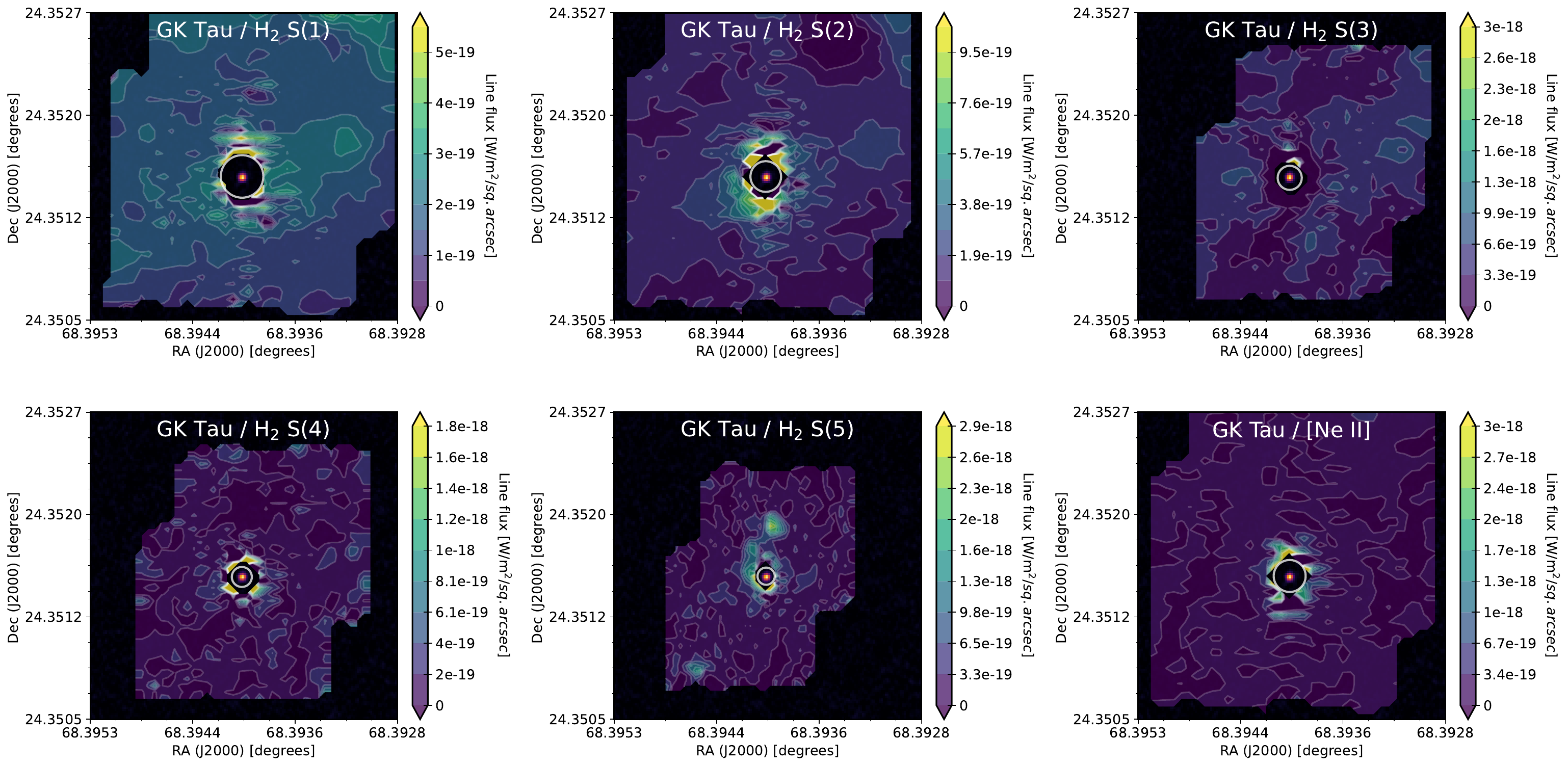}
\caption{\label{fig:gktau}}
\end{figure*}

\begin{figure*}[h]
\centering
\includegraphics[width=18cm]{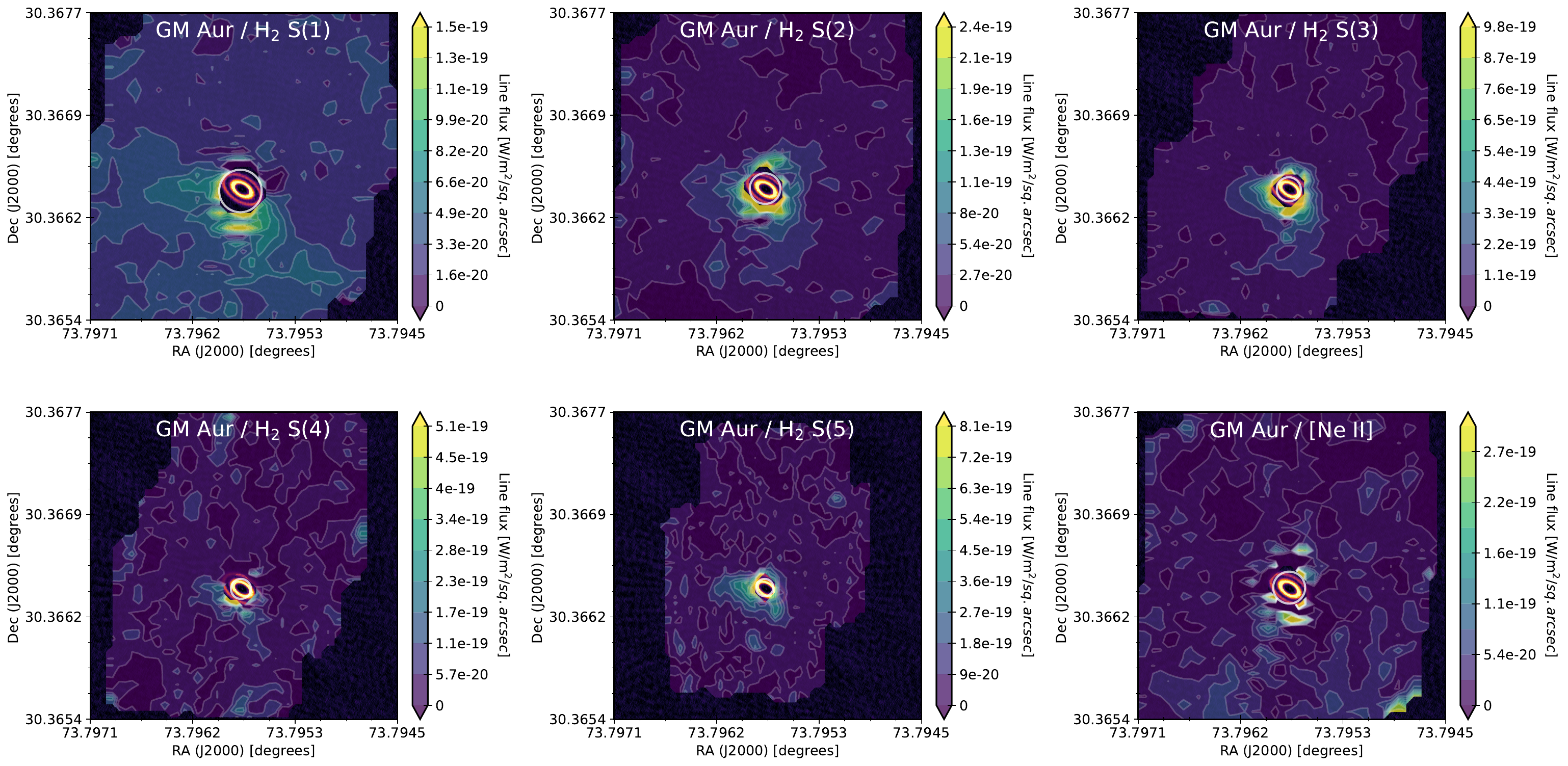}
\caption{\label{fig:gmaur}}
\end{figure*}

\newpage

\begin{figure*}[h]
\centering
\includegraphics[width=18cm]{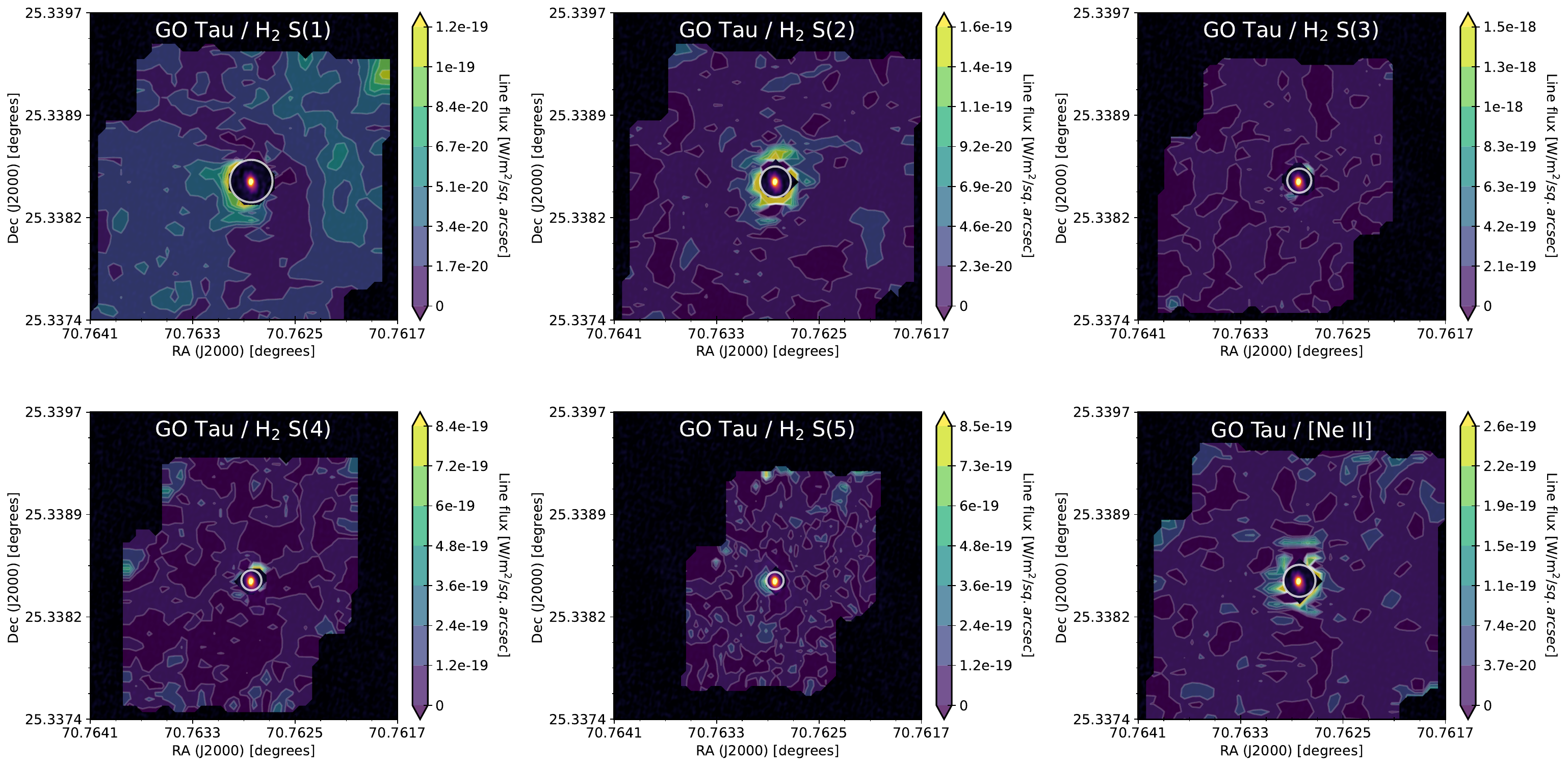}
\caption{\label{fig:gotau}}
\end{figure*}

\begin{figure*}[h]
\centering
\includegraphics[width=18cm]{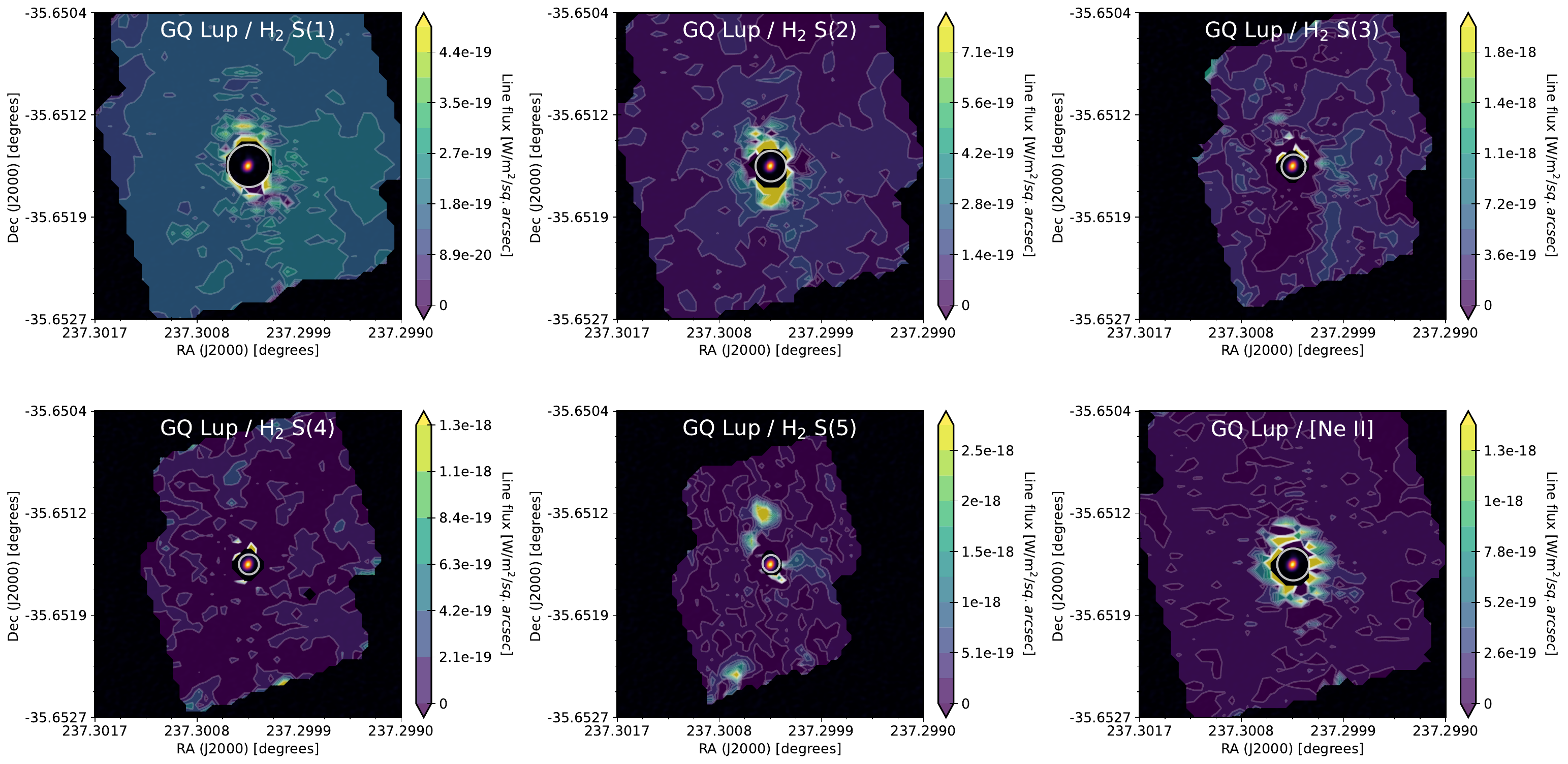}
\caption{\label{fig:gqlup}}
\end{figure*}

\newpage

\begin{figure*}[h]
\centering
\includegraphics[width=18cm]{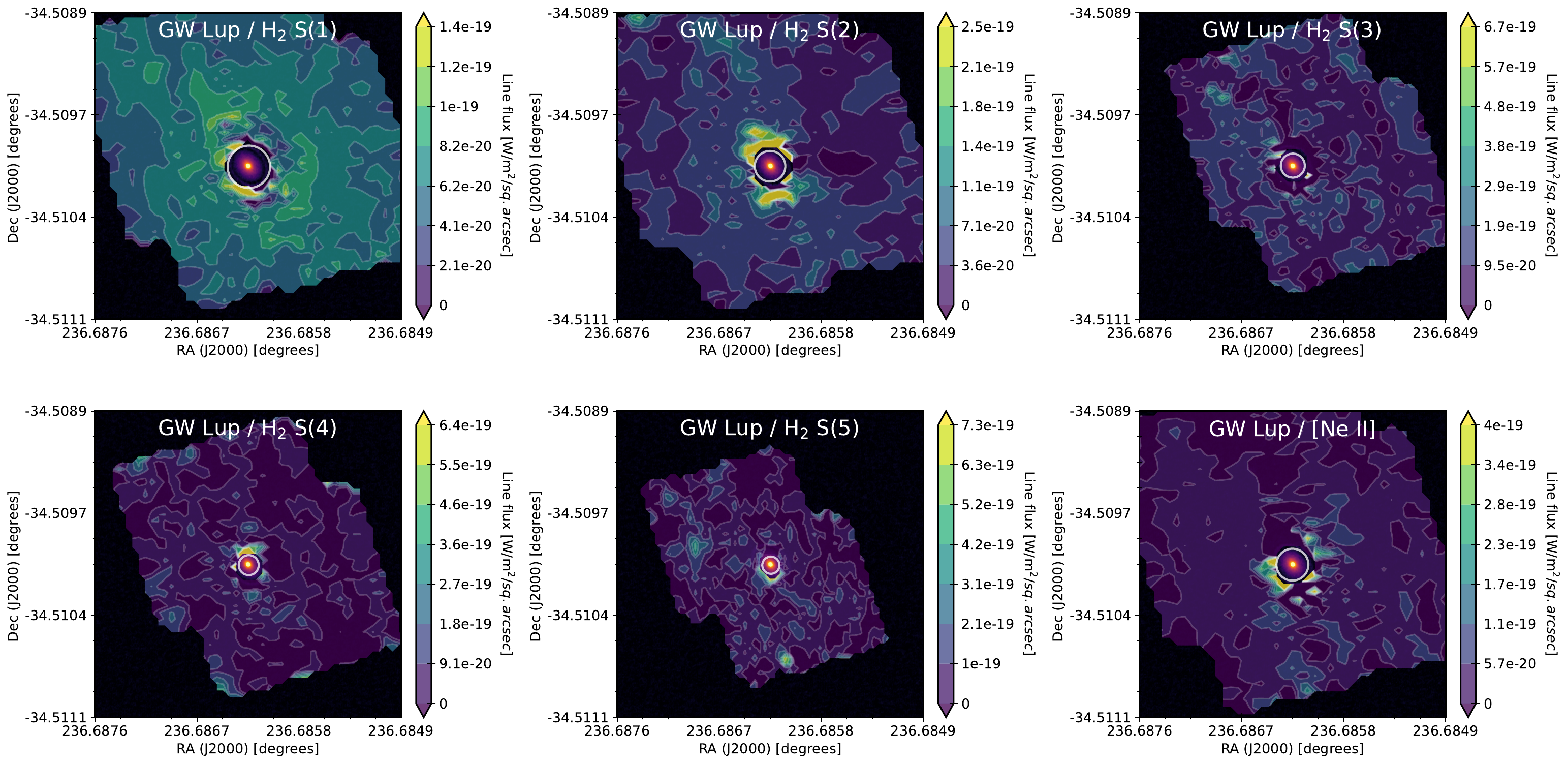}
\caption{\label{fig:gwlup}}
\end{figure*}

\begin{figure*}[h]
\centering
\includegraphics[width=18cm]{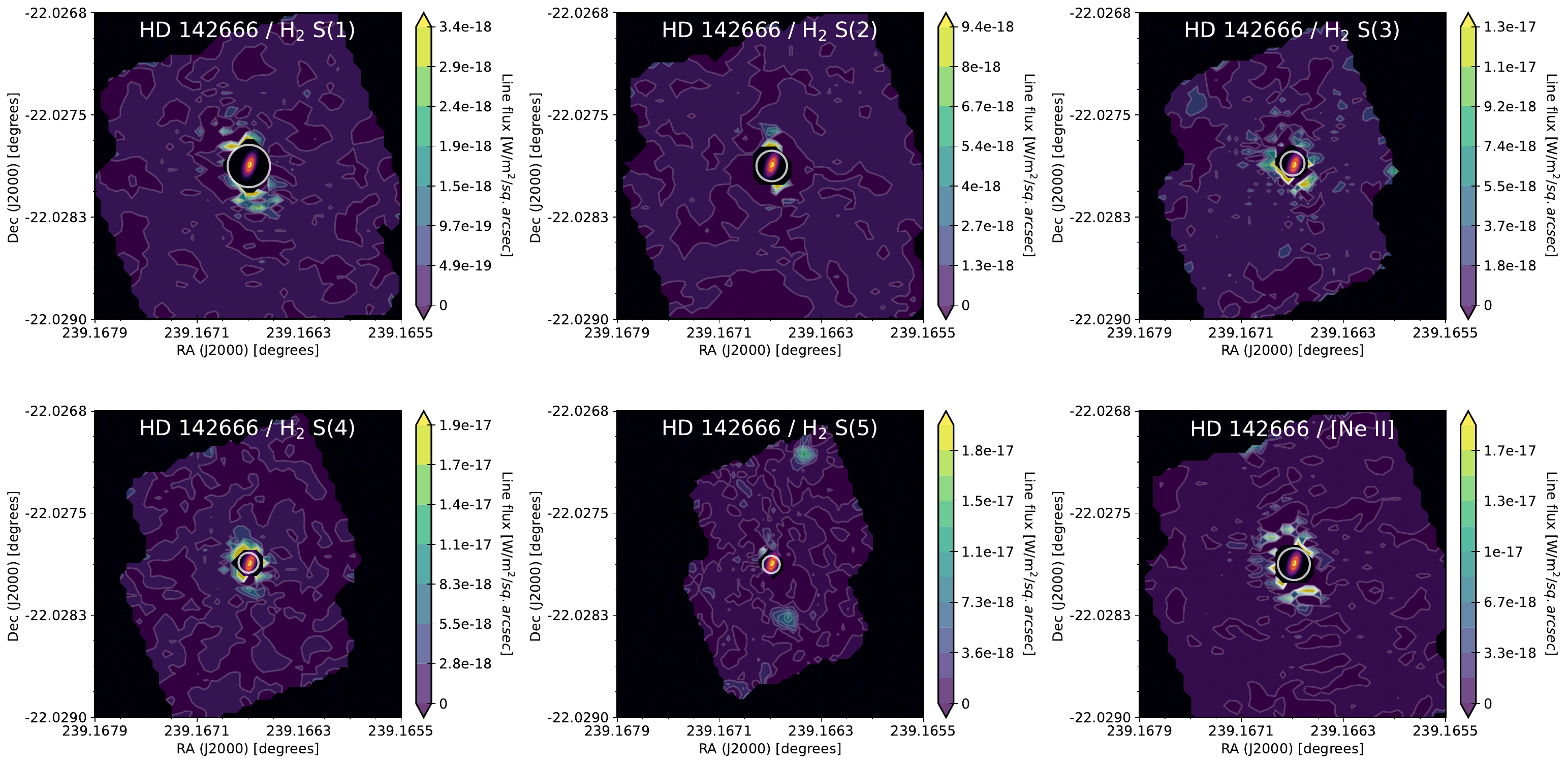}
\caption{\label{fig:hd142666}}
\end{figure*}

\newpage

\begin{figure*}[h]
\centering
\includegraphics[width=18cm]{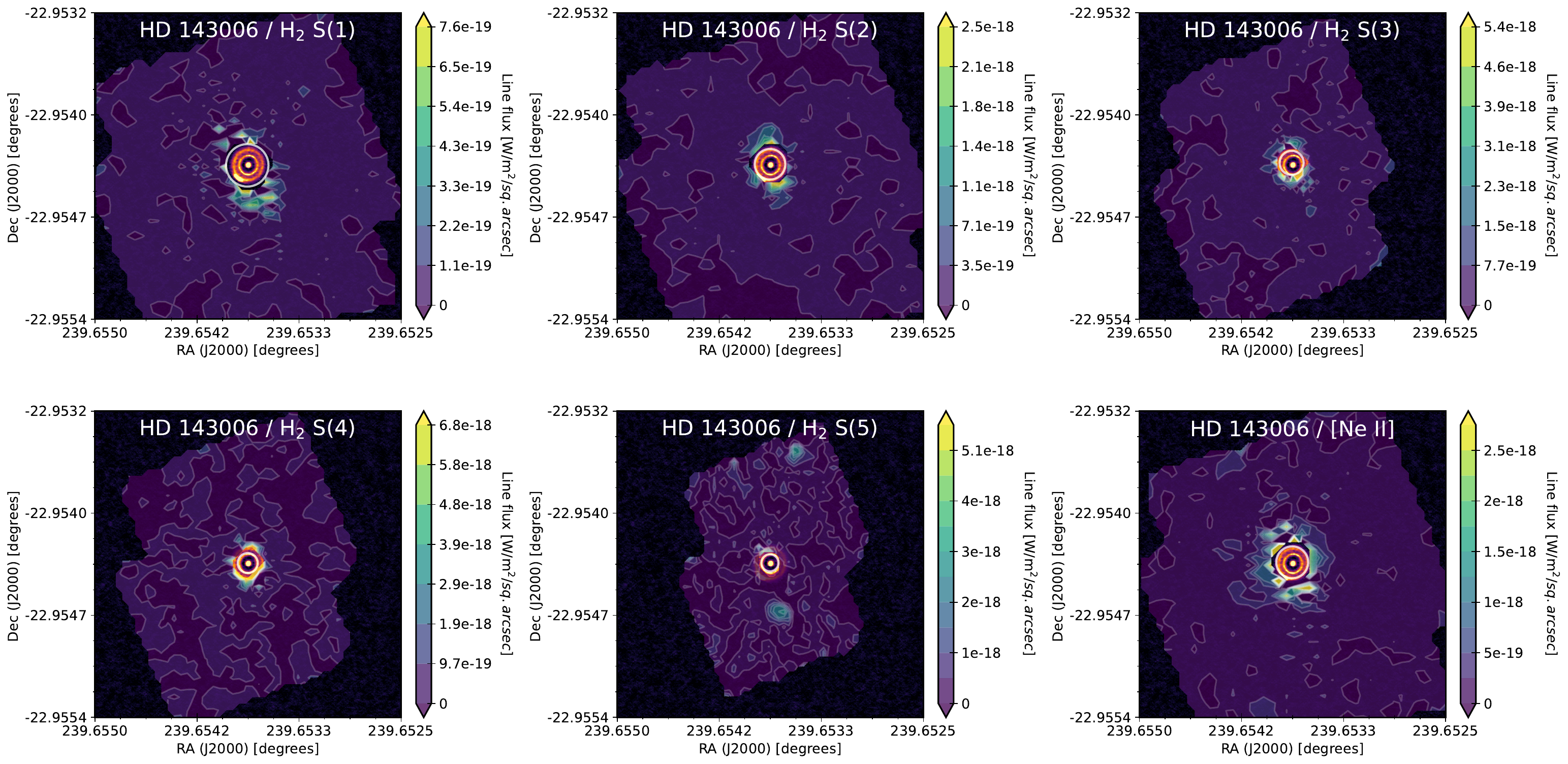}
\caption{\label{fig:hd143006}}
\end{figure*}

\begin{figure*}[h]
\centering
\includegraphics[width=18cm]{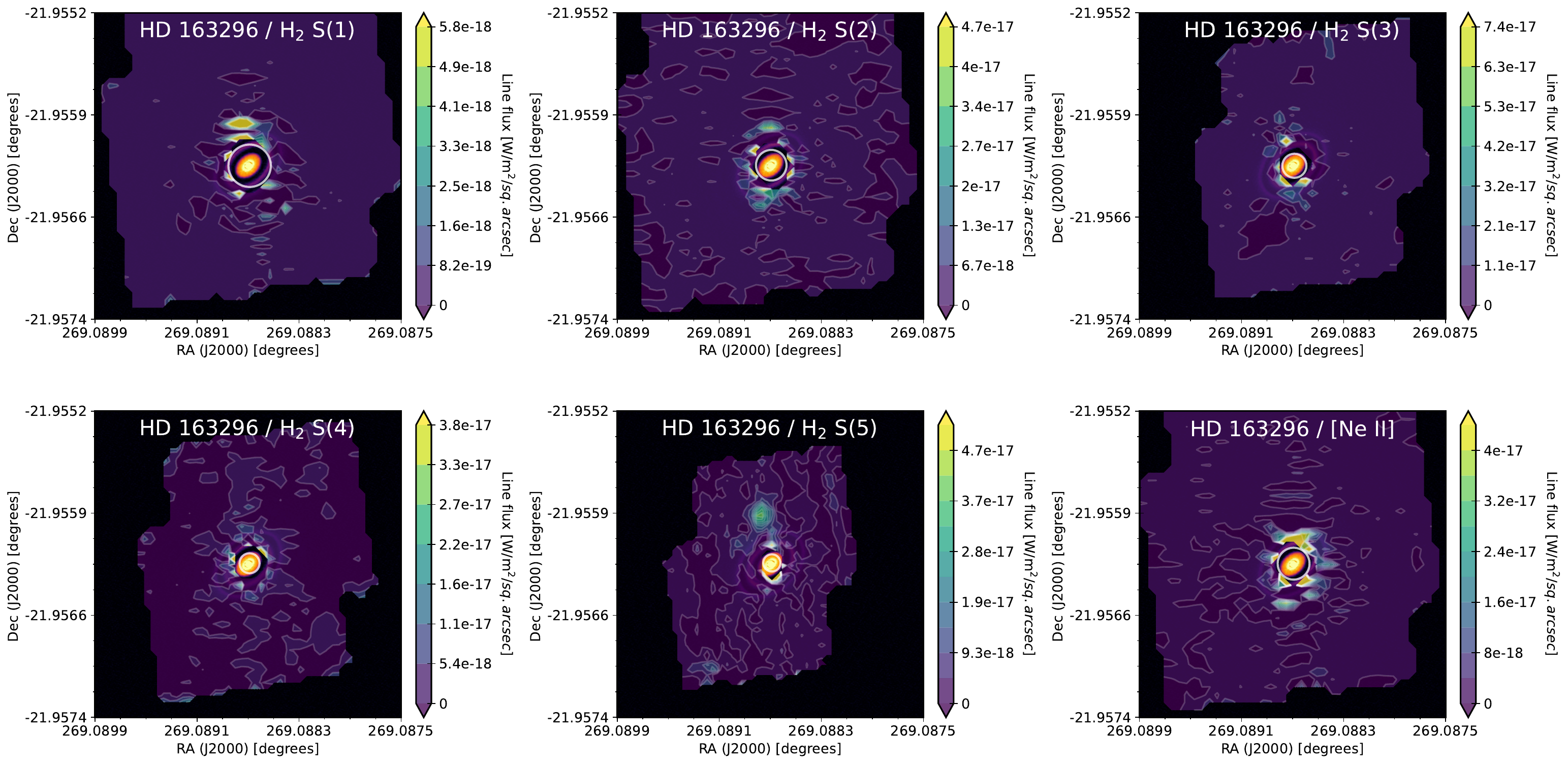}
\caption{\label{fig:hd163296}}
\end{figure*}

\newpage

\begin{figure*}[h]
\centering
\includegraphics[width=18cm]{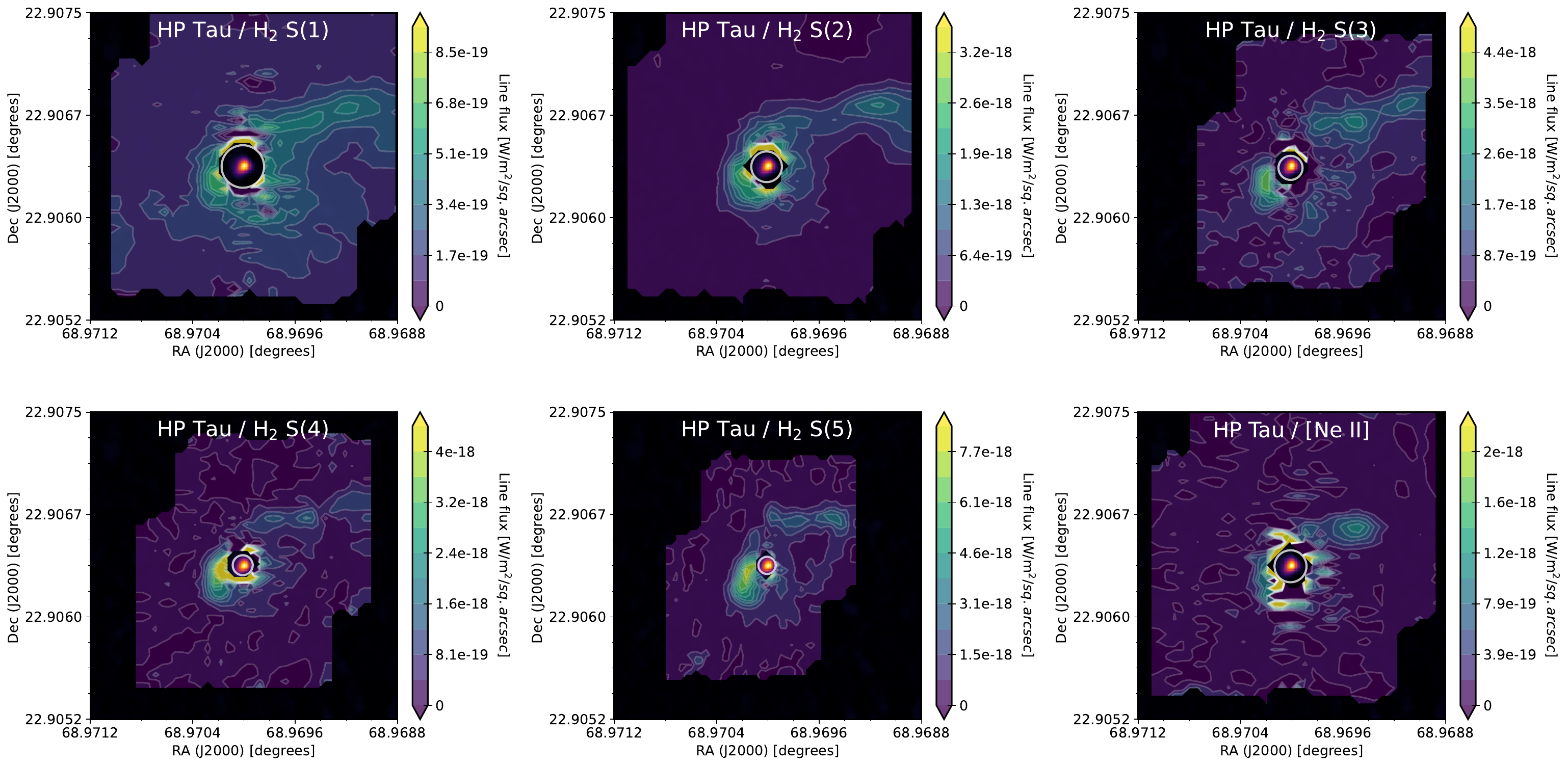}
\caption{\label{fig:hptau}}
\end{figure*}

\begin{figure*}[h]
\centering
\includegraphics[width=18cm]{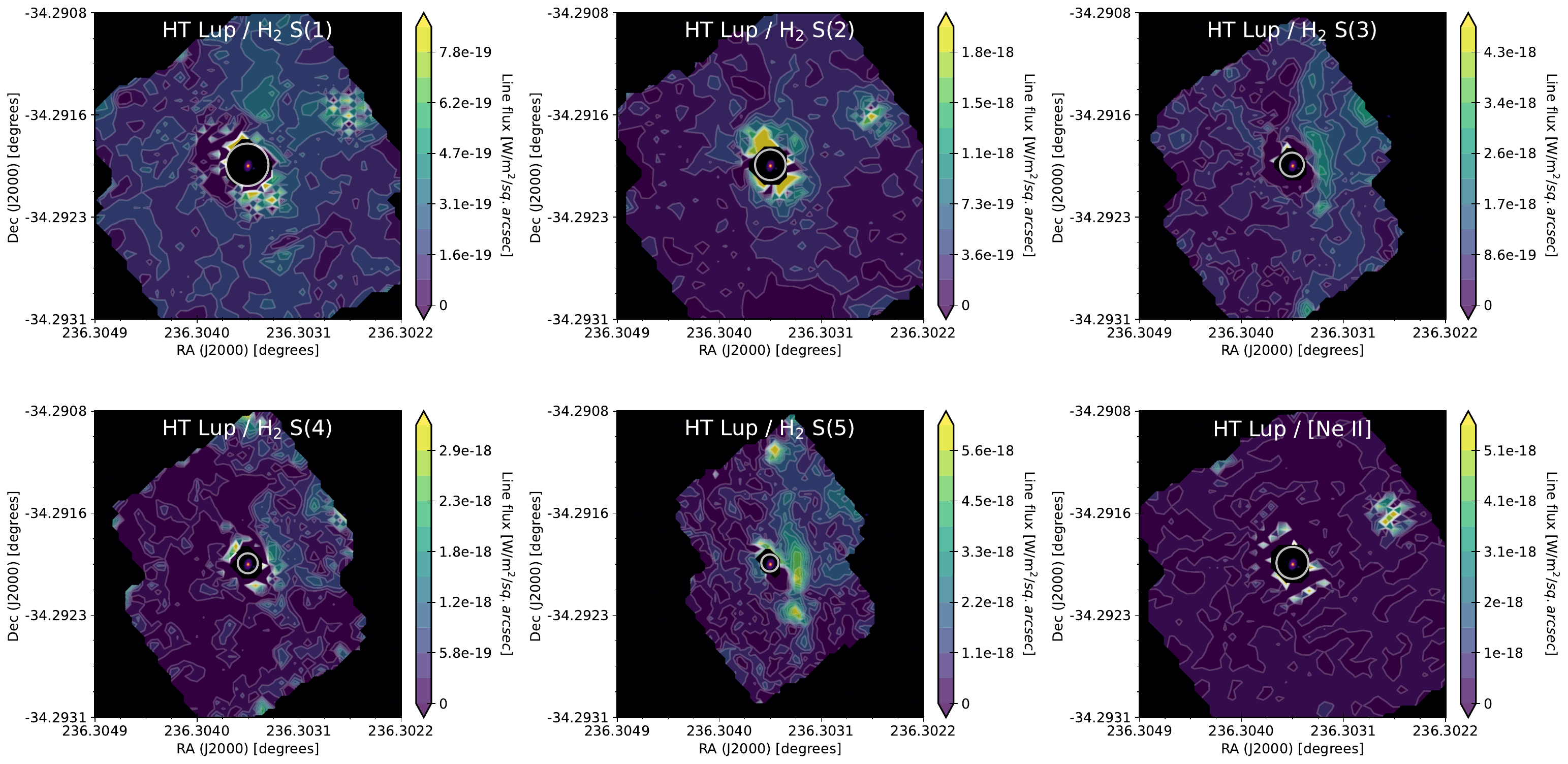}
\caption{\label{fig:htlup}}
\end{figure*}

\newpage

\begin{figure*}[h]
\centering
\includegraphics[width=18cm]{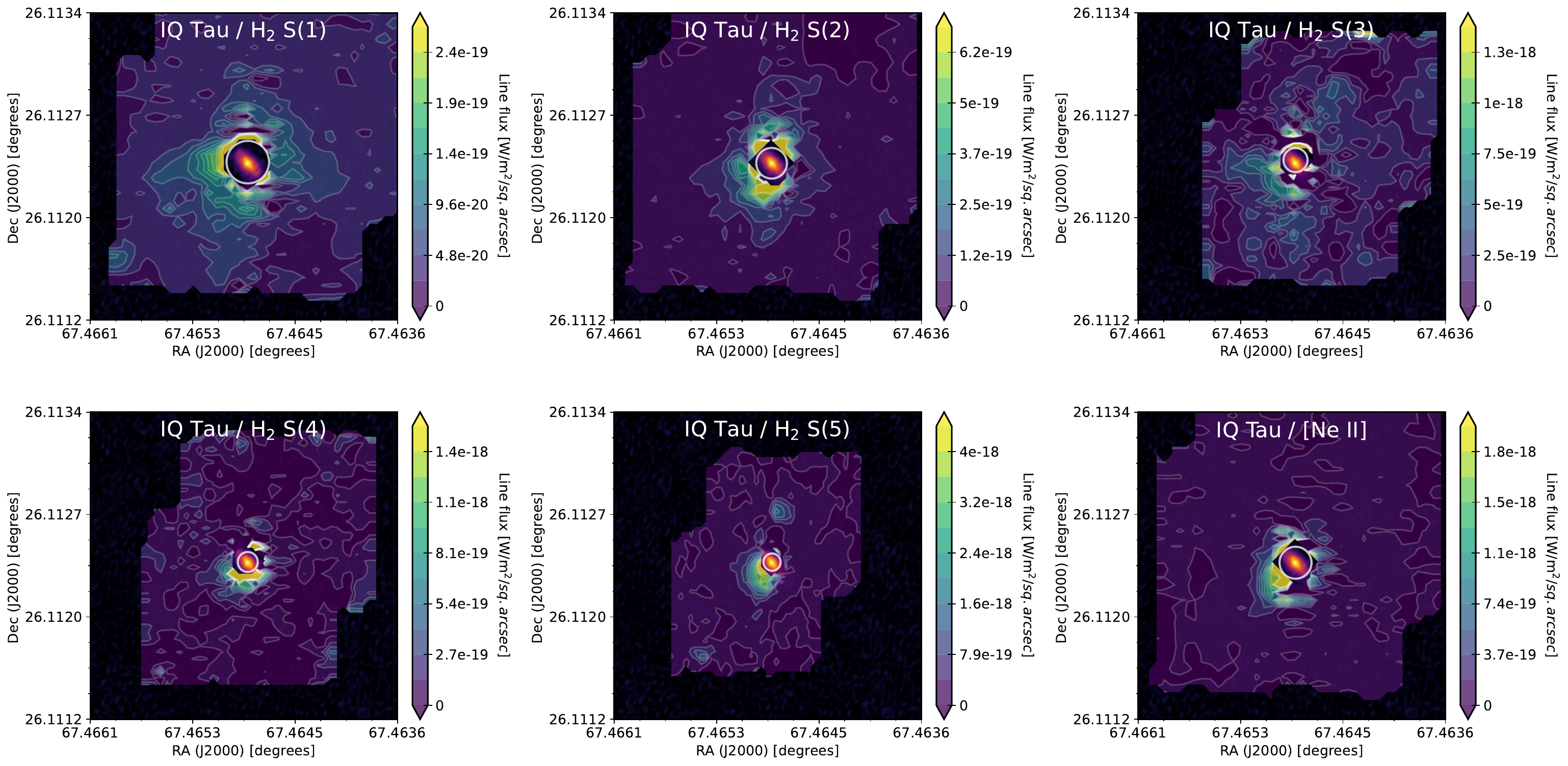}
\caption{\label{fig:iqtau}}
\end{figure*}

\begin{figure*}[h]
\centering
\includegraphics[width=18cm]{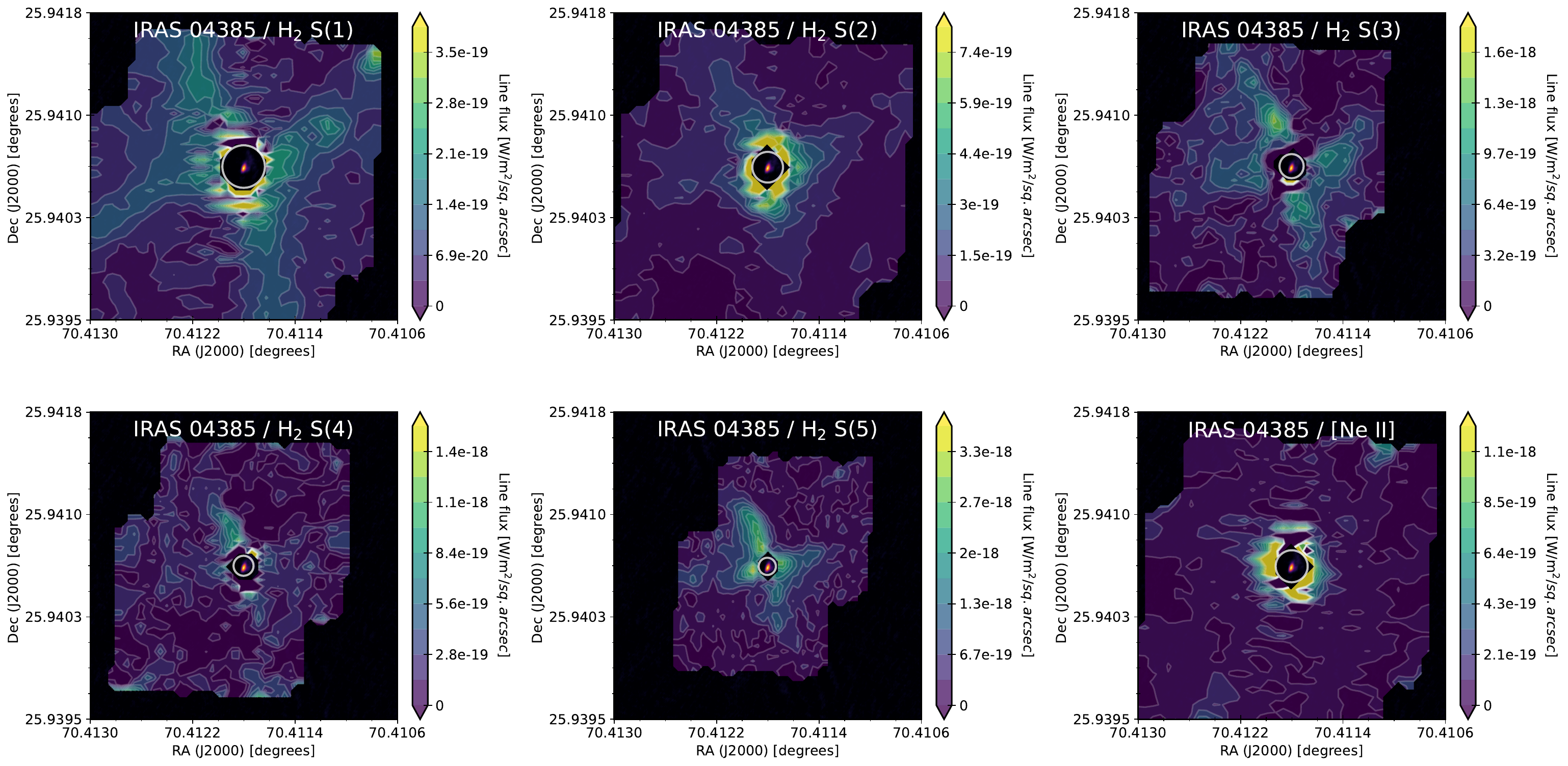}
\caption{\label{fig:iras04385}}
\end{figure*}

\newpage

\begin{figure*}[h]
\centering
\includegraphics[width=18cm]{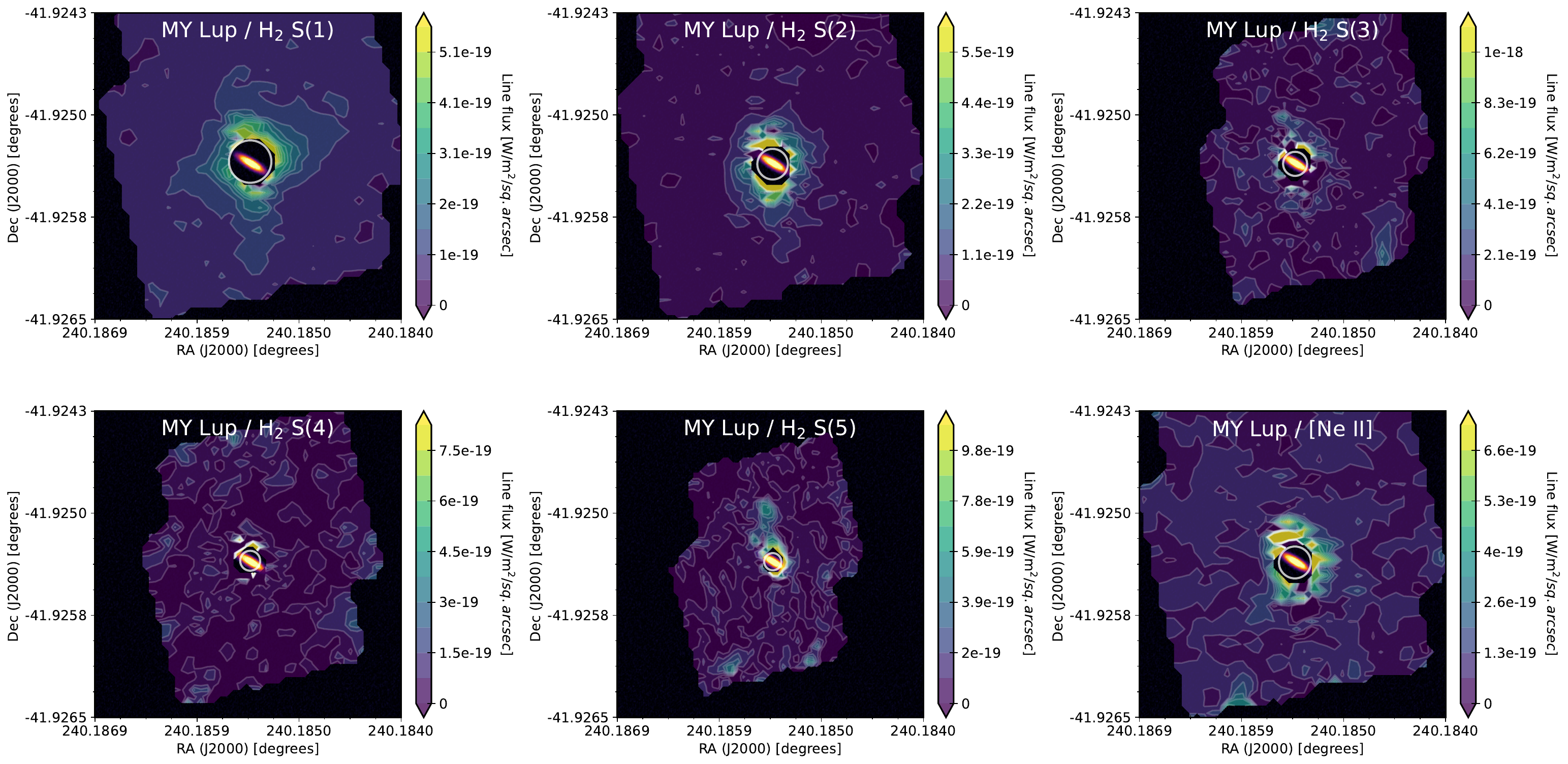}
\caption{\label{fig:mylup}}
\end{figure*}

\begin{figure*}[h]
\centering
\includegraphics[width=18cm]{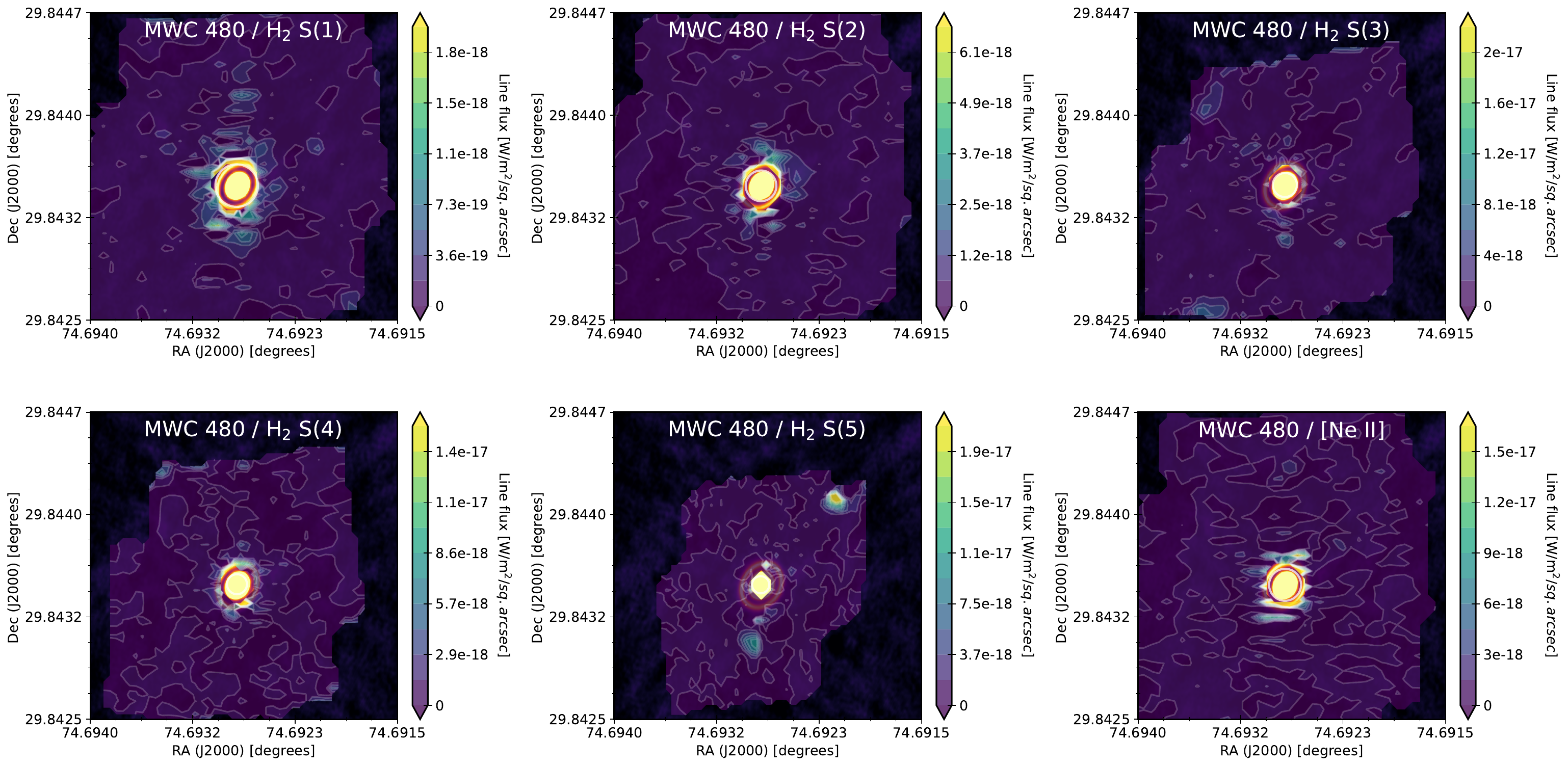}
\caption{\label{fig:mwc480}}
\end{figure*}

\newpage

\begin{figure*}[h]
\centering
\includegraphics[width=18cm]{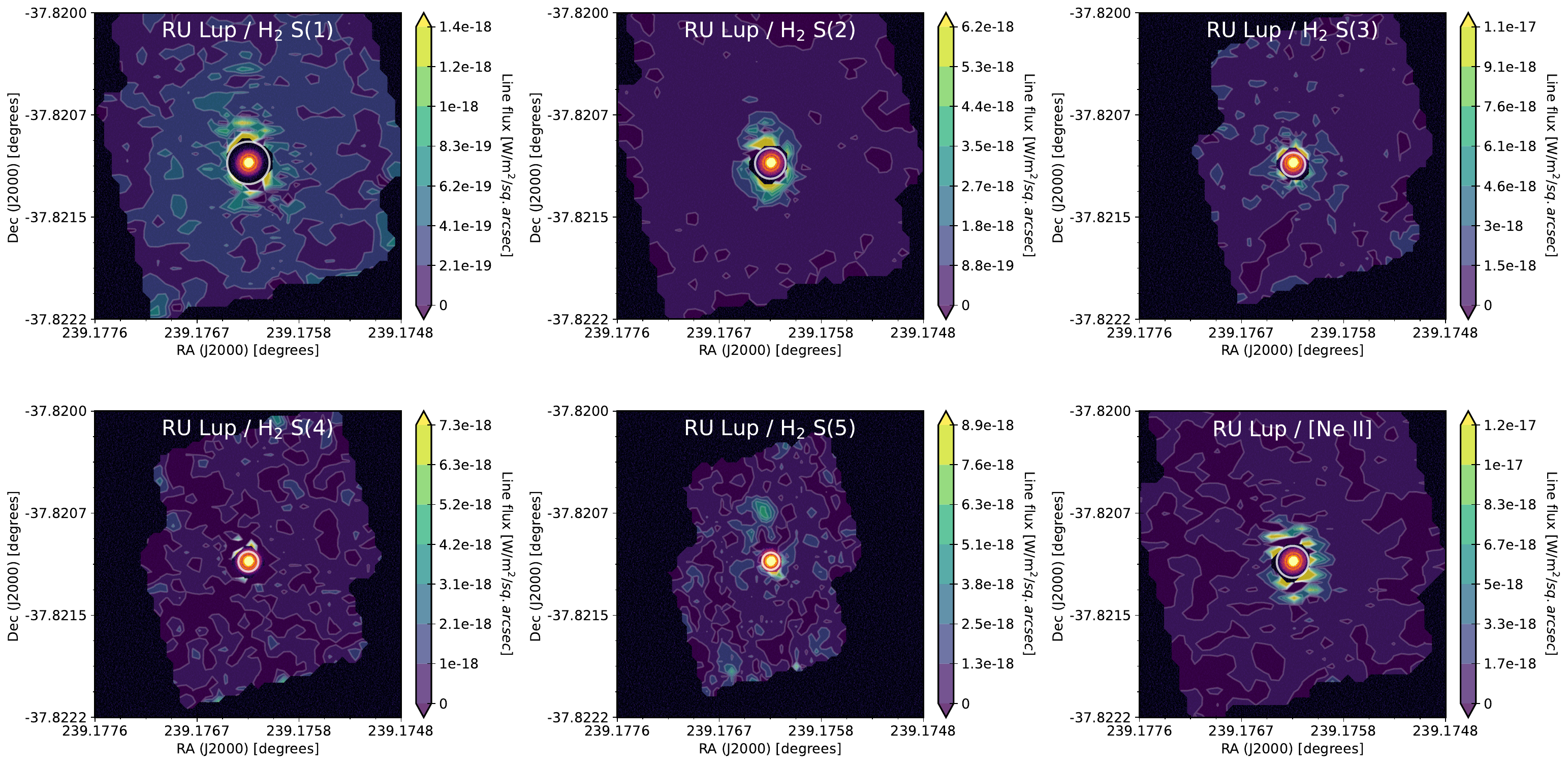}
\caption{\label{fig:rulup}}
\end{figure*}

\begin{figure*}[h]
\centering
\includegraphics[width=18cm]{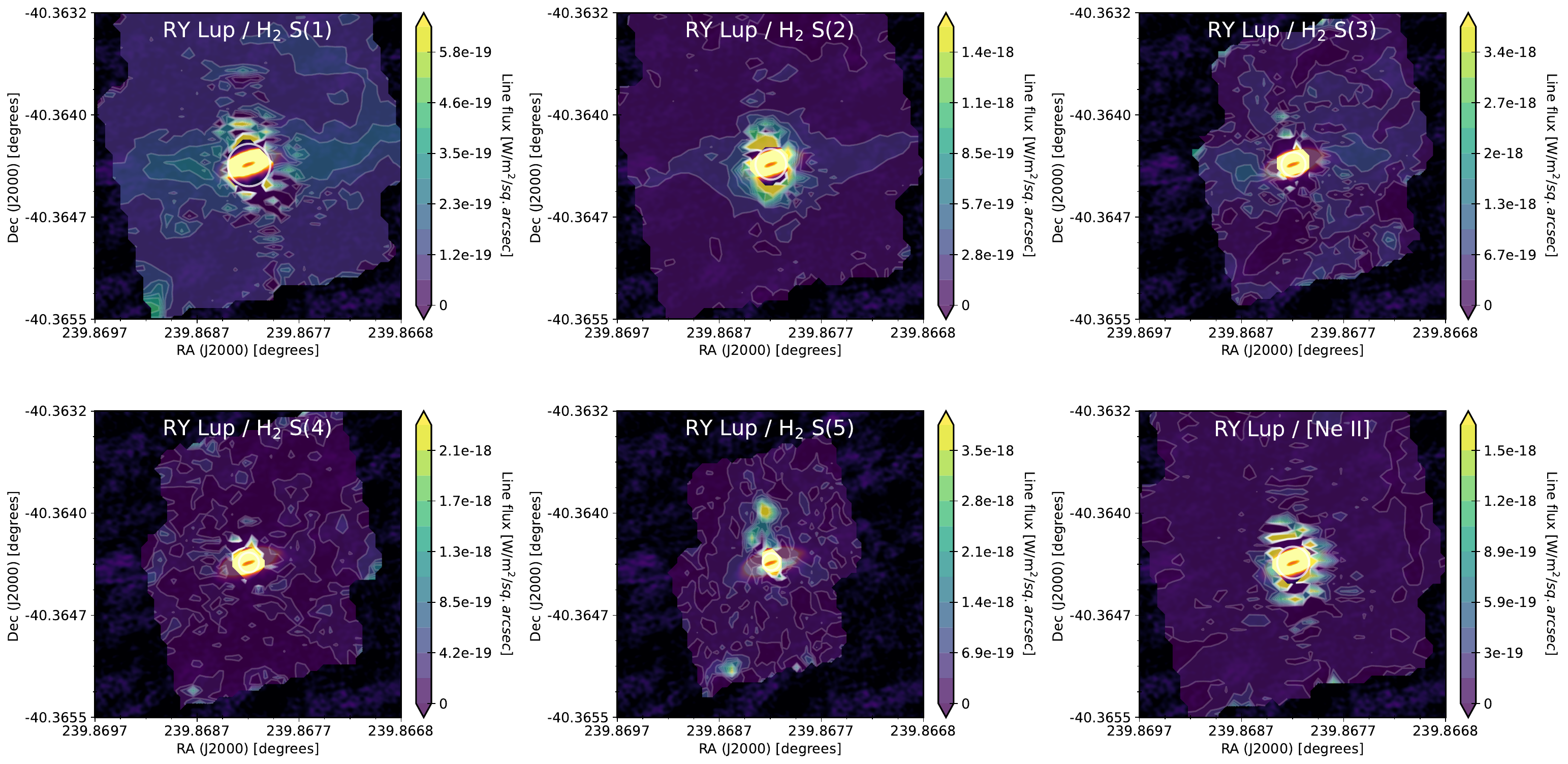}
\caption{\label{fig:rylup}}
\end{figure*}

\newpage

\begin{figure*}[h]
\centering
\includegraphics[width=18cm]{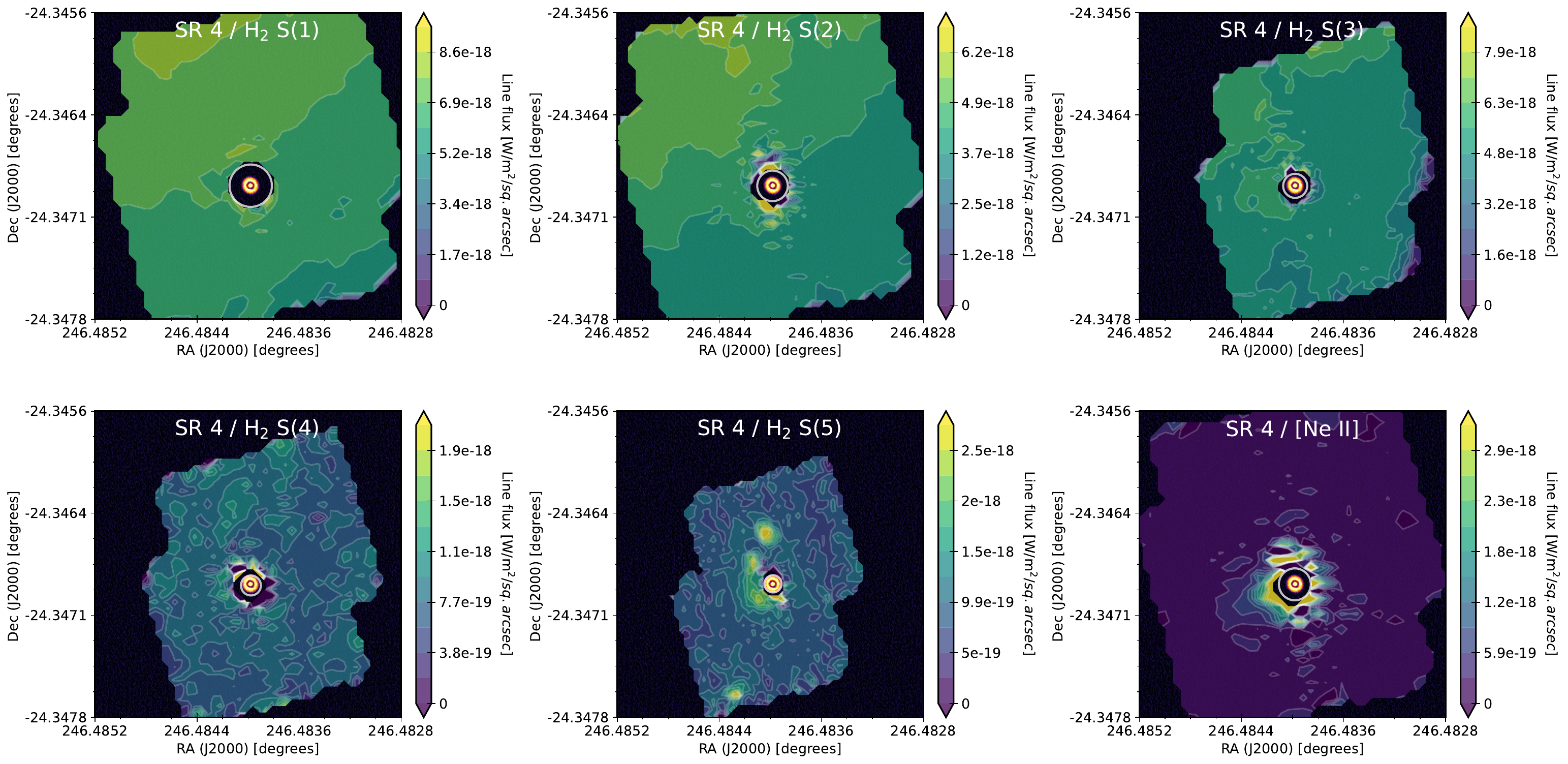}
\caption{\label{fig:sr4}}
\end{figure*}

\begin{figure*}[h]
\centering
\includegraphics[width=18cm]{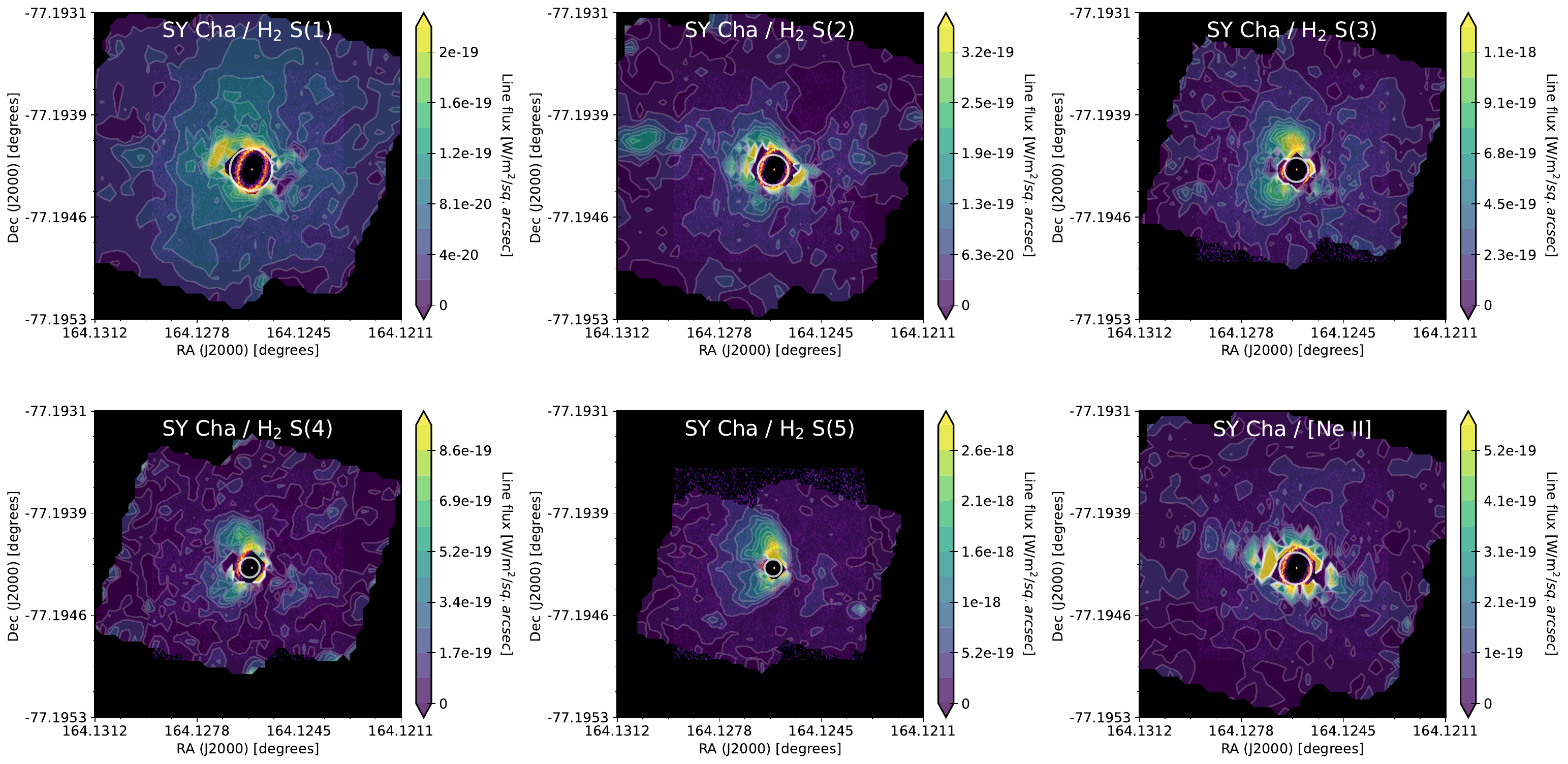}
\caption{\label{fig:sycha}}
\end{figure*}

\newpage

\begin{figure*}[h]
\centering
\includegraphics[width=18cm]{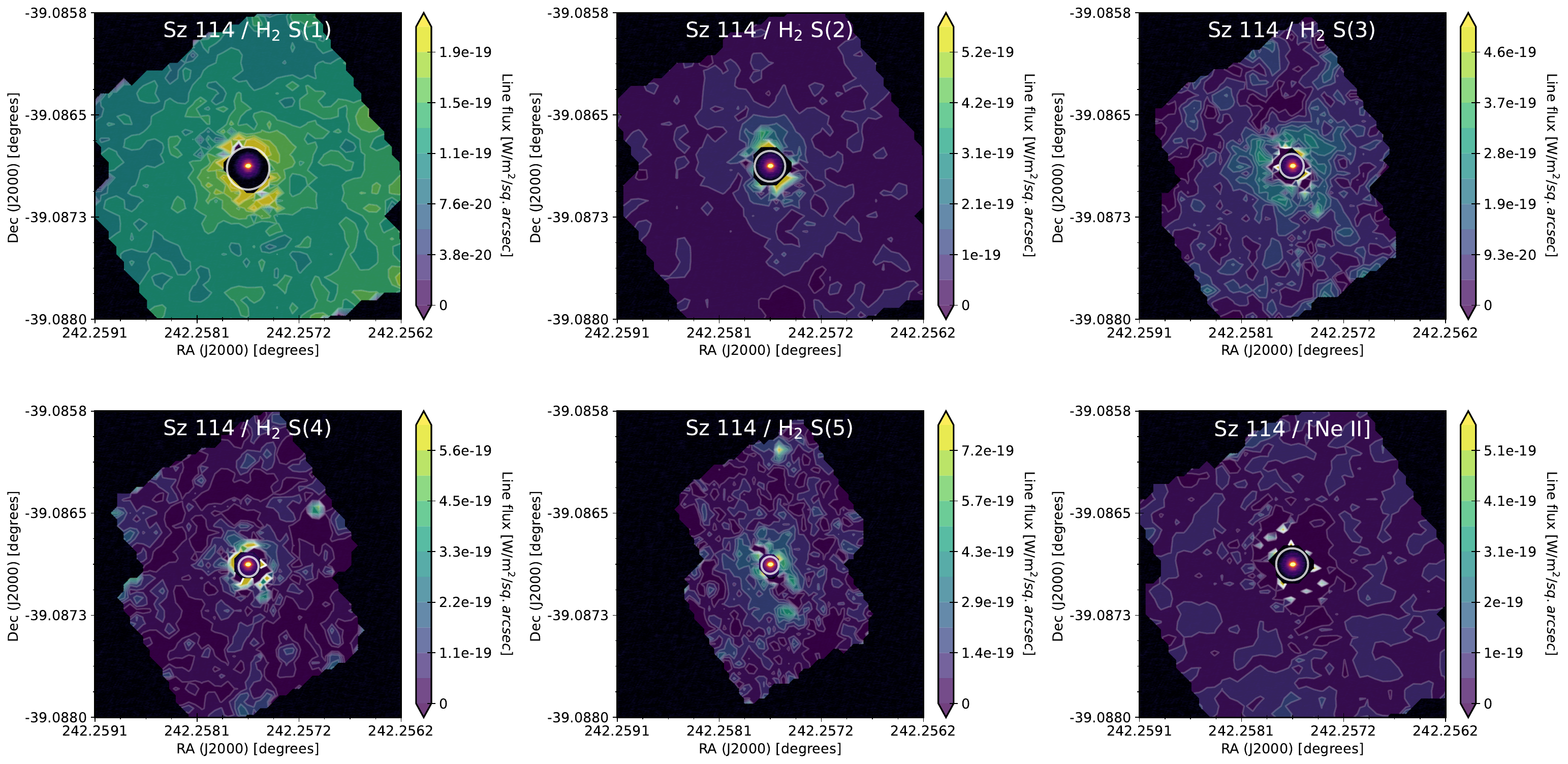}
\caption{\label{fig:sz114}}
\end{figure*}

\begin{figure*}[h]
\centering
\includegraphics[width=18cm]{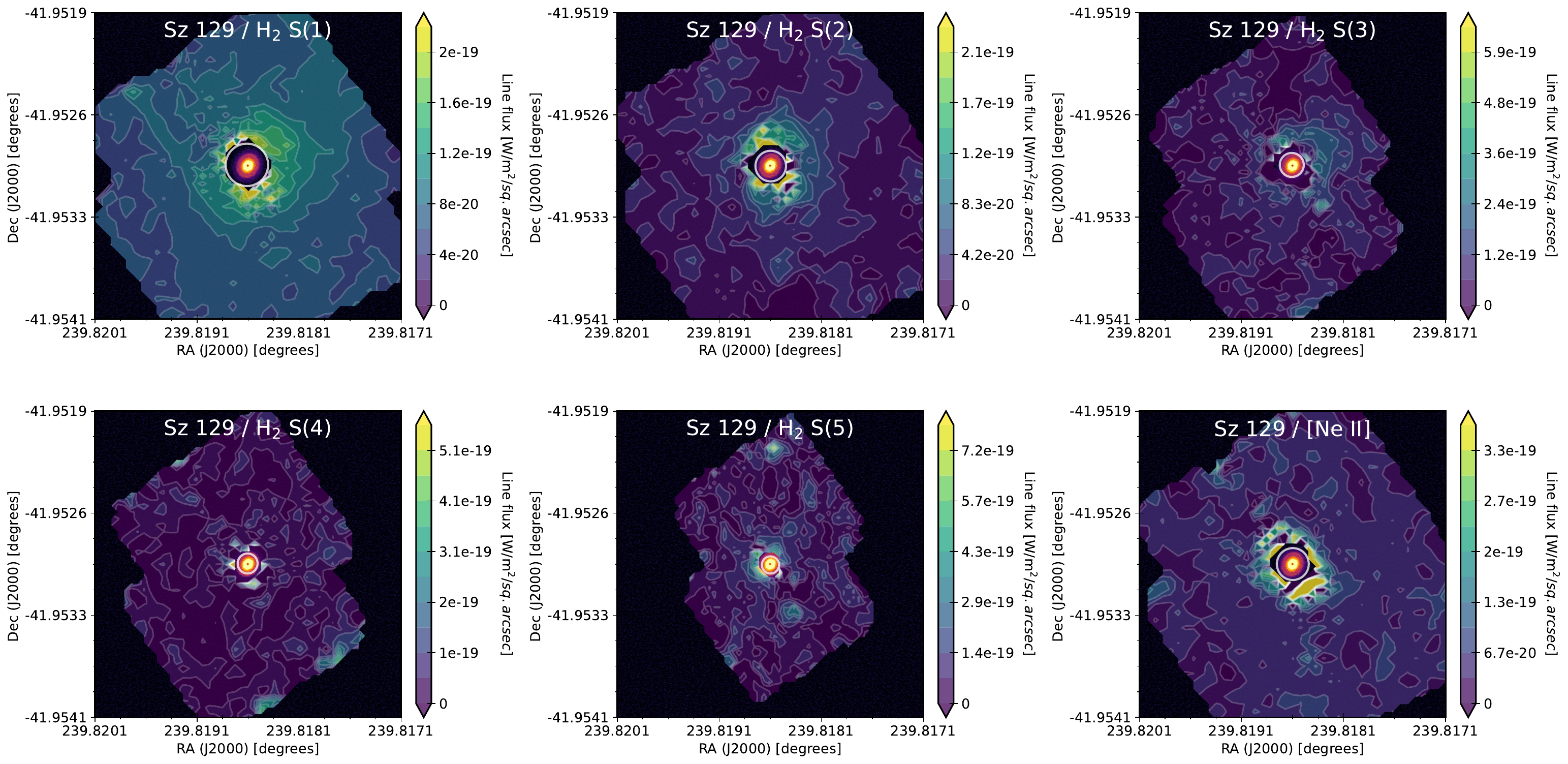}
\caption{\label{fig:sz129}}
\end{figure*}

\newpage

\begin{figure*}[h]
\centering
\includegraphics[width=18cm]{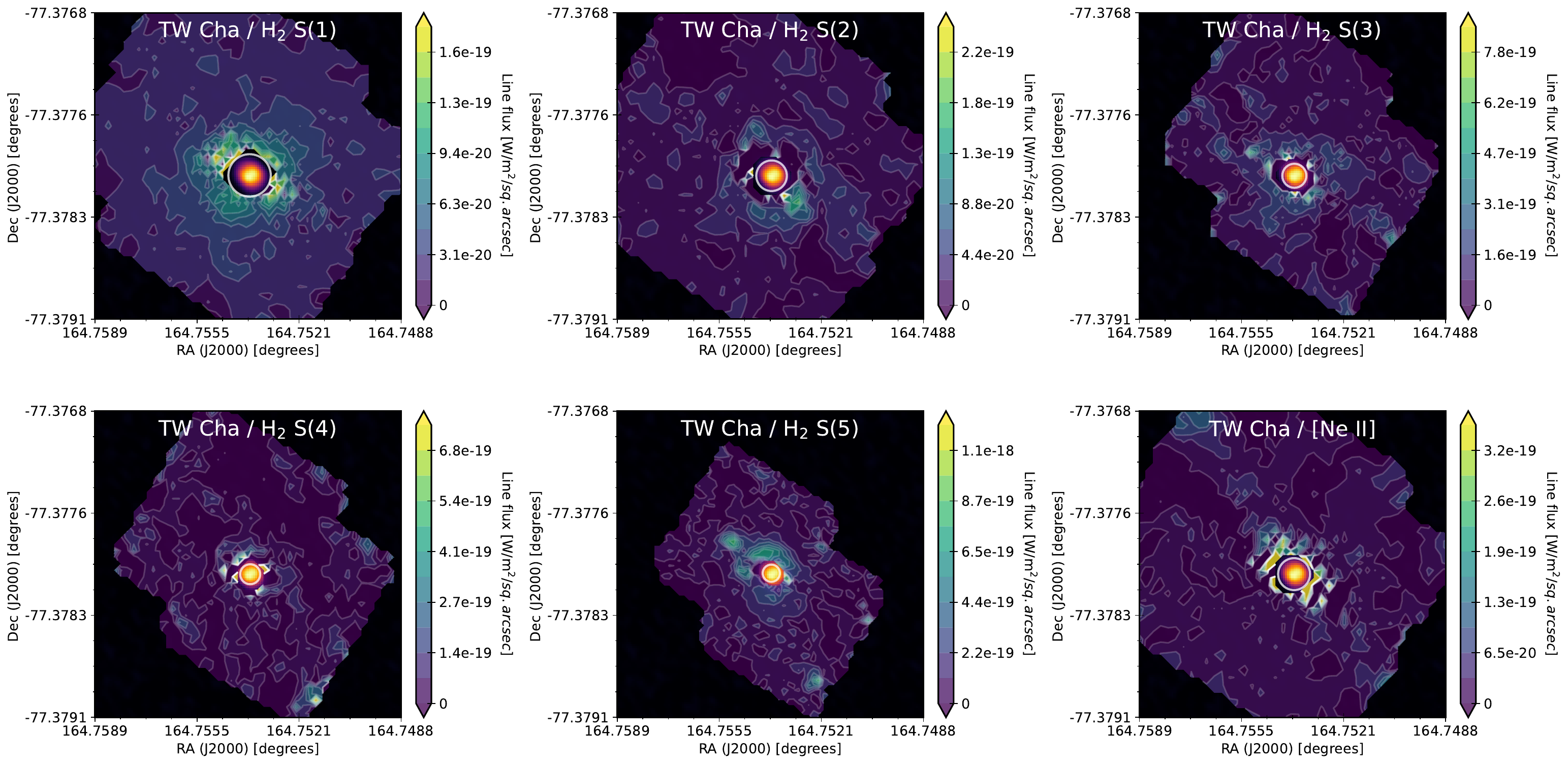}
\caption{\label{fig:twcha}}
\end{figure*}

\begin{figure*}[h]
\centering
\includegraphics[width=18cm]{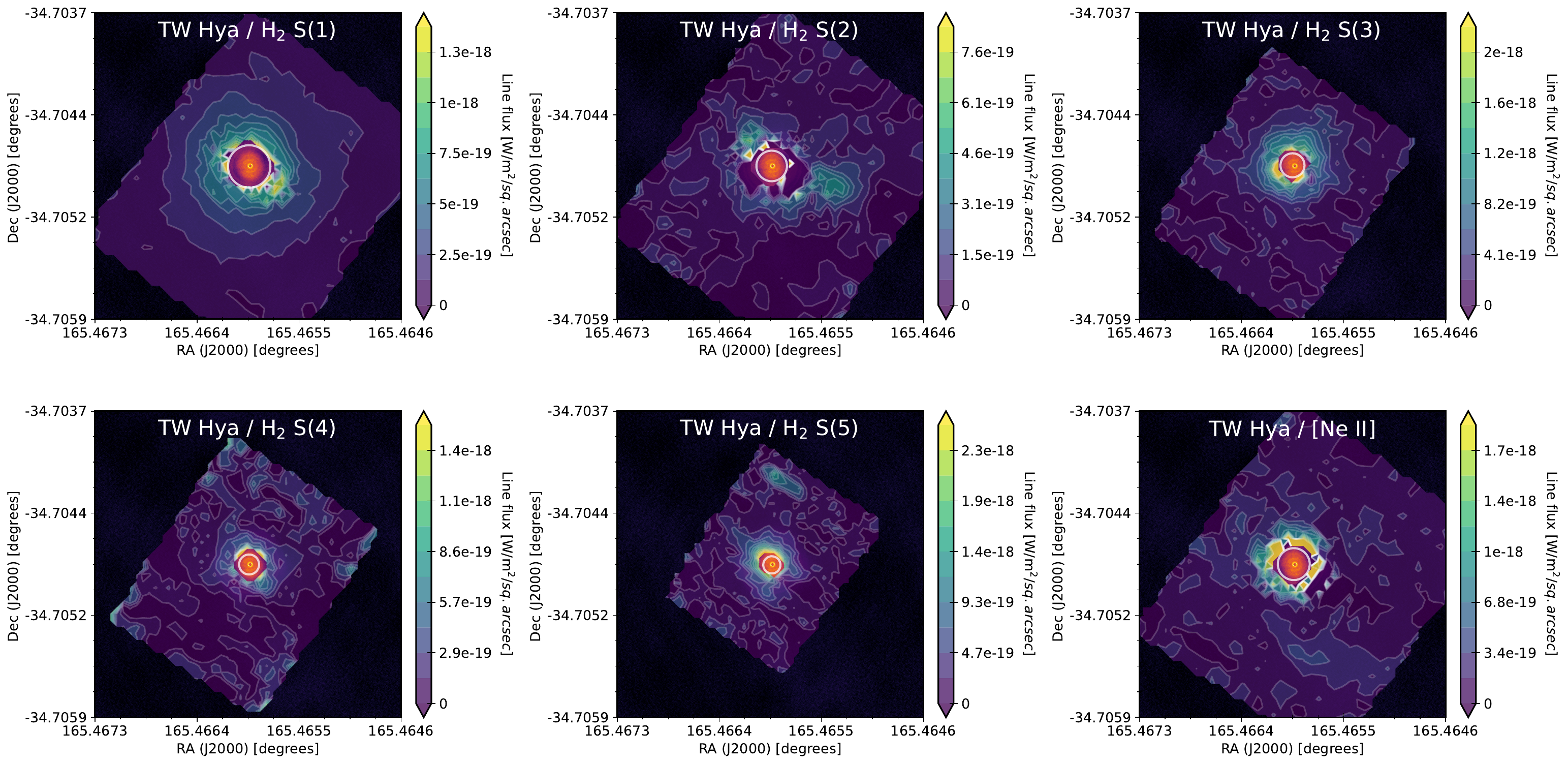}
\caption{\label{fig:twhya}}
\end{figure*}

\newpage

\begin{figure*}[h]
\centering
\includegraphics[width=18cm]{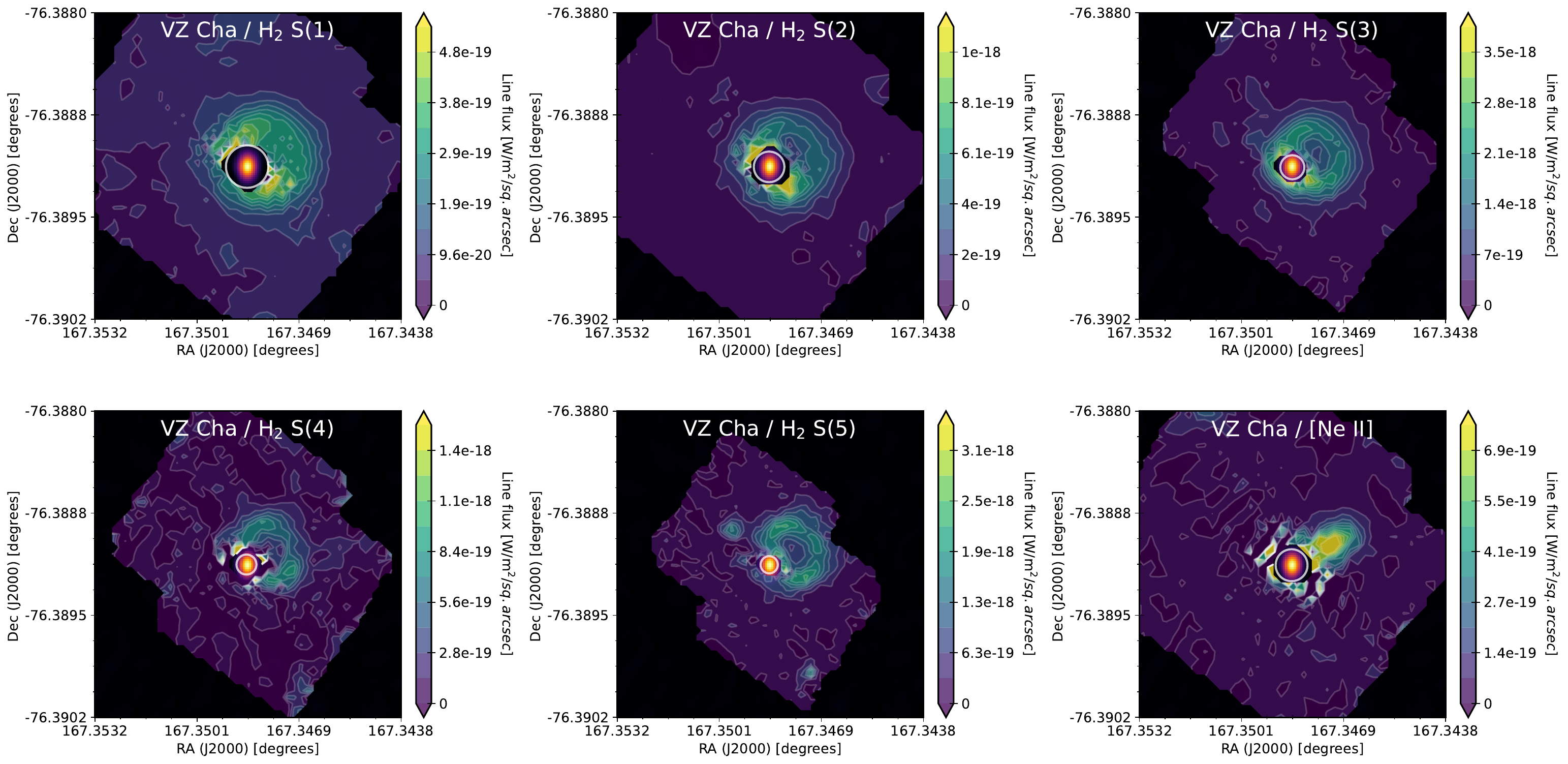}
\caption{\label{fig:vzcha}}
\end{figure*}

\begin{figure*}[h]
\centering
\includegraphics[width=18cm]{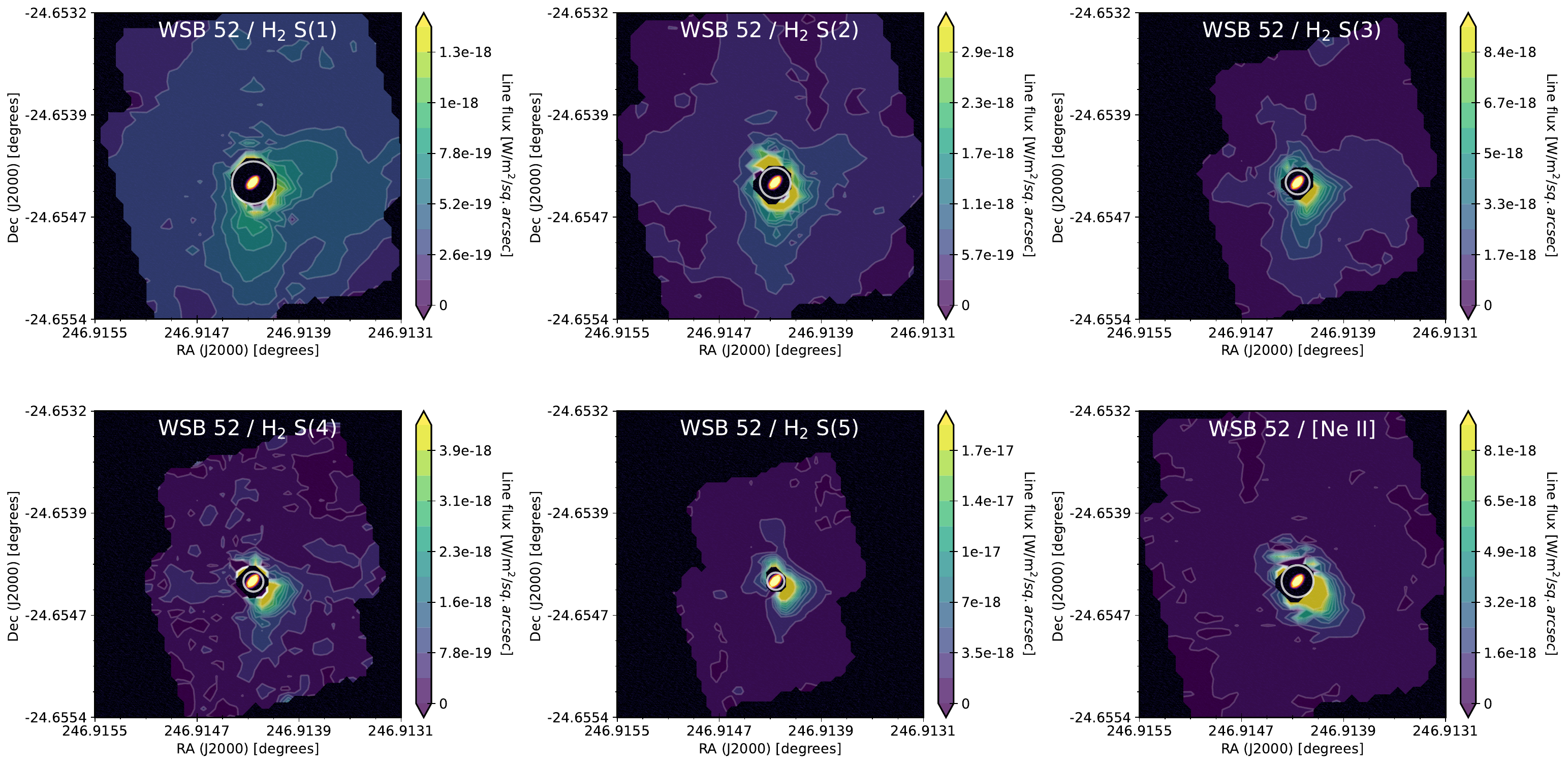}
\caption{\label{fig:wsb52}}
\end{figure*}

{\section{Rotation diagrams}
This section of the appendix contains the rotation diagrams for each of our targets.  
\begin{figure*}[h]
\centering
\includegraphics[width=0.385\linewidth]{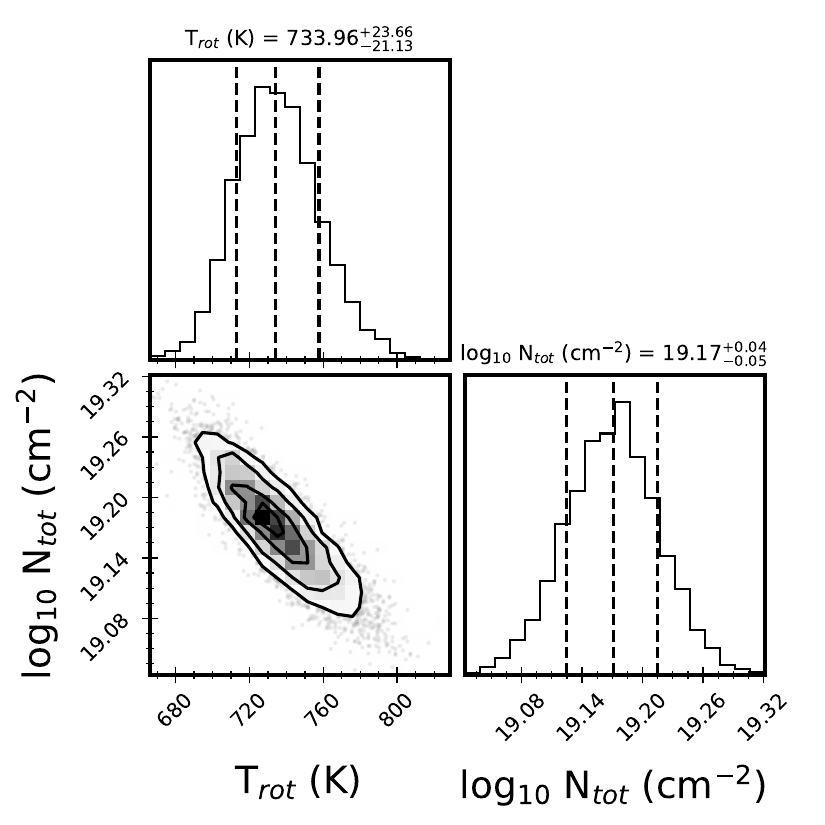}\includegraphics[width=0.385\linewidth]{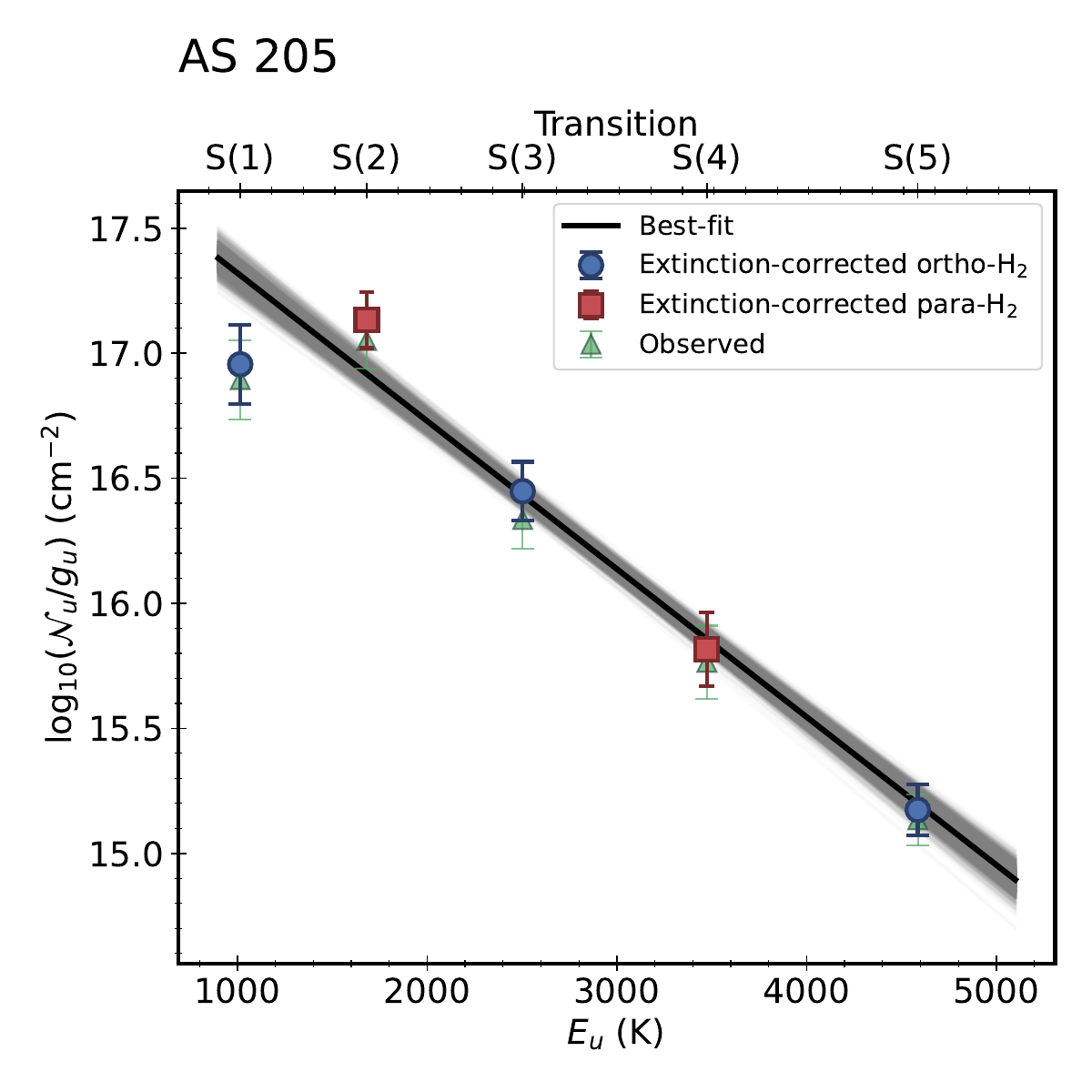}

\includegraphics[width=0.385\linewidth]{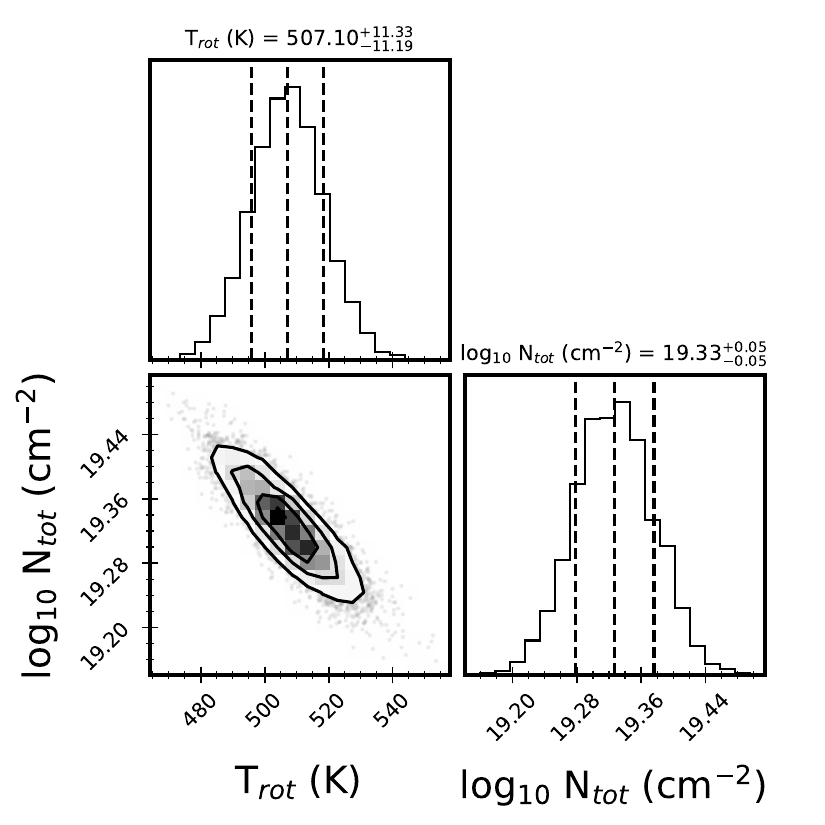} \includegraphics[width=0.385\linewidth]{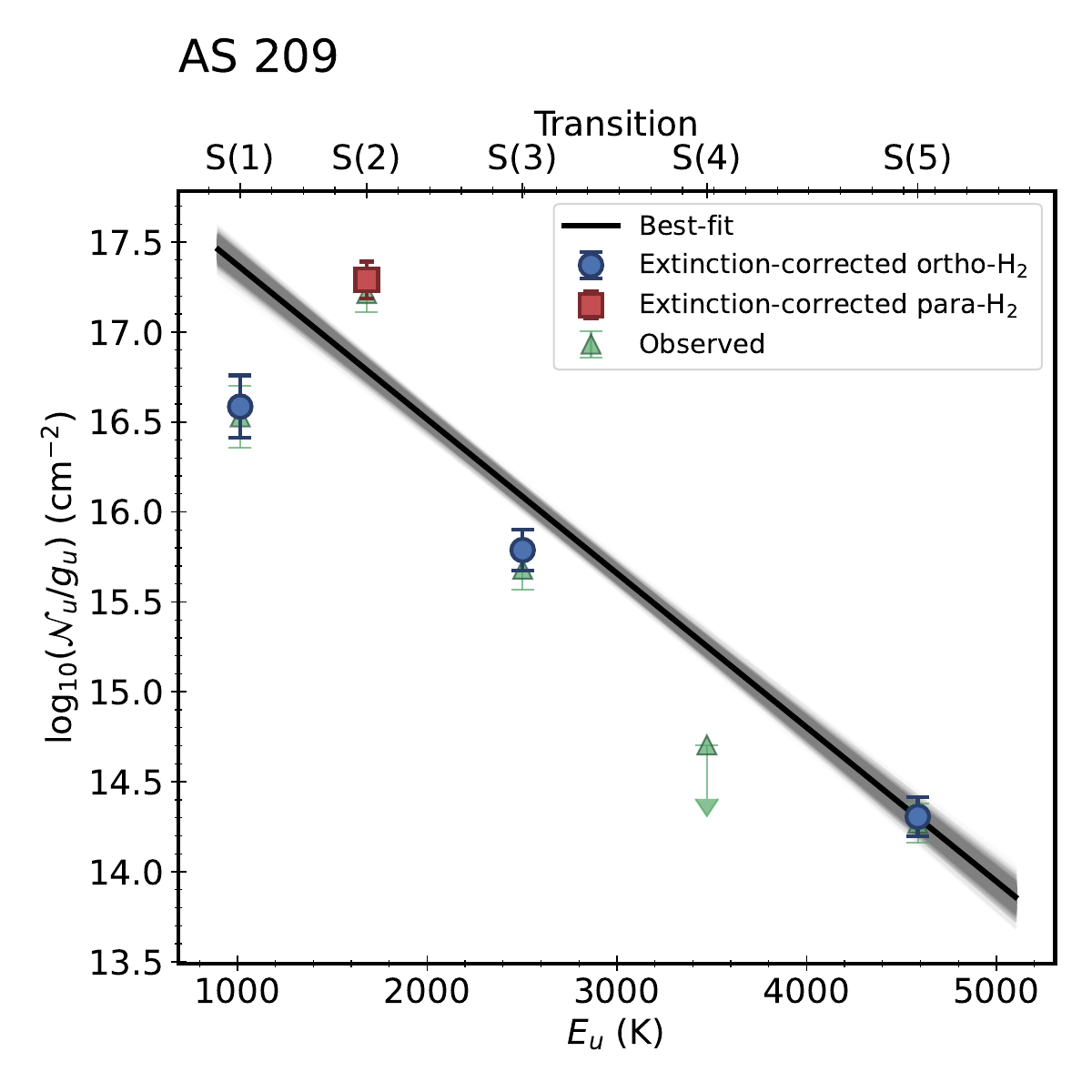} 

\includegraphics[width=0.385\linewidth]{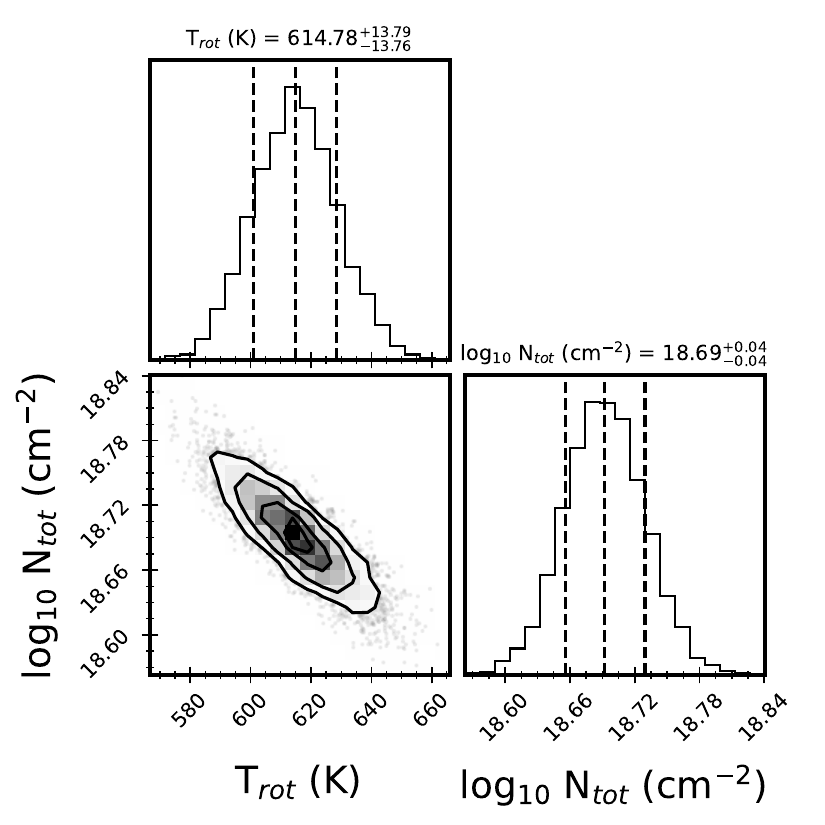}\includegraphics[width=0.385\linewidth]{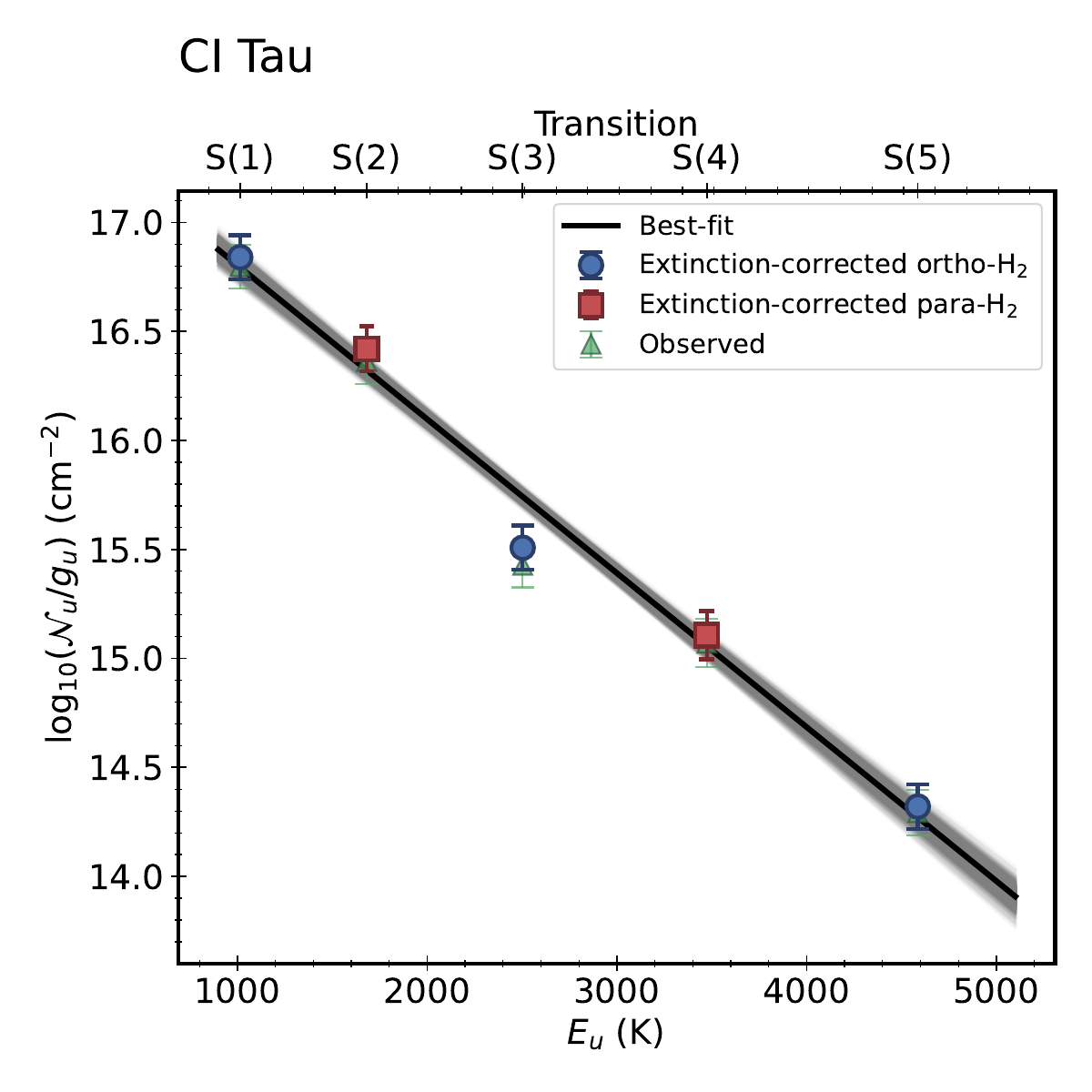}

\caption{(Left) Corner plot showing the posterior distributions and best-fit parameters from the MCMC fit to the rotation diagram. (Right) Rotation diagram with observed data points in green and extinction-corrected points in blue for ortho-H$_2$ and red for para-H$_2$. The black line denotes the best-fit model, and the shaded grey region represents the uncertainty range derived from the MCMC fitting. \label{rotfig1}} 
\end{figure*}

\begin{figure*}[h]
\centering
\includegraphics[width=0.385\linewidth]{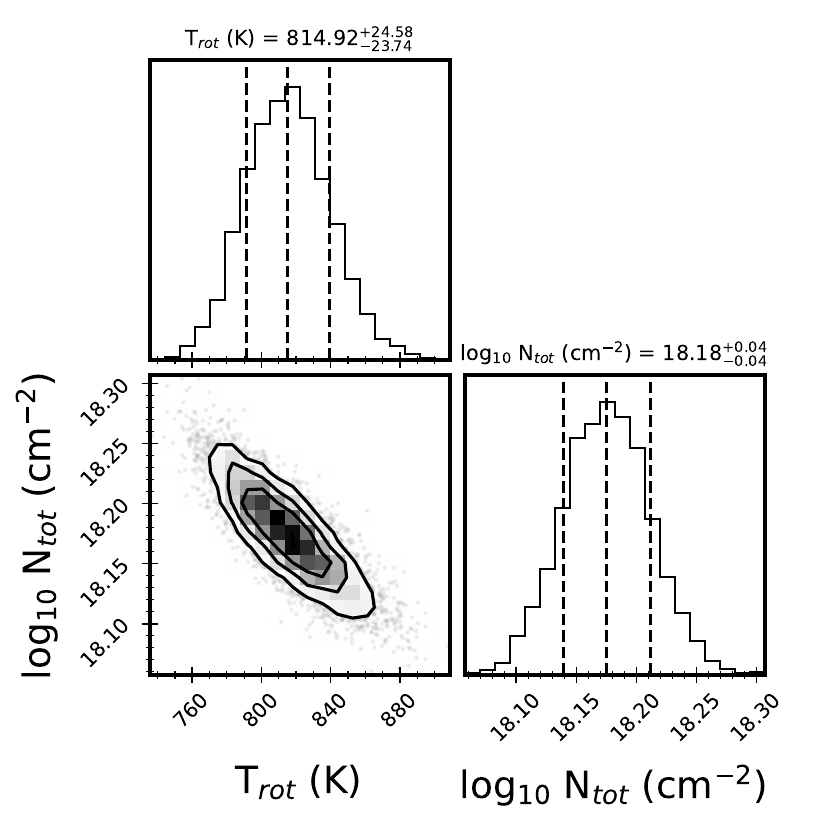}\includegraphics[width=0.385\linewidth]{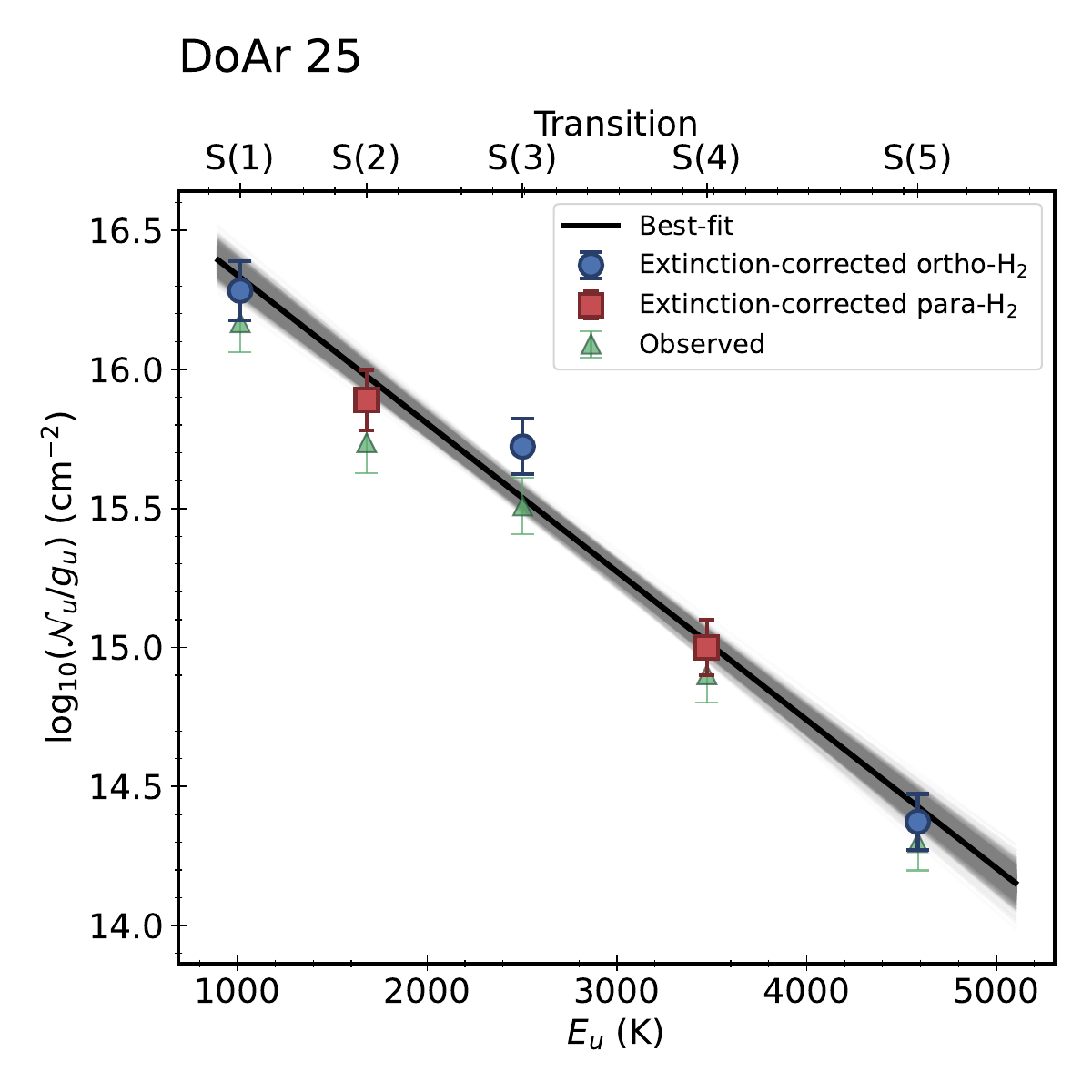}

\includegraphics[width=0.385\linewidth]{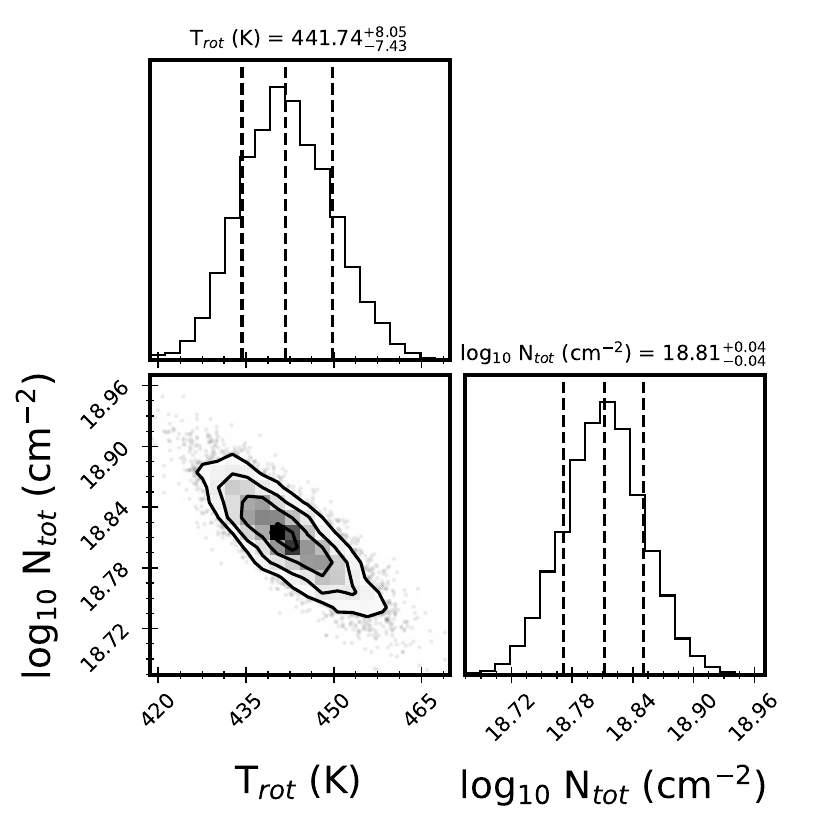}\includegraphics[width=0.385\linewidth]{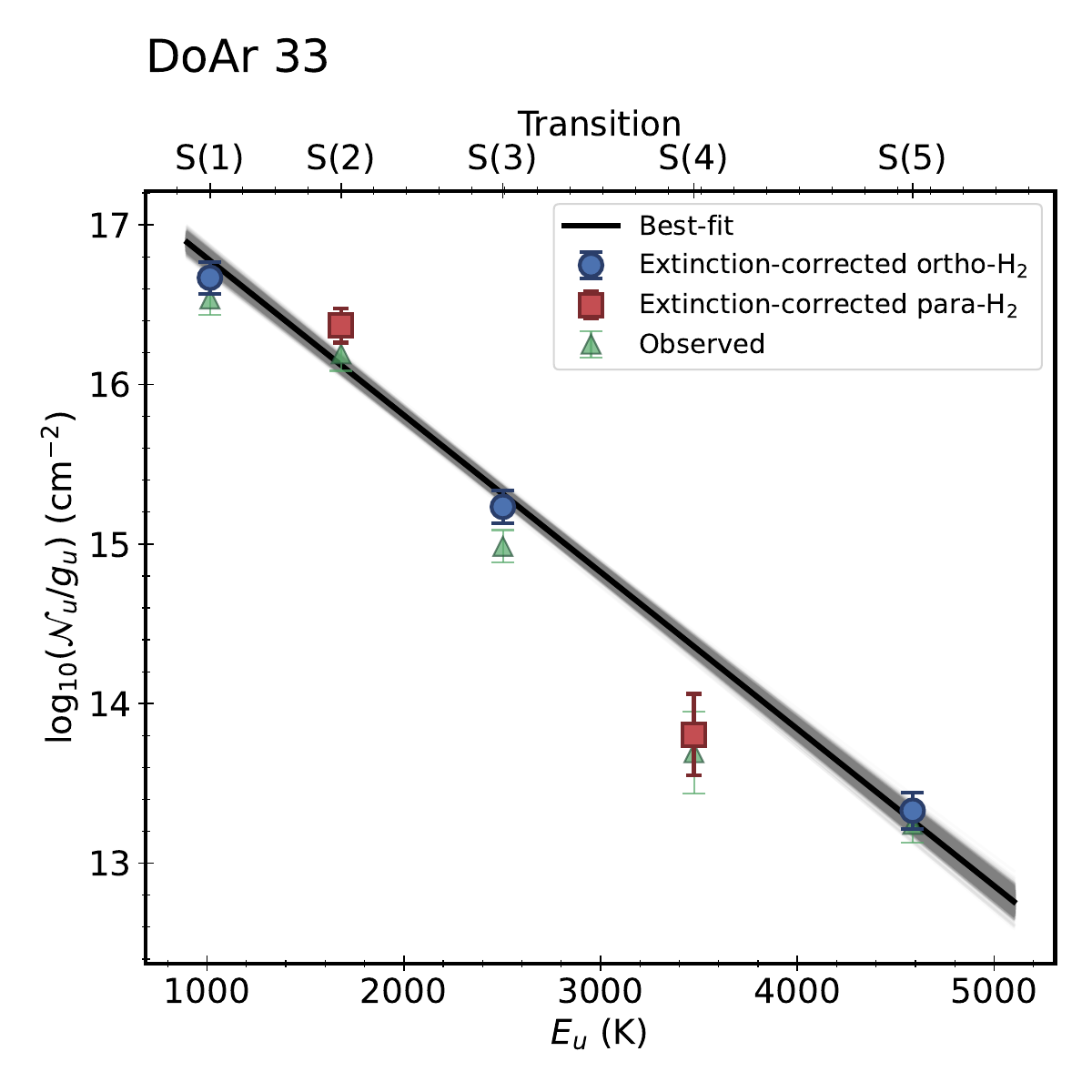}

\includegraphics[width=0.385\linewidth]{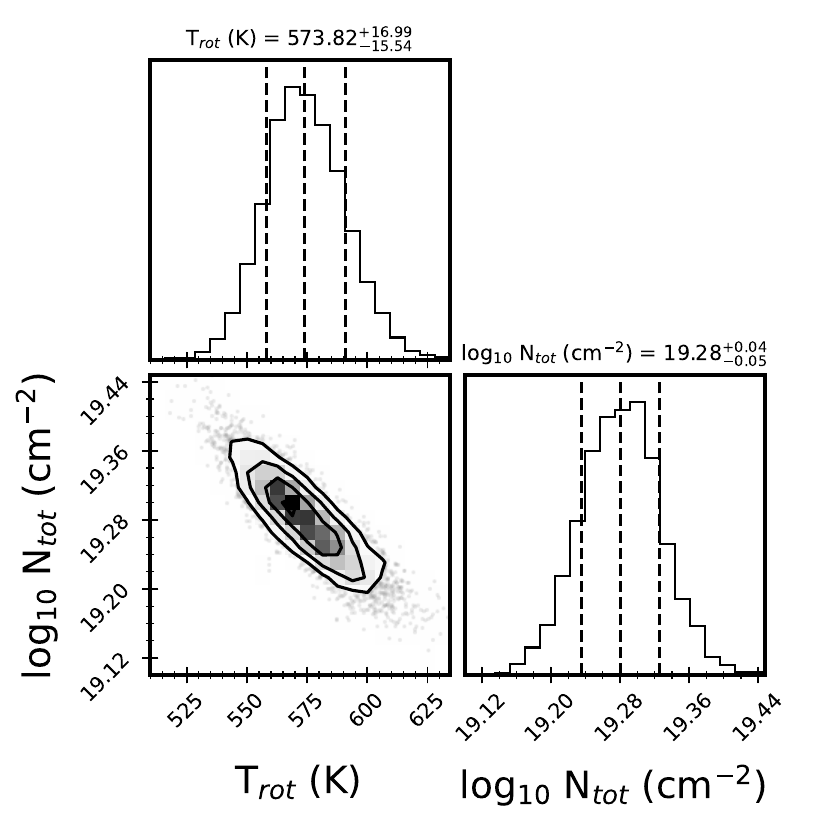} \includegraphics[width=0.385\linewidth]{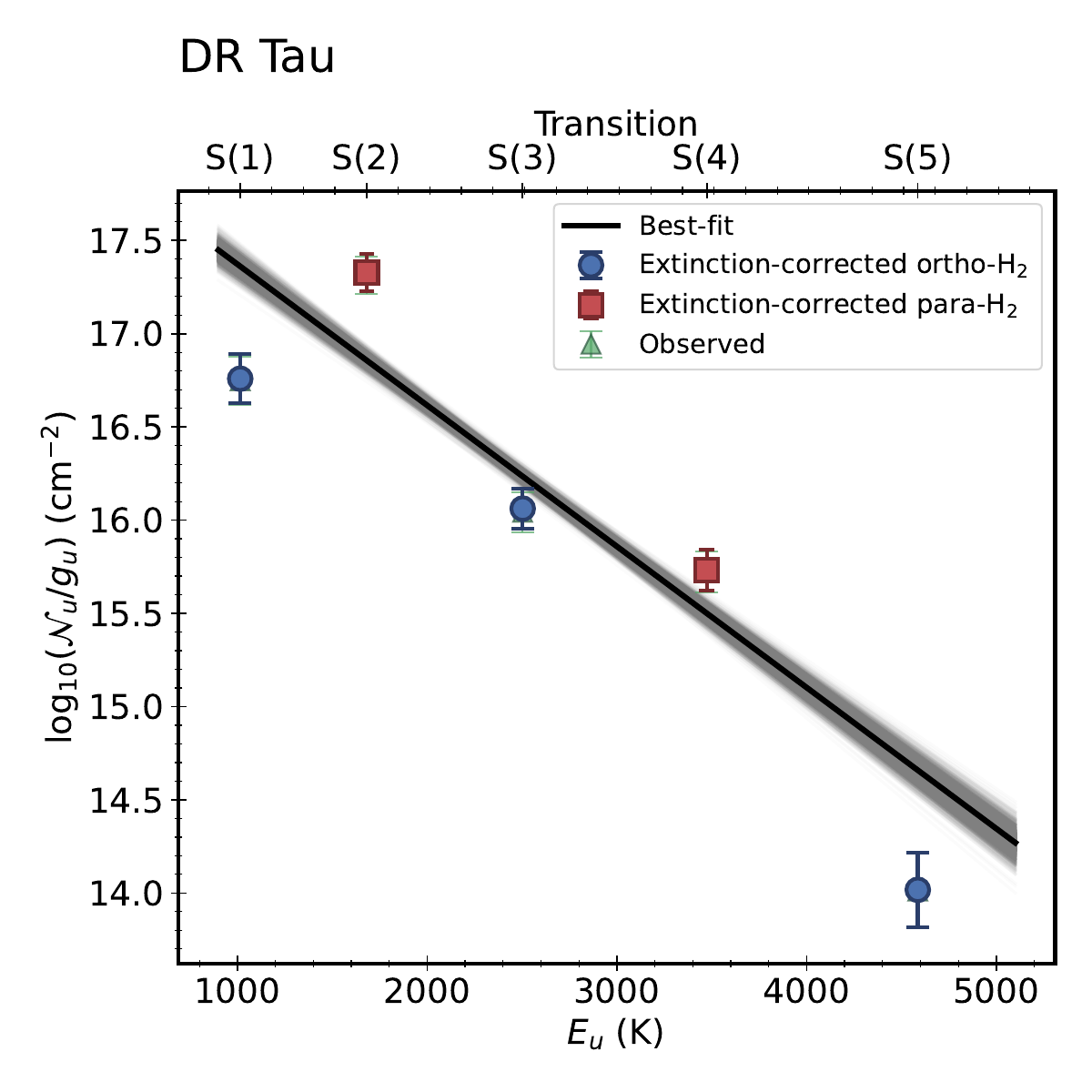} 

\caption{Left) Corner plot showing the posterior distributions and best-fit parameters from the MCMC fit to the rotation diagram. (Right) Rotation diagram with observed data points in green and extinction-corrected points in blue for ortho-H$_2$ and red for para-H$_2$. The black line denotes the best-fit model, and the shaded grey region represents the uncertainty range derived from the MCMC fitting.\label{rotfig2}} 
\end{figure*}

\begin{figure*}[h]
\centering
\includegraphics[width=0.385\linewidth]{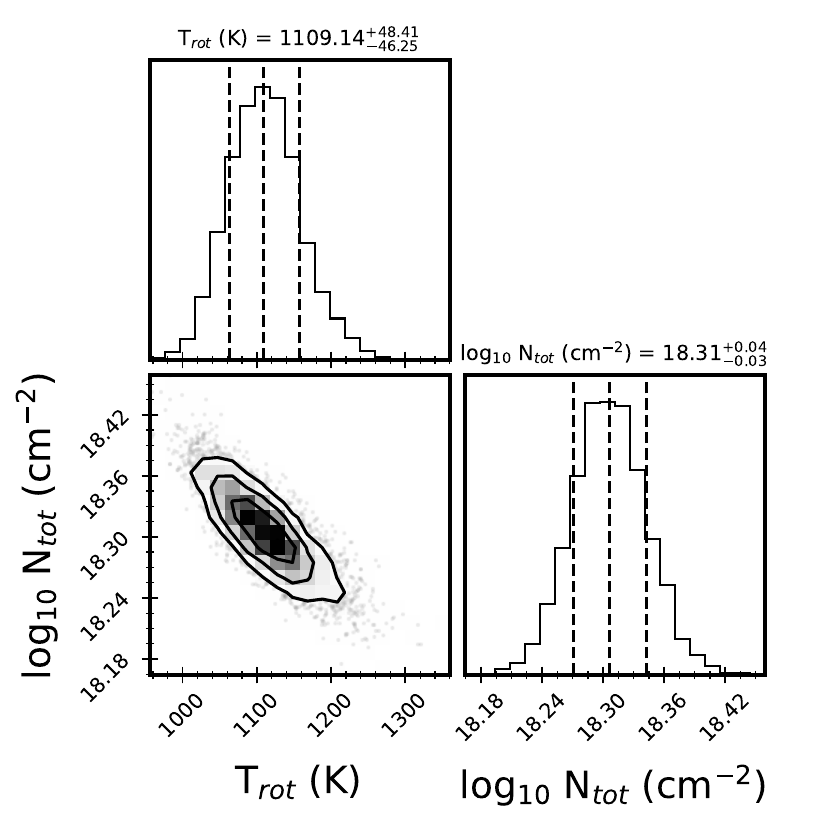}\includegraphics[width=0.385\linewidth]{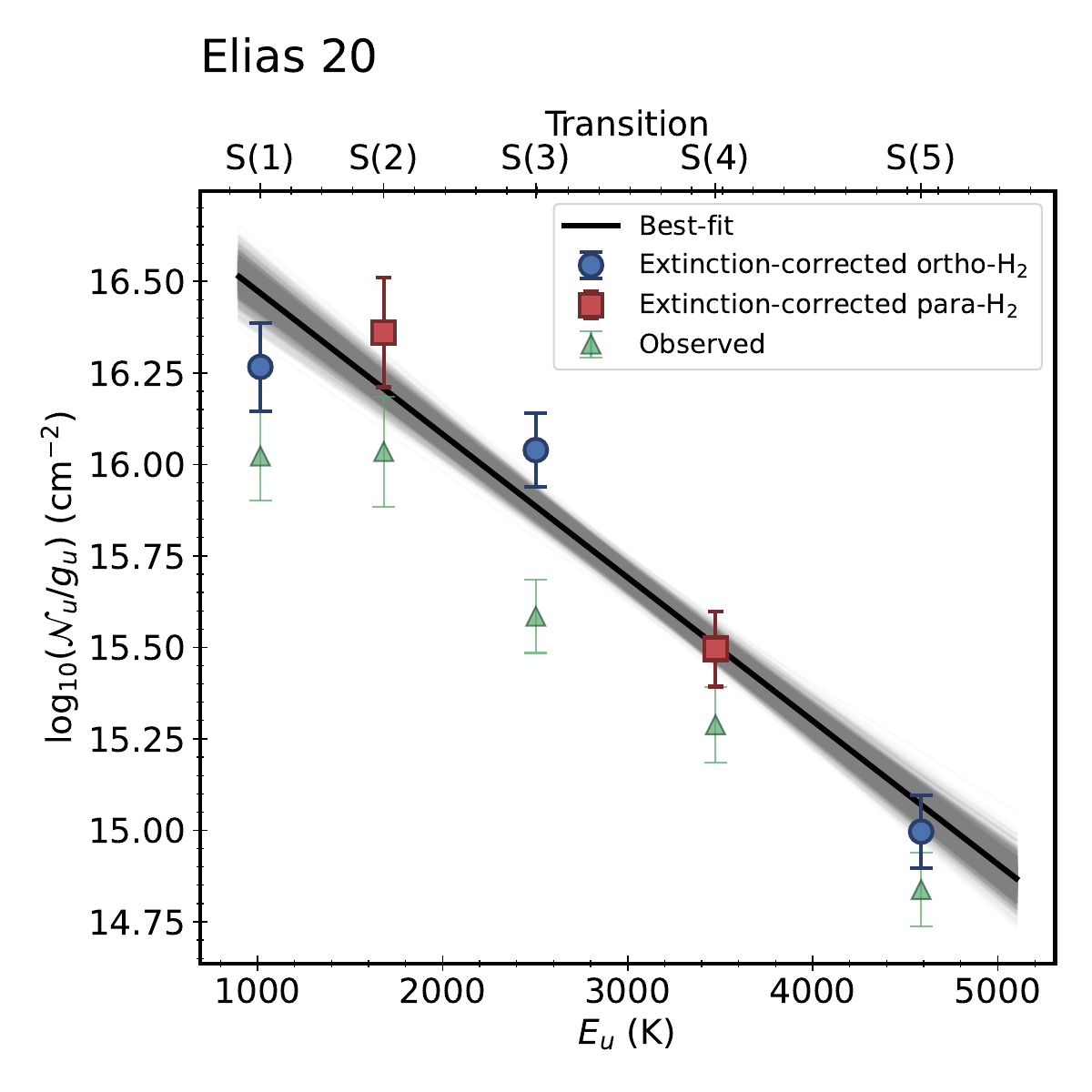}

\includegraphics[width=0.385\linewidth]{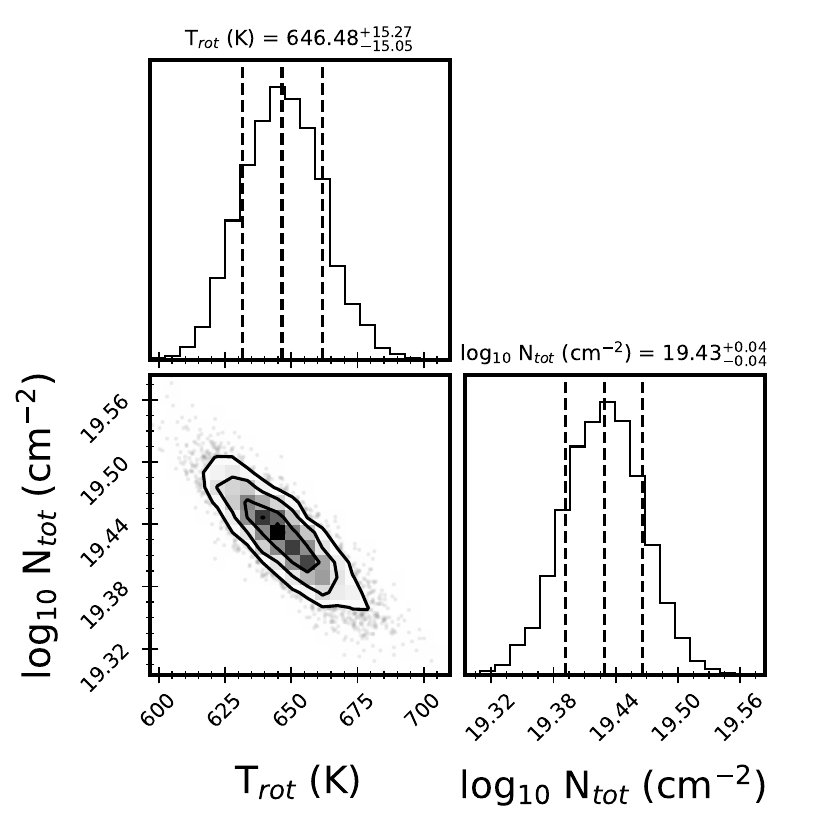} \includegraphics[width=0.385\linewidth]{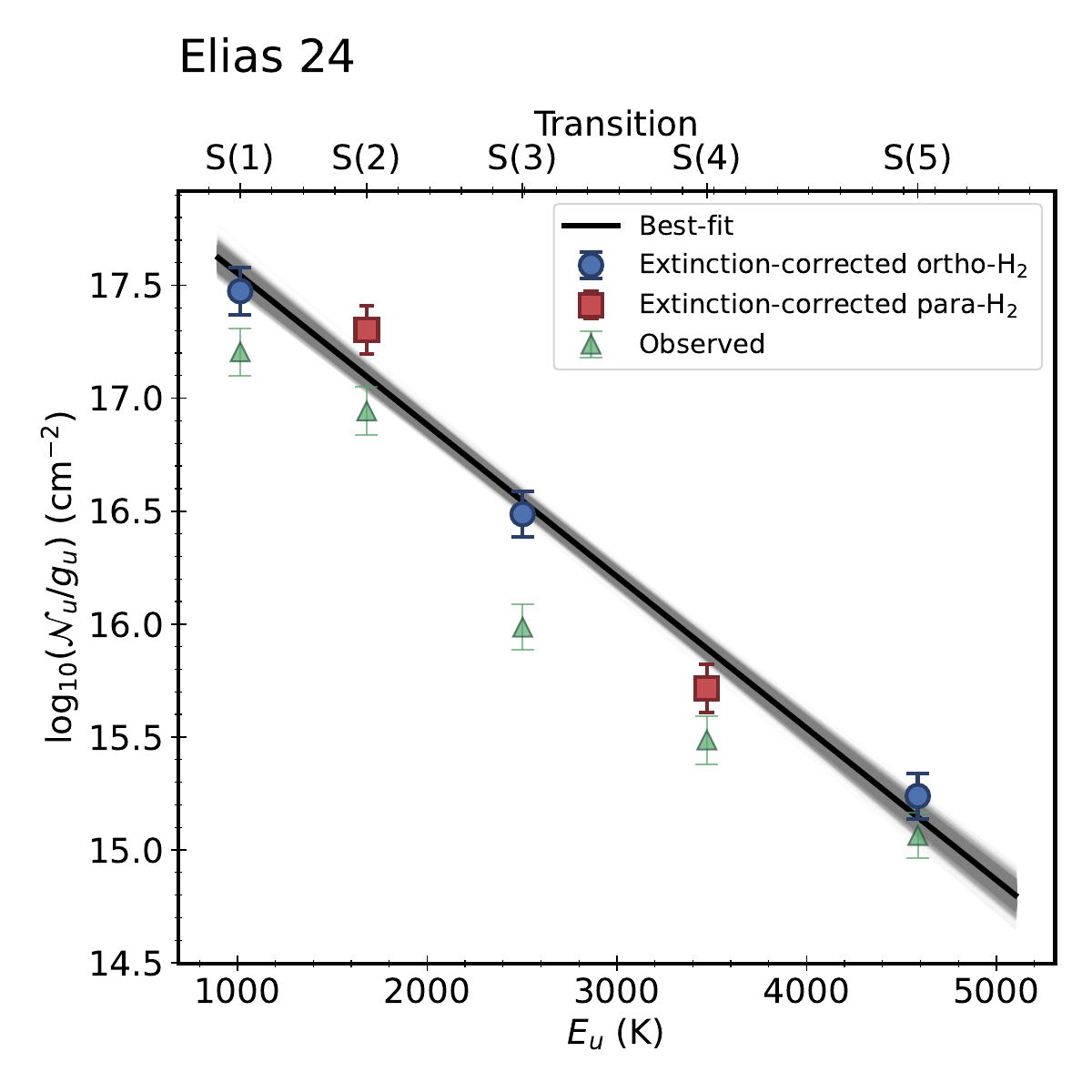} 

\includegraphics[width=0.385\linewidth]{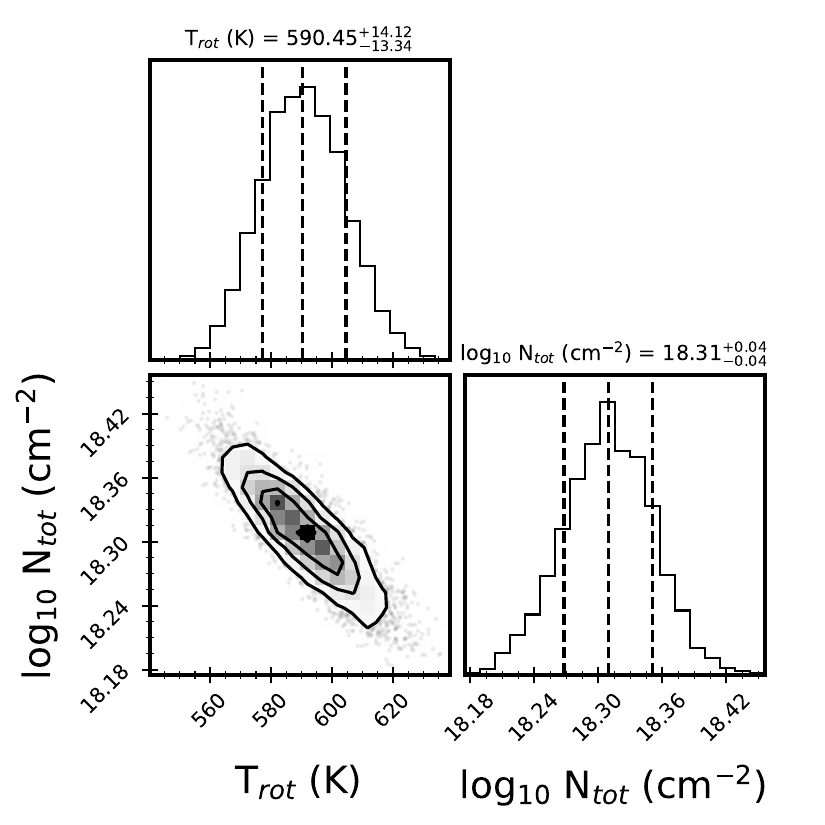}\includegraphics[width=0.385\linewidth]{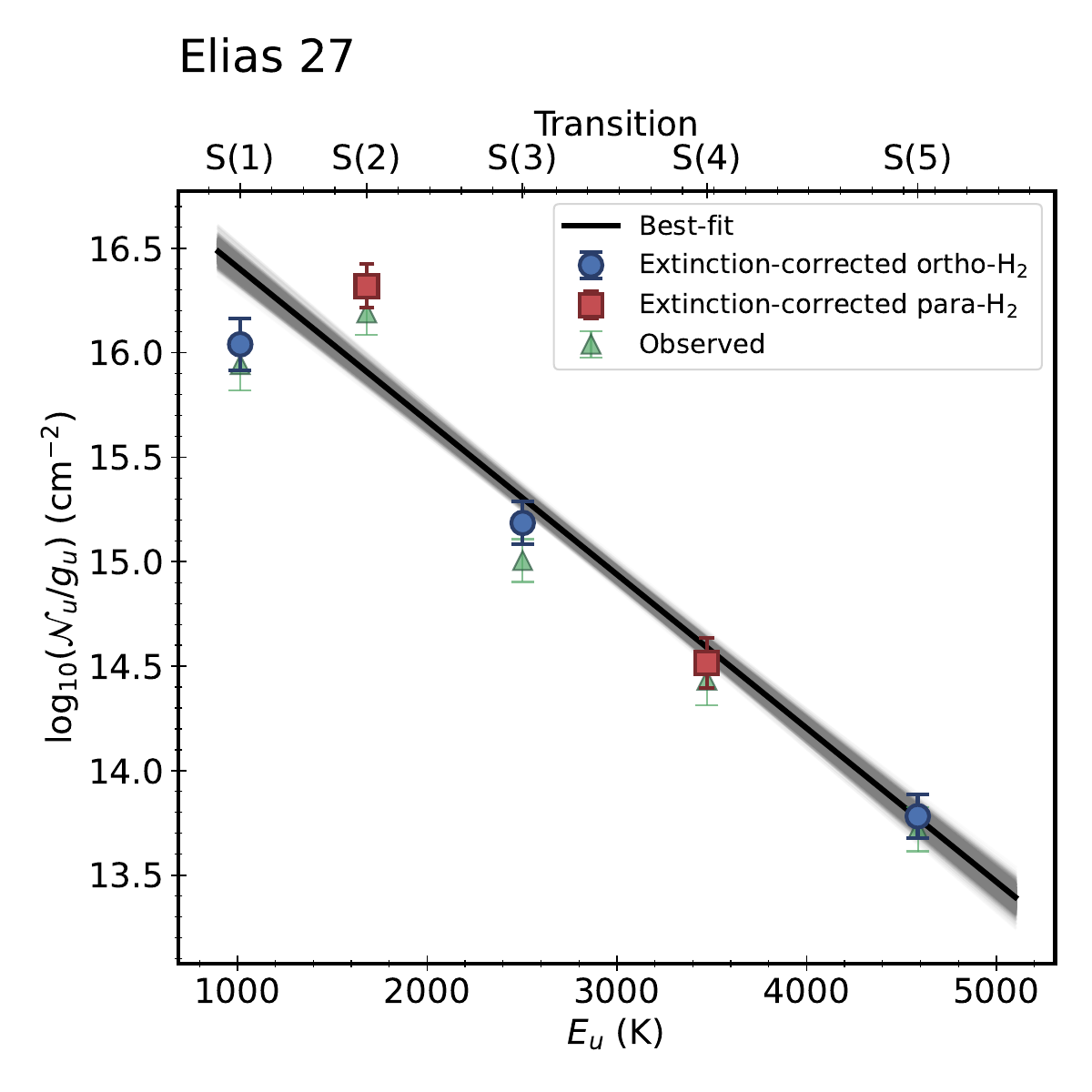}

\caption{Left) Corner plot showing the posterior distributions and best-fit parameters from the MCMC fit to the rotation diagram. (Right) Rotation diagram with observed data points in green and extinction-corrected points in blue for ortho-H$_2$ and red for para-H$_2$. The black line denotes the best-fit model, and the shaded grey region represents the uncertainty range derived from the MCMC fitting.\label{rotfig3}} 
\end{figure*}

\begin{figure*}[h]
\centering
\includegraphics[width=0.385\linewidth]{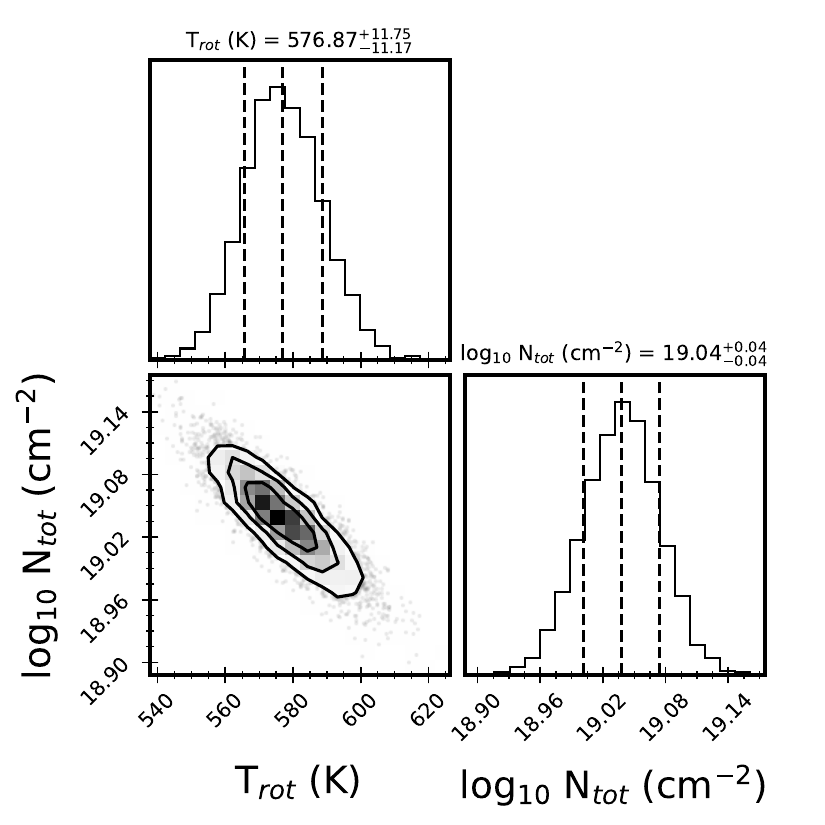}\includegraphics[width=0.385\linewidth]{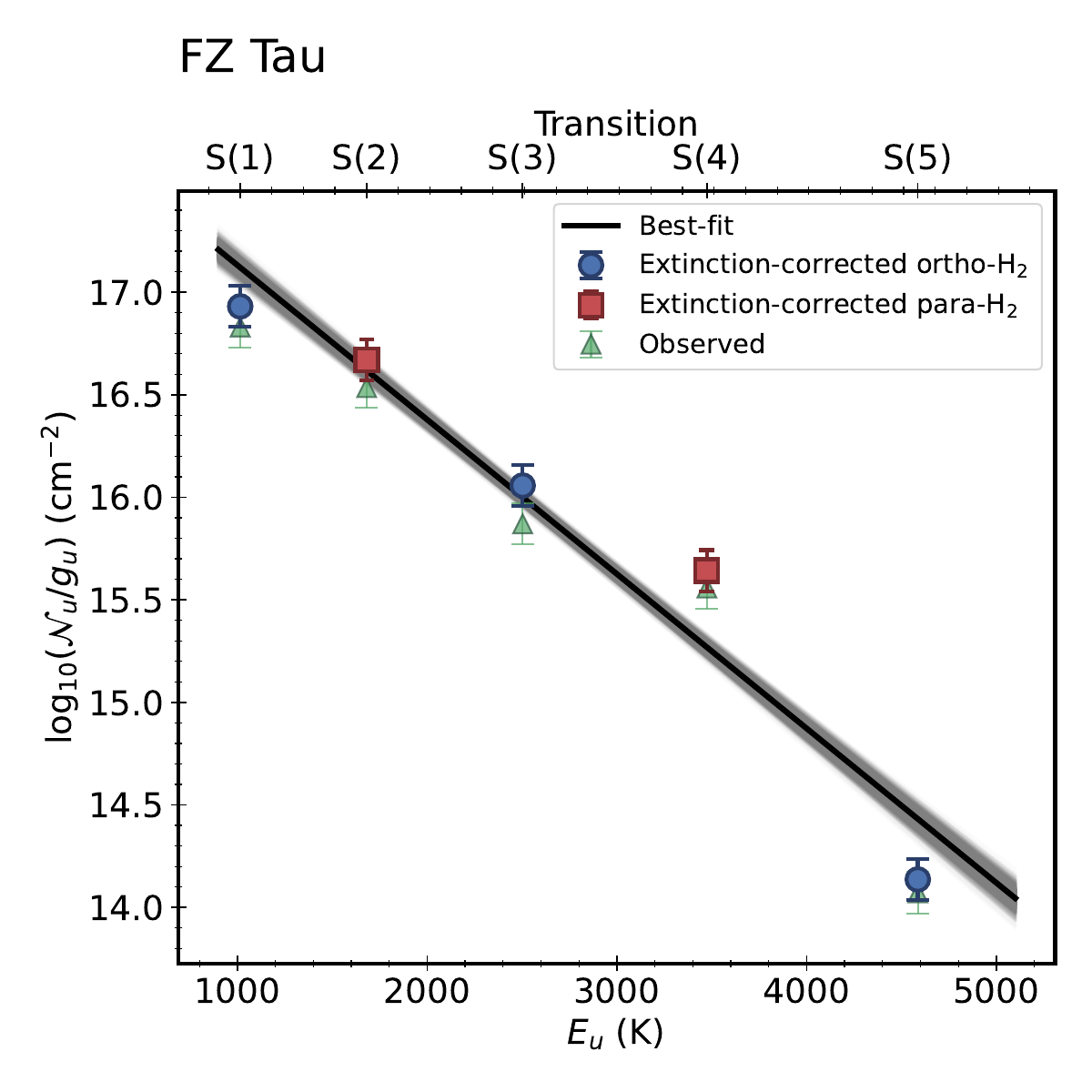}

\includegraphics[width=0.385\linewidth]{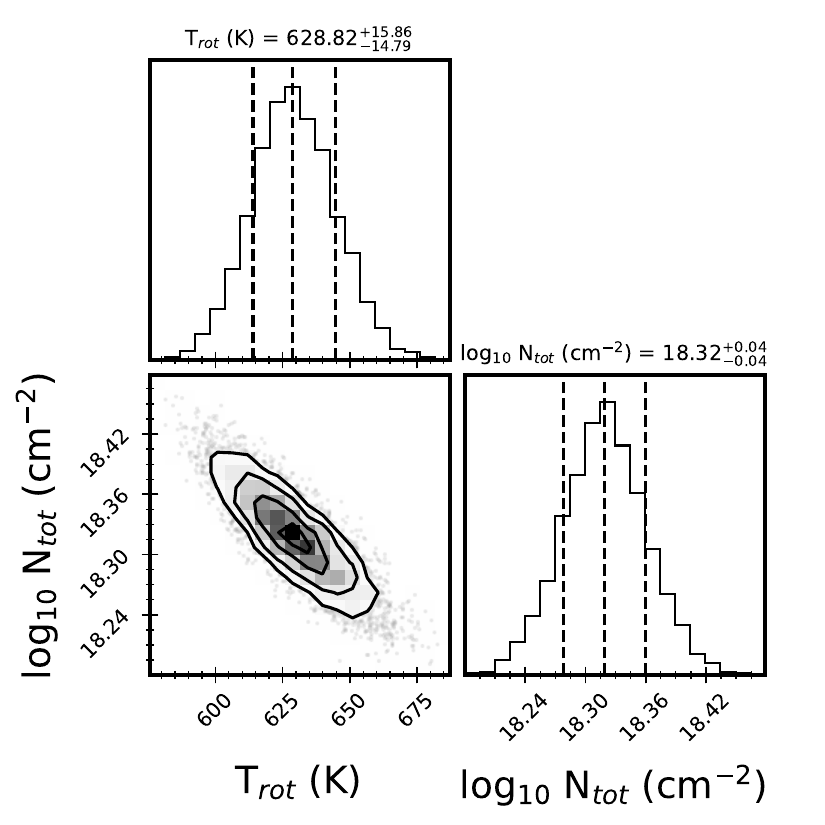}\includegraphics[width=0.385\linewidth]{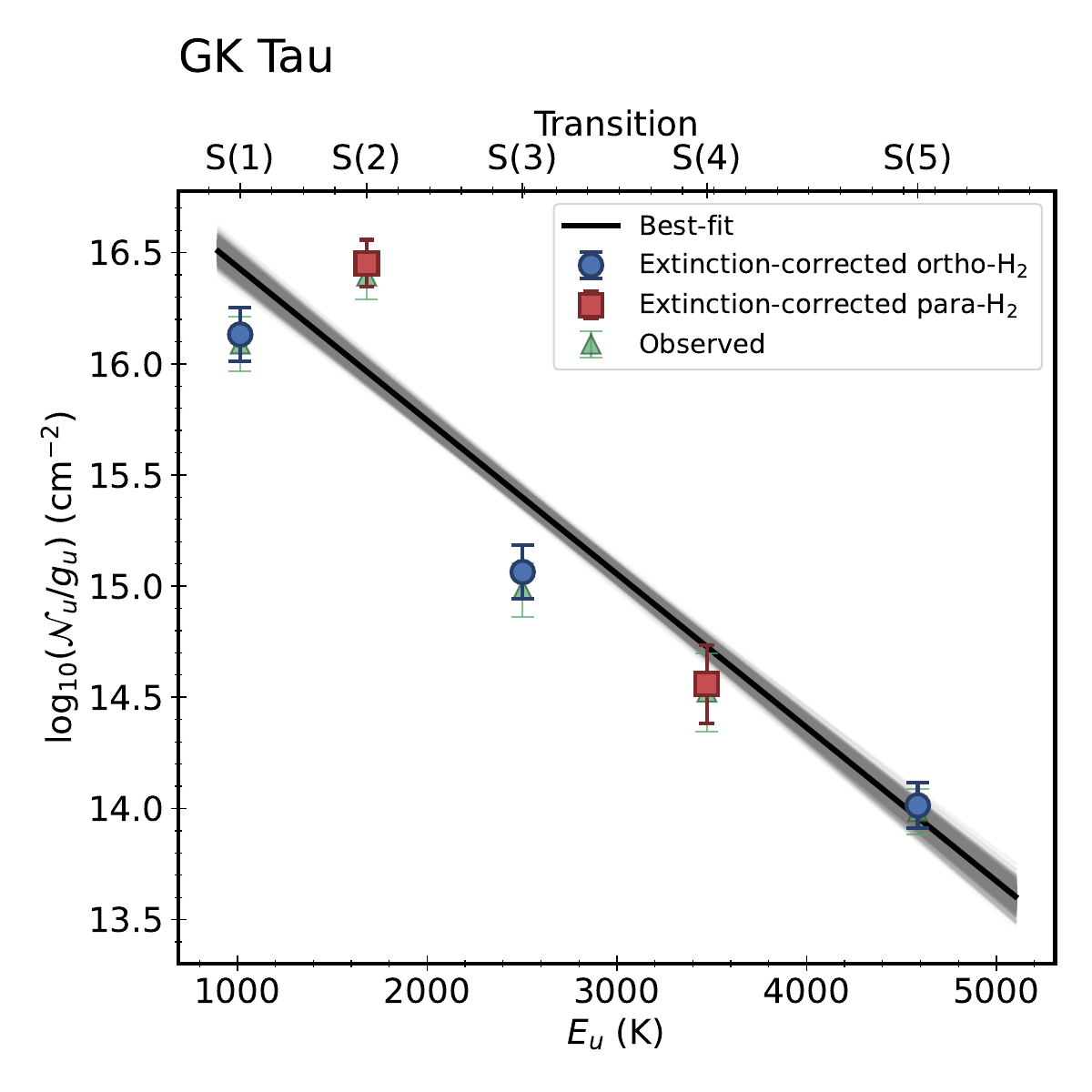}

\includegraphics[width=0.385\linewidth]{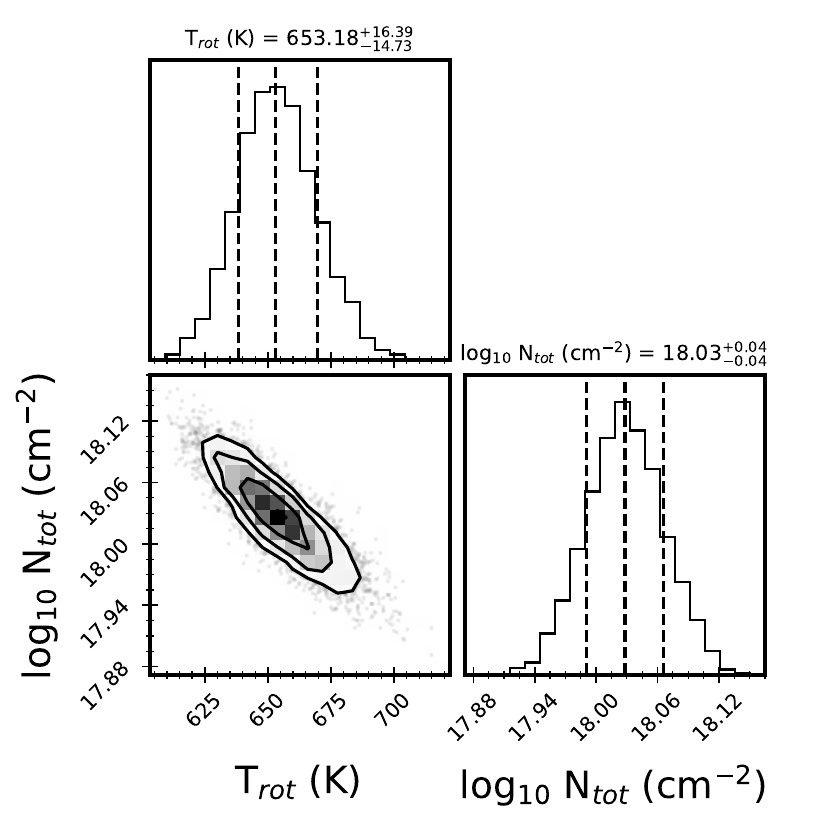}\includegraphics[width=0.385\linewidth]{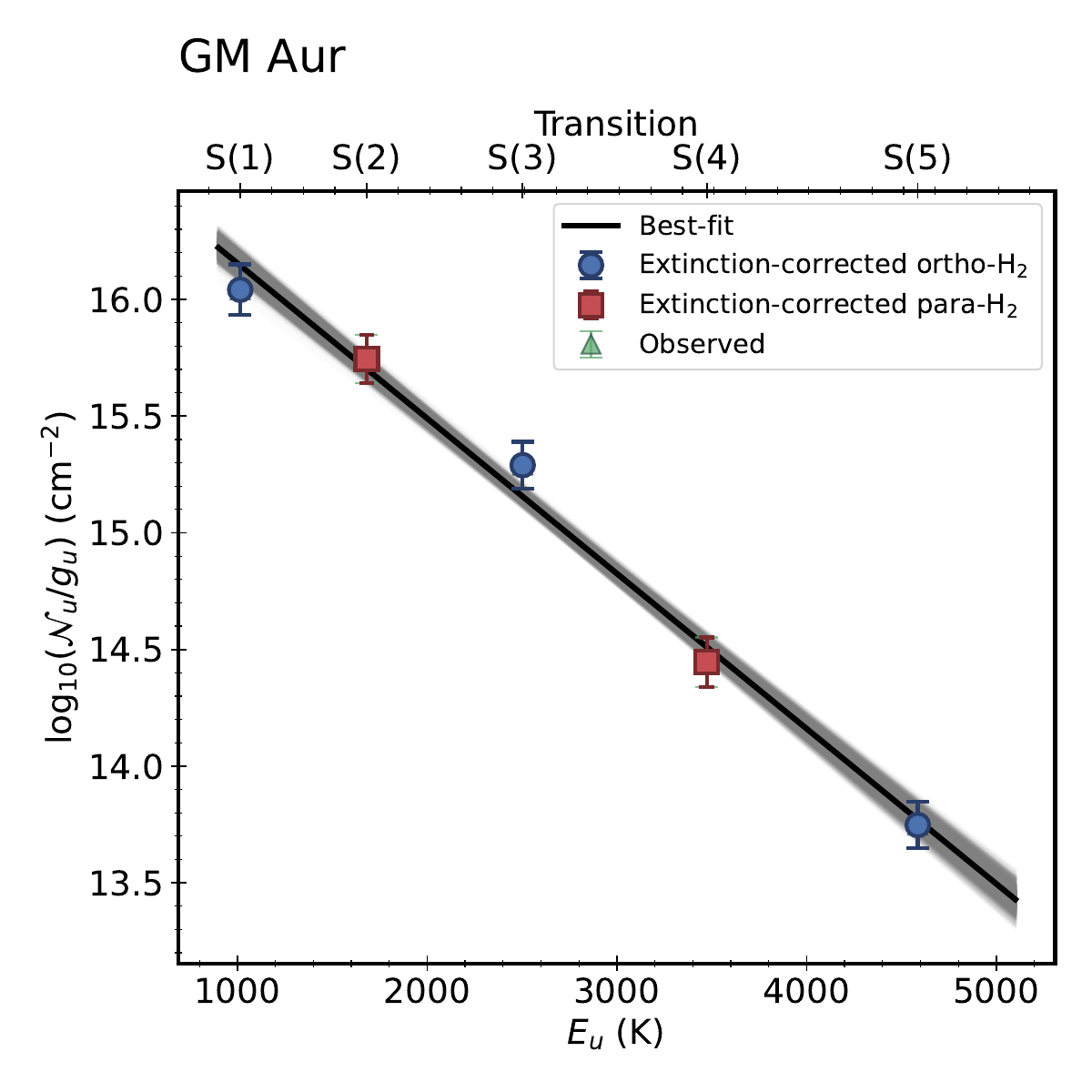}
\caption{Left) Corner plot showing the posterior distributions and best-fit parameters from the MCMC fit to the rotation diagram. (Right) Rotation diagram with observed data points in green and extinction-corrected points in blue for ortho-H$_2$ and red for para-H$_2$. The black line denotes the best-fit model, and the shaded grey region represents the uncertainty range derived from the MCMC fitting.\label{rotfig4}} 
\end{figure*}

\begin{figure*}[h]
\centering
\includegraphics[width=0.385\linewidth]{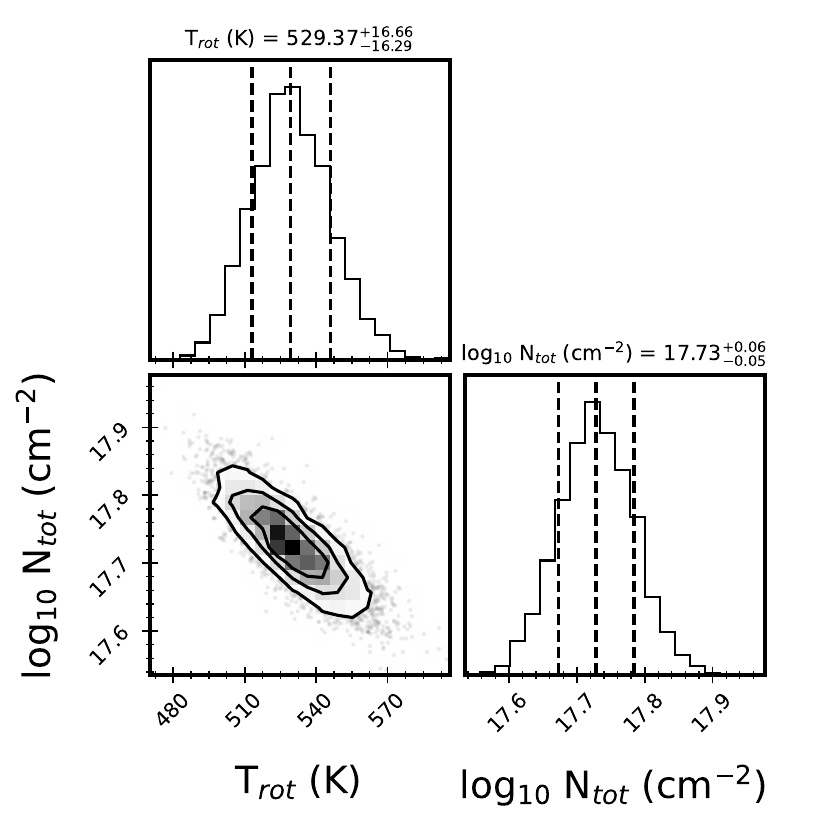}\includegraphics[width=0.385\linewidth]{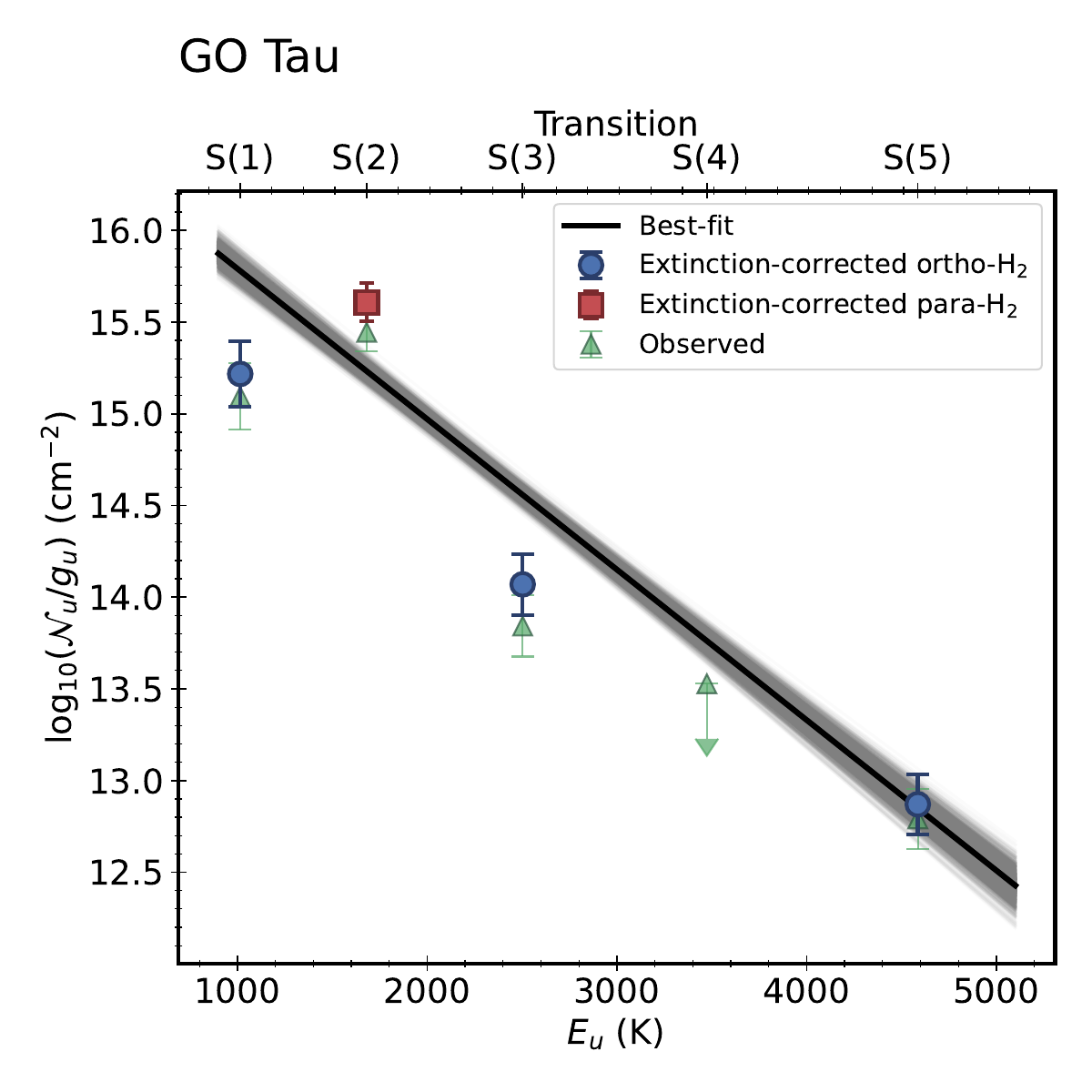}
\includegraphics[width=0.385\linewidth]{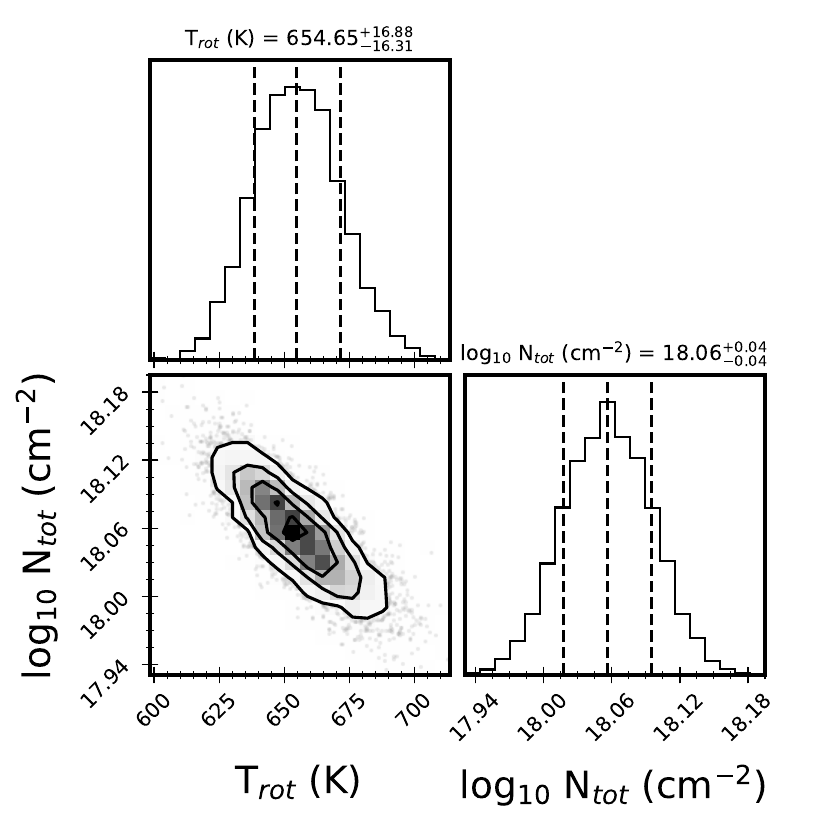}\includegraphics[width=0.385\linewidth]{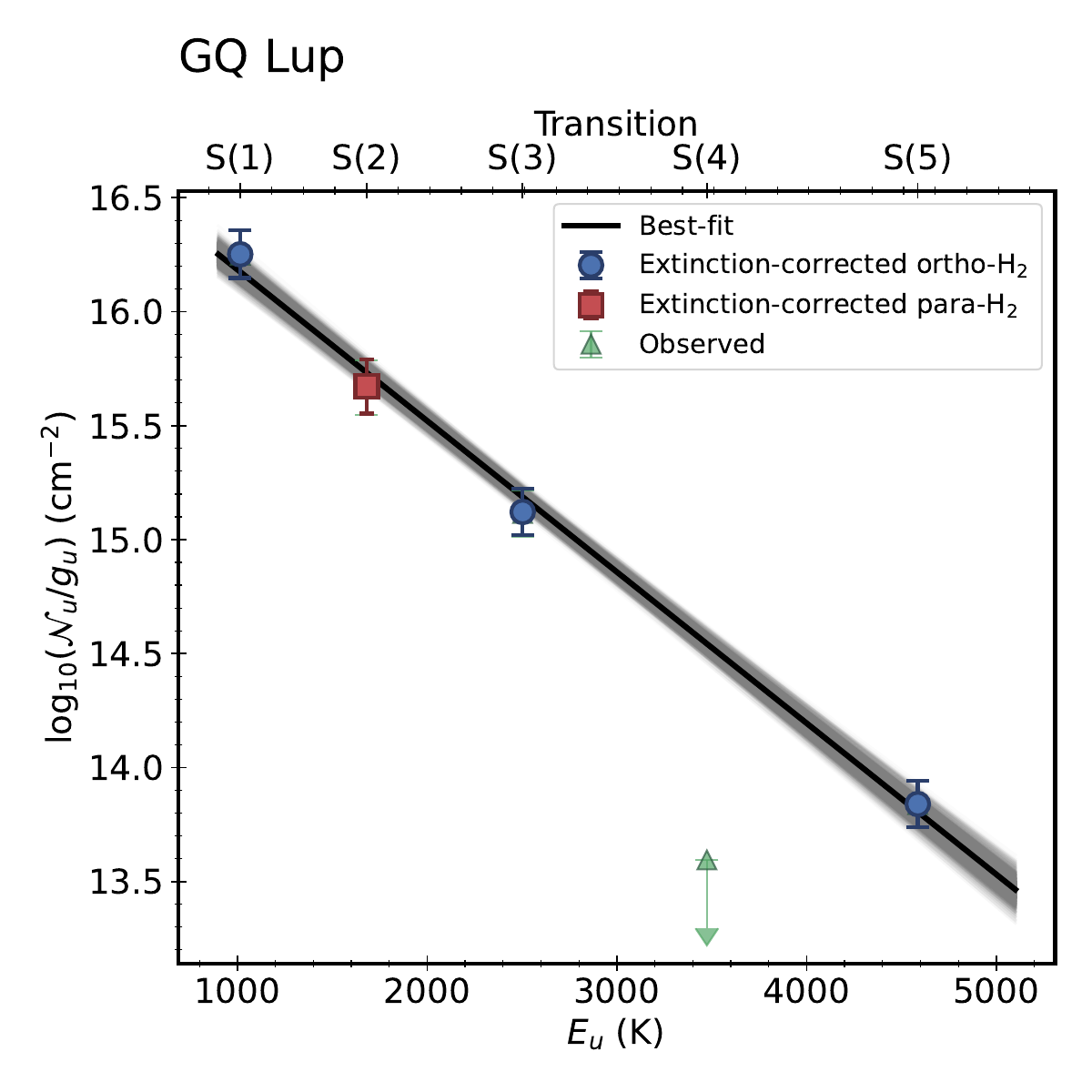}

\includegraphics[width=0.385\linewidth]{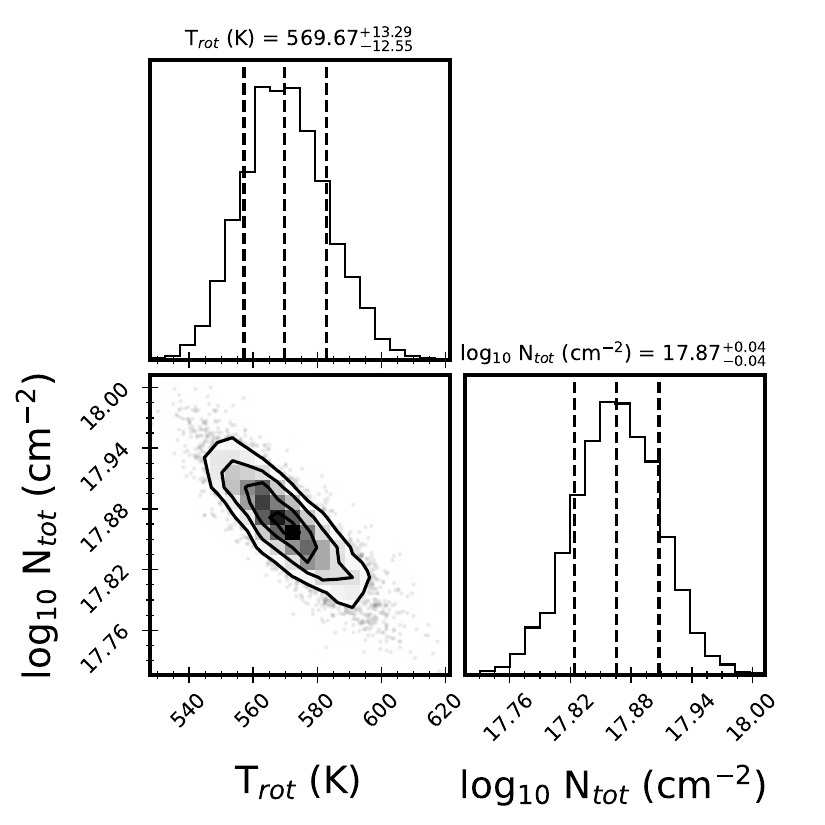}\includegraphics[width=0.385\linewidth]{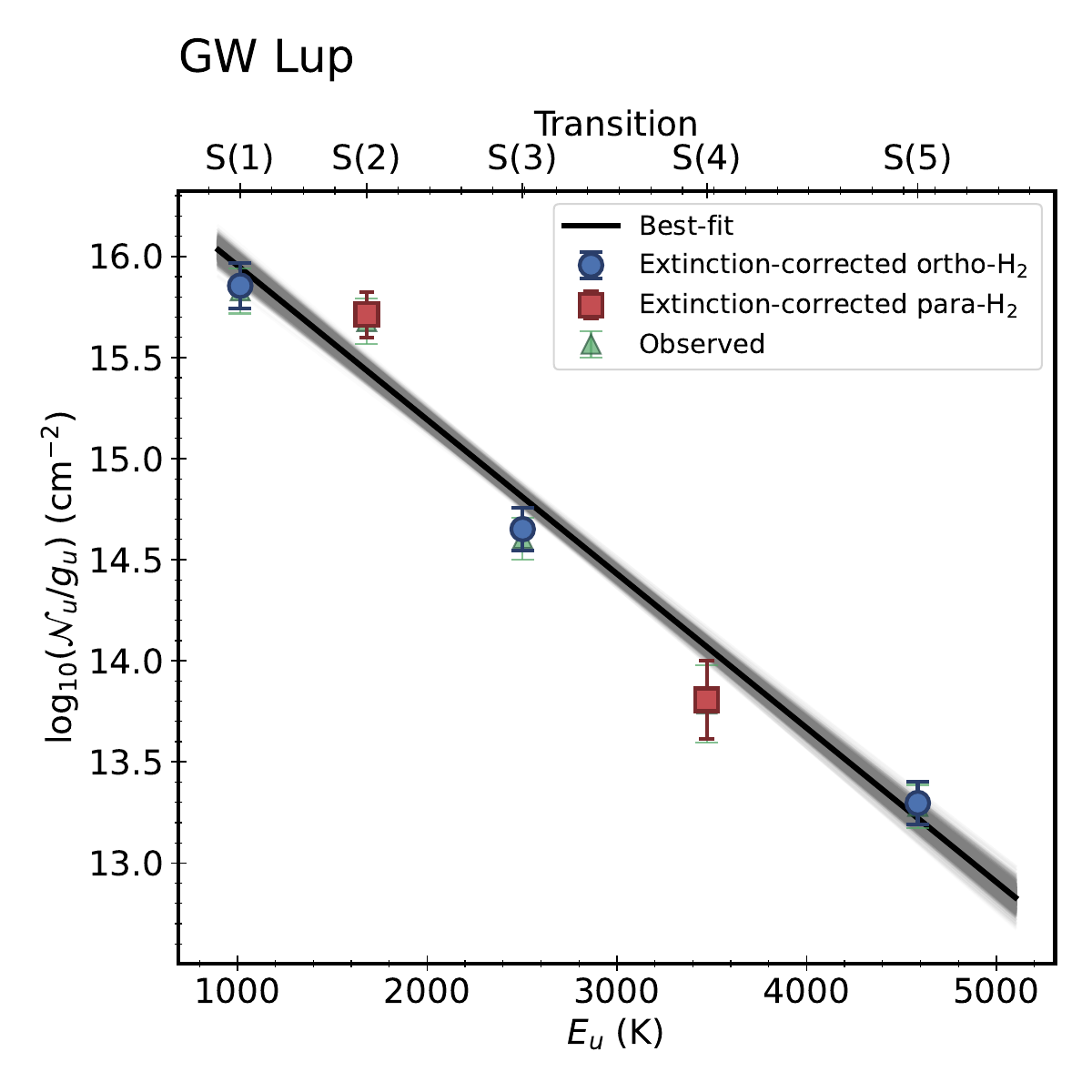}

\caption{Left) Corner plot showing the posterior distributions and best-fit parameters from the MCMC fit to the rotation diagram. (Right) Rotation diagram with observed data points in green and extinction-corrected points in blue for ortho-H$_2$ and red for para-H$_2$. The black line denotes the best-fit model, and the shaded grey region represents the uncertainty range derived from the MCMC fitting. \label{rotfig5}} 
\end{figure*}

\begin{figure*}[h]
\centering
\includegraphics[width=0.385\linewidth]{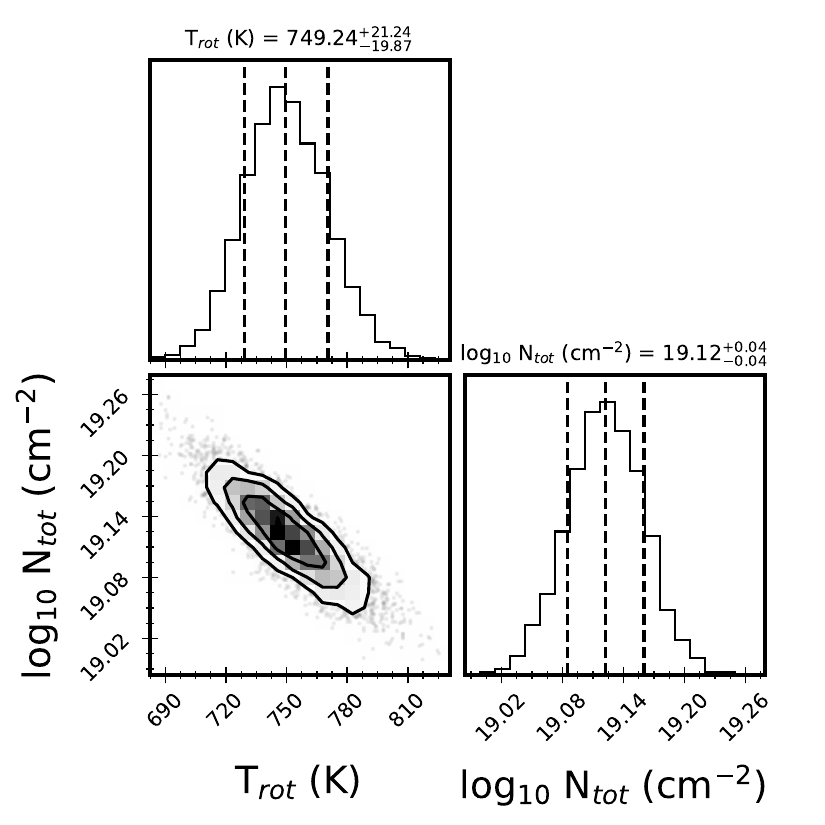}\includegraphics[width=0.385\linewidth]{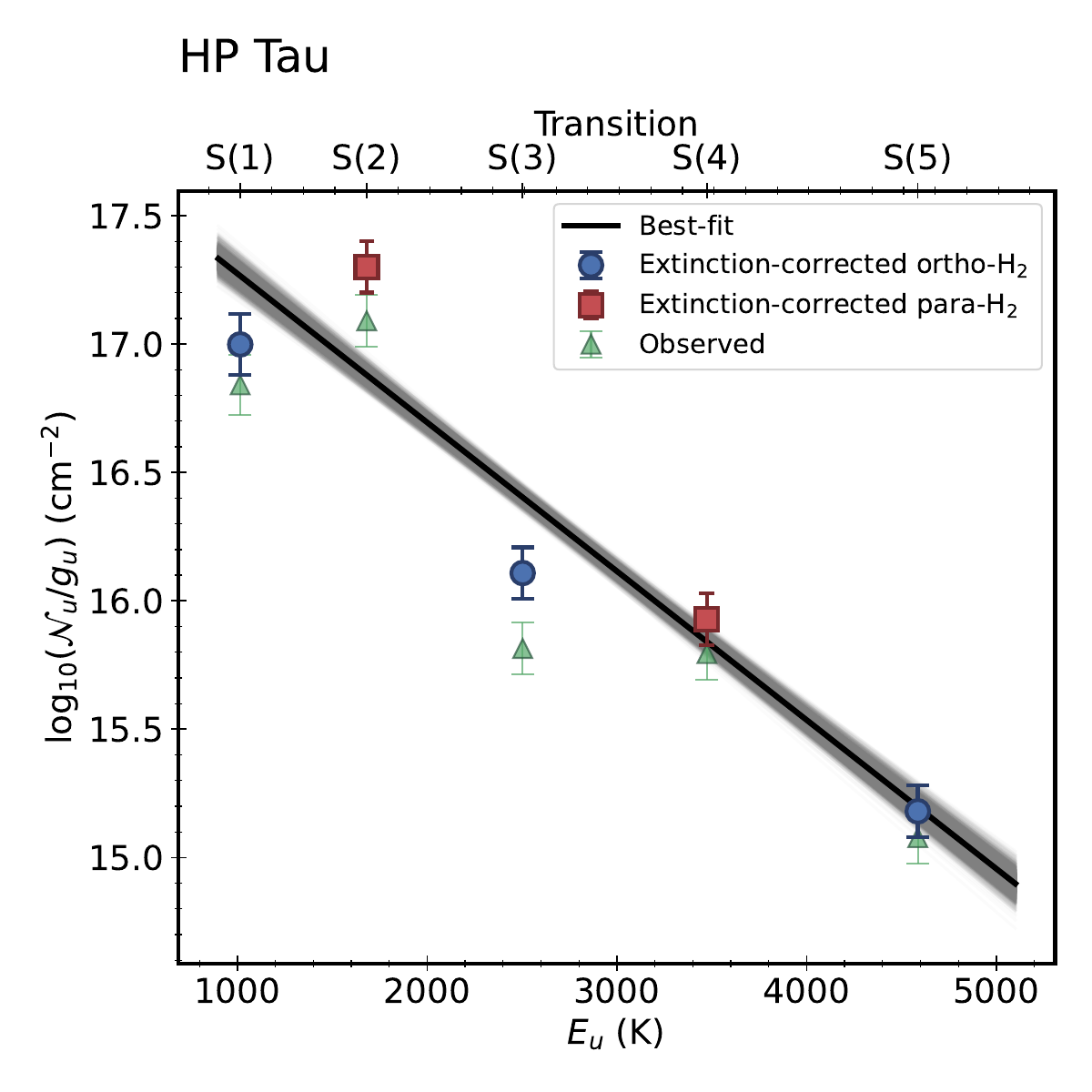}

\includegraphics[width=0.385\linewidth]{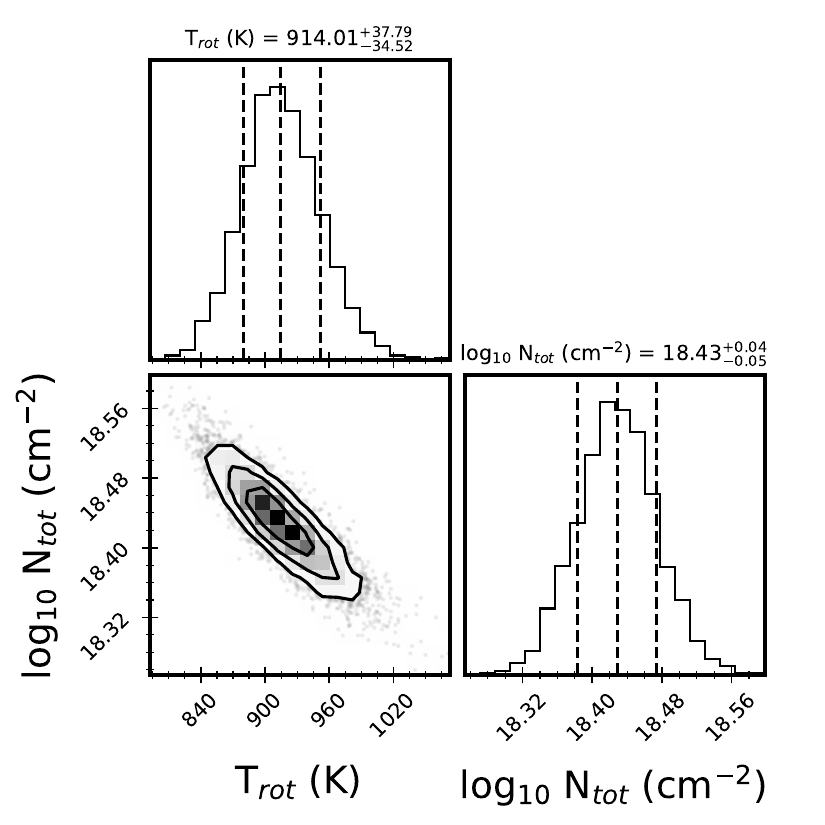}\includegraphics[width=0.385\linewidth]{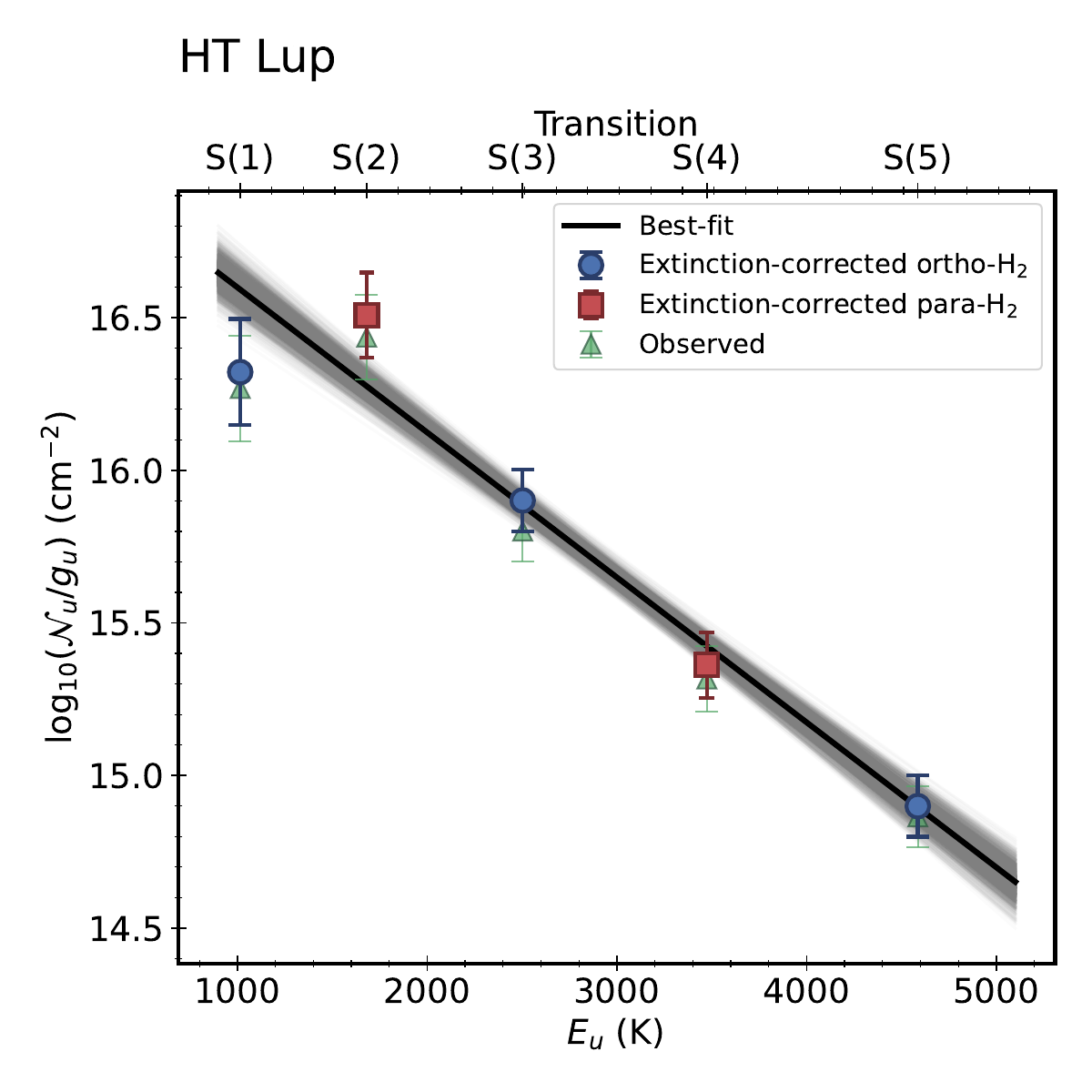}

\includegraphics[width=0.385\linewidth]{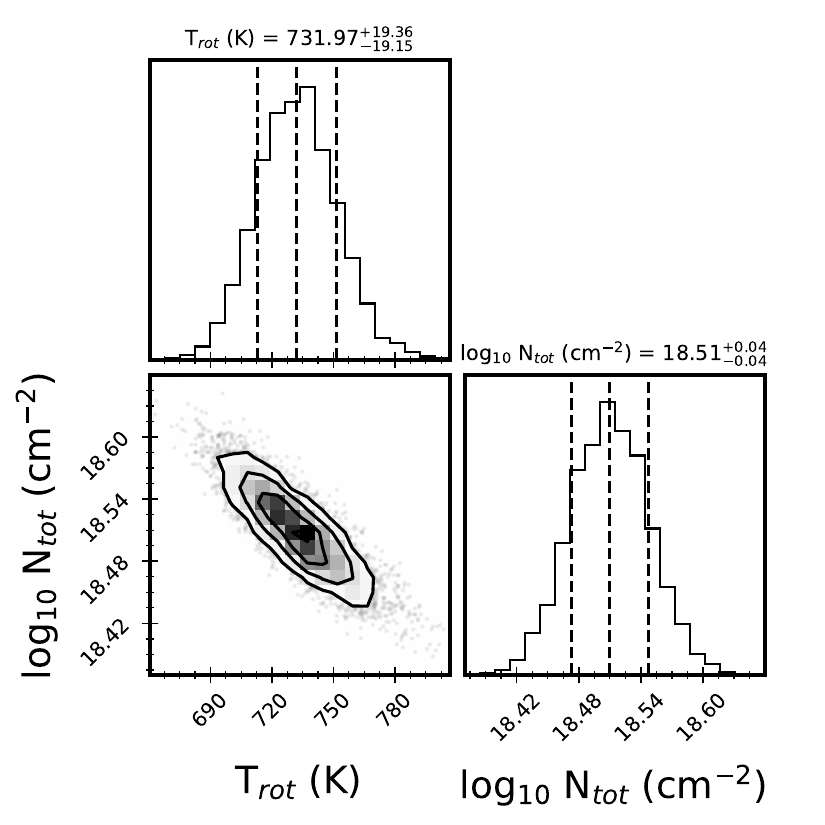}\includegraphics[width=0.385\linewidth]{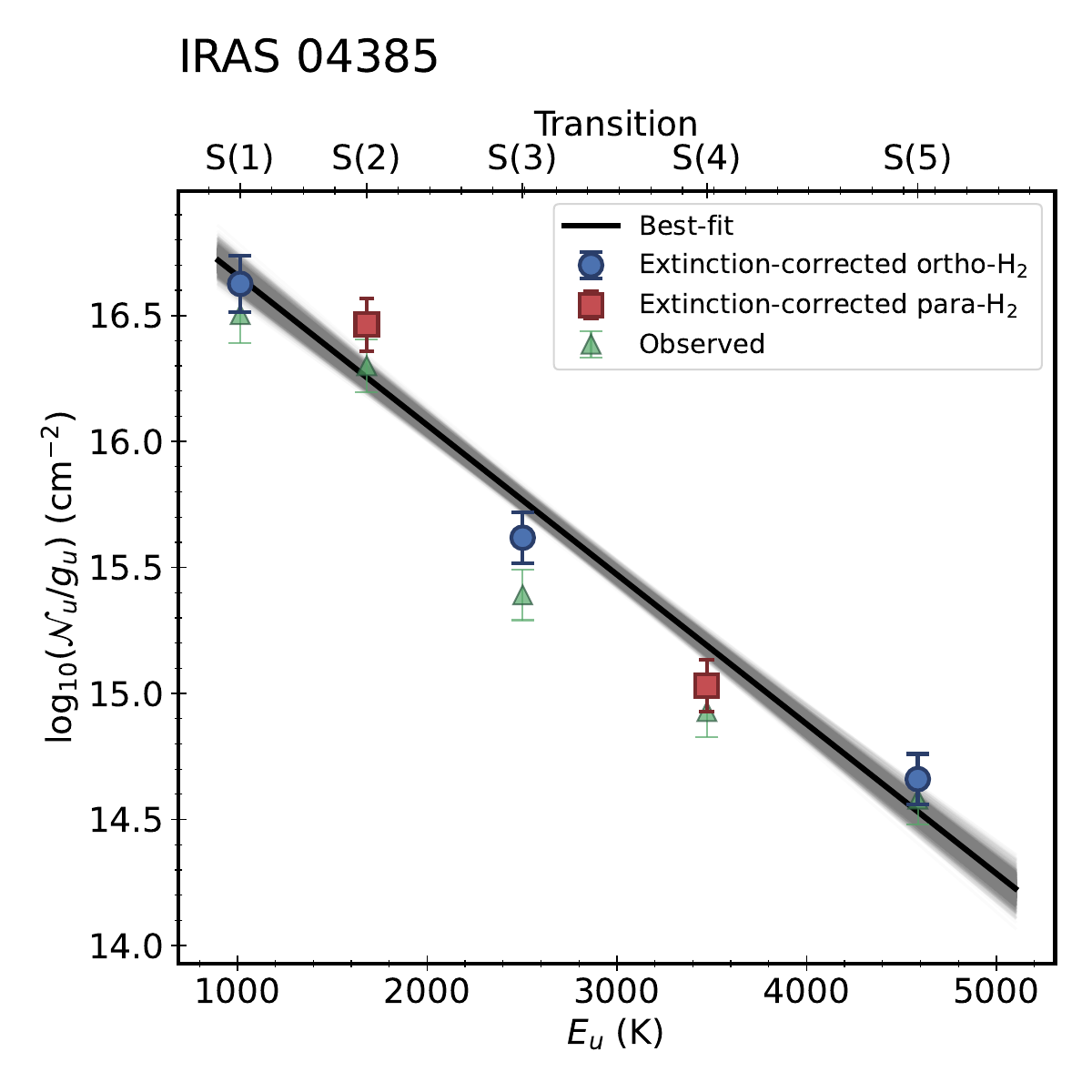}
\caption{Left) Corner plot showing the posterior distributions and best-fit parameters from the MCMC fit to the rotation diagram. (Right) Rotation diagram with observed data points in green and extinction-corrected points in blue for ortho-H$_2$ and red for para-H$_2$. The black line denotes the best-fit model, and the shaded grey region represents the uncertainty range derived from the MCMC fitting. \label{rotfig6}} 
\end{figure*}

\begin{figure*}[h]
\centering
\includegraphics[width=0.385\linewidth]{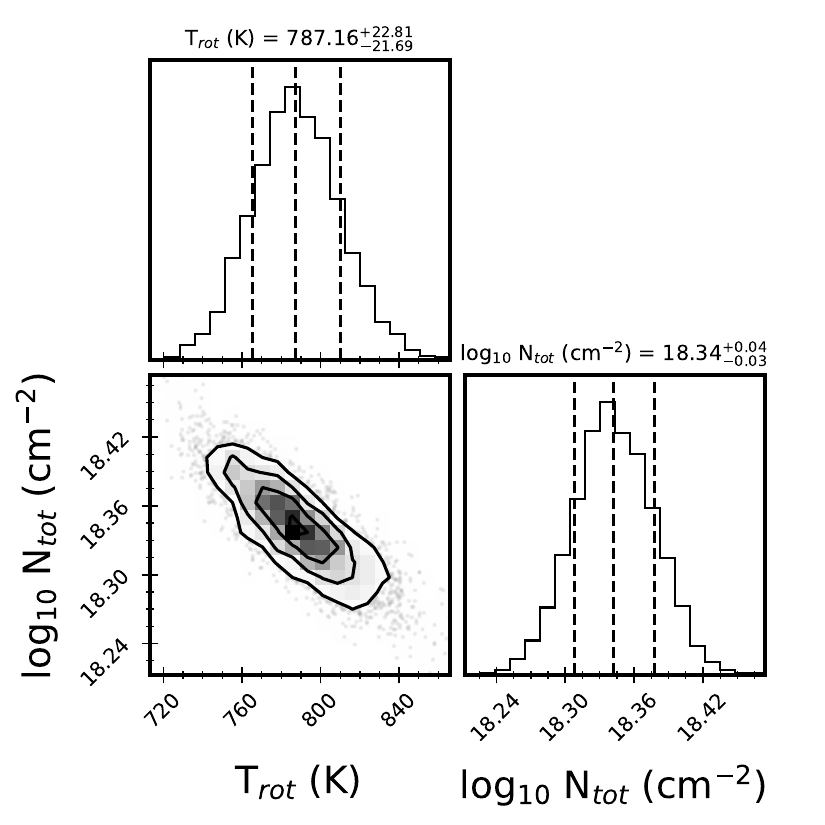}\includegraphics[width=0.385\linewidth]{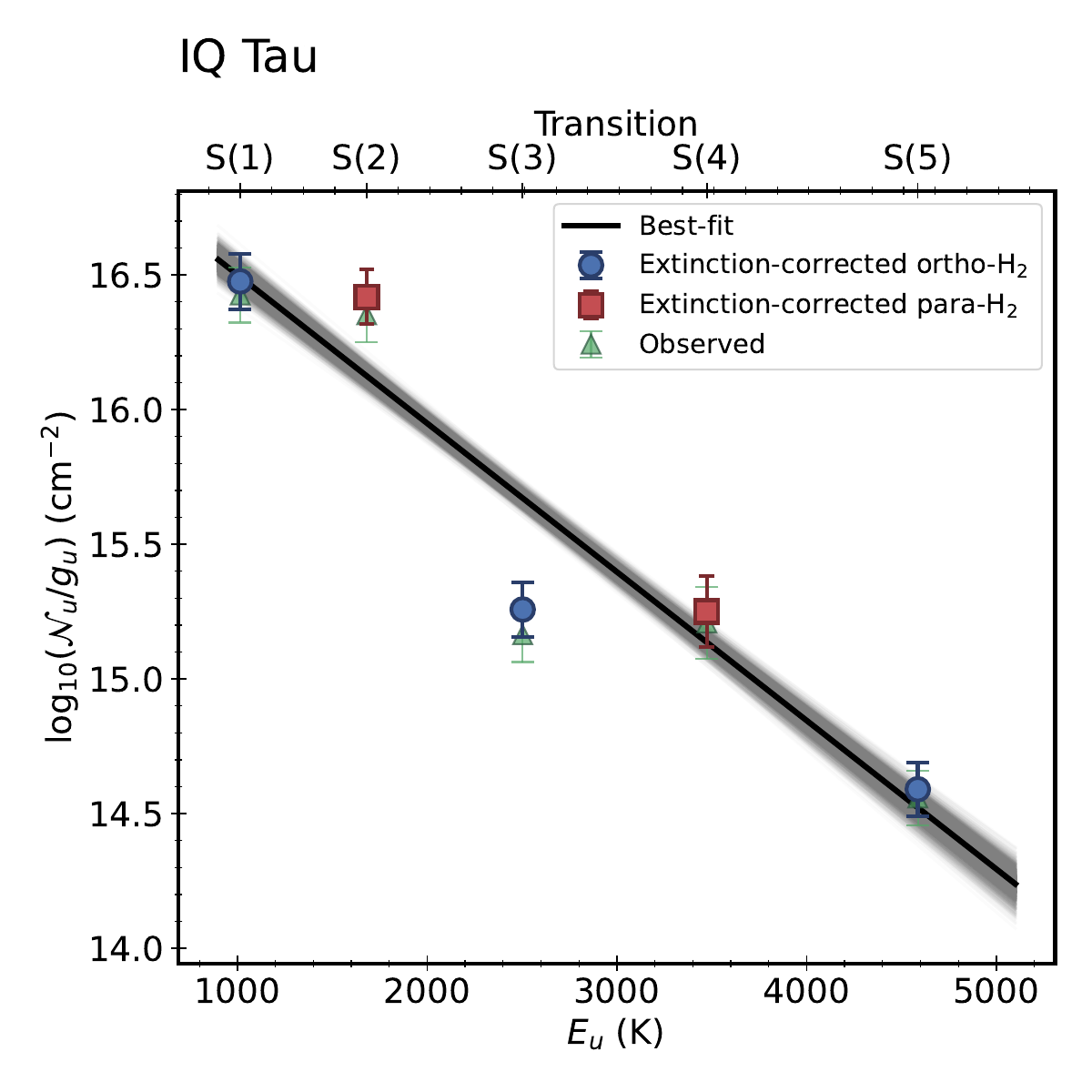}

\includegraphics[width=0.385\linewidth]{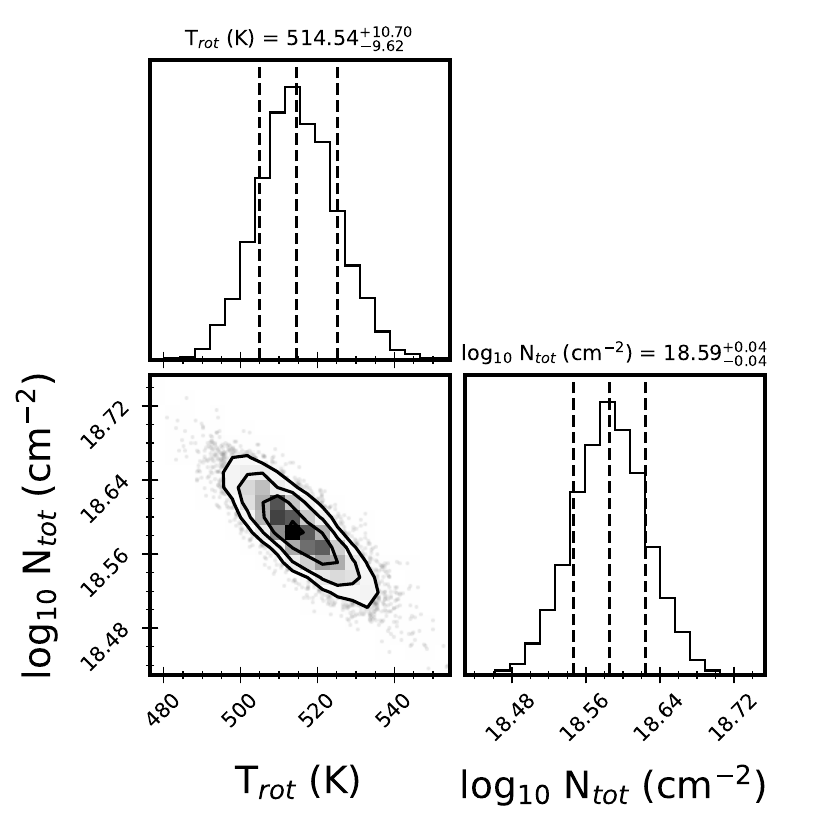}\includegraphics[width=0.385\linewidth]{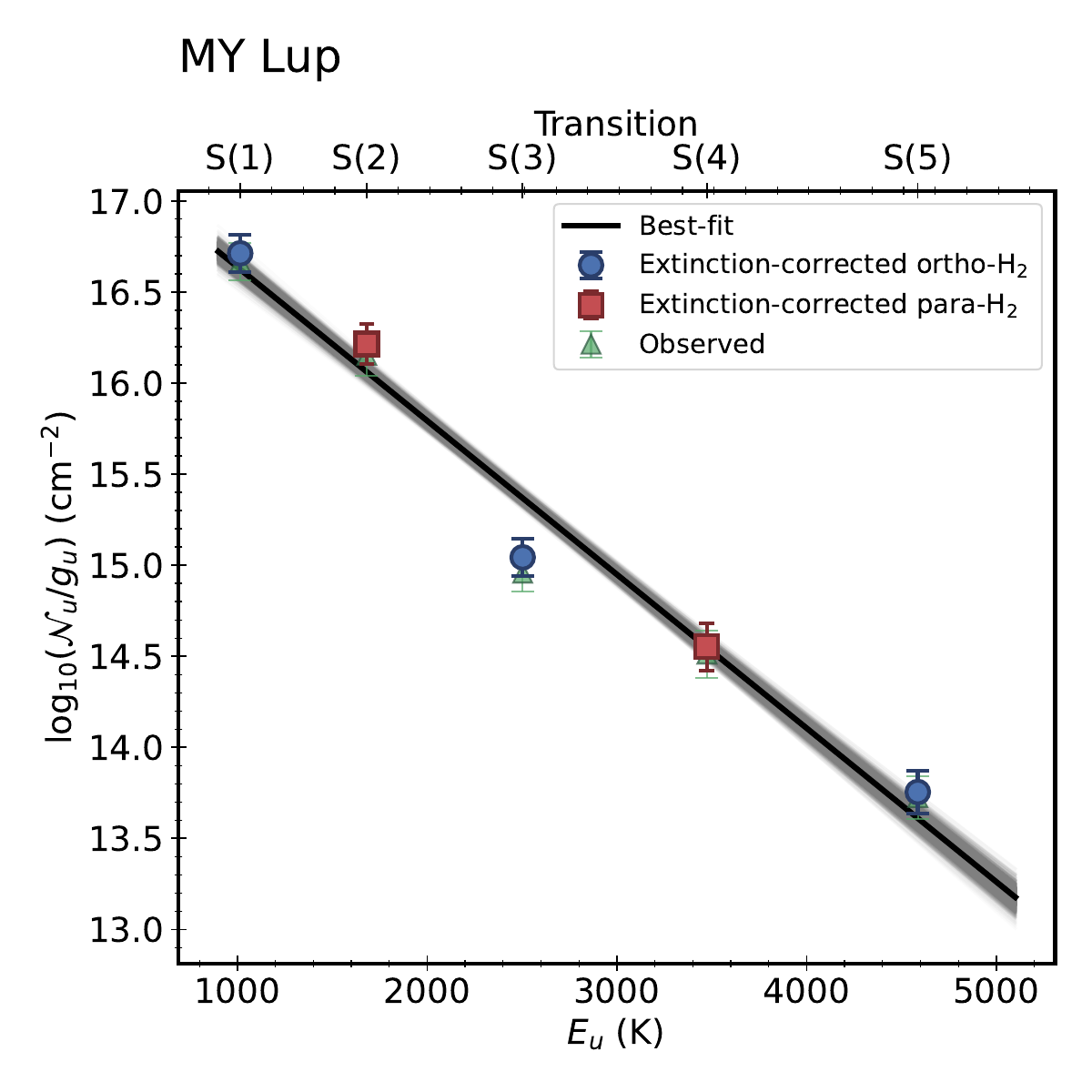}

\includegraphics[width=0.385\linewidth]{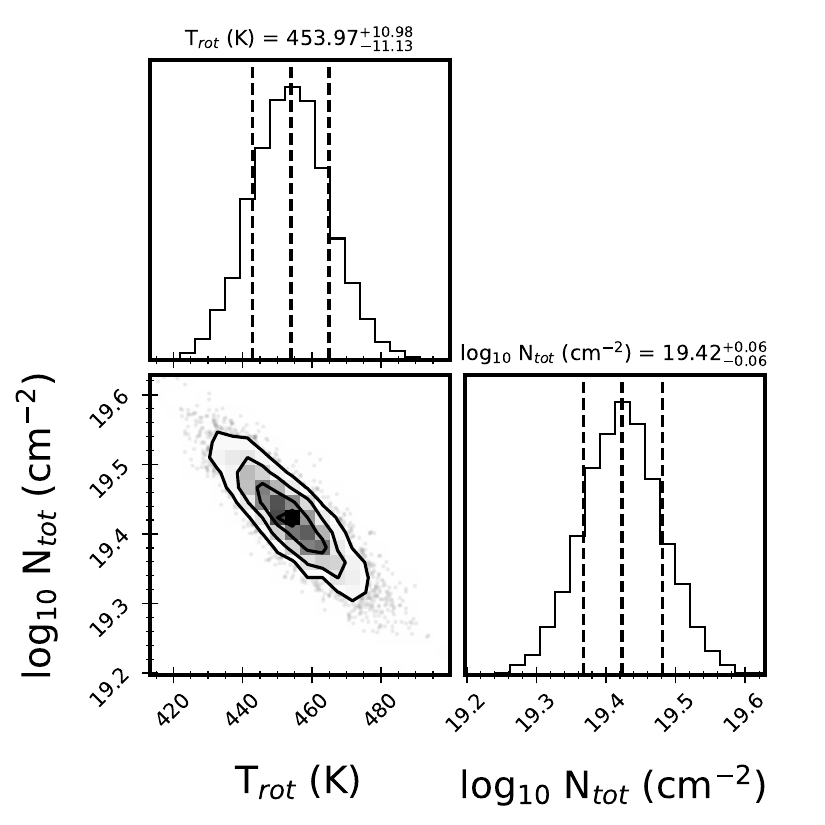}\includegraphics[width=0.385\linewidth]{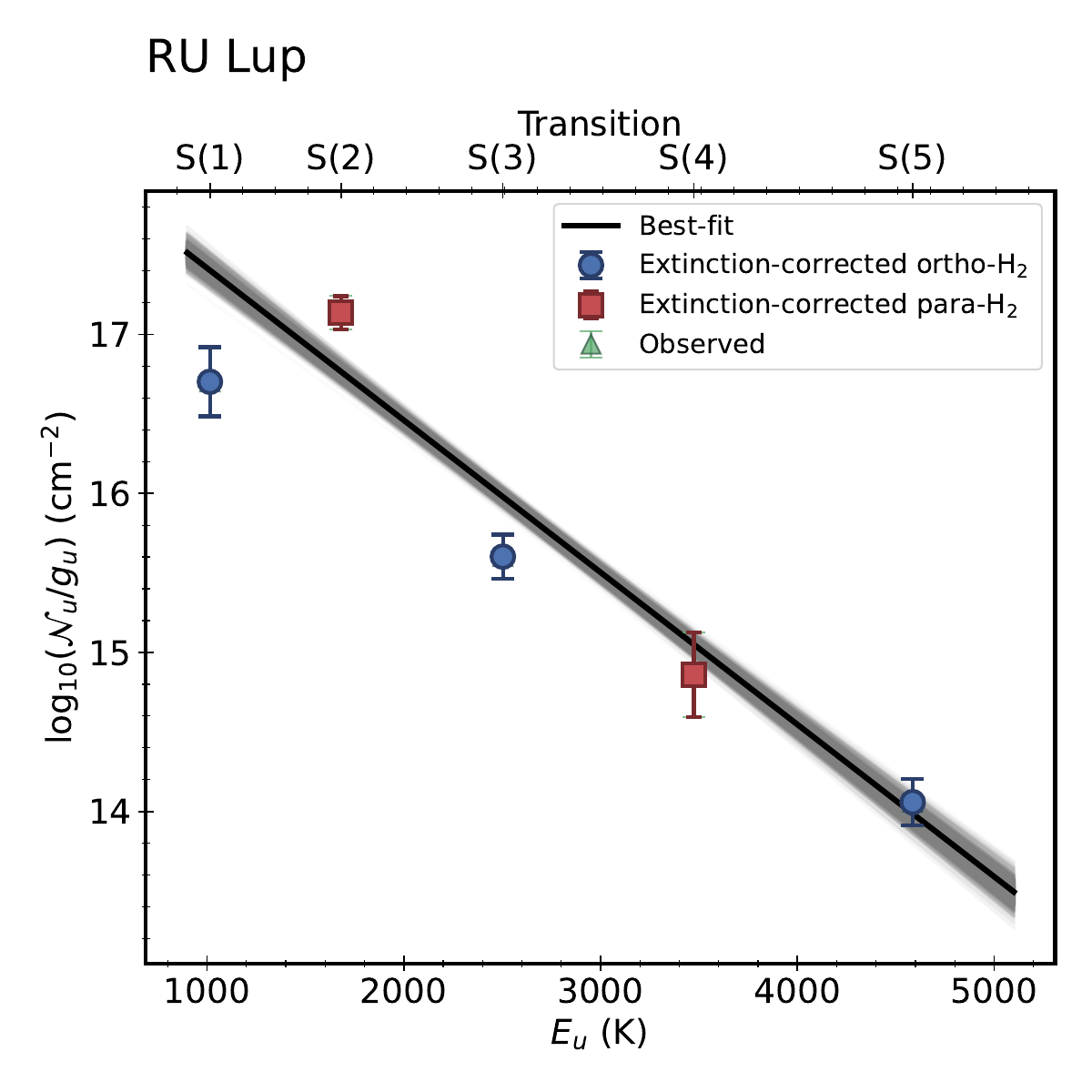}

\caption{Left) Corner plot showing the posterior distributions and best-fit parameters from the MCMC fit to the rotation diagram. (Right) Rotation diagram with observed data points in green and extinction-corrected points in blue for ortho-H$_2$ and red for para-H$_2$. The black line denotes the best-fit model, and the shaded grey region represents the uncertainty range derived from the MCMC fitting.\label{rotfig7}} 
\end{figure*}
\begin{figure*}[h]
\centering
\includegraphics[width=0.385\linewidth]{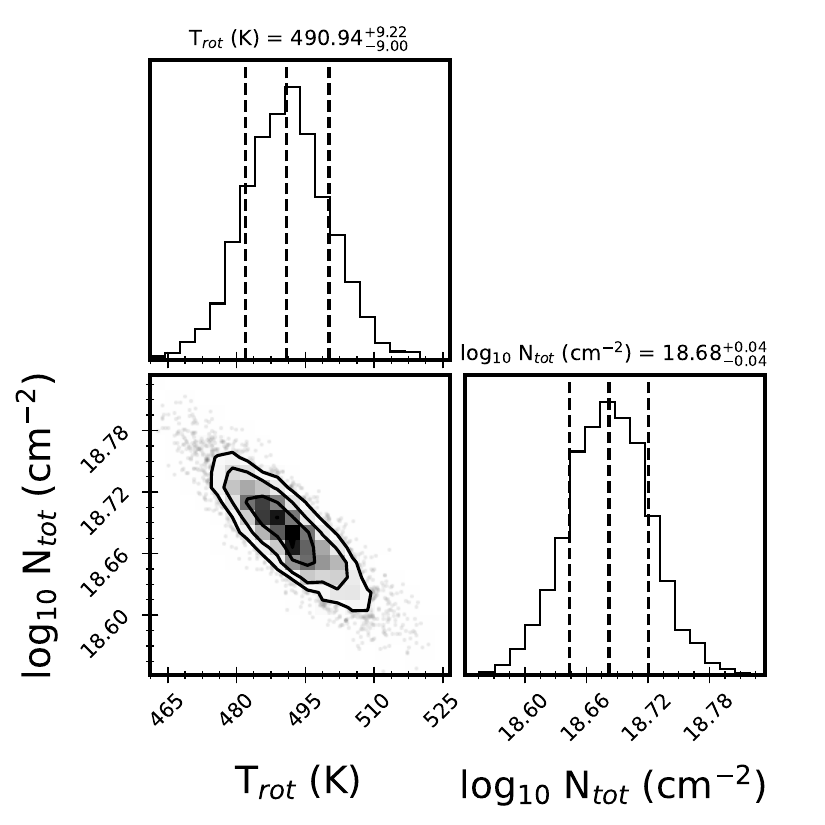}\includegraphics[width=0.385\linewidth]{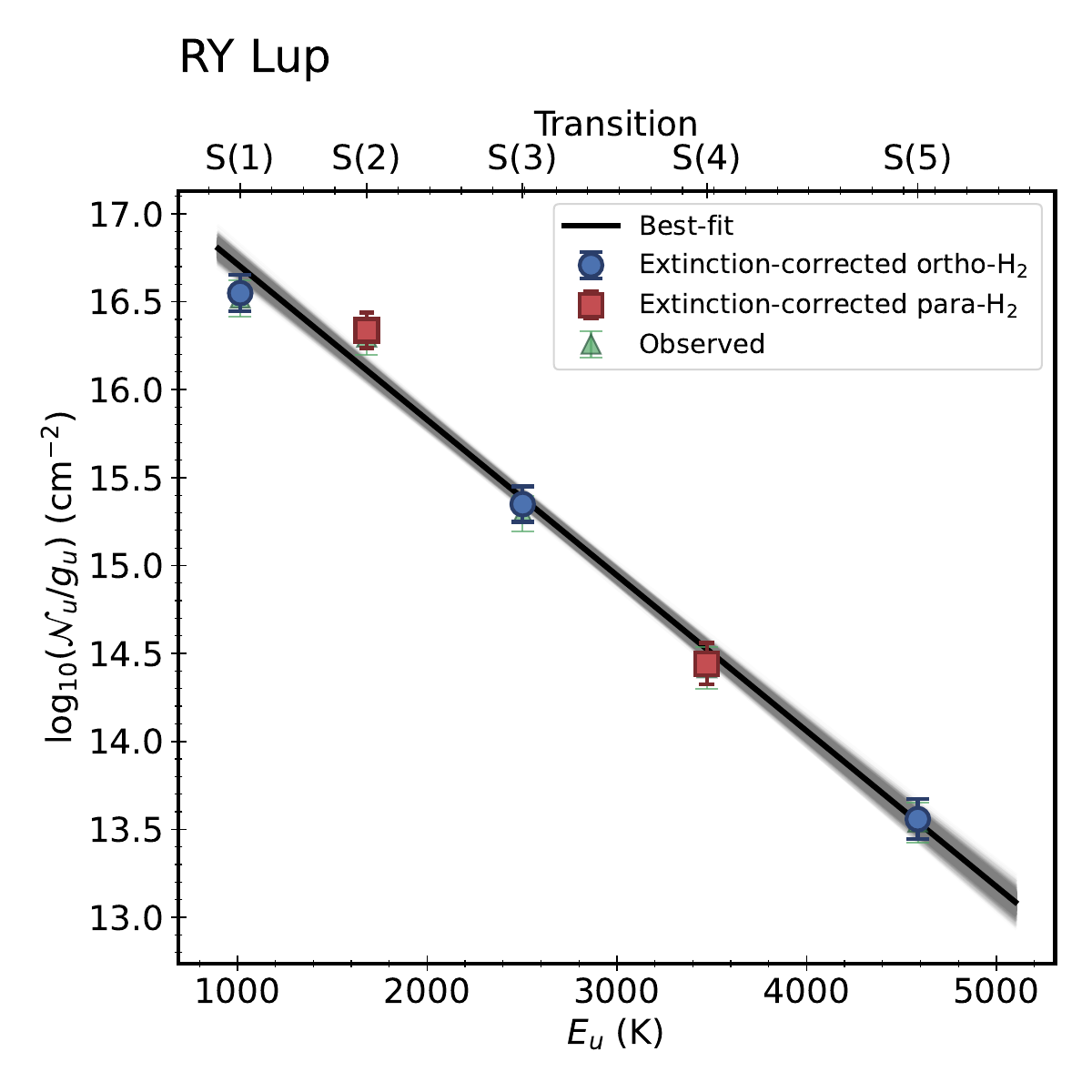}

\includegraphics[width=0.385\linewidth]{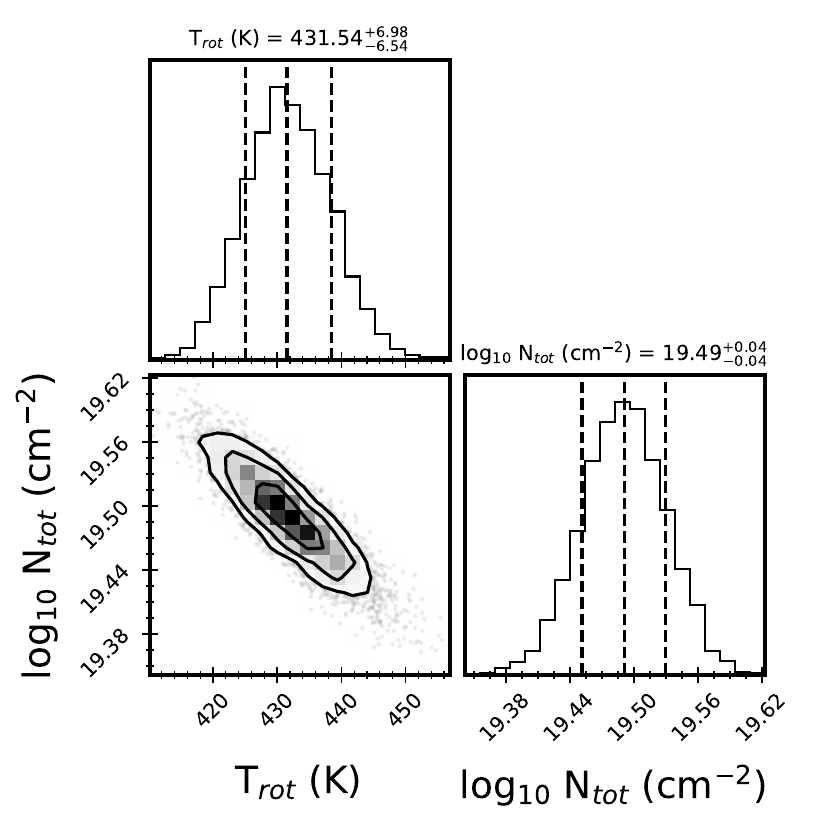}\includegraphics[width=0.385\linewidth]{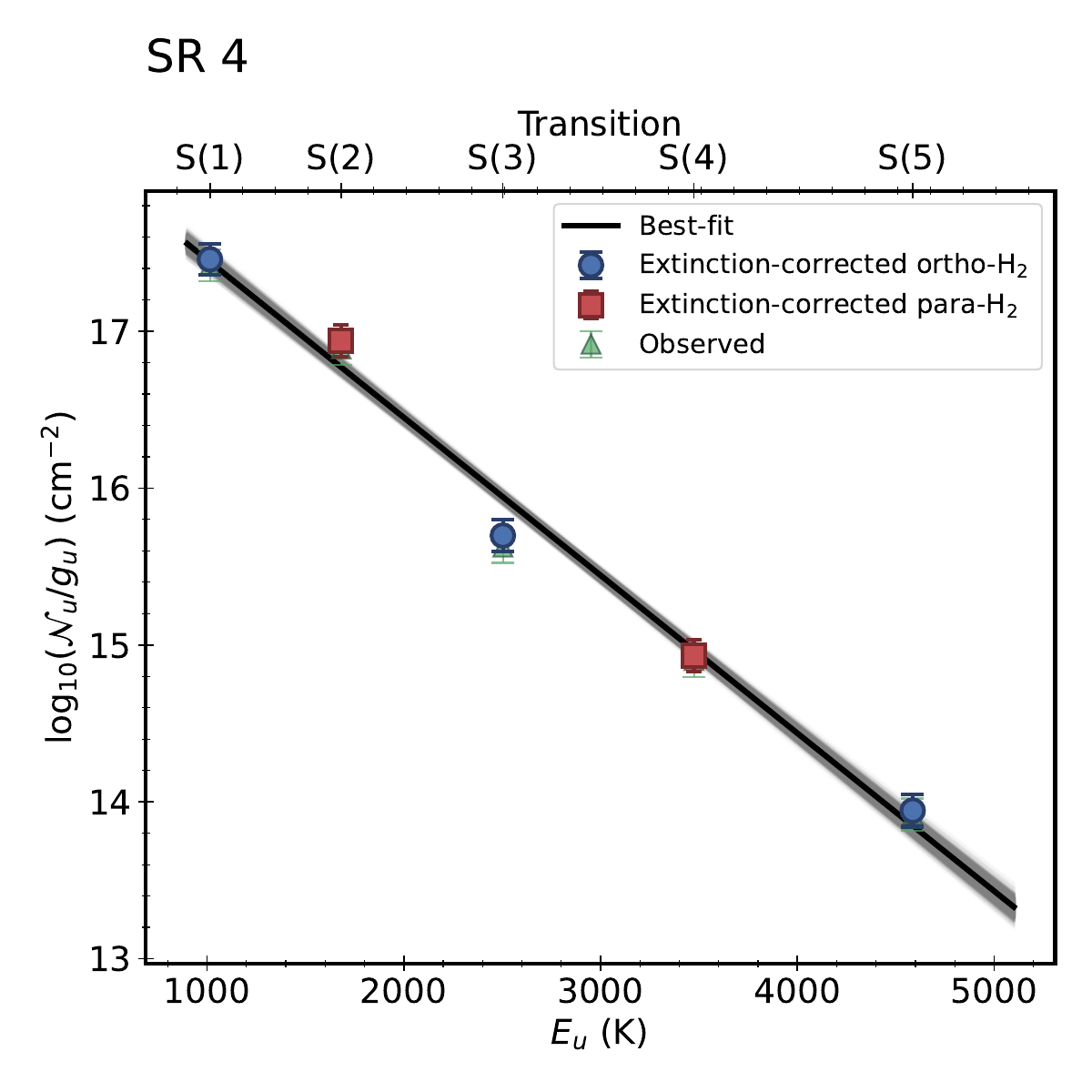}

\includegraphics[width=0.385\linewidth]{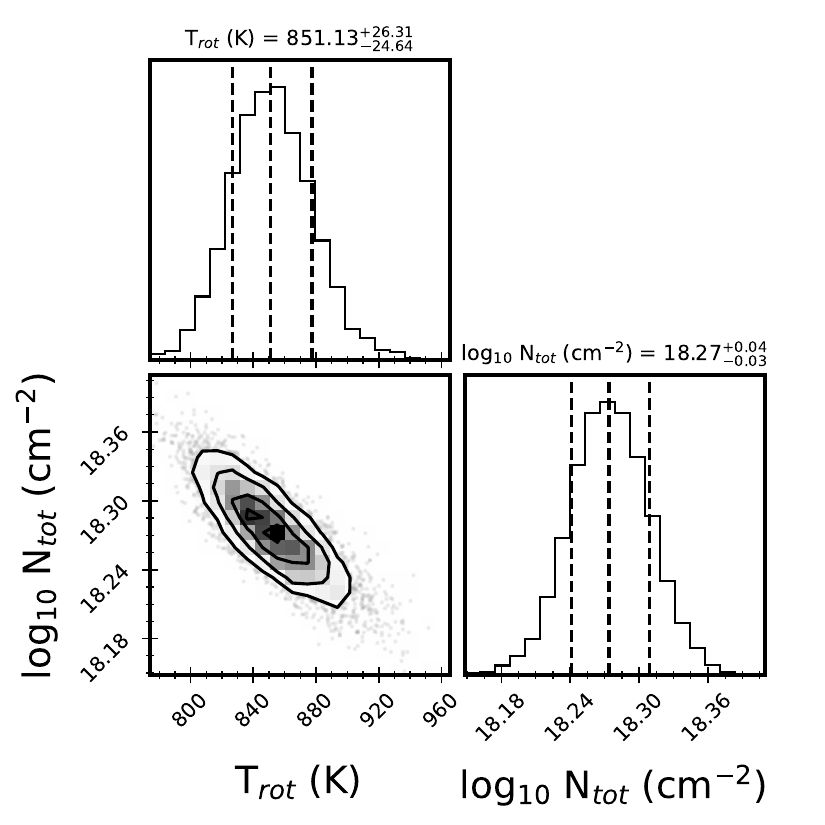}\includegraphics[width=0.385\linewidth]{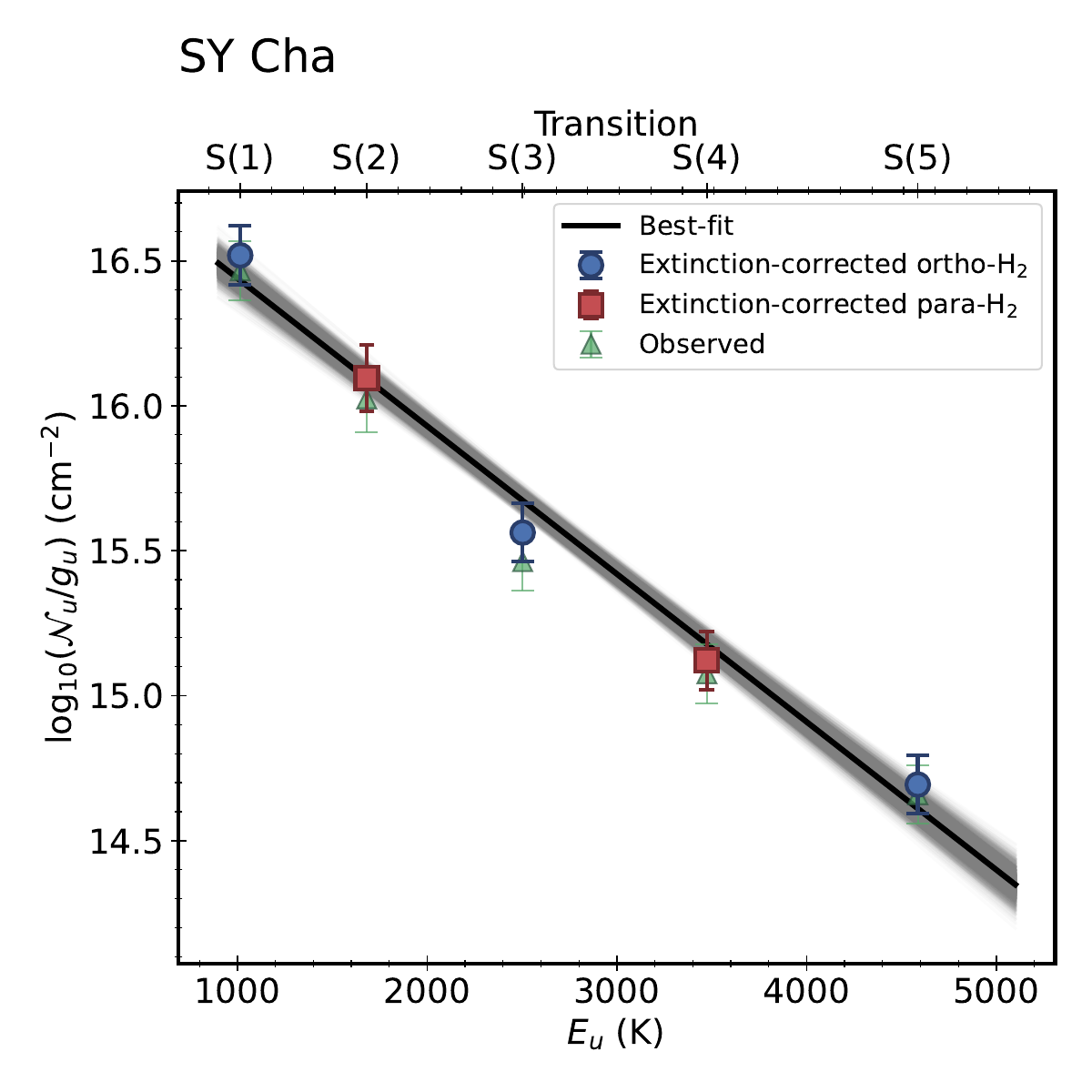}
\caption{Left) Corner plot showing the posterior distributions and best-fit parameters from the MCMC fit to the rotation diagram. (Right) Rotation diagram with observed data points in green and extinction-corrected points in blue for ortho-H$_2$ and red for para-H$_2$. The black line denotes the best-fit model, and the shaded grey region represents the uncertainty range derived from the MCMC fitting.\label{rotfig8}} 
\end{figure*}

\begin{figure*}[h]
\centering
\includegraphics[width=0.385\linewidth]{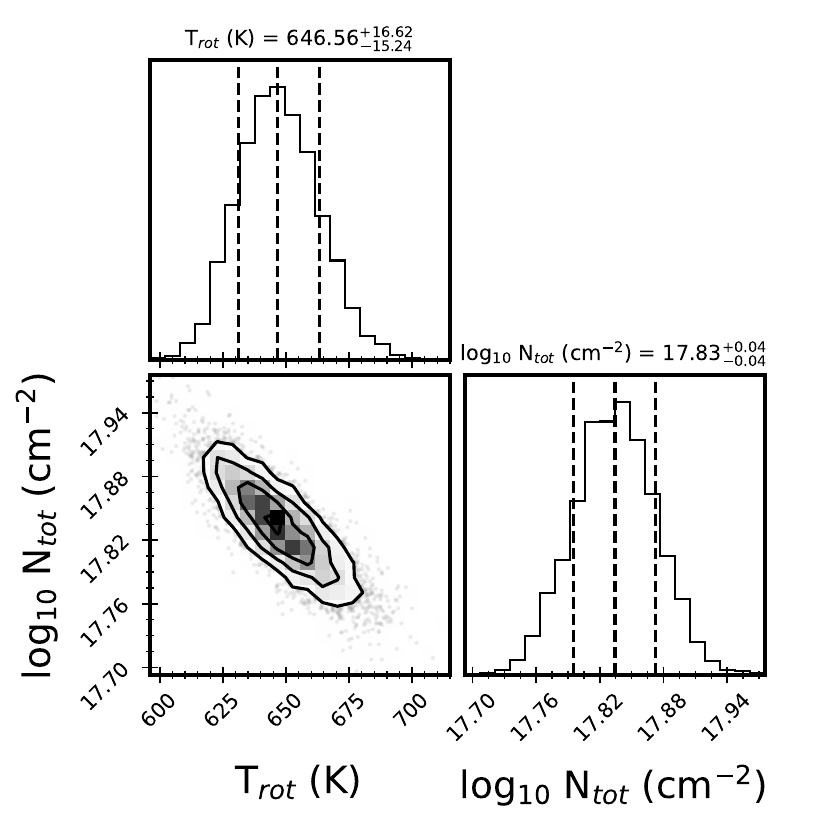}\includegraphics[width=0.385\linewidth]{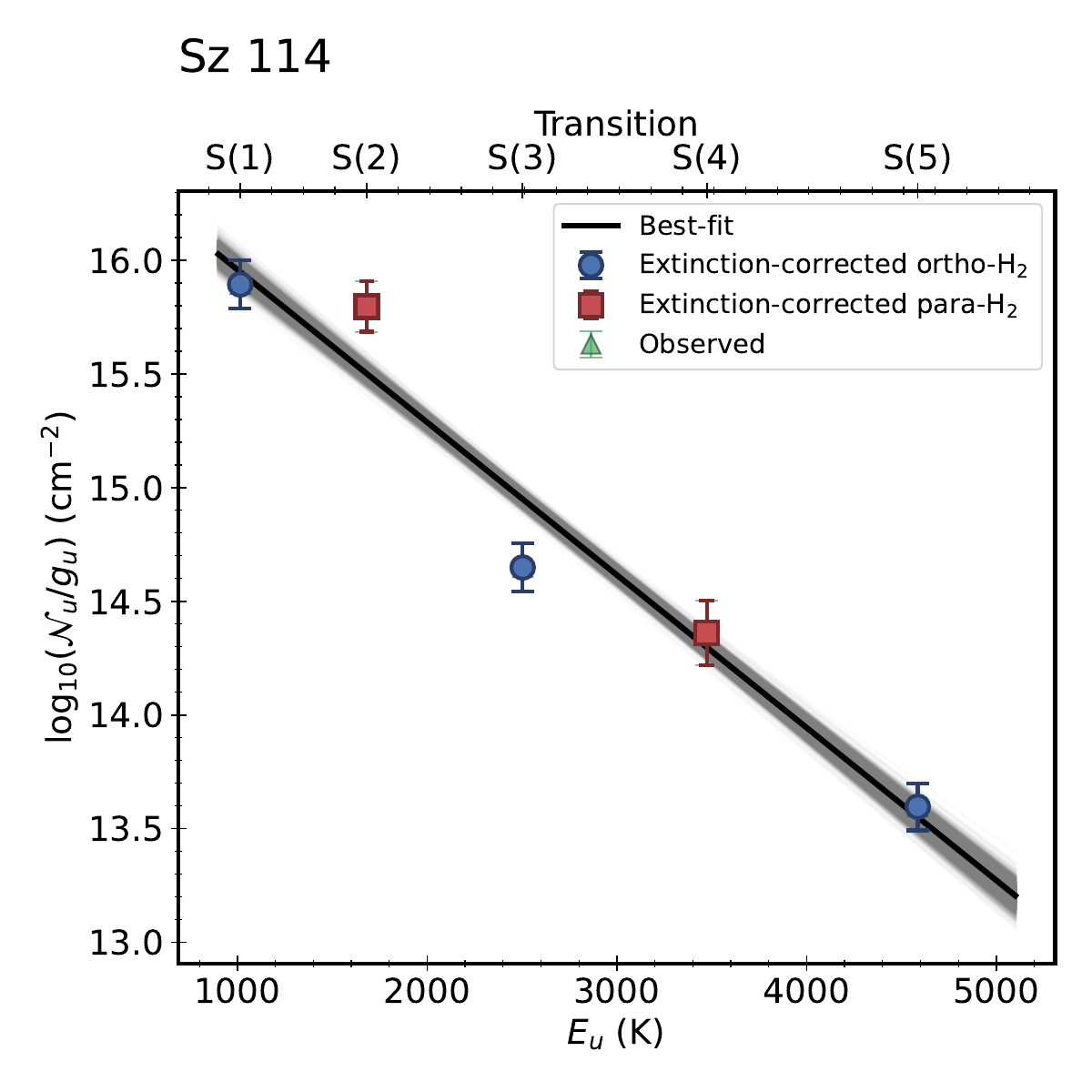}

\includegraphics[width=0.385\linewidth]{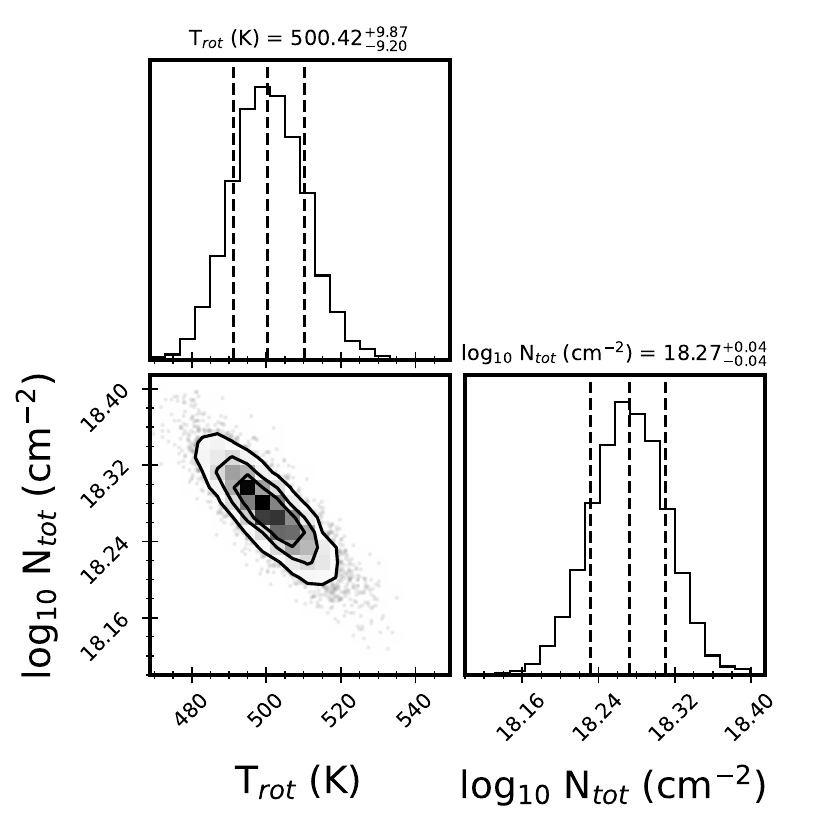}\includegraphics[width=0.385\linewidth]{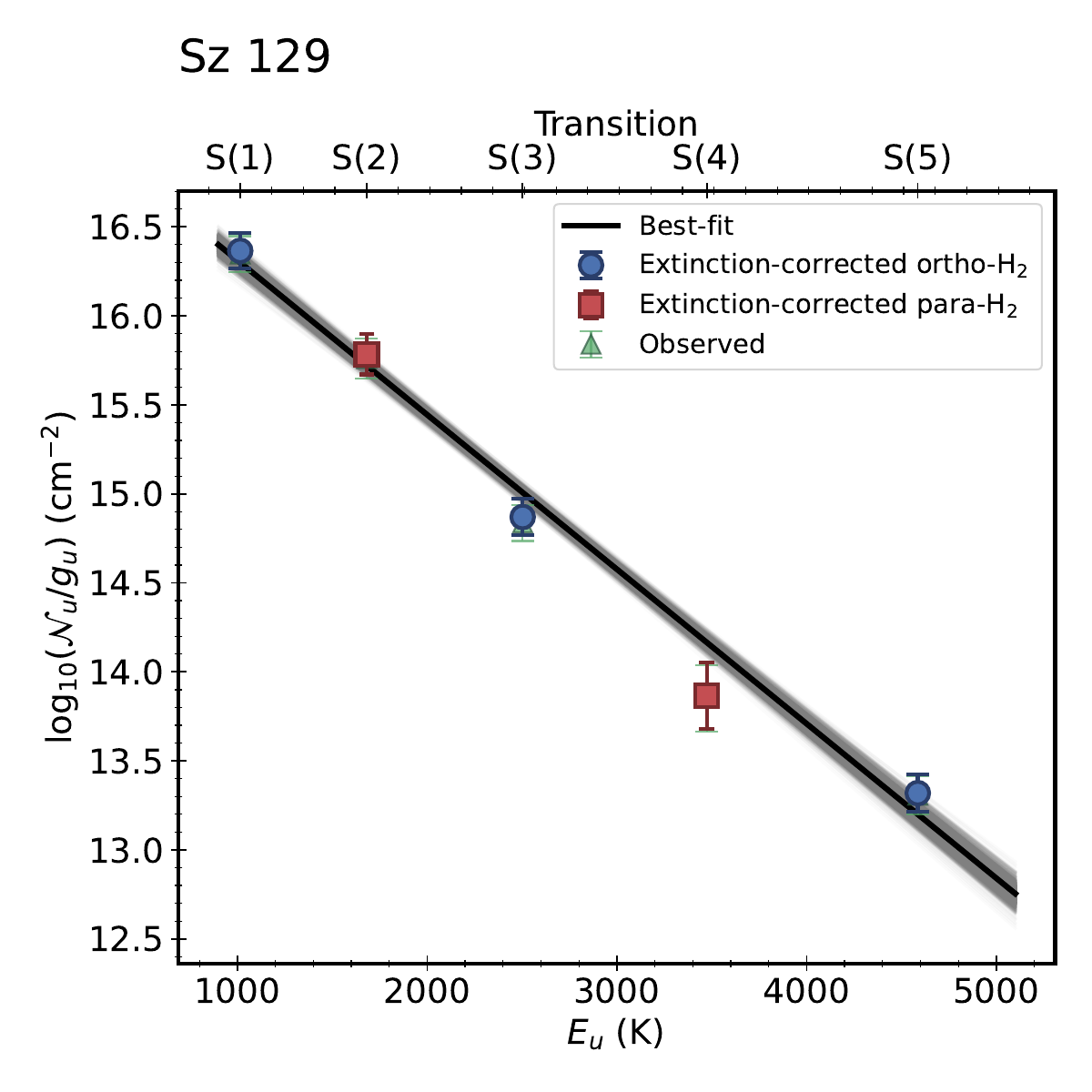}

\includegraphics[width=0.385\linewidth]{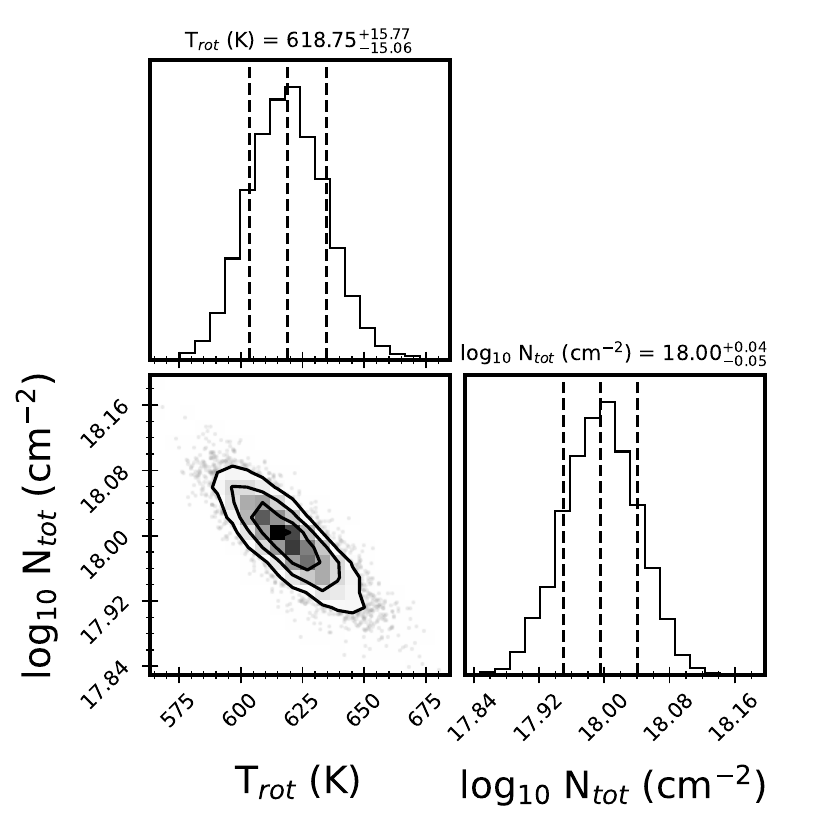}\includegraphics[width=0.385\linewidth]{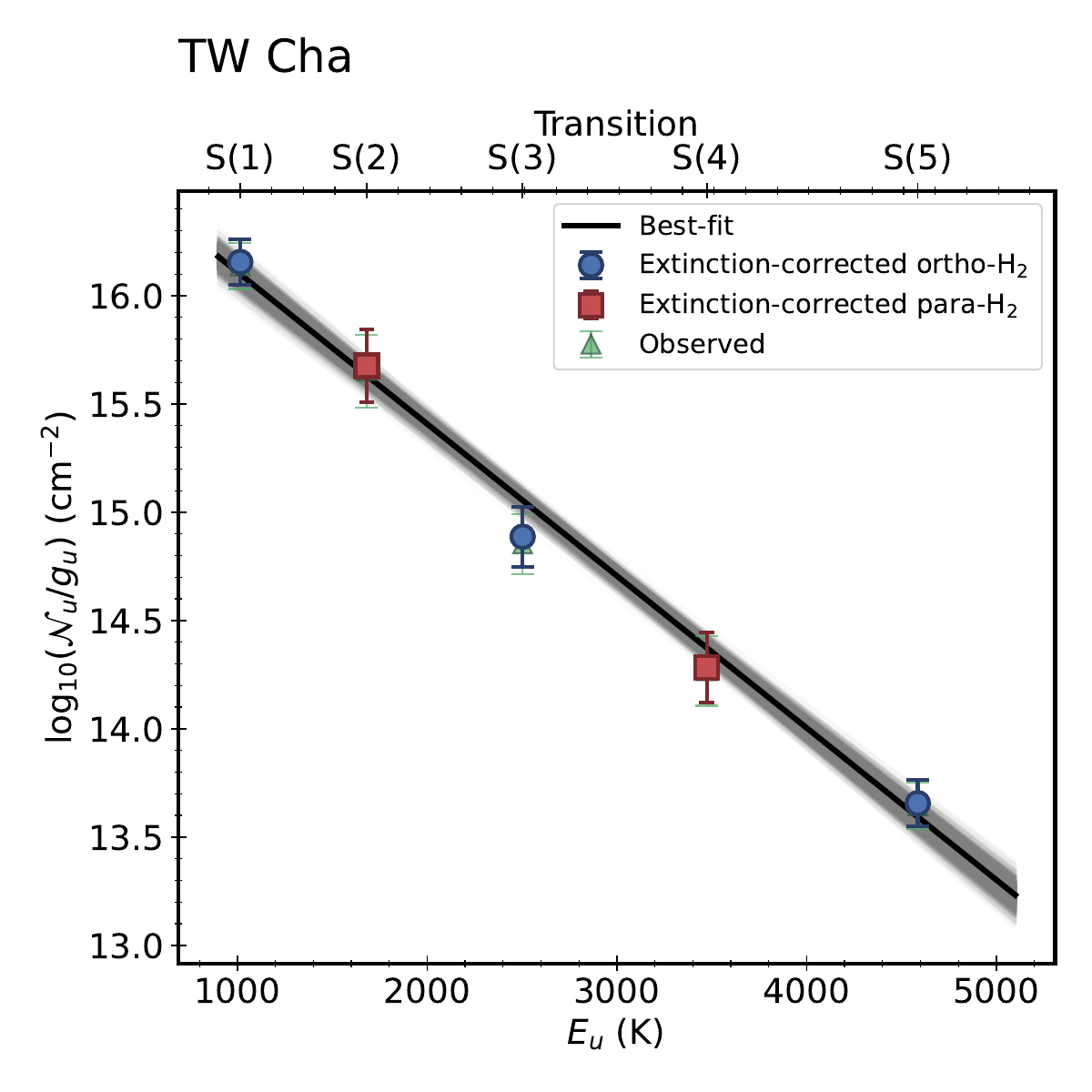}

\caption{Left) Corner plot showing the posterior distributions and best-fit parameters from the MCMC fit to the rotation diagram. (Right) Rotation diagram with observed data points in green and extinction-corrected points in blue for ortho-H$_2$ and red for para-H$_2$. The black line denotes the best-fit model, and the shaded grey region represents the uncertainty range derived from the MCMC fitting.\label{rotfig9}} 
\end{figure*}

\begin{figure*}[h]
\centering
\includegraphics[width=0.385\linewidth]{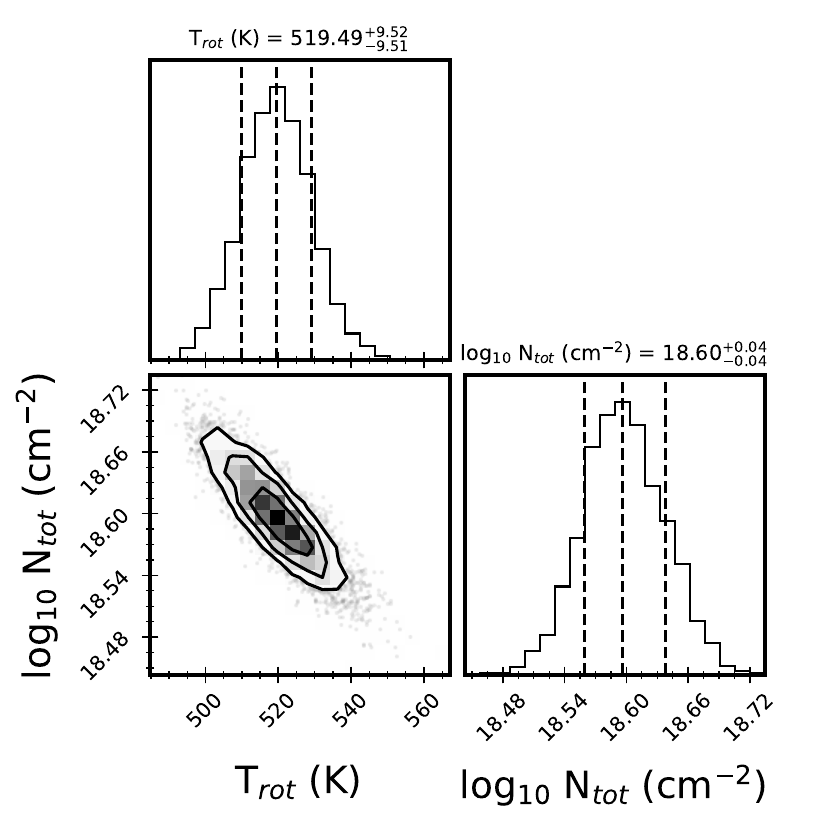}\includegraphics[width=0.385\linewidth]{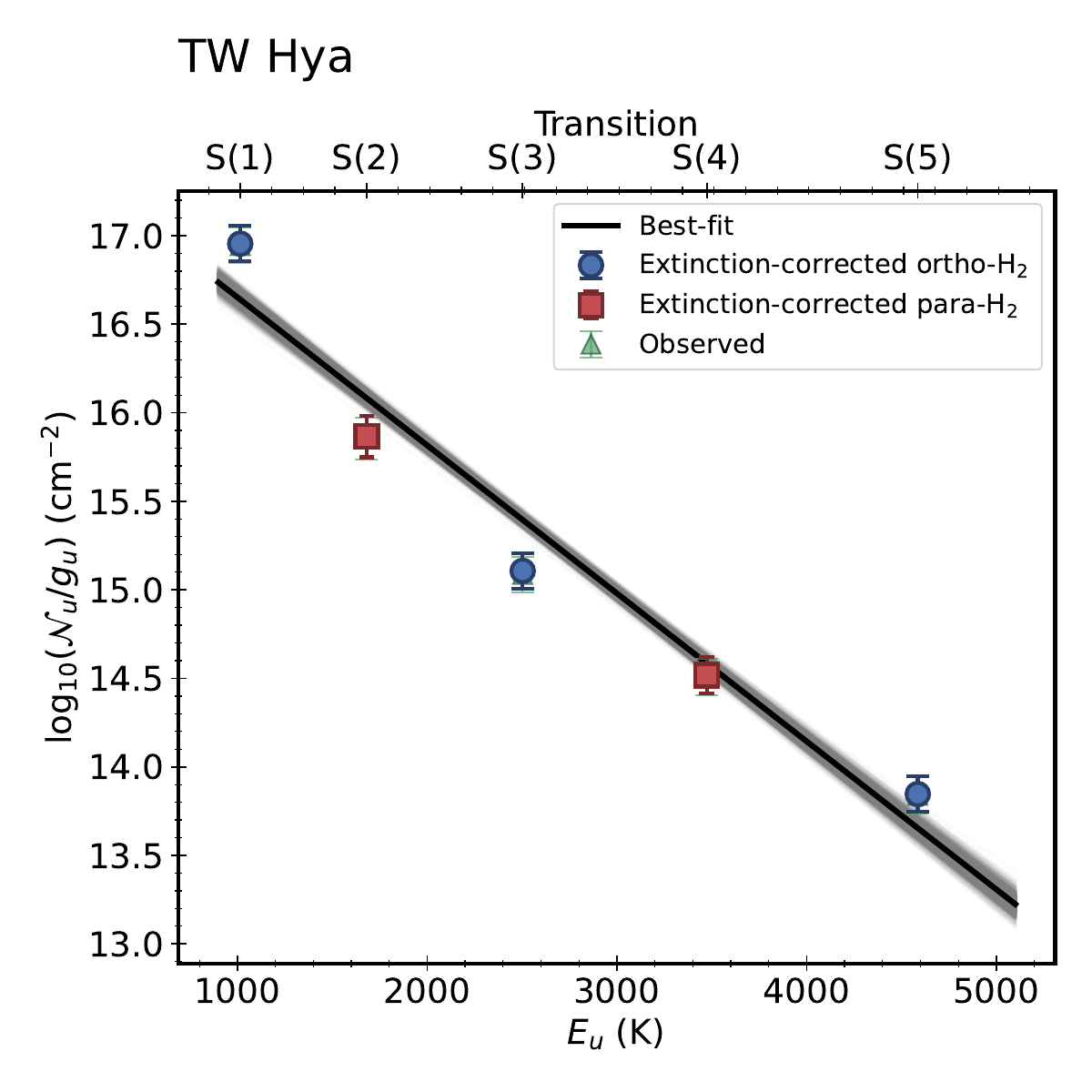}

\includegraphics[width=0.385\linewidth]{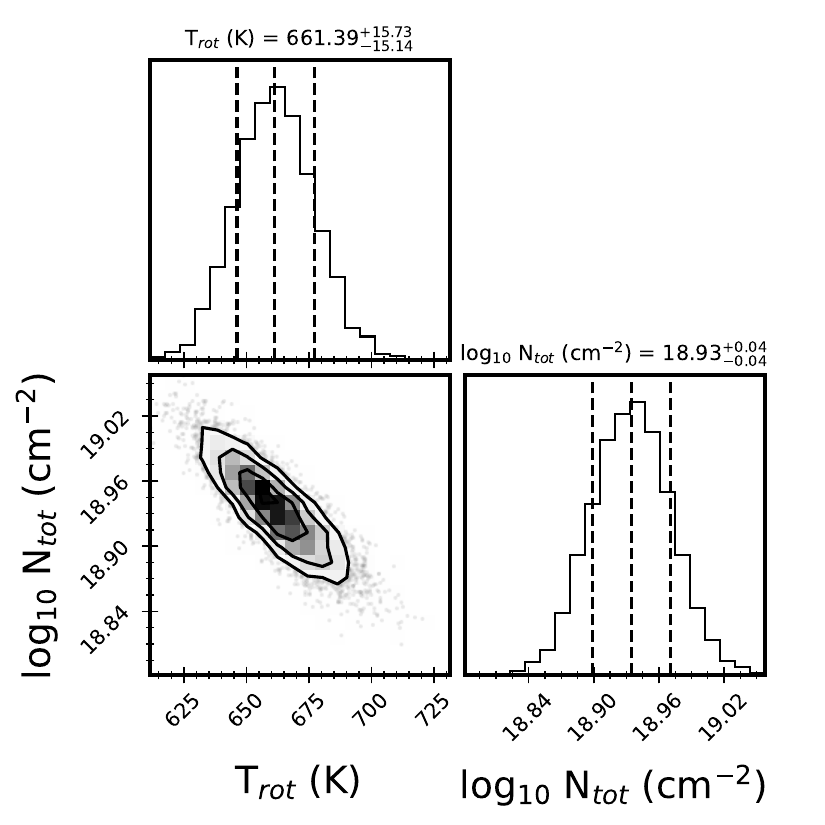}\includegraphics[width=0.385\linewidth]{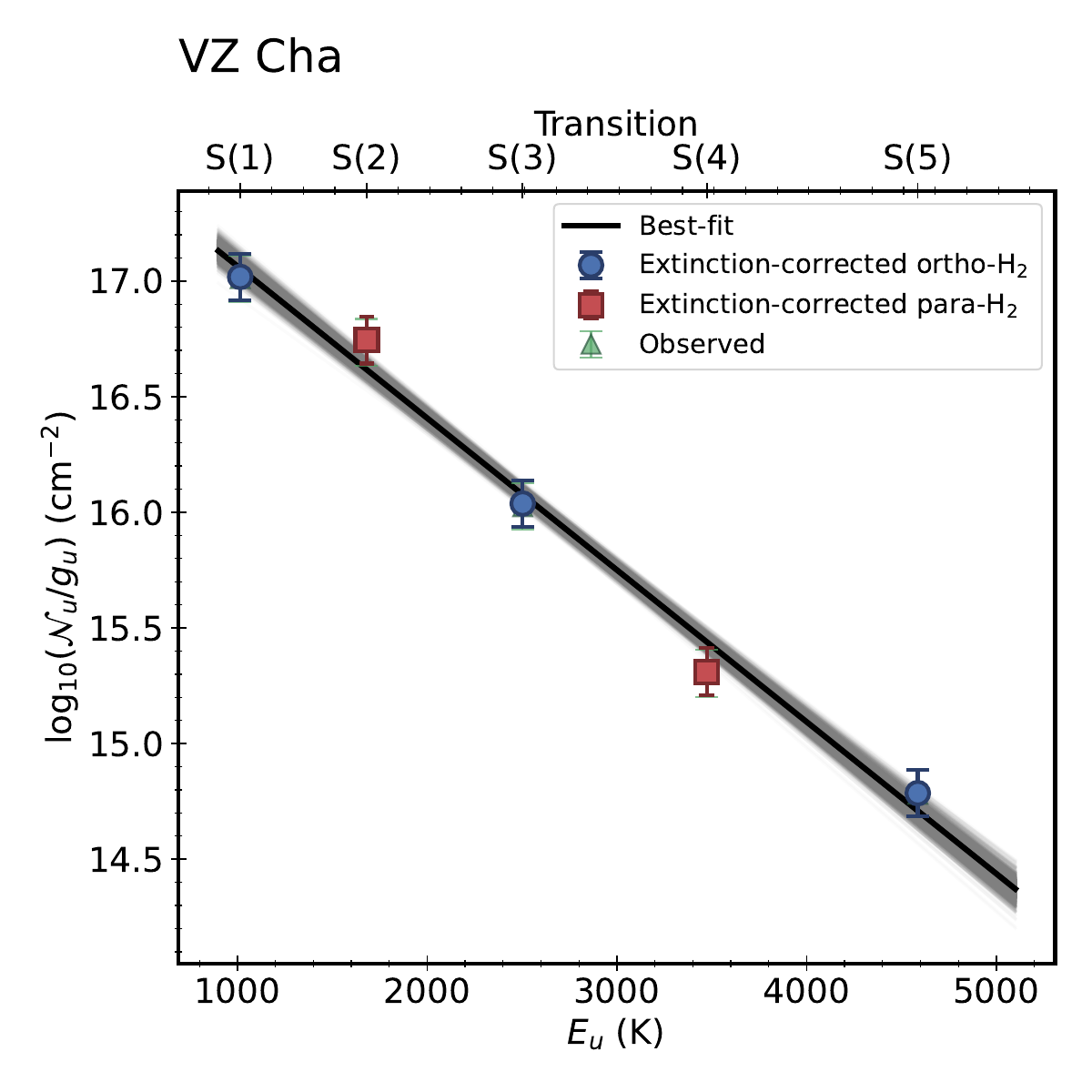}

\includegraphics[width=0.385\linewidth]{wsb52_corner.pdf}\includegraphics[width=0.385\linewidth]{wsb52_rotdiag.pdf}
\caption{Left) Corner plot showing the posterior distributions and best-fit parameters from the MCMC fit to the rotation diagram. (Right) Rotation diagram with observed data points in green and extinction-corrected points in blue for ortho-H$_2$ and red for para-H$_2$. The black line denotes the best-fit model, and the shaded grey region represents the uncertainty range derived from the MCMC fitting.\label{rotfig10}} 
\end{figure*}
}

\end{document}